\pdfoutput=1

\documentclass[12pt]{iopart}

\def\eg{{\em e.g.}}
\def\ie{{\em i.e.}}

\newcommand{\beq}{\begin{equation}}
\newcommand{\eeq}{\end{equation}}
\newcommand{\bea}{\begin{eqnarray}}
\newcommand{\eea}{\end{eqnarray}}
\newcommand{\ltsim}{\raisebox{-4pt}{$\,\stackrel{\textstyle <}{\sim}\,$}}
\newcommand{\gtsim}{\raisebox{-4pt}{$\,\stackrel{\textstyle >}{\sim}\,$}}

\newcommand{\sqrtsNN}{\sqrt{s_{\rm \scriptscriptstyle NN}}}
\newcommand{\Raa}{{R_{\rm AA}}}

\newcommand{\GeV}{\mathrm{GeV}}
\newcommand{\TeV}{\mathrm{TeV}}
\newcommand{\fm}{\mathrm{fm}}
 
\newcommand{\cm}{\mathrm{cm}}

\newcommand{\s}{\mathrm{s}}

\newcommand{\mb}{\mathrm{mb}}
\newcommand{\mub}{\mathrm{\mu b}}
\newcommand{\nb}{\mathrm{nb}}
\newcommand{\pT}{p_{\rm T}}

\usepackage{cite}
\usepackage{graphicx}
\usepackage{epstopdf}

\begin{document}

\topical[Heavy-Flavor in Heavy-Ion Collisions]
{Open Heavy Flavor in QCD Matter and in Nuclear Collisions}

\author{Francesco Prino}
\address{INFN Torino, Via Pietro Giuria 1, I-10125 Torino, Italy}
\ead{prino@to.infn.it}

\author{Ralf Rapp}
\address{Cyclotron Institute and Department of Physics and Astronomy, 
Texas A{\&}M University, College Station, TX 77843-3366, USA}
\ead{rapp@comp.tamu.edu}

\begin{abstract}
We review the experimental and theoretical status of open heavy-flavor (HF) production 
in high-energy nuclear collisions at RHIC and LHC. We first overview the theoretical 
concepts and pertinent calculations of HF transport in QCD matter, including perturbative 
and non-perturbative approaches in the quark-gluon plasma, effective models in hadronic 
matter, as well as implementations of heavy-quark (HQ) hadronization. This is followed 
by a brief discussion of bulk evolution models for heavy-ion collisions and initial 
conditions for the HQ distributions which are needed to calculate HF spectra in 
comparison to observables. We then turn to a discussion of experimental data that 
have been collected to date at RHIC and LHC, specifically for the nuclear suppression 
factor and elliptic flow of semileptonic HF decays, $D$ mesons, non-prompt $J/\psi$ 
from $B$-meson decays, and $b$-jets. 
Model comparisons to HF data are conducted with regards to 
extracting the magnitude, temperature and momentum-dependence of HF transport 
coefficients from experiment.
\end{abstract}

\maketitle

\tableofcontents

\section{Introduction}
\label{sec_intro}

\subsection{QCD Matter and Heavy-Ion Collisions}
The investigation of strongly interacting matter under extreme conditions of
temperature and energy density is at the forefront of modern nuclear physics
research. 
On the one hand, this research improves our understanding of the hot medium 
that prevailed in the first few microseconds of the early universe, as well as
of the interior of neutron stars where a compression of several times normal 
nuclear matter density is realized at small temperatures. On the other hand, 
it remains a formidable challenge to map out the phase structure of Quantum 
Chromodynamics (QCD) as part of the standard model of elementary particle physics, 
based on the many-body physics of quarks and gluons and/or suitable effective 
degrees of freedom. In particular, the asymptotic freedom of the QCD coupling 
strength suggests that hadronic matter, which prevails at low temperatures
and densities, will change into a quark-gluon plasma (QGP) at high temperatures. 

Collision experiments of heavy atomic nuclei at high energies provide the 
unique opportunity to create and study the properties of QCD matter in the
laboratory. Due to the transient nature of the medium produced in these 
reactions, which explodes with large collective velocities, its systematic 
investigation is not easy. Nevertheless, progress has 
been made over about 30 years of ultrarelativistic heavy-ion collisions 
(URHICs) due to a concerted effort of experiment and theory. The basic 
picture that has emerged for the reaction dynamics is that of an expanding 
fireball which rapidly reaches near local equilibrium within about 1\,fm/$c$
after the initial collision, a pressure-driven expansion for about 10-15\,fm/$c$ 
and a subsequent rather sudden freezeout into free-streaming hadrons which 
are observed in the detectors, 
cf.~Refs.~\cite{Landolt23,Akiba:2015jwa,Braun-Munzinger:2015hba} for recent 
overviews. Analyses of bulk-hadron momentum distributions provide a snapshot 
of the fireball at its break-up stage, revealing a ``thermal freezeout" at a 
temperature, $T_{\rm fo}\simeq100$\,MeV, and an average collective expansion 
velocity of more than half the speed of 
light~\cite{Abelev:2008ab,Abelev:2012wca}. 
The analysis of abundances 
of different hadron species, the so-called hadro-chemistry, reveals a higher 
temperature, $T_{\rm ch}\simeq 160$\,MeV, characterizing the ``chemical 
freezeout" of the hadronic system. This temperature is close to the 
pseudo-critical temperature of the chiral transition computed in lattice-QCD, 
$T_{\rm pc}=155\pm10$\,MeV~\cite{Borsanyi:2010bp,Bhattacharya:2014ara}, 
although the transition from partons to hadrons probably occurs over an 
extended temperature window.  Electromagnetic radiation from the medium 
further supports the formation of matter at and above the transition;
in particular, dilepton invariant-mass spectra show that the $\rho$-meson,
as a ``prototype" light hadron, strongly broadens in hadronic matter and
dissolves in the vicinity of $T_{\rm pc}$~\cite{Rapp:2013nxa}.

A fundamental question that remains open thus far is how the quark-hadron 
transition emerges from the underlying in-medium QCD force, and how the 
latter determines the transport properties of the QGP. The heavy charm and 
bottom quarks ($Q=c,b$) play a special role in this context. In the present 
review we will elaborate on the physics of open heavy-flavor (HF) probes 
in URHICs, focusing on recent developments pertaining to theoretical modeling, 
the first round of LHC data as well as the most recent RHIC data (see, \eg, 
Ref.~\cite{Rapp:2009my,Averbeck:2013oga,Andronic:2015wma} for recent reviews). 
While the physics of the open 
HF sector bears intimate connetions to the hidden HF (quarkonium) sector, 
which are increasingly exploited, we will here concentrate on the former and 
refer to recent overviews on in-medium quarkonium physics in 
Refs.~\cite{BraunMunzinger:2009ih,Kluberg:2009wc,Rapp:2008tf,Mocsy:2013syh,Andronic:2015wma}.

\subsection{Diagnostic Potential of Heavy-Flavor Particles}
\label{ssec_dia}
The masses of charm and bottom quarks (as well as hadrons containing them)  
are much larger than both the QCD scale parameter and the pseudo-critical
temperature of the QCD phase transition, 
$m_{c,b} \gg \Lambda_{\rm QCD}, T_{\rm pc}$. 
In the vacuum, this has long been recognized as an effective expansion tool 
to formulate a potential theory of the fundamental QCD force. The potential
approach was originally motivated by phenomenological descriptions of quarkonium 
spectroscopy in vacuum, but is now accurately confirmed by lattice QCD 
(lQCD) calculations~\cite{Bali:2000vr}.   

In the context of QCD matter studies in heavy-ion collisions, the large
heavy-quark (HQ) mass, $m_Q$, together with flavor conservation 
in strong interactions, has several important implications:
\\
{\bf (a)} $Q\bar Q$ production is essentially restricted to the primordial 
nucleon-nucleon collisions in the reaction, since the production threshold 
is much higher than the typical medium temperatures.
\\
{\bf (b)} Since the HQ mass is much larger than the pseudo-critical
temperature, charm and bottom quarks retain their ``identity" through the 
hadronization transition in URHICs; this renders them an excellent probe of
hadronization mechanisms down to small momenta, \ie, whether they pick up a 
light anti-/quark from the surrounding medium or fragment independently.  
\\
{\bf (c)} The typical momentum exchange of HF particles with the heat bath, 
$\vec q^{\,2} \simeq T^2$, is parametrically small compared to the thermal 
HF momentum, $p_Q^{\rm th} \simeq \sqrt{2m_Q T}$; HF particles thus execute 
a ``Brownian motion" with many, relatively small momentum kicks from the 
medium.
\\
{\bf(d)} The typical thermal energy transfer on the HF particle is 
parametrically small compared to the momentum transfer, 
$q_0^2\sim{\vec q}^{\,4}/m_Q^2\ll\vec q^{\,2}$; therefore, the 4-momentum 
transfer $q^2=q_0^2-\vec q^{\,2}$ is essentially spacelike, which is 
characteristic of potential-type interactions. 
\\
{\bf (e)} The thermal relaxation time of HF particles, 
$\tau_Q\simeq\tau_{\rm th}m_Q/T$, is much longer than the thermalization 
time, $\tau_{\rm th}$, of the bulk medium; thus, in connection with item (a) 
and the possibility that $\tau_Q$ is comparable to (or longer than) the 
fireball lifetime, $\tau_{\rm FB}$, HF spectra in heavy-ion collisions 
retain a memory of their interaction history. If so, their finally observed 
modifications can serve as a gauge of the HF coupling strength to the 
medium. 
\\
As a further benefit for phenomenology, various quantities which characterize the 
interactions of heavy flavor in medium can be computed in thermal lattice QCD, 
\eg, the HF diffusion coefficient~\cite{Banerjee:2011ra,Kaczmarek:2014jga}, 
HQ susceptibilities~\cite{Petreczky:2008px} and mesonic correlation 
functions~\cite{Bazavov:2014cta}. Even if these quantities are not directly 
applicable to experiment, they can serve as valuable constraints for model 
calculations which provide a bridge to experimental observables~\cite{Riek:2010py}. 
The above features provide for a promising framework to determine the basic QCD 
force in the medium and its emergent phenomena of HF transport in QCD matter, 
by combining lattice QCD, model calculations and the phenomenological analysis 
of experimental data in a controllable way.

For sufficiently high momenta, $p_Q\gg m_Q$, heavy quarks will behave as light particles,
transitioning into the regime of energy loss dominated by gluon radiation. It currently
remains an open question at which momenta this transition occurs, whether it coincides
with a transition from non-perturbative to perturbatively calculable mechanisms, and 
how the quark mass effect establishes itself prior to reaching the ultrarelativistic 
regime.

 \subsection{Brief Outline}
The remainder of this review is organized as follows. 

In Sec.~\ref{sec_theo} we review general frameworks for the theoretical 
description of HF propagation through QCD matter and their implementations into 
the phenomenology of heavy-ion collisions.
We start by recalling the basic elements of HF transport approaches, in particular 
simulations based on the Boltzmann and relativistic Langevin equations, 
and critically discuss their virtues and limitations as well as respective 
regimes of applicability (Sec.~\ref{ssec_trans}). 
We then survey and compare different mechanisms of HF transport through QCD matter 
including perturbative and non-perturbative elastic HQ interactions in the QGP, 
perturbative radiative HQ energy loss, heavy-meson scattering in hadronic matter, 
analyses using thermal lQCD, as well as transport through the quark-hadron transition 
and pre-equilibrium phases of heavy-ion collisions (Sec.~\ref{ssec_coeff}). 
This is followed by a discussion of further ingredients needed to make 
realistic contact with data (Sec.~\ref{ssec_bulk}), pertaining to the initial 
conditions for the HQ spectra and the space-time evolution of the bulk medium 
formed in heavy-ion collisions.

In Sec.~\ref{sec_exp} we summarize the current status of experimental HF 
measurements in light- and heavy-ion collisions and their theoretical interpretation 
through model comparisons. We first survey the relevant accelerator facilities, 
experimental tools and types of HF observables (Sec.~\ref{ssec_scope}), followed by   
a presentation of the current data on HF production in $pp$ collisions and their 
modification in proton/deuteron-nucleus ($p$A/$d$A) collisions (Sec.~\ref{ssec_element}).
The compilation of HF data in nucleus-nucleus (AA) systems 
(Sec.~\ref{ssec_AAdata}) is organized into 
three parts, dedicated to HF decay lepton, charmed-hadron,
and beauty production, each of them reviewing the results
on the modification of transverse-momentum ($p_T$) spectra and 
the elliptic flow in Au-Au and Cu-Cu collisions at RHIC and 
Pb-Pb collisions at the LHC.
The discussion of the data is accompanied by model comparisons which we attempt
to translate into qualitative and quantitative information about the mechanisms
and magnitudes of HF interactions and transport coefficients in QCD matter
(Sec.~\ref{ssec_impl}).  

We summarize and give an outlook in Sec.~\ref{sec_sum}.

\section{Theoretical Descriptions of Heavy Flavor in Medium}
\label{sec_theo}

As alluded to in the introduction, the motion of HF particles (both quarks 
and hadrons) through the QCD medium at temperatures relevant for URHICs is 
akin to a Brownian motion, where a heavy {\em probe} is injected into a 
background medium of light particles. The subsequent diffusion process of the 
probe particle is governed by its coupling to the medium, schematically given by
an average displacement squared,   
\beq
\langle \vec r^{\,2} \rangle = (2d) D_s t \ 
\eeq  
($t$: time). The transport properties of the medium are encoded in the spatial 
diffusion coefficient, $D_s$ (the prefactor $2d$, where $d$ is the number of 
spatial dimensions, is a convention). A small value of $D_s$ characterizes a strong 
coupling: frequent rescattering limits the spatial dispersion of the Brownian
particle. The spatial diffusion coefficient is directly related to the thermal 
relaxation (or equilibration) time, $\tau_{\rm eq}$, of the heavy particle via
\beq
\tau_{\rm Q} = \frac{m_Q}{T} D_s \ .  
\label{Ds}
\eeq
This relation makes explicit the ``time delay" in the HF thermalization process 
by the HQ mass to temperature ratio, ${m_Q}/{T}$, and further suggests 
that $D_s$ is a generic medium property. 
When scaling $D_s$ by the thermal wavelength of the medium, 
$\lambda_{\rm th} = 1/2\pi T$, one obtains a dimensionless quantity which has 
been suggested to be proportional to the widely discussed ratio of viscosity 
to entropy density of the medium~\cite{Moore:2004tg,Rapp:2009my}, 
\beq 
D_s (2\pi T) \propto \frac{\eta}{s} (4\pi) \ .
\eeq
The numerical coefficient in this relation is, however, not unique, varying, 
\eg, from $\sim$1 in strongly-coupled conformal field theory 
(CFT)~\cite{Herzog:2006gh,Gubser:2006bz,CasalderreySolana:2006rq} 
to 2/5 in kinetic theory for a weakly coupled ultrarelativistic 
gas~\cite{Danielewicz:1984ww}.   

In the remainder of this section we first discuss two of the most commonly 
employed transport approaches to HF propagation in QCD matter, namely relativistic 
Langevin processes and Boltzmann simulations (Sec.~\ref{ssec_trans}). In 
particular, we revisit their regimes of applicability in relation to both the 
large mass limit of the HF particle and the nature of the background medium. We 
then turn to microscopic calculations of HF transport coefficients in QCD matter 
(Sec.~\ref{ssec_coeff}), encompassing elastic and inelastic interactions in both 
perturbative and nonperturbative approaches, for both QGP (Secs.~\ref{sssec_qgp} 
and \ref{sssec_qgp-rad}) and hadronic matter (Sec.~\ref{sssec_had}). This will be 
followed by a discussion of lQCD computations (Sec.~\ref{sssec_lqcd}), 
the spatial HF diffusion coefficient (Sec.~\ref{sssec_Ds}), hadronization 
mechanisms and their relation to diffusion processes (Sec.~\ref{sssec_hadro}), 
and pre-equilibrium as well as mean-field effects on HQ propagation 
(Sec.~\ref{sssec_pre-mean}). 
Finally, we discuss further model components needed for quantitative phenomenology 
of HF data in URHICs (Sec.~\ref{ssec_bulk}), focusing on properties of the bulk 
medium evolution pertinent to HF transport. Another component, namely initial 
conditions for the HQ spectra, will be discussed in more detail in connection
with $pp$ and $p$A data in Sec.~\ref{ssec_element}.
Throughout Sec.~\ref{sec_theo} we will adopt natural units with $\hbar=c=1$.

\subsection{Frameworks for Heavy-Flavor Transport}
\label{ssec_trans}

In kinetic theory, the starting point for describing the motion of HF particles
through QCD matter is the Boltzmann equation. Since there is no principal 
difference between HF hadrons in hadronic matter and heavy quarks in QGP, 
we focus our formal 
discussion in this section on the latter. The space-time evolution of the HQ 
phase space distribution function, $f_{\mathrm{Q}}$, is then governed by
the integro-differential equation
\begin{equation}
\label{boltz}
\left [ \frac{\partial}{\partial t} + \frac{\vec{p}}{E_p}
  \frac{\partial}{\partial \vec{x}} + \vec{F}
  \frac{\partial}{\partial{\vec{p}}} \right ]
f_{\mathrm{Q}}(t,\vec{x},\vec{p}) = C[f_{\mathrm{Q}}] \ , 
\label{Boltz}
\end{equation}
where $E_p=\sqrt{m_Q^2+\vec{p}^{\,2}}$ denotes the HQ on-shell energy (reflecting 
the inherently classical nature of the Boltzmann equation). In a static medium in 
equilibrium the distribution function approaches the Boltzmann distribution, 
$f_{\mathrm{Q}} = d_Q \exp(-E_p/T)$, where $T$ is the temperature and $d_Q$=6 the 
HQ degeneracy. 

The space-time evolution of $f_{\mathrm{Q}}$ is generated by the first two terms 
on the left-hand side ($lhs$) of Eq.~(\ref{Boltz}). The third term represents 
the effects due to a force $\vec{F}$ induced by an external (or mean) field, \eg, 
(chromo-) electric and/or magnetic fields as could be relevant in the early phases 
of a heavy-ion collision. Examples of those will be discussed in 
Sec.~\ref{sssec_pre-mean}. 

The right-hand side ($rhs$) of the Boltzmann equation is the collision integral. 
For $2\to2$ scattering of a heavy quark off thermal partons it takes the form 
\begin{eqnarray}
C[f_Q]= & \frac{1}{2 E_p} \int\tilde{dk}\int\tilde{dp'}\int\tilde{dk'} 
 \frac{1}{d_Q} \sum\limits_{m=q,\bar q, g} |\mathcal{M}_{Qm}|^2  (2 \pi)^4 
\delta^{(4)}(p+k-p'-k') 
\nonumber\\
 & \qquad \qquad \qquad \qquad 
\times [f_Q(E_{p'}) f_{m}(\omega_{k'}) - f_Q(E_p) f_{m}(\omega_k) ] \ .
\label{coll}
\end{eqnarray}
where $\tilde{dk}\equiv d^3k/(2\omega_k(2\pi)^3)$ are the standard Lorentz-invariant 
phase space differentials and $\mathcal{M}_{Qm}$ is the HQ-parton scattering amplitude 
with $\vec{p}$ ($\vec{p}'$) and $\vec{k}$ ($\vec{k}'$) the 
3-momenta of the incoming (outgoing) heavy quark $Q$ and medium parton $m$ (quark, antiquark 
or gluon), respectively; the $f_{m}$ are the latter's thermal distribution functions 
and $\omega_{k,k'}$ their on-shell energies (quantum effects can be included through 
the use of Fermi/Bose distribution
functions and final-state blocking/enhancement factors, and are implicit in
field-theoretical evaluations of the scattering amplitude). In a dilute medium, 
the collision integral can be approximated by using particle cross sections. For 
large scattering rates the definition of asymptotic states becomes problematic
and a formulation in terms of scattering probabilities is preferable. The inclusion 
of radiative processes is more challenging, especially if interference effects 
for subsequent scatterings become significant, due to the inherent off-shell 
nature of these processes, see, \eg, Refs.~\cite{Gossiaux:2012cv,Uphoff:2014hza} 
for recent work in the HQ context. As is well known, electromagnetic Bremsstrahlung 
off heavy fermions is suppressed by the fourth power of their mass. It is currently 
an open question at which HQ momenta and QGP temperatures radiative processes take 
over from elastic ones. We will return to this question below. 

As discussed in the introduction, for large quark masses and moderate 
temperatures, the typical momentum transfer from the heat bath to the heavy 
quark is parametrically small, $\vec{q}^{\,2} \ll \vec p^{\,2}$ where 
$\vec{q}=\vec{p}-\vec{p\,}'=\vec{k}'-\vec{k}$.  In addition, the energy transfer 
is further suppressed, $q_0^2 \ll \vec{q}^{\,2}$. These conditions imply that 
the heavy quark is undergoing soft but incoherent collisions characteristic 
for Brownian motion. The scattering rate in the collision integral
can then be expanded in powers of the momentum transfer. Carrying this out to
second order, the Boltzmann equation can be transformed into the Fokker-Planck 
equation, 
\begin{equation}
\frac{\partial}{\partial t} f_Q(t,\vec{p}) = \frac{\partial}{\partial p_i}
\left \{ A_{i}(\vec{p\,}) f_Q(t,\vec{p\,})+\frac{\partial}{\partial
    p_j}[B_{ij}(\vec{p\,}) f_Q(t,\vec{p\,})] \right \} \ , 
\label{FP}
\end{equation}
which now is only a differential equation for the HQ phase space distribution
function. Its key ingredients are the transport parameters $A$ and $B$. In a
medium in (local) thermal equilibrium they simplify to three a priori 
independent coefficients, 
\begin{eqnarray}
A_i(\vec{p\,}) &=& A(p) p_i \ , 
\\
B_{ij}(\vec{p\,}) &=& B_0(p) P_{ij}^{\perp}(\vec{p\,}) +
B_1(p) P_{ij}^{\parallel}(\vec{p\,})\ ,
\label{AB}
\end{eqnarray}
characterizing momentum friction and diffusion, respectively, of the propagating 
heavy quark ($P_{ij}^{\perp}=\delta_{ij}-p_ip_j/\vec{p\,}^2$ and 
$P_{ij}^{\parallel}=p_ip_j/\vec{p\,}^2$ are the standard 3D transverse and 
longitudinal projectors, respectively). They correspond to the first and 
second moments of the thermally weighted scattering amplitude,
\begin{eqnarray}
A(p) &=& \left\langle {1-\frac{\vec{p} \cdot \vec{p\,}'}{\vec{p\,}^2}} \right\rangle \ , 
\\
B_{0}(p) &=& \frac{1}{4} \left\langle{\vec{p\,}'^{2}-\frac{(\vec{p\,}'\cdot \vec{p\,})^2}
{\vec{p\,}^2}} 
\right\rangle \ , \\
B_{1}(p) &=& \frac{1}{2} \left\langle{\frac{(\vec{p\,}'\cdot \vec{p\,})^2}{\vec{p\,}^2} 
 - 2 \vec{p\,}'\cdot  \vec{p} + \vec{p\,}^2} \right\rangle \ , 
\end{eqnarray}  
where the definition of the brackets follows 
from the collision term in the Boltzmann equation, 
\begin{eqnarray}
\left\langle X(\vec{p}') \right\rangle &=& \frac{(2 \pi)^4}{2E_pd_Q} \int \tilde{dk} \tilde{dk'} 
\tilde{dp'}
\nonumber \\
&&\qquad\qquad \times\sum_{m}|\mathcal{M}_{Qm}|^2\delta^{(4)}(p+k-p'-k')f_{m}(\omega_k) X(\vec{p}') \ .
\label{X}
\end{eqnarray}
In the non-relativistic limit of 3-momentum independent transport coefficients, 
$\gamma\equiv A(p)={\rm const}$ and $D_p\equiv B_0(p) = B_1(p)={\rm const}$, the Fokker-Planck 
equation takes the simpler form
\begin{equation}
\frac{\partial}{\partial t} f_Q(t,p) = \gamma
\frac{\partial}{\partial p_i} [p_i f_Q(t,p)] 
              + D_p \Delta_{\vec{p\,}} f_Q(t,p) \   
\end{equation}
($\Delta_{\vec{p\,}}$: 3-momentum Laplace operator). For large times, this equation 
has the solution
\begin{equation}
f_Q(t,p) = \left (\frac{2 \pi D_p}{\gamma} \right )^{3/2} \exp\left
  (-\frac{\gamma \vec{p\,}^2}{2 D_p} \right ) \ .  
\end{equation}
Matching this to the non-relativistic equilibrium limit one obtains the
Einstein relation (or dissipation-fluctuation theorem),
\begin{equation}
D_p = m_Q \gamma T \ , 
\label{einstein}
\end{equation}
highlighting the intimate relation between momentum friction and diffusion, and 
their role in imprinting the temperature of the heat bath on the HQ distribution.

In URHICs, the Fokker-Planck equation can be implemented to simulate HF motion
through QCD matter via a Langevin process. The latter consists of a ``deterministic" 
drag and ``stochastic" diffusion part, defined via momentum and position updates
as 
\begin{eqnarray}
dp_j &=& -\Gamma(p,T)  p_j dt + \sqrt{dt} C_{jk} \rho_k 
\nonumber \\ 
dx_j &=& \frac{p_j}{E} dt \ . 
\label{langevin}
\end{eqnarray} 
The $C$'s are uniquely related to the diffusion coefficients,
\begin{equation}
C_{jk} = \sqrt{2B_0(p)} P_{jk}^\perp(\vec{p\,}) + 
\sqrt{2B_1(p)} P_{jk}^\parallel(\vec{p\,}) \ , 
\end{equation} 
and are weighted with a random Gaussian noise distribution, 
$P(\vec{\rho\,})=\exp(-\vec{\rho\,}^2/2)/(2\pi)^{3/2}$. The precise 
relation of $\Gamma$ to the friction coefficient,
$\Gamma(p)=A(p,T) + {\cal O}(T/m_Q)$, depends on the numerical scheme 
to carry out the Langevin process~\cite{Koide:2011yy,He:2013zua}. 
In the so-called post-point scheme, the relativistic version of
the Einstein relation takes the form $B_0(p)=B_1(p)=TE_p\Gamma(p)$.
The latter is not automatically satisfied for microscopically calculated 
transport coefficients, but rather becomes an important tool to ensure 
their consistency~\cite{Moore:2004tg,vanHees:2004gq} and, consequently, the 
proper equilibrium limit when implementing them into Langevin simulations. 
In both perturbative~\cite{Moore:2004tg} and 
non-perturbative approaches~\cite{vanHees:2004gq} for calculating the transport
coefficients from underlying microscopic interactions, the Einstein relation
was found to be rather well satisfied for the friction and transverse diffusion
coefficient. Significant deviations were found to develop for the longitudinal 
coefficient toward higher momenta, presumably due to the specific kinematics of
the scattering processes (\eg, pQCD scattering is typically strongly peaked at 
forward angles). To ensure the correct equilibrium limit in Langevin simulations 
of the Fokker-Planck equation, $B_1$ is then corrected for by its value obtained 
from the Einstein relation~\cite{vanHees:2005wb,Gossiaux:2006yu,Das:2010tj} in 
terms of the friction coefficient; alternatively, the latter is readjusted by the 
diffusion coefficients $B_{0,1}$~\cite{Moore:2004tg,Akamatsu:2008ge,Alberico:2011zy}.  

One may wonder how accurate the Fokker-Planck approximation due to the expansion 
in the momentum transfer is, especially for charm quarks at high temperatures.
This question has recently been revisited by comparing the results of HQ Boltzmann 
and Langevin simulations using leading-order pQCD cross section for heavy-light 
parton scattering~\cite{Scardina:2014gxa}. It was found that for charm quarks 
with $m_c=1.3$\,GeV in a heat bath of $T=0.4$\,GeV, appreciable deviations of 
several 10's of percent develop in the momentum spectra after an evolution time 
of $\Delta t=2$\,fm. For bottom quarks with $m_b=4.2$\,GeV, the deviations 
are below $\sim$5\% even after $\Delta t=6$\,fm, and also for $c$ quarks 
at $T=0.2$\,GeV the agreement is much improved. This suggests that for ratios 
$M_Q/T\gtsim6$ the Langevin approximation is quite reliable. In fact, if the 
dissipation-fluctuation theorem is enforced for the pQCD transport coefficients 
in the Langevin simulation, the discrepancies with the Boltzmann results are 
further reduced by up to a factor of $\sim$2~\cite{Scardina:2014gxa}.   
 
Another feature of the Boltzmann equation that is not easily implemented in
Langevin simulations is that of off-equilibrium effects in the surrounding medium.
In the Boltzmann equation, such effects on the medium parton distribution 
functions can be treated on the same footing as for the HQ distribution, as done,
\eg, in Refs.~\cite{Uphoff:2012gb,Das:2015ana,Song:2015sfa}. On the other hand, 
the simulations of the bulk medium in the Boltzmann approach usually rely on 
quasi-particles with vanishing widths. In the Langevin approach, the underlying
transport coefficients can be calculated with finite-temperature field theory
techniques, where the inclusion of quantum effects through the use of broad 
spectral functions for the bulk medium constituents is in principal straightforward. 
The Langevin approach is thus better suited for situations where the medium is 
strongly coupled (as implicit in hydrodynamic medium evolutions), as long as the 
``Brownian" particle is still a reasonably well-defined quasi-particle, \ie, with 
an energy much larger than its width, $E_p\gg \Gamma_Q(p)$ (here, $\Gamma_Q$ denotes
the scattering rate, not equilibration rate, characterizing the width of
the quasi-particle peak in the HQ spectral function).

\subsection{Calculations of Heavy-Flavor Transport}
\label{ssec_coeff}

\subsubsection{Quark-Gluon Plasma I: Elastic Interactions} 
\label{sssec_qgp}
\hspace{2cm}

Early evaluations of HQ transport coefficients in deconfined matter were carried out 
using perturbative Born diagrams for cross sections off thermal partons, with a Debye 
screening mass $m_D=gT$ introduced as an infrared regulator into the (dominant) 
$t$-channel gluon exchange propagator~\cite{Svetitsky:1987gq}, 
\begin{equation}
D_g=\frac{1}{(t-m_D^2)} \ ,
\label{Dg-schem}
\end{equation}
where $t$ is the 4-momentum transfer in the scattering. Subsequently, these 
calculations have been routinely repeated to serve as a benchmark, and will be 
referred to as ``schematic LO pQCD" in the following. Even for a coupling constant 
as large as $\alpha_s=g^2/4\pi=0.4$, 
which produces appreciable total cross sections of a few millibarns, the predominantly 
forward scattering results in friction coefficients for charm (bottom) quarks reaching 
about $A(p$=$0)\simeq12(5)$\,MeV, with a weak momentum dependence. The corresponding 
thermal relaxation times, $\tau_Q\simeq$~15-20(40)\,fm/$c$ (translating into a spatial 
diffusion coefficient $D_s(2\pi T)\simeq30$), are quite a bit longer than the typical 
QGP lifetime of about 5--10\,fm/$c$ in central heavy-ion collisions at 
RHIC and LHC energies~\cite{Aamodt:2011mr}. When first RHIC data on HF spectra and 
elliptic flow in Au-Au collisions at $\sqrt{s_{\rm NN}}$=200\,GeV were 
published~\cite{Adler:2005xv,Adare:2006nq}, 
it became clear that much shorter thermalization times are required to account for 
the strong medium modifications implied by these data (and subsequently by LHC data). 
A non-perturbative resonance model~\cite{vanHees:2004gq,vanHees:2005wb}, which gave 
a fair description of these first data, led to an initial estimate of the diffusion 
coefficient of $D_s(2\pi T)\simeq$~4-6.   

Several developments in the perturbative framework have been carried out since then. 
In Refs.~\cite{Peshier:2008bg,Gossiaux:2008jv}, the schematic screening in 
eq.~(\ref{Dg-schem}) has been improved following the methods of 
Ref.~\cite{Braaten:1991we} 
by matching a hard-thermal loop calculation of the gluon self-energy for small momentum 
transfers, $t$, to a perturbative (unscreened) calculation at large $t$.
This effectively leads to a reduction in the screening of the gluon 
exchange propagator, which may be represented as 
\begin{equation}
D_g=\frac{1}{t-r m_D^2} \ .
\end{equation}
Numerically, the effective reduction parameter turns out to be quite small, 
$r\simeq0.2$, which produces a substantial increase (by about a factor of 2) 
of the energy loss of a heavy quark for a typical strong coupling constant 
of $\alpha_s$$\simeq$0.2. 
It has furthermore been argued~\cite{Peshier:2008bg,Gossiaux:2008jv} that the 
running of the QCD coupling constant ought to be accounted for down to soft 
scales, rising up to values of $\alpha_s$$\simeq$1 for small $|t|$. When 
combining the running coupling with the reduced Debye screening, the charm-quark 
friction coefficient (or thermalization rate) increases by a factor of 5-10 over 
the schematic fixed-coupling/-Debye mass scheme at small charm-quark 
momenta~\cite{Gossiaux:2008jv}. 
A matching of the HTL calculations at low $|t|$ to pQCD scattering at high $|t|$
has also been utilized in the calculations of the transport coefficients by the
Torino group (or POWLANG approach)~\cite{Alberico:2011zy,Alberico:2013bza}, but
with a fixed Debye mass as obtained from weak-coupling calculations and a 
coupling constant running with temperature (not with momentum
transfer)~\cite{Kaczmarek:2005ui}; in particular, the calculated longitudinal
diffusion coefficient has been used to fix the friction coefficient via 
the Einstein relation. This leads to a significantly smaller thermalization rate
at zero charm-quark momentum, but also to a different momentum dependence which
is initialy rising with $p$. 

In a spirit similar to the implementation of 
Refs.~\cite{Peshier:2008bg,Gossiaux:2008jv}, the HQ-medium interactions have 
been calculated in the 
dynamical quasiparticle model using elastic Born cross sections with infrared 
enhanced coupling constant~\cite{Berrehrah:2013mua}. In this calculation the HQ 
transport is coupled to a bulk medium whose equation of state is consistent with 
lQCD. The results for the charm-quark relaxation rate in this approach are smaller 
by about a factor of $\sim$2-3 compared to the HTL-inspired scheme with running 
coupling and reduced Debye mass. These transport coefficients have been subsequently 
employed in the PHSD tranport model~\cite{Song:2015sfa,Song:2015ykw}.      
 
A large increase in the HQ thermalization rate raises the question of higher order
corrections. In Ref.~\cite{CaronHuot:2007gq} the convergence of the perturbative 
expansion for the HQ transport coefficient has been scrutinized. It was found that 
even for a strong coupling constant as small as $\alpha_s$=0.03 (corresponding to 
$g_s$=0.6) the NLO correction is around a factor of 2 (and further growing for 
larger values of $\alpha_s$). This gives little hope for a rigorous perturbative 
evaluation of HQ transport, but rather calls for non-perturbative methods. 
For example, it would be of interest to carry out a ladder resummation of the 
infrared-augmented one-gluon exchange interaction and check for the appearance of 
bound-state solutions.

A large running coupling toward small momentum transfers in the one-gluon exchange 
interaction corresponds to a strong Coulomb force on the heavy quark at large 
distances. However, in the QCD vacuum, the long-distance part of the potential 
between a static color charge and its anti-charge is characterized by a linearly 
increasing ``string" interaction, presumably induced by gluonic condensates. 
The pertinent Cornell potential, 
\begin{equation}
V(r) = -\frac{4}{3} \frac{\alpha_s}{r} + \sigma r \ , 
\end{equation}
was originally inferred from a successful description of the vacuum charmonium 
and bottomonium spectra~\cite{Eichten:1979ms} and is now well established from 
lQCD~\cite{Bali:2000vr}. A key question in QGP research is how this fundamental
force changes in the medium. Lattice-QCD computations of the free energy at finite
temperature indicate that remnants of the confining force survive in the QGP,
possibly up to 2\,$T_c$~\cite{Bazavov:2014kva}. Thus, string-like interactions
are likely to play an important role for HQ interactions when approaching
$T_c$ from above. As elaborated in Sec.~\ref{ssec_dia}, an extension of the 
potential model to finite temperature may provide an opportunity to include 
non-perturbative interactions in the description of open and hidden HF 
particles in QCD matter at moderate temperatures. This can be realized in the 
thermodynamic $T$-matrix formalism, which has been widely used in other contexts, 
such as the nuclear many-body problem\cite{Brockmann:1990cn}, electromagnetic 
plasmas~\cite{Redmer:1997}, or cold atomic gases~\cite{Pantel:2014jfa}. In 
Ref.~\cite{Riek:2010fk} the $T$-matrix formalism has been deployed to 
selfconsistently evaluate HQ and quarkonium properties in the QGP. 
With additional relativistic (magnetic) 
corrections to the static potential~\cite{Brambilla:2003nt}, and an approximate
description of vacuum spectroscopy, a comprehensive treatment of HF properties in
the QGP can be carried out. Euclidean correlators, susceptibilities and spectral 
functions can be calculated and tested against lQCD results, as well as transport
properties which can be implemented into URHIC phenomenology.
In particular, the in-medium $T$-matrix for heavy-light parton scattering can be 
straightforwardly implemented to calculate HQ transport coefficients using 
eq.~(\ref{X})~\cite{vanHees:2007me,Riek:2010fk,Huggins:2012dj}. A critical input 
to this approach is the two-body interaction kernel, \ie, the HQ potential. Ideally, 
one could take it directly from lQCD. However, lQCD computations rather provide 
the free ($F_{Q\bar Q}$) and internal ($U_{Q\bar Q}$) energies, which strictly 
speaking are not potentials. However, they have been used to bracket the uncertainty 
in the potential definition; the internal energy leads to much stronger interactions 
and is clearly preferred over the free energy when carrying out 
quarkonium~\cite{Zhao:2010nk,Liu:2010ej,Zhao:2011cv} and open HF 
phenomenology~\cite{He:2014cla} at RHIC and LHC energies. Most notably, when using 
the free energy, the calculated suppression of the $Y(1S)$ state is larger than 
observed~\cite{Emerick:2011xu,Strickland:2011aa}, and, pertinent to the present 
topic, the calculated elliptic flow of open HF observables is much too 
small~\cite{He:2016}. 
\begin{figure}[!t]
\begin{center}
\includegraphics[width=0.6\textwidth]{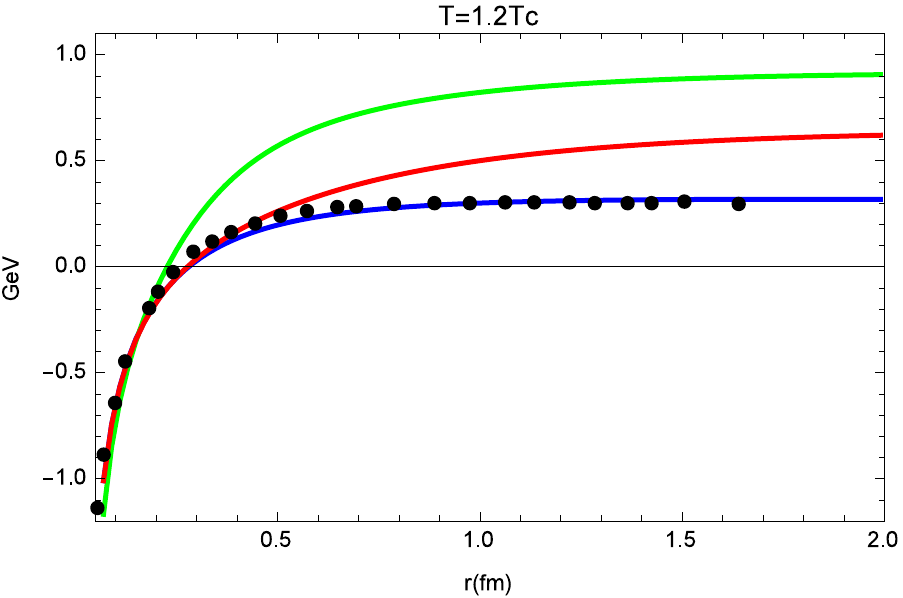}
\end{center}
\caption{Fit of a $T$-matrix calculation of the free energy (blue line) to lQCD 
data~\cite{Kaczmarek:2007pb} (black points) at $T$=240\,MeV, and the resulting 
internal energy (green line) and in-medium input potential (red line). 
Figure taken from Ref.~\cite{Liu:2015ypa}.}
\label{fig_pot}
\end{figure}
Recent progress in determining the underlying HQ potential
has been made by directly calculating the static free and internal energies from
the $T$-matrix~\cite{Liu:2015ypa}. It turns out that large imaginary parts in the
single HQ selfenergies as well as in the potential induce substantial deviations 
from the free energy (the latter emerges as potential in the limit of small imaginary 
parts, \eg, in the weak-coupling limit). Remarkably, the in-medium potential, $V$, 
gives rise to a long-range (non-Coulombic) force which is neither present in $F$ 
nor in $U$, cf.~Fig.~\ref{fig_pot}. This has important consequences for the HQ transport
coefficient, which for low momenta turns out to be larger than when $U$ is assumed as
potential. These results are not unlike the ones found in Ref.~\cite{Bazavov:2014kva}        
where a 2-parameter ansatz for the spectral function with a functional form
adopted from HTL perturbation theory was used for the extraction of the potential.
On the other hand, in Ref.~\cite{Burnier:2014ssa} the (real part of the) potential 
turns out to be close to the free energy. More work is needed to clarify the
differences in these findings. 

An in-medium Cornell potential is also underlying the collisional dissociation 
approach for HF propagation in the QGP adopted in the model by 
Vitev {\it et al.}~\cite{Sharma:2009hn}: the propagating heavy quarks
can form in-medium $D$- and $B$-meson bound states which subsequently dissociate again,
and so on. As in Refs.~\cite{vanHees:2004gq,vanHees:2007me}, the presence of such mesonic
correlations in the QGP has been found to be instrumental in generating sufficient 
suppression of the HF spectra at RHIC, and later LHC~\cite{Sharma:2009hn}. 

\begin{figure}[!t]
\includegraphics[width=0.57\textwidth]{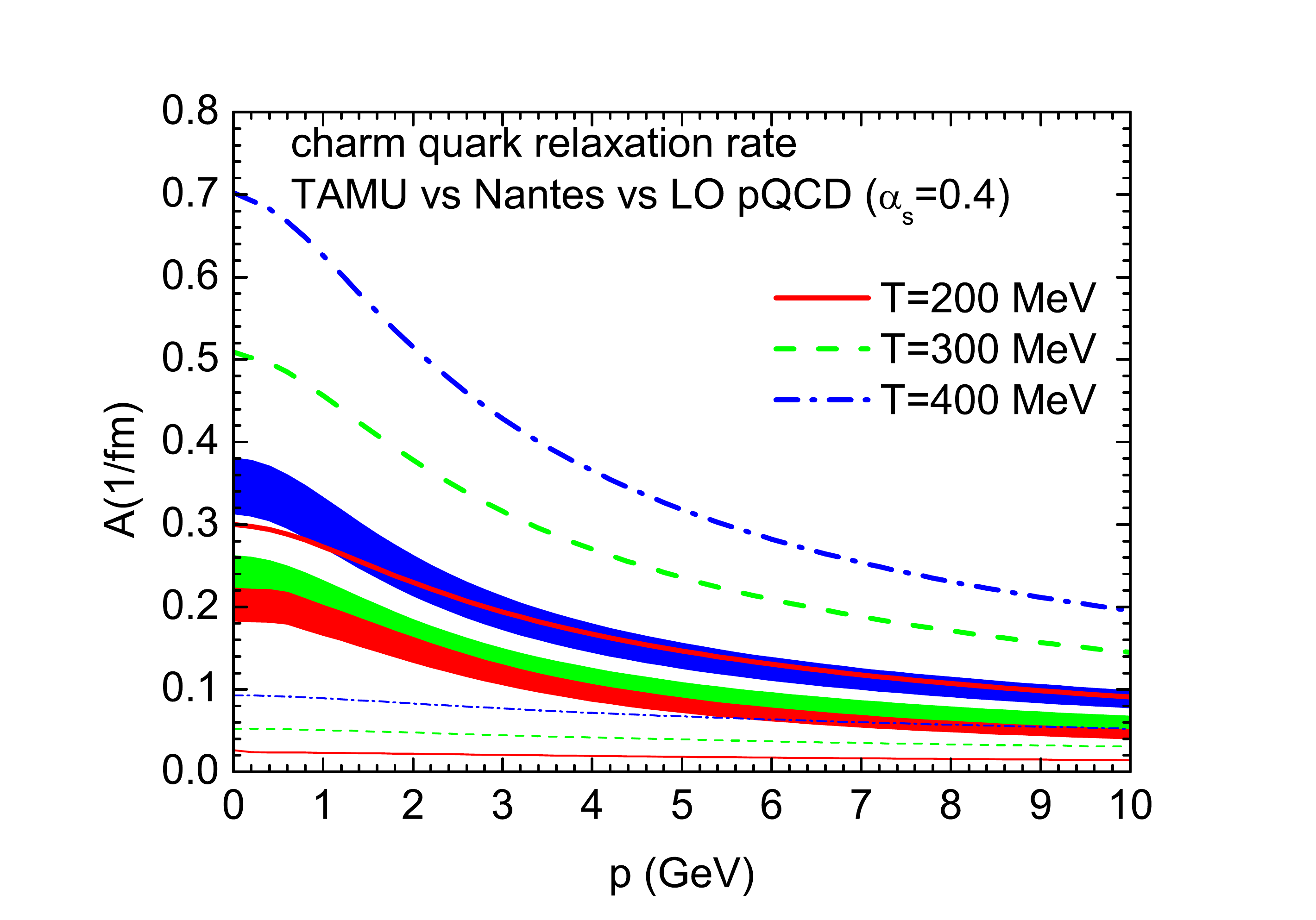}
\hspace{-1.3cm}
\includegraphics[width=0.57\textwidth]{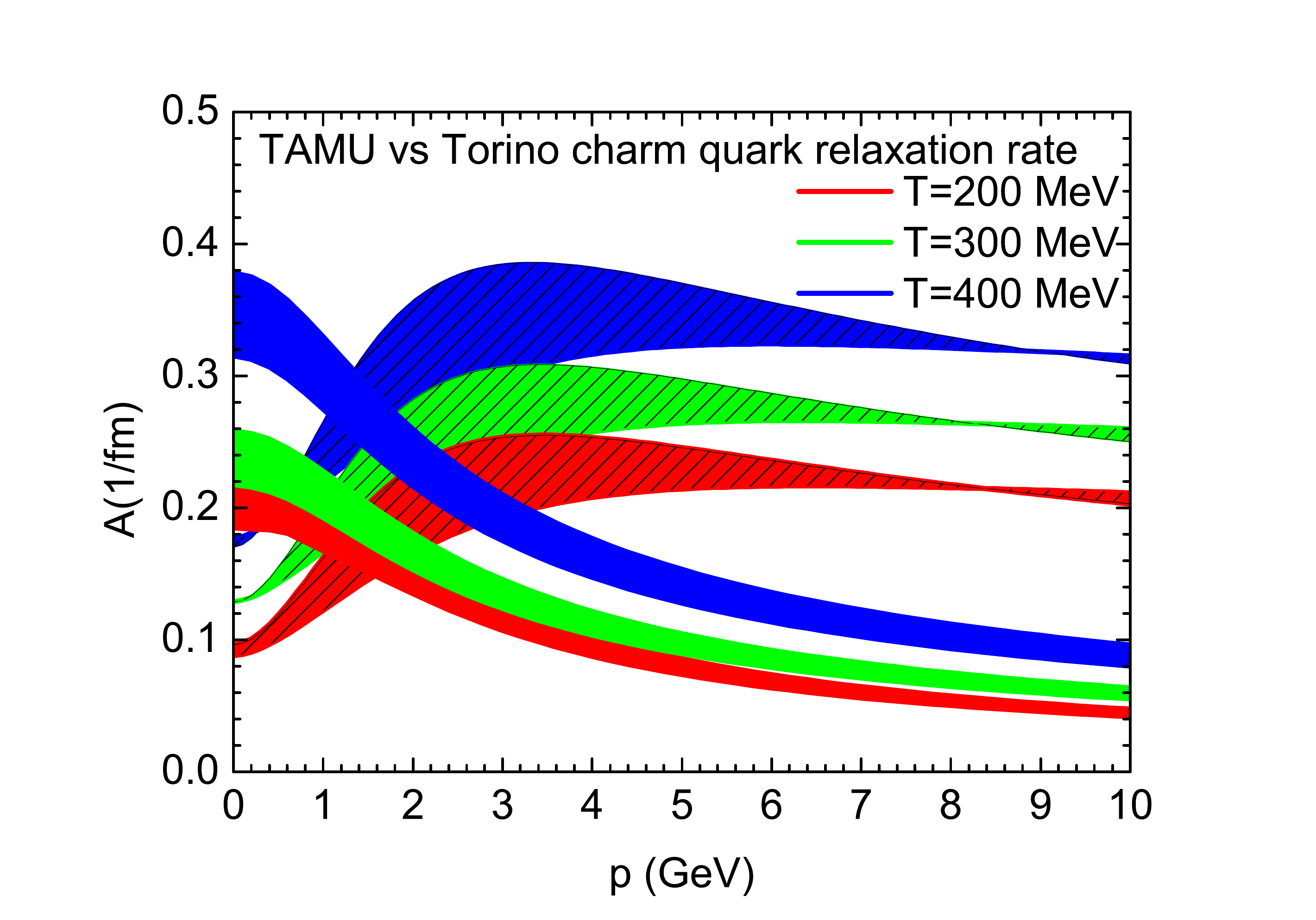}
\caption{Charm-quark friction coefficient calculated from different underlying
elastic interactions in the QGP. Left panel: schematic LO pQCD calculations with
fixed Debye mass and coupling constant (thin lines), pQCD Born calculations with 
reduced Debye mass (from HTL/pQCD matching) and running coupling
constant~\cite{Gossiaux:2008jv} (thick lines), and 
in-medium $T$-matrix calculations for HQ-parton scattering using internal energies
from thermal lQCD as potential~\cite{Riek:2010fk,Huggins:2012dj} (bands with a
20\% uncertainty range due to different lQCD inputs).
Right panel: pQCD calculations in the HTL-pQCD matched approach (Torino) used in 
POWLANG where the friction coefficient has been inferred from the longitudinal 
diffusion coefficient~\cite{Alberico:2011zy,Alberico:2013bza} (shaded
bands with uncertainty due to different matching scales), compared
to the same $T$-matrix results as in the left panel.}
\label{fig_A}
\end{figure}
The momentum dependence of the charm-quark friction coefficient (or thermalization 
rate) is summarized in Fig.~\ref{fig_A} for several of the above discussed approaches, 
as a function of 3-momentum and for several temperatures. The schematic LO pQCD 
approach~\cite{Svetitsky:1987gq} with fixed coupling ($\alpha_s$=0.4) and Debye mass 
gives relatively small values of around 0.05/fm, with a weak momentum dependence. 
In the HTL approach of the Torino group (as implemented in the POWLANG HF transport 
approach 
for URHICs)~\cite{Alberico:2011zy,Alberico:2013bza}, much larger values are found
(by factor of $\sim$3-4 at low $p$), which 
is in part due the improved treatment of the screening and in part due to the procedure 
of inferring $A(p)$ from the longitudinal momentum diffusion coefficient, $B_1(p)$ 
(also, the charm-quark mass is $m_c$=1.3\,GeV). In particular, the latter produces a 
momentum dependence of the friction coefficient which increases with $p$ before 
approximately leveling off. 
On the contrary, the increase in soft HQ interactions due to a reduced screening 
mass and running coupling constant at low momentum 
transfers~\cite{Peshier:2008bg,Gossiaux:2008jv} (as 
implemented in the MC@sHQ+EPOS~\cite{Gossiaux:2010yx,Nahrgang:2013xaa} and 
BAMPS~\cite{Uphoff:2011ad,Uphoff:2012gb} HF transport approach) leads to a 
marked enhancement of the friction coefficient at low charm-quark momenta, by up to 
an order of magnitude over the schematic LO pQCD benchmark, while at high momenta it  
falls below the HTL approach of Refs.~\cite{Alberico:2011zy,Alberico:2013bza}. 
The $T$-matrix approach (as used in the TAMU HF transport 
approach~\cite{He:2011qa,He:2014cla}) also predicts a marked low-momentum 
enhancement of the thermalization rate, but generated by near-threshold resonances 
as a consequence of the ladder resummation.  At high momenta it tends toward the 
schematic pQCD results.  The $T$-matrix results have a slightly different temperature 
dependence~\cite{vanHees:2007me,Riek:2010fk} than the pQCD-like approaches, with a 
relatively larger interaction strength near $T_c$, falling off at higher $T$ as the 
resonances dissolve. This is even more pronounced with the newly derived 
potential~\cite{Liu:2015ypa} where the low-momentum friction coefficient  
{\em decreases} with temperature.

\subsubsection{Quark-Gluon Plasma II: Radiative Interactions}
\label{sssec_qgp-rad}
\hspace{2cm}  

Radiative ($2\to3$) scattering processes are suppressed compared to elastic
scattering with increasing mass of the incoming particle(s). For example, in
electrodynamics, the radiative energy loss of a muon is parametrically suppressed
relative to an electron by the fourth power of the mass ratio, $(m_e/m_\mu)^4$. The
reason is the suppression of the energy relative to the momentum transfer,
$q_0^2 \sim \vec q^{\,4}/m_Q^2 \ll \vec q^{\,2}$ (as discussed in the introduction), 
which renders the emission of an on-shell gluon (or photon) with $q^2=0$ 
unfavorable, at least as long as the 3-momentum transfer is small. 
Within QCD, the pertinent suppression of forward-angle gluon emission was termed 
the ``dead cone" effect~\cite{Dokshitzer:2001zm}; subsequently, it was aruged that
medium-induced gluon radiation can fill this dead cone, while a depletion persists
for large energies of the emitted gluon~\cite{Armesto:2003jh,Armesto:2005iq}.    
For sufficiently energetic heavy quarks, the mass effect is expected `
to cease and the radiative energy loss to take over from the elastic one, since the 
energy tends to be radiated in large quantities per scattering event. 
On the other hand, at the level
of the nuclear modification factor in URHICs, also elastic scatterings with
typically smaller energy loss can lead to a significant suppression effect due to
the rather steeply falling initial spectra. Even for light-parton jet quenching,
the quantitative role of elastic scattering has not been settled yet; \eg, in
a pQCD thermal-field theory framework, the elastic contribution to the nuclear
modification factor, $R_{\rm AA}$ (see Sec.~\ref{sec:expobs} for its definition),
of pions remains significant at high $p_T$, despite the much smaller
energy loss~\cite{Qin:2007rn}. 

Heavy-flavor observables at intermediate and high $p_T$ thus provide an excellent 
tool to map out the transition from a predominantly elastic- to radiative-scattering 
regime (and, in principle, as a function of temperature). The availability of two 
different HQ masses (charm and bottom) adds a further handle to determine this 
transition and test underlying mechanisms. Toward this end, several strategies 
have been pursued thus far. In 
Refs.~\cite{vanHees:2005wb,vanHees:2007me,He:2011qa,Alberico:2011zy,Lang:2012cx,Alberico:2013bza,Song:2015sfa} (UrQMD, POWLANG, PHSD and TAMU models),
an absolute determination of the elastic contribution is attempted to infer the
onset of radiative contributions by deviations from the experimental data beyond a
certain $p_T$. Alternatively, in the MC@sHQ+EPOS~\cite{Gossiaux:2008jv,Gossiaux:2010yx}, Vitev {\it et al.}~\cite{Sharma:2009hn}, BAMPS~\cite{Uphoff:2012gb,Uphoff:2014hza}, and Cao {\it et al.}/Duke~\cite{Cao:2013ita,Cao:2015hia} models,
both elastic and radiative contributions are included in a simultaneous best fit to 
data.  Early on it was already realized that radiative energy loss alone, which was 
able to describe light-hadron suppression at RHIC, is insufficient to account for 
the suppression of HF decay electrons at RHIC~\cite{Wicks:2007am}.

The evaluation of HQ radiative energy loss has thus far been mostly based on pQCD. 
One starts from the elastic heavy-light parton scattering diagrams, augmented 
with the radiation of a gluon, albeit in different approximation schemes. In 
Ref.~\cite{Djordjevic:2009cr}, within the HTL framework, the radiative energy loss 
rate for a heavy quark has been expanded in the number of HQ scattering events, \ie, 
in its inverse mean-free path. This is also known as an opacity expansion which, 
in particular, accounts for a finite size of the QCD medium, as originally 
developed for light-parton jet quenching~\cite{Gyulassy:2000er}. The extension
to heavy quarks~\cite{Djordjevic:2009cr} includes recoil effects of the plasma
partons and the possibility that the initial heavy quark is produced off its mass
shell and goes on-shell by radiating a gluon. The approach used in 
MC@sHQ+EPOS~\cite{Gossiaux:2010yx,Aichelin:2013mra} and 
BAMPS~\cite{Uphoff:2014hza} utilizes the Born 
diagrams for elastic parton scattering with a radiated gluon~\cite{Gunion:1981qs} 
extended to a heavy quark in a regime of ``intermediate" energies where the mass 
effect can be argued to largely suppress coherence effects known to be important 
for light partons. In this framework the radiative $2\to3$ cross section can be 
approximately cast into a factorized form of an elastic cross section times a 
probability of radiating an extra gluon ($P_g\propto \alpha_s$), schematically 
given as
\begin{equation}
d\sigma(Qp\to Qpg) \simeq d\sigma(Qp\to Qp) \ P_g  \ \ \quad (p=q,\bar q, g)  \ . 
\end{equation}
To account for interference effects, in particular for a suppression of gluon 
emission if the latter is characterized by a formation time longer than the
time between subsequent HQ scatterings, a practical prescription has been 
implemented in the BAMPS model~\cite{Uphoff:2014hza}. It amounts to vetoing the 
gluon emission process if the HQ mean-free path, $\lambda$, is shorter than a 
fraction $x_{\rm LPM}$ of the gluon formation time, $\tau_{\rm form}$, \ie,  
$\lambda < x_{\rm LPM} \tau_{\rm form}$, with 0~$<$~$x_{\rm LPM}$~$<$~1 a free 
parameter. A different implementation of coherence effects is used in the 
MC@sHQ+EPOS model~\cite{Nahrgang:2013saa}, where a suppression factor has 
been deduced assuming 
that multiple scatterings during the formation time can be represented by a 
single ``effective" scattering center with modified formation length. Further 
quantitative differences in these two implementations include the choice of 
charm-quark mass, $m_c$=1.5\,GeV~\cite{Nahrgang:2013saa} 
vs.~1.3\,GeV~\cite{Uphoff:2014hza}, and the use of finite vs.~zero mass for 
the radiated gluons, respectively.


\subsubsection{Hadronic Matter}
\label{sssec_had}
\hspace{2cm}  

Investigations of $D$-meson properties in hadronic matter have been carried out
in several approaches to date, \eg, QCD sum rules~\cite{Hilger:2008jg,Suzuki:2015est} 
and many-body theory~\cite{Fuchs:2004fh,Lutz:2005vx,Tolos:2007vh}. Until rather 
recently, their focus was mostly on $D$-meson spectral properties, including mass 
shifts and in-medium broadening, while transport properties did not receive much 
attention. In the year 2011 a series of works were conducted to investigate
the latter.

\begin{figure}[!t]
\begin{center}
\includegraphics[width=0.6\textwidth]{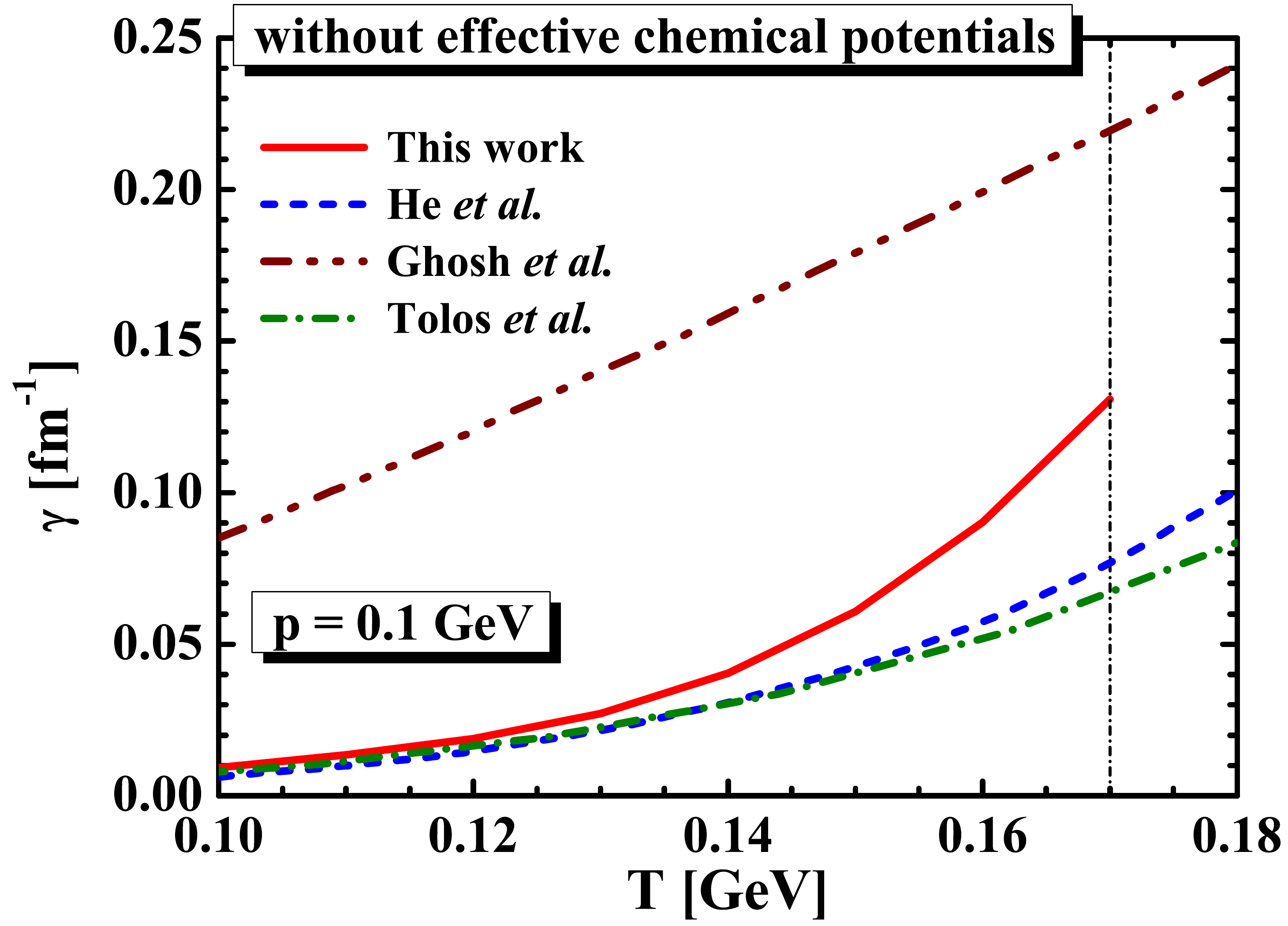}
\end{center}
\caption{Thermal relaxation rate of low-momentum $D$-mesons in hadronic matter
in chemical equilibrium, computed in the approaches of Refs.~\cite{He:2011yi}
(dashed line), \cite{Ghosh:2011bw} (dash-double-dotted line), \cite{Tolos:2013kva} 
(dash-dotted line) and \cite{Ozvenchuk:2014rpa} (solid line). 
Figure taken from Ref.~\cite{Ozvenchuk:2014rpa}.}
\label{fig_gamD}
\end{figure}

In Ref.~\cite{Laine:2011is} heavy-meson chiral perturbation theory was used to 
calculate the $D$-meson diffusion constant in a low-temperature pion gas using 
the $\pi$-$D$ scattering amplitude to lowest order in the couplings (Born 
amplitudes).  When extrapolated up to a temperature of $T$=100\,MeV (which 
might be beyond the applicability of the approximations), a relaxation rate 
of $\gamma_D$$\simeq$0.05/fm for zero-momentum $D$ mesons was found. 
In Ref.~\cite{He:2011yi} scattering amplitudes of $D$ mesons off 
pseudoscalar ($\pi$, $K$, $\eta$) and vector mesons ($\rho$, $\omega$, $K^*$) 
as well as baryons and anti-baryons ($N$, $\bar N$, $\Delta$, $\bar \Delta$) 
were adopted from existing calculations based on effective lagrangians. 
The resummation (unitarization) underlying these amplitudes led to 
significantly smaller values for the $p$=0 relaxation rate, by about a factor 
of 10 at $T$=100\,MeV (where essentially only pions contribute). However, when
extrapolated to temperatures around $T_c$, the $D$-meson relaxation rate 
increases to $\sim$0.08/fm, or a relaxation time of $\sim$12~fm/$c$, which
is comparable to the fireball lifetime in URHICs. The pertinent spatial 
diffusion coefficient amounts to $D_s (2\pi T)\simeq 6$, not far from 
estimates on the QGP side within the $T$-matrix formalism~\cite{Riek:2010fk}. 
In Ref.~\cite{Ghosh:2011bw} a similar set of hadronic scattering partners
was employed with the amplitudes treated in Born approximation. At 
$T$=100\,MeV the relaxation rate was calculated to be rather large, 
$\gamma_D$$\simeq$0.08/fm, although the temperature dependence is more mild
than in Ref.~\cite{Laine:2011is},  cf.~Fig.~\ref{fig_gamD}.
Furthermore, using unitarized amplitudes with interaction vertices from 
HQ chiral perturbation theory, a pion gas result of $\gamma_D$$\simeq$0.005/fm 
was found in Ref.~\cite{Abreu:2011ic}, consistent with Ref.~\cite{He:2011yi}. 
A more complete calculation~\cite{Tolos:2013kva} including scattering off the 
light pseudoscalar octet as well as nucleons and $\Delta$'s led to good agreement 
with the full results of Ref.~\cite{He:2011yi}, cf.~Fig.~\ref{fig_gamD}. 
This suggests that the use of resummed 
amplitudes, as well as excited states in hadronic matter, is critical to 
arrive at reliable results.  

The effects of higher excited hadronic states were estimated in 
Ref.~\cite{Ozvenchuk:2014rpa} using the idea~\cite{He:2011yi} that 
$D$-meson cross sections for scattering off light mesons ($m$) and baryons ($b$) 
can be approximated by $\sigma_{Dm}$=7-10\,mb and $\sigma_{Db}$=10-15\,mb, 
respectively. 
These values, motivated by constituent light-quark counting arguments, were 
found to reproduce the results of the microscopic calculations for $\gamma_D$ 
in Ref.~\cite{He:2011yi} reasonably well. Up to $T$=130\,MeV the contribution 
of excited states is found to be negligible, but it increases the $D$-meson 
relaxation rate by up to $\sim$60\% at $T$=170\,MeV, cf.~Fig.~\ref{fig_gamD}. 

The transport of $B$ mesons in hadronic matter was also studied in due course. 
In Ref.~\cite{Das:2011vba} Born amplitudes off pseudoscalar mesons from 
heavy-meson chiral perturbation theory where employed, leading to a small 
relaxation rate of $\gamma_B$$\simeq$0.001/fm at $T$=100\,MeV (the $D$-meson rate 
was found to be only slightly larger in this framework). In the unitarized calculation 
of Ref.~\cite{Abreu:2012et}, a significantly larger value was obtained, close to
$\gamma_B$=0.003/fm, almost a factor of 2 smaller than for $D$-mesons in the same 
framework~\cite{Abreu:2011ic}. A more complete calculation~\cite{Torres-Rincon:2014ffa} 
including $B^*$ resonances and scattering off $N$ and $\Delta$'s gave similar results 
at $T$=100\,MeV, but a significant increase of $\gtsim$50\% for $T$$\ge$140\,MeV.

\subsubsection{Lattice QCD}
\label{sssec_lqcd} 
\hspace{2cm}  

The extraction of transport coefficients in lQCD is rather challenging, but
progress is being made. 
For the HF case one usually computes the euclidean $Q\bar Q$ correlation 
function in the vector channel, transforms it to the spectral function 
in Minkowski space (requiring an inverse integral transform), and then
takes the zero-energy limit to extract the friction coefficient and from
it the spatial diffusion coefficient, $D_s$.
This has been carried out in quenched QCD~\cite{Ding:2012sp}, yielding a 
result of $D_s (2\pi T)\simeq 2\pm1$ without a significant temperature
dependence over the range $T$=1.4-3\,$T_c$.
Alternatively, in the static limit the diffusion coefficient can be extracted
from the color-electric field correlator which has been simulated in 
quenched QCD in Ref.~\cite{Banerjee:2011ra}. Here the computed diffusion
coefficient turned out to be $D_s (2\pi T)\simeq6\pm2$ for $T$=1.1-1.5\,$T_c$ 
and possibly increasing at higher $T$. 
A more recent quenched lQCD extraction based on the electric-field correlator
method gives $D_s (2\pi T)\simeq$\,3.7-7 for $T$=1.5\,$T_c$~\cite{Francis:2015daa}, 
consistent with the previous analysis~\cite{Banerjee:2011ra}.

Important questions remain about the microscopic mechanisms underlying the 
rather small values of the spatial HF diffusion coefficient, $D_s$. Detailed 
comparisons to model calculations are one way to gain insights, which will be
done in Sec.~\ref{sssec_Ds} in connection with Fig.~\ref{fig_Ds}. 
Alternatively, within lQCD, information about the effective degrees of 
freedom of the QGP at given temperature can be obtained from generalized 
susceptibilities. These can be used to correlate the (conserved) charm quantum 
number with other conserved quantum numbers (\eg, baryon charge or strangeness),
to test whether the prevalent degrees of freedom carrying charm are 
mesonic/diquark-like, or if they are quark-like. This has been scrutinized in a recent
work~\cite{Mukherjee:2015mxc}, and it was found that hadronic degrees of 
freedom play a significant role in contributing to the partial charm-generated 
pressure above $T_c$, up to at least 
$T\simeq1.2T_c\simeq$\,200\,MeV. This finding is quite consistent with the 
picture that broad $D$-meson resonances gradually dissolve with
increasing temperature of the QGP, as predicted within the $T$-matrix 
approach~\cite{Mannarelli:2005pz,vanHees:2007me,Riek:2010fk}. It is 
furthermore quite intriguing that the corresponding $T$-matrix calculations 
of the HQ diffusion coefficient~\cite{vanHees:2007me,Riek:2010fk} are 
comparable to the lQCD results discussed above.

\subsubsection{Comparison of Spatial Diffusion Coefficients}
\label{sssec_Ds}
\hspace{2cm}

\begin{figure}[!t]
\begin{center}
\includegraphics[width=0.7\textwidth]{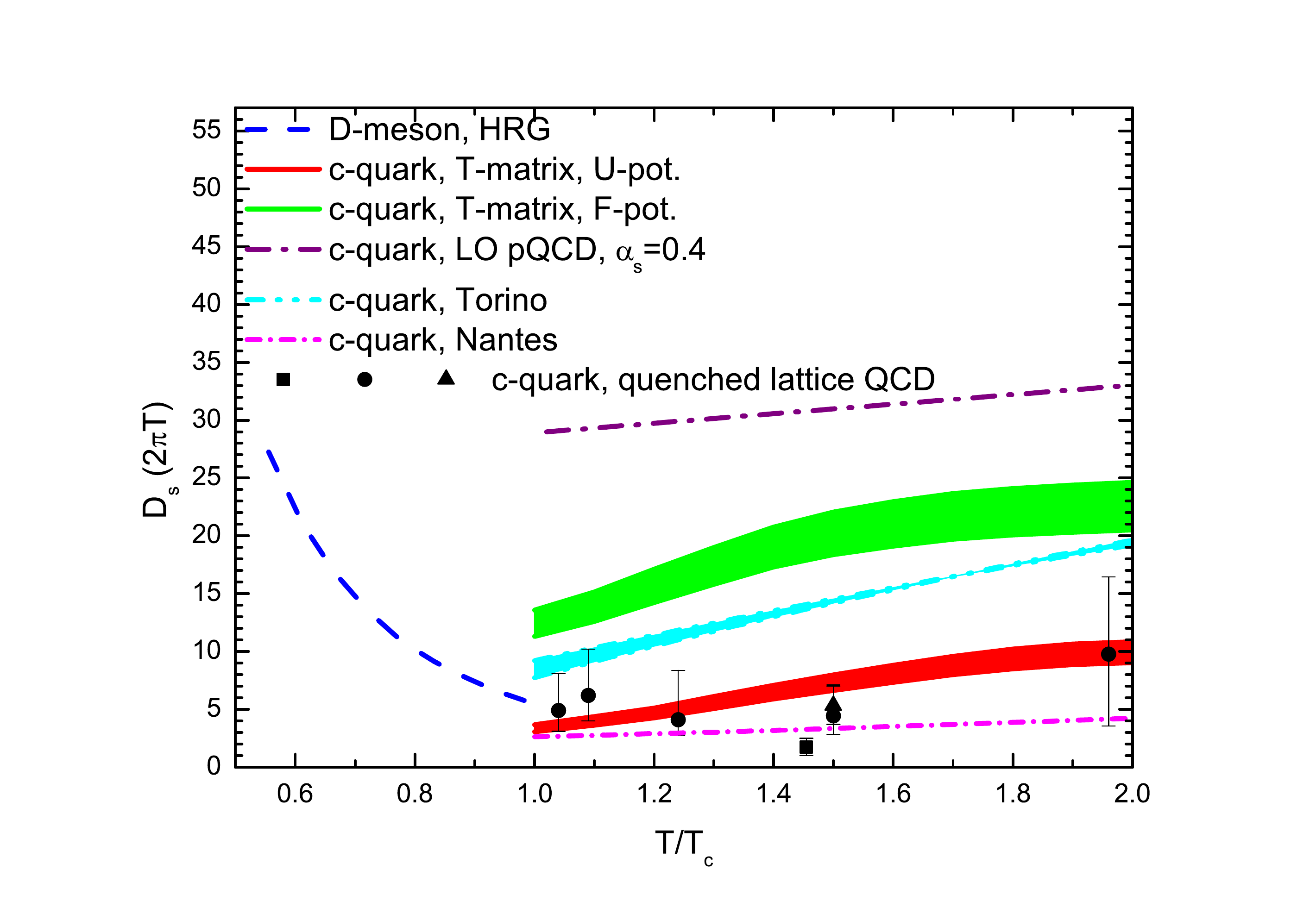}
\end{center}
\caption{Charm-quark diffusion coefficients from quenched lQCD (circles~\cite{Banerjee:2011ra},
squares~\cite{Ding:2012sp}, and triangle~\cite{Francis:2015daa}) compared to model
calculations based on different elastic interactions in the QGP (corresponding to
the $A$($p$=0) limit in Fig.~\ref{fig_A}): $T$-matrix calculations with either free (green
band) or internal energy (red band) as potential~\cite{Riek:2010fk,Huggins:2012dj}, pQCD
Born calculations from HTL/pQCD matching using a reduced Debye mass and running coupling
(Nantes~\cite{Peshier:2008bg,Gossiaux:2008jv}, pink dash-dotted line) or with perturbative
Debye mass and fixed coupling (Torino~\cite{Alberico:2011zy,Alberico:2013bza}, cyan
band), as well as schematic LO pQCD with fixed coupling and Debye mass
$m_D$=$gT$ (purple dash-dotted line). The blue-dashed line below $T_c$ is a calculation of
$D$-meson diffusion in hadronic matter from elastic scattering off various mesons
and anti-/baryons~\cite{He:2011yi}.
}
\label{fig_Ds}
\end{figure}

In Fig.~\ref{fig_Ds} we present a summary of results for the spatial diffusion 
coefficient, $D_s$, scaled by the thermal wavelength. We compare the 
quenched lQCD results from Sec.~\ref{sssec_lqcd} to various model calulations
of elastic charm-quark scattering in the QGP as discussed in Sec.~\ref{sssec_qgp},
corresponding to the $p\to0$ limit of the friction coefficients displayed
in Fig.~\ref{fig_A}. The schematic LO pQCD calculations with fixed coupling constant
($\alpha_s$=0.4) and Debye mass ($m_D$=$gT$), as well as the $T$-matrix calculations
with the free energy as potential, are significantly above the lQCD values. On the
other hand, the $T$-matrix results with the internal energy as potential, the Torino 
and Nantes HTL/pQCD calculations, as well as the infrared-enhanced pQCD implementation 
into the dynamical quasiparticle model~\cite{Berrehrah:2013mua} (not shown), are 
largely within the range of the quenched lQCD data.
The $T$-matrix and Torino results show a significant temperature dependence, while
the lQCD results are currently inconclusive on this aspect. Also shown is the
calculation of the $D$-meson diffusion coefficient in hadronic matter based
on effective interactions~\cite{He:2011yi}. As mentioned earlier, its temperature
dependence and values extrapolated to the pseudo-critical region are suggestive
for a continuous transition into the quark-based calculations, together with
a shallow minimun structure around $T_{\rm pc}$~\cite{He:2011yi,He:2012df}.
We also note that calculations of $D_s$ based on the bottom-quark friction
coefficients~\cite{vanHees:2004gq,Gossiaux:2008jv,Riek:2010fk,Alberico:2011zy,Huggins:2012dj}
give similar results to the charm-quark ones, within 20\% for most of
the approaches; in other words, the charm- and bottom-quark friction 
coefficients, $A$($p$=0), differ by approximately the mass ratio $m_b/m_c$. 
Since the masses are divided out in converting $A$($p$=0) to $D_s$, the 
latter can indeed serve as a reasonably universal measure of the (HQ) 
interaction strength in the QGP.

The diffusion coefficient has also been computed in the strong-coupling limit
of conformal field theories (CFTs) by using the AdS/CFT correspondence 
principle. The result for the HQ drag (or friction) coefficient in a 
Super-Yang Mills (SYM) plasma with $N_c$ fundamental charges has been worked 
out as~\cite{Herzog:2006gh,Gubser:2006bz,CasalderreySolana:2006rq}
\begin{equation}
\gamma_Q^{\rm SYM}=\frac{\pi\sqrt{\lambda}T_{\rm SYM}^2}{2m_Q} \ , 
\end{equation}
which turns out to be proportional to the square root of the 't Hooft coupling,
$\lambda=g_{\rm SYM}^2N_c$, highlighting its nonperturbative nature.
Interestingly, the AdS/CFT friction coefficient (thermalization rate) exhibits the 
factor $T/m_Q$ characteristic for the time delay in the thermalization of a Brownian 
particle. Several caveats arise in converting this result into an estimate for the QCD
plasma~\cite{Gubser:2006qh}. When rescaling the temperature of the SYM plasma
to match the degrees of freedom (energy density) of the QCD plasma, and matching
the coupling constant to an $\alpha_s$ estimated from the lQCD HQ free energies,
one obtains $\gamma_Q=CT^2/m_Q$ with $C$\,$\simeq$\,1.5-2.5, yielding
$\gamma_Q\simeq$\,0.3-0.5\,/fm at $T$=250\,MeV.
The pertinent diffusion coefficient (not shown in Fig.~\ref{fig_Ds}) turns out 
to be rather similar to the pQCD results with running coupling and reduced Debye
mass~\cite{Peshier:2008bg,Gossiaux:2008jv} (Nantes, pink dash-dotted line in
 Fig.~\ref{fig_Ds}).

\subsubsection{Hadronization}
\label{sssec_hadro}
\hspace{2cm}  

The diffusion of heavy flavor in the QGP and hadronic phase of URHICs needs to be 
interfaced by a transition of the degrees of freedom, from charm and bottom quarks 
to HF hadrons. Unlike in a macroscopic modeling of the bulk evolution, \eg, with 
hydrodynamics, where this transition is encoded in the equation of state, the 
hadronization mechanism for HF transport requires a microscopic treatment. This is 
both a challenge and an opportunity to learn about this fundamental process in QCD. 

In the vacuum, the common procedure to hadronize quarks at high momentum is the use
of an empirical fragmentation function, $D^{h}_q(z)$, which describes the probability
distribution of producing a hadron of momentum $p_h$ from a parent quark of momentum
$p_q$, with the momentum fraction $z=p_h/p_q$. Different quark species (as well as 
gluons) are modeled with different distributions which, once fixed (say, in $e^+e^-$
annihilation), are assumed to be universal for other collision systems. For HQ 
fragmentation a widely adopted form is the Peterson fragmentation 
function~\cite{Peterson:1982ak}, 
\begin{equation}
D^H_Q(z;\epsilon_Q)=\frac{N}{z [1- (1/z) -\epsilon_Q/(1-z)]^2}  \ , 
\end{equation}   
where $N$ is a normalization factor to ensure $\int D(z)dz=1$. The parameter 
$\epsilon_Q$ controls the hardness of the distribution, with a maximum
at $z\simeq1-2\epsilon_Q$ and a width of $\sim$$\epsilon_Q$. A standard choice
is $\epsilon_c$=0.04 and $\epsilon_b=0.005$, in line with the expected behavior
of $\epsilon_Q\propto 1/m_Q^2$\cite{Peterson:1982ak}. 
An alternative, more involved  modeling of the HQ fragmentation function has been 
developed in connection with fixed-order-next-to-leading-logarithm (FONLL) 
calculations of HQ production~\cite{Cacciari:1998it,Cacciari:2005rk}, where the 
Peterson framework does not straightforwardly apply.

At low momentum, the fragmentation picture breaks down, and other mechanisms must 
set in. In particular, a recombination of heavy quarks with surrounding light 
anti-/quarks has been suggested~\cite{Kartvelishvili:1980uz,Hwa:1994uha} and found
to account for flavor asymmetries in $D$-meson production in elementary hadronic
reactions, especially at forward rapidities through recombination with valence 
anti-/quarks of the projectile~\cite{Braaten:2002yt,Rapp:2003wn}. 
The recombination picture is even more compelling within a QGP cooling through 
the transition temperature, where an ample abundance of thermal light quarks and 
anti-quarks provides a natural source to color-neutralize heavy quarks. 
Initial works have implemented this idea using an instantaneous approximation 
using 3D space-momentum Wigner functions to convert charm quarks into $D$-mesons 
(or charmonia) with comoving light quarks~\cite{Lin:2003jy,Greco:2003vf}. Despite 
the fact that the major momentum fraction is carried by the heavy quark, a large 
effect on the $D$-meson elliptic flow in URHICs 
at low and intermediate momenta was found, as being imprinted by recombination 
with light quarks. One expects this mechanism to reach out to higher 
transverse momenta than for light hadrons since, at comparable momenta, 
the velocity of heavy quarks is smaller than for light quarks thus facilitating 
recombination with comoving thermal partons.  
A more rigorous implementation of the recombination mechanism at low and 
intermediate $p_T$ should account for energy conservation (\ie, be formulated 
as a {\em rate}) and satisfy thermal equilibrium as the correct long-time limit. 
This is particularly important in situations with significant space-momentum 
correlations as is the case in non-central heavy-ion collisions with a 
collectively expanding partonic source with elliptic flow~\cite{Molnar:2004zj}. 
Such correlations are not accounted for in coalescence formalisms with global 
(momentum-space only) or factorized momentum- and coordinate-space distribution 
functions (such as usually done with 3D Wigner distributions). 
In the ``resonance recombination model" (RRM) the hadronization of 
heavy quarks is described by scattering off thermal light quarks into broad 
$D$-meson like resonances~\cite{Ravagli:2007xx,Ravagli:2008rt}, which by 
construction conserves 4-momentum and recovers the thermal-equilibrium limit. 
In the HF context the RRM was implemented on a hydrodynamic hypersurface with 
elliptic flow~\cite{He:2011qa}; it was shown that a HQ distribution in local 
equilibrium, with an elliptic flow as dictated by the
hydrodynamic velocity fields, indeed maps into a $D$-meson distribution
in local equilibrium with the correct mesonic elliptic flow\footnote{For example, 
for sufficiently large radial flow, a single charm-quark distribution 
with strictly positive $v_2^c(p_t)$ implemented into RRM gives rise to a charmonium 
distribution with $v_2^{\psi}(p_T)$ dipping into negative values at small $p_T$, 
reflecting the well-known mass effect in the $v_2$ behavior of heavy particles.}.      
Utilizing the heavy-light quark $T$-matrix interactions, which generate resonances 
close to $T_c$, to compute the HQ recombination rates within 
RRM~\cite{He:2011qa,He:2012df} puts the hadronization process on the same footing 
as the non-perturbative HQ diffusion processes (we recall in passing that the 
confining force plays an essential role in these interactions). In this picture, 
hadronization is simply a manifestation of the increasing strength of HQ interactions 
with the partonic medium as $T_c$ is approached from above.  

A quantitative treatment of HQ recombination processes needs to allow for the
possibility of forming higher excited hadrons, in particular $D^*$ mesons, 
$\Lambda_c$ baryons and $D_s$ mesons in the charm sector (and likewise for 
bottom). Even if these particles are not measured, they can have a significant 
impact on the $D$-meson abundance through depleting the charm quarks available 
for $D$-meson recombination, the so-called ``chemistry effect" (hadrons 
containing more than one anti-/charm quark, including double-charm baryons 
and charmonia, give small corrections in this context as their abundance only 
constitutes up to a few percent of the total charm yield). For example, the 
fraction of $D_s$ and $\Lambda_c$ could be significantly enhanced in URHICs 
relative to $pp$ collisions in the presence of coalescence processes.

A ``hybrid" in-medium hadronization scheme, which embeds fragmentation into an 
environment of thermal partons, has recently been implemented into the POWLANG 
transport approach~\cite{Beraudo:2014boa}. Here, the hadronization of a heavy 
quark propagating in a hydrodynamic background is carried out once it enters a 
fluid cell which has cooled down to a pseudo-critical temperature of 
$T_{\rm pc}$=155\,MeV. The thermal light- and strange-quark distributions in the 
rest frame of the fluid cell are sampled and boosted into the lab frame where a 
string is constructed by joining the endpoints of the heavy- and light-quark 
positions.  These configurations are then passed to PYTHIA 6.4~\cite{Sjostrand:2006za} 
to simulate the fragmentation of the string into hadrons. In this way the HF 
fragmentation process inherits collective flow of the expanding medium.
By the nature of the fragmentation process, where the strings, once formed,
are hadronized as in vacuum, the HF hadro-chemistry remains rather similar to 
that in the vacuum; in particular, it does not induce a sizeable enhancement 
of the ${D_s^+}/D$ or $\Lambda_{c}^+/D$ ratios relative to $pp$ collisions. 

In practice, the question arises how to disentangle the effects of the QGP 
phase, hadronization (via coalescence) and hadronic phase from HF observables
in URHICs. In Ref.~\cite{He:2012df} it has been suggested that heavy-strange
mesons ($D_s$ and $B_s$) can help in this respect, based on two main ideas. 
First, the suppression of strangeness in elementary ($pp$) collisions is
lifted in heavy-ion collisions reflected in the near-equilibrium level 
of strange-particle production (quantified by a strangeness fugacity 
$\gamma_s$$\simeq$~1). Consequently, if HF recombination is effective, it
would manifest itself as an increase in the $D_s/D$ (or $B_s/B$) ratio in
AA relative to $pp$ collisions. Moreover, the $p_T$ dependence of this ratio 
could reveal how far out coalescence mechanisms are operative, especially 
since the small mass difference between strange and non-strange mesons renders 
a splitting in their $R_{\rm AA}$ due to collective flow a small effect. 
Second, it is expected that hadronic rescattering is much weaker for $D_s$ 
than for $D$ mesons, based on the picture that the coupling to the heat 
bath is mostly driven by the light-quark content of the hadron (the 
approximate constituent light-quark scaling of the $D$-meson interactions
with light hadrons referred to in Sec.~\ref{sssec_had} is one indication
thereof; another one is the apparent earlier kinetic decoupling of 
multi-strange hadrons in the hadronic phase of URHICs, see, \eg, 
Refs.~\cite{He:2011zx,Takeuchi:2015ana} for recent works on this issue). 
This implies that the difference between the $v_2$ of $D$ and $D_s$ mesons
can serve as a measure of the elliptic flow imprinted on the $D$ during
the hadronic phase, and thus serve as a measure of its hadronic transport
coefficient. It furthermore turns out that the hadronic interactions do not 
have a large effect on the $R_{\rm AA}$ of the $D$-mesons, presumably due 
to a compensation between the dropping temperature and increasing flow of 
the medium. This would be fortunate as to conserve the comparison of the 
$D$- and $D_s$-meson $R_{\rm AA}$'s as a measure of coalescence processes.  

\begin{figure}[!t]
\begin{minipage}{0.33\linewidth}
\vspace{-0.2cm}
\includegraphics[width=1.12\textwidth]{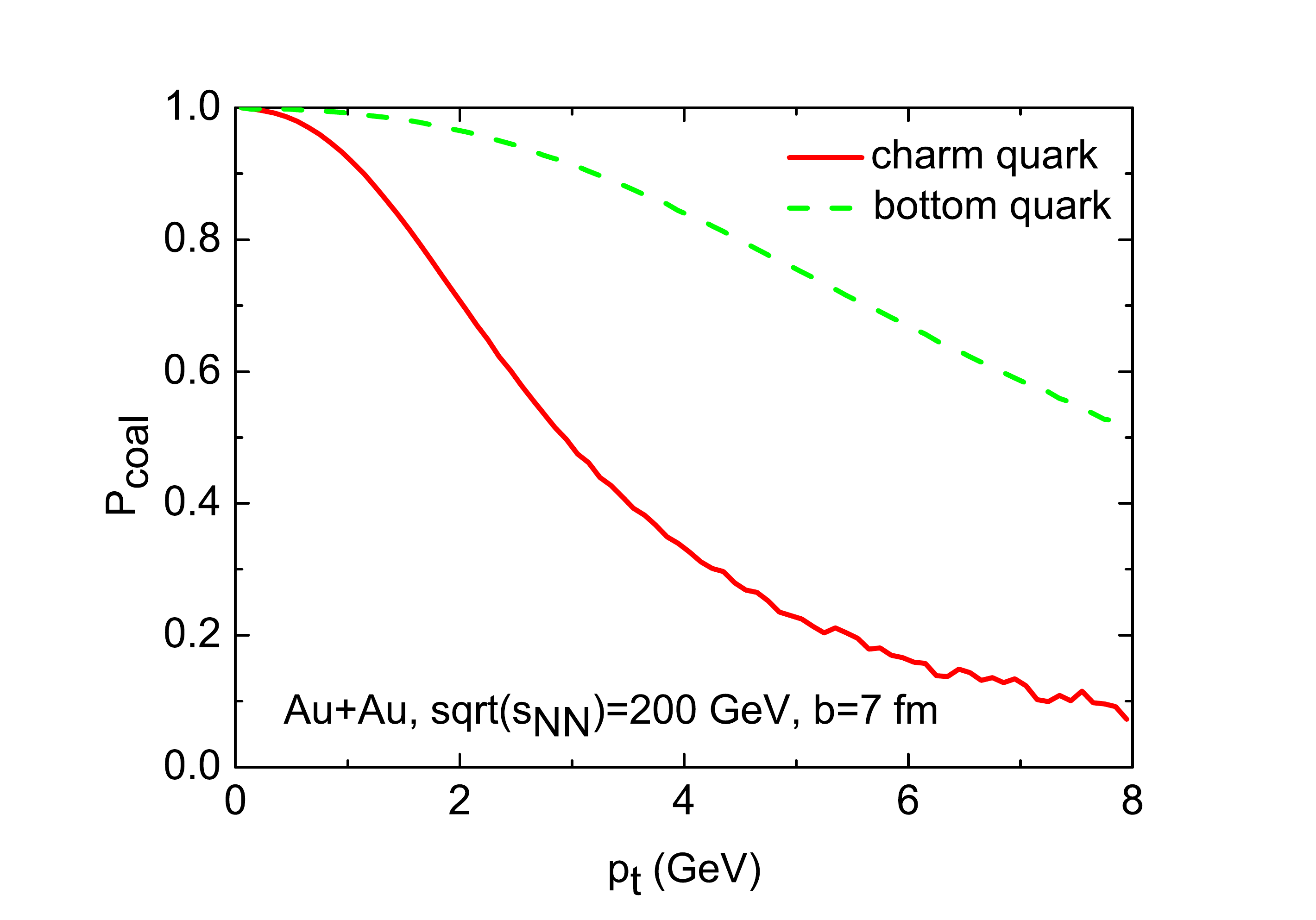}
\end{minipage}
\begin{minipage}{0.33\linewidth}
\hspace{0.1cm}
\includegraphics[width=0.9\textwidth]{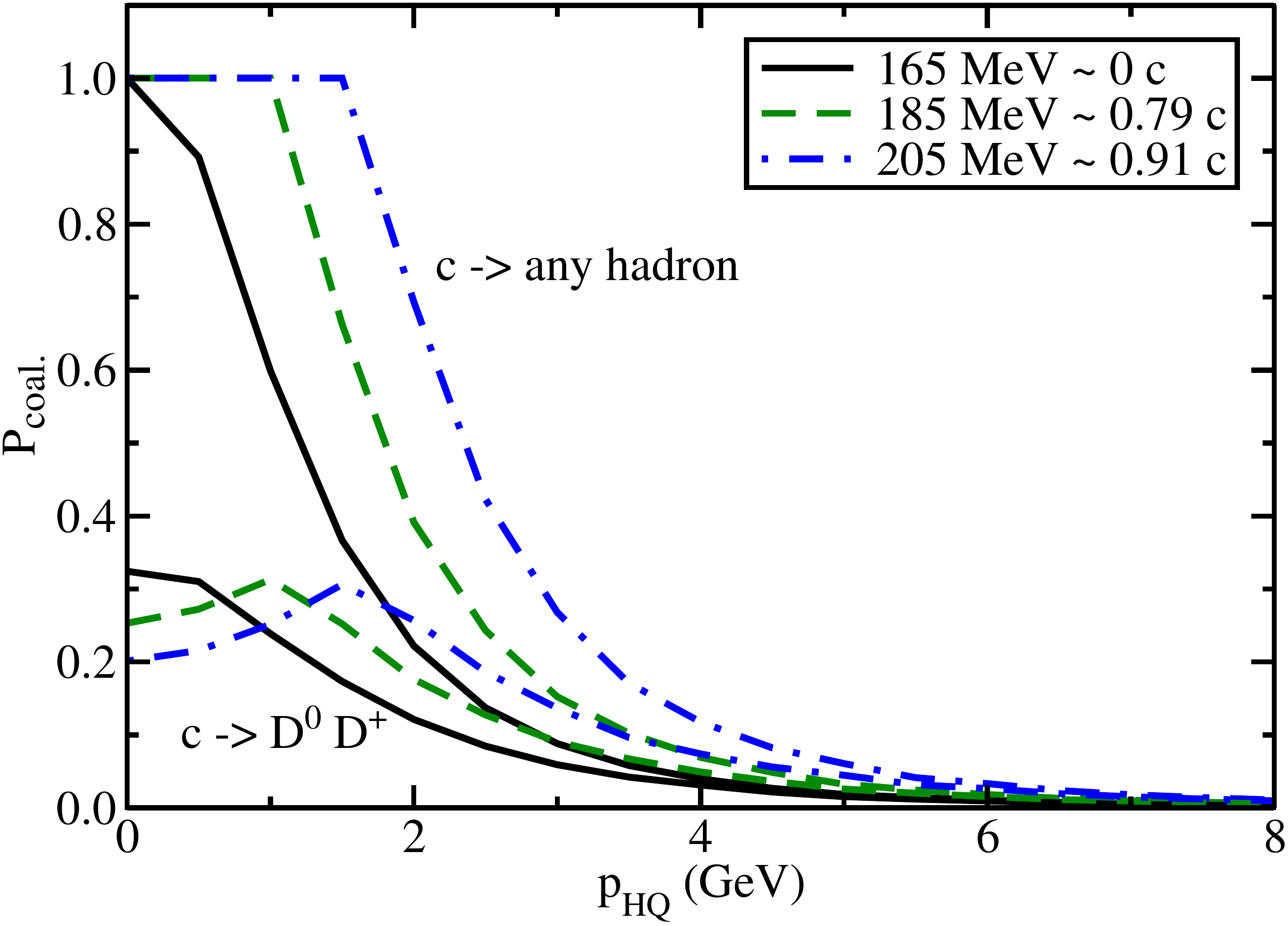}
\end{minipage}
\begin{minipage}{0.33\linewidth}
\vspace{-0.2cm}
\includegraphics[width=1.1\textwidth,height=0.76\textwidth]{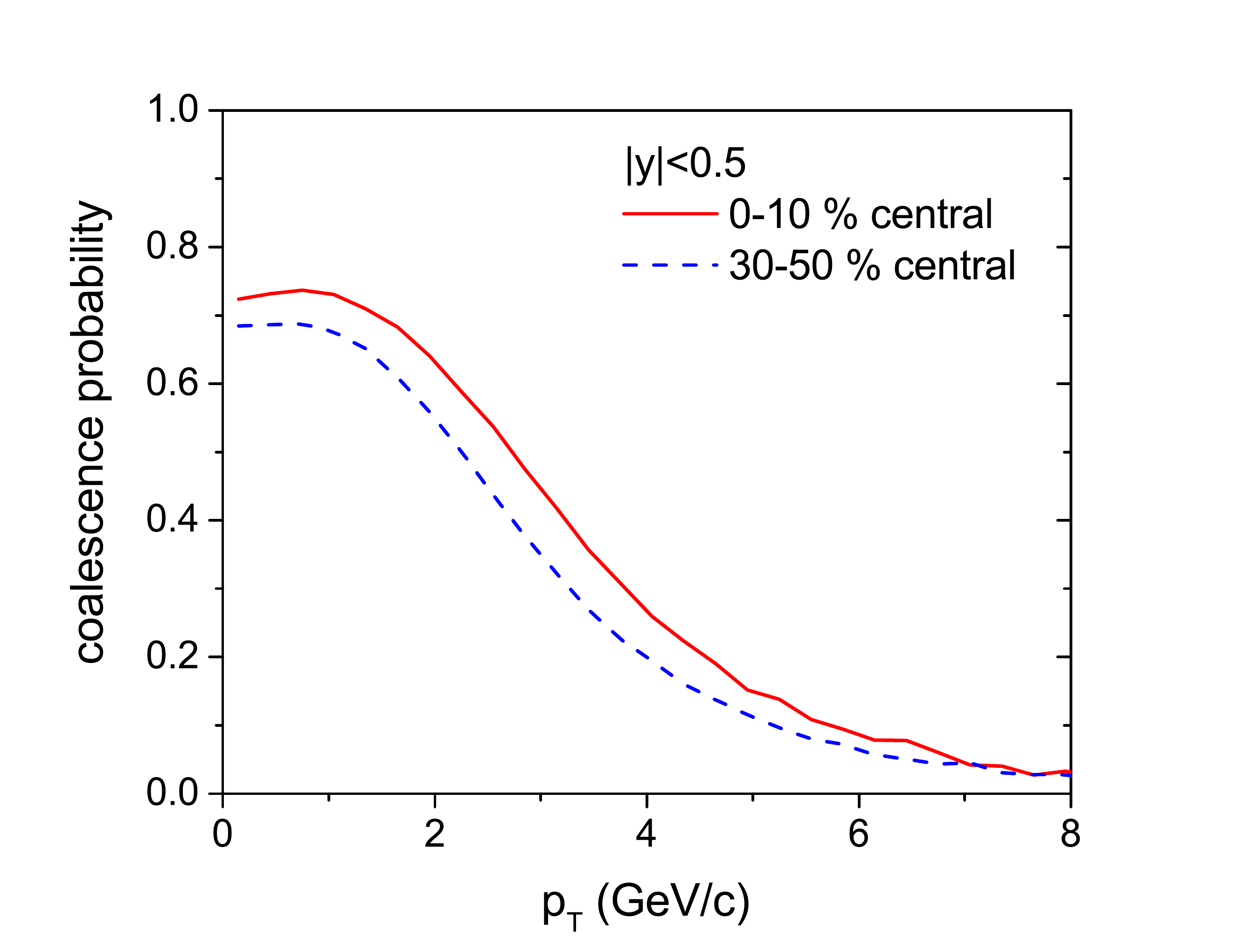}
\end{minipage}
\caption{Coalescence probabilities of charm quarks in heavy-ion collisions as a
function of their (transverse) momentum in the lab frame using
(a) the resonance recombination model~\cite{Ravagli:2007xx} within the TAMU
transport approach~\cite{He:2011qa} at RHIC (left panel, for $c$ and $b$ quarks),
(b) an instantaneous coalescence model with Wigner functions~\cite{Oh:2009zj}
within the Duke transport approach~\cite{Cao:2013ita} (middle panel), and
(c) a stochastic sampling of Wigner functions within the PHSD transport
approach~\cite{Song:2015sfa,Song:2015ykw} (right panel).}
\label{fig_Pcoal}
\end{figure}
We finish this section by illustrating the results for the coalescence 
probabilities of charm quarks as a function of their (transverse) momentum in 
URHICs for some of the implementations employed 
in the literature, cf.~Fig.~\ref{fig_Pcoal}. The three examples are based on rather 
different ingredients. The TAMU model~\cite{He:2011qa,He:2014cla} (left panel) 
utilizes RRM~\cite{Ravagli:2007xx} with mesonic resonance rates for charm quarks 
of mass $m_c$=1.7-1.8\,GeV interacting with constituent thermal light quarks on 
a hydro hypersurface at $T$=170\,MeV, assuming a thermal hadro-chemistry. 
The Duke model~\cite{Cao:2013ita} employs an instantaneous Wigner function 
approach~\cite{Oh:2009zj} for charm quarks of mass $m_c$=1.27\,GeV with constituent 
thermal light quarks at $T$=165\,MeV where transverse-flow effects are simulated 
through an effective medium temperature; a hadro-chemistry is accounted for by 
coalescence into the ground and first excited states of $D$ mesons and charmed 
baryons. In the PHSD transport model~\cite{Song:2015sfa,Song:2015ykw}, a Wigner 
function coalescence is implemented by a stochastic sampling of the local 
environment with charm quarks of mass $m_c$=1.5\,GeV, as given by the light-parton 
phase space distributions; the sampling is carried out when the PHSD bulk medium 
evolves through energy densities around $\epsilon_{\rm pc}$$\simeq$0.5\,GeV/fm$^3$,  
representing the pseudo-transition region; also here the hadro-chemistry is
accounted for through the inclusion of higher charm resonance states.   
The calculated coalescence probabilites in all three approaches share basic 
features, such as a maximum at low momentum and decreasing over a momentum 
range of a few times the HQ mass. The normalization of $P_{\rm coal}(p_t\to0)$ 
has been put to one in the TAMU and Duke models, while in the PHSD model it 
is below one and significantly depends on the assumed hadron radius in the 
Wigner function. A more rigorous way to determine the absolute normalization 
remains to be developed. The drop of $P_{\rm coal}(p_t)$ with increasing momentum
is strongest in the Duke model, while it is weaker for PHSD and TAMU, possibly 
caused (at least in part) by the larger charm-quark masses used in the latter 
two models. We also note that other features of the coalescence process have
significant impact on experimental observables, \eg,  how the HQ momentum 
translates into a hadron momentum. In the Duke model, the net effect of 
heavy-light coalescence on $D$-meson observables turns out to be quite small, 
while is noticeable in PHSD, and still larger in the TAMU model. More detailed 
studies are needed to unravel the origin of these differences.

\subsubsection{Pre-Equilibrium and Mean-Field Transport}
\label{sssec_pre-mean}
\hspace{2cm}

In this section we return to the role that a ``mean-field" induced force could
play in the propagation of heavy quarks in URHICs. The canonical environment 
for such a formulation are the earliest, off-equilibrium phases in the collision,
where a high occupancy of gluonic modes is suggestive for a field description, 
or, from a slightly different viewpoint, a description in terms of a strongly 
coupled fluid without quasipaticle structure. 

The latter approach has been adopted in Ref.~\cite{Chesler:2013urd}, by employing
the AdS/CFT duality to study the force needed to drag a heavy quark at constant
velocity through an environment of two colliding (longitudinally Gaussian and 
transversely uniform) sheets of energy. When comparing the required force in this
highly off-equilibrium medium to that of the equilibrium limit in the ${\cal N}$=4 
super-Yang-Mills plasma, it was found that the results are quite similar (within
a few 10's of percent) if evaluated at the same energy densities (or, equivalently,
at the same ``pseudo"-temperatures extracted from different diagonal elements
of the off-equilibrium stress-energy tensor). This suggests that the effects of
the initial off-equilibrium phases on HQ diffusion in URHICs can be reasonably 
well approximated by the standard transport approach with coefficients extrapolated 
from the equilibrium theory at corresponding energy densities. In addition, a 
characteristic time delay in the impact of the initial medium on the HQ motion 
was found, on the order of 0.1-0.2~fm/$c$, which could serve as the absolute 
starting time for the transport simulation.   

The gluon saturation approach has been followed in Ref.~\cite{Das:2015aga}, 
by converting the gluonic field configurations obtained within the IP-glasma 
model~\cite{Schenke:2013dpa} into off-equilibrium quantum-statistical gluon 
distribution functions, associated with a typical timescale of the inverse 
saturation momentum, $\tau_{\rm glasma}\sim 1/Q_s \sim$~0.1-0.2\,fm/$c$ (not 
unlike the delay time found in the AdS/CFT approach discussed in the previous 
paragraph). This distribution has then been inserted into eq.~(\ref{X}) to 
evaluate the HQ transport coefficients using the schematic LO pQCD matrix 
elements. The findings parallel those of the AdS/CFT approach, in that the 
HQ drag coefficient turns out to be very similar to the one in equilibrium 
provided the same total gluon density is enforced. It turns out to be larger 
than for a chemically equilibrated QGP at the same parton number (or energy) 
density, primarily due to the larger color factor in the pQCD HQ-gluon scattering 
amplitude relative to HQ-quark scattering.  

A genuine mean-field contribution can also occur in later stages of the fireball 
evolution, \eg, due to an effective in-medium quark mass, $m_Q^*(T)$, when the 
quark propagates through regions with non-zero temperature gradient. The HQ mass 
is generally expected to increase with decreasing temperature of the QGP, as 
borne out of both perturbative and non-perturbative calculations (see, \eg, 
Ref.~\cite{Mocsy:2013syh}). Thus, for a heavy quark propagating through an 
expanding fireball where the temperature decreases both with time and in outward 
direction, its 3-momentum is reduced due the pertinent mean-field force. This 
effect has been studied in relativistic Langevin simulations using the transport 
coefficients of Ref.~\cite{Riek:2010fk} and found to soften the momentum spectrum 
while slightly reducing the elliptic flow, although both effects are quantitatively 
small (a few percent)~\cite{He-priv}. Generally speaking, for large HQ scattering 
rates, corresponding to large imaginary parts in the self-energy, real parts tend 
to be rather structureless (by means of a dispersion integral) and the corresponding 
mean-field forces small. Repulsive forces on heavy quarks due to a vector mean-field
in the QGP have been studied in the PHSD approach~\cite{Song:2015ykw} and found to 
induce a hardening of the final $D$-meson spectra, enhancing the ``flow bump" in 
their $\Raa$ and shifting it to slightly higher $\pT$.

\subsection{Bulk Medium Evolution Models for URHICs}
\label{ssec_bulk}
To implement the HF transport properties as discussed in the preceding
section into the phenomenology of URHICs, two additional ingredients are
required. The first one concerns the initial conditions for the HQ phase
space distribution. Here one usually adopts pQCD calculations fit to
experimental $p_T$ spectra of HF particles in elementary $pp$ collisions
(augmented with suitable fragmentation functions). In an AA collision,
their spatial distribution is usually assumed to follow the $NN$
collision profile, while the $p_T$ distributions are possibly modified
due to nuclear modifications in the parton distribution functions
(and constrained by $p$A data). A more detailed discussion of these
issues will be given in Sec.~\ref{ssec_element} below.


The second ingredient beyond the HQ interactions with the medium
is the space-time evolution of the latter. The Fokker-Planck/Langevin
framework, with transport coefficients depending on temperature
(possibly chemical potentials) and 3-momentum in the rest frame of
the medium, is naturally implemented into hydrodynamic (or fireball) 
models based on local thermal equilibrium. A minimal requirement for
state-of-the-art evolution models for HF transport simulations is the 
description of light-hadron ($\pi$, $K$, $p$) multiplicities, $p_T$ 
spectra and their elliptic-flow coefficient, $v_2(p_T)$, at least 
approximately and up to a typical transverse momentum of $p_T\simeq 2$\,GeV, 
which encompasses 
$\sim$90\% or more of the produced bulk particles. However, even
within this constraint, an appreciable uncertainty in the space-time 
evolution of the temperature and the collective flow field (in 
particular the elliptic flow) remains. 
For example, an initial study of this issue~\cite{Gossiaux:2011ea}, 
comparing charm-quark Langevin simulations within the original Kolb-Heinz 
(KH) 2+1D hydrodynamic model~\cite{Kolb:2003dz} and a parameterized 
expanding thermal fireball model, revealed significant deviations between 
the results for the charm-quark $v_2$ at the end of the quark-hadron 
mixed phase (using the same transport coefficients). A large part of the 
discrepancy originated from the different build-up of the radial and 
elliptic flow in the two evolution models, proceeding significantly slower 
(``softer") in the KH hydro. Retuning the fireball to agree with the inclusive 
light-quark $v_2$ of the KH hydro at the end of the mixed phase leads to 
much closer agreement for the resulting charm-quark $v_2$~\cite{He:2011qa}
(see also Ref.~\cite{Alberico:2011zy} for a comparison of HF spectra
from different hydro evolutions). 
Typical components of a ``soft" hydro evolution include a standard initial 
participant profile for the entropy density, vanishing initial-flow 
velocity at the thermalization time, 
and an equation-of-state (EoS) with a quark-hadron mixed-phase. Among 
the consequences of a ``soft" expansion are that the bulk-$v_2$ builds 
up rather slowly and receives sizable contributions from the hadronic phase 
(30-40\%), that multi-strange particles need to freeze out well inside 
the hadronic phase, and that the pion HBT radii disagree with experiment. 
All of these features are modified if the medium evolution is made more 
explosive, by introducing pre-equilibrium 
flow~\cite{Kolb:2002ve,Broniowski:2008qk,Gale:2012rq,vanderSchee:2013pia}, 
a realistic lQCD EoS with cross-over transition, a more compact 
initial-density profile and a finite shear viscosity. This resolves the 
HBT puzzle~\cite{Pratt:2008qv,Pratt:2009hu}, makes the bulk-$v_2$ saturate 
close to the 
hadronization transition, and generates sufficient radial flow to allow 
multi-strange hadrons to kinetically decouple at chemical 
freezeout~\cite{He:2011zx}. A ``harder" hydro evolution has significant 
consequences for HF observables. Most notably, the flow bump in the 
charm-quark (and subsequent $D$-meson) $R_{\rm AA}$ is shifted to higher 
$p_T$, and the pertinent $v_2(p_T)$ tends to be larger for $p_T\gtsim2$\,GeV, 
both because the bulk-$v_2$ is available from earlier times on (giving 
charm quarks more time to pick it up) and the collectivity of the charm 
quarks is pushed out to higher $p_T$.  
It is interesting to note that an ``explosive" medium evolution is an 
important ingredient to understand the rather large $v_2$ observed for direct 
photons at RHIC (and possibly LHC)~\cite{vanHees:2011vb,vanHees:2014ida}. 
A harder hydro evolution with improved initial conditions and lQCD-based EoS
was also implemented in Refs.~\cite{Nahrgang:2013xaa,Nahrgang:2013saa}. Also
here the flow bump in the $D$-meson $R_{\rm AA}$ at RHIC showed a notable 
shift to higher $p_T$~\cite{Nahrgang:2014ila} over previous results with 
the same HQ transport coefficients~\cite{Gossiaux:2010yx}.

Alternative to macroscopic bulk evolutions based on local thermal 
equilibrium, semi-classical microscopic transport simulations have been 
employed~\cite{Zhang:2005ni,Molnar:2006ci,Uphoff:2011ad,Uphoff:2012gb,Uphoff:2014hza,Das:2015ana,Song:2015sfa,Song:2015ykw}, 
or combinations thereof within so-called ``hybrid" models~\cite{Lang:2012cx}.
In the BAMPS model~\cite{Uphoff:2012gb}, for example, the evolution of the
partonic system is terminated at the transition temperature, where quarks 
and gluons are fragmented into hadrons. This model approximately describes 
the measured bulk-hadron $v_2$, suggestive for a rather explosive QGP 
expansion, but it predicts a much softer flow bump in the $D$-meson 
$R_{\rm AA}$ at RHIC than the simulations based on a ``hard-hydro" 
evolution (cf.~upper left panel in Fig.~\ref{fig:Raav2DLHCRHIC}). 
In the PHSD transport simulations~\cite{Song:2015sfa}, the $D$-meson 
$R_{\rm AA}$ at RHIC turns out to be rather close to the results from 
``hard-hydro" evolutions.

\section{Experimental Results and Comparison to Models}
\label{sec_exp}

\subsection{Experimental Facilities and Techniques}
\label{ssec_scope}

\subsubsection{Accelerators}

\hspace{2cm}

Heavy-ion collisions are a unique way to explore the phase diagram of QCD 
matter in the laboratory and study the properties of the QGP.
The first fixed-target experiments with collisions of light nuclei at 
ultra-relativistic energies started in 1986 at the BNL AGS and at the CERN SPS.
Heavy nuclei (Au and Pb) became available in the 90's at both facilities.
Since the year 2000, heavy-ion colliders are in operation, first at BNL 
(Relativistic Heavy Ion Collider, RHIC, first Au--Au collisions in 2000) 
and ten years later at CERN (LHC, first Pb--Pb collisions in 2010).

The AGS and SPS fixed-target programs enabled the study of heavy-ion reactions 
at center-of-mass ($CM$) energies of $\sqrtsNN=$~1-20~GeV per nucleon pair.
In the heavy-ion program at the CERN SPS, open HF production was 
not measured directly. Indirect measurements based on the contribution 
of simultaneous semi-muonic decays of correlated  ${D\overline{D}}$ 
pairs to the dimuon invariant-mass spectra  were carried out in Pb--Pb 
and In--In collisions at $\sqrtsNN=17.2~\GeV$ by the NA50 and NA60 
collaborations~\cite{Abreu:2000nj,Arnaldi:2008er}.

The Relativistic Heavy Ion Collider (RHIC) began operation in 2000 with the 
capability of colliding nuclei from deuterons to Au at $CM$ energies of up 
to 200\,GeV per nucleon pair. Over the 15 years of data taking to date, several 
collision systems were studied at different collision energies. The luminosity 
for full energy Au--Au collisions was increased over the years from the
design value of $2 \cdot 10^{26}~\cm^{-2}~\s^{-1}$ to about
$5 \cdot 10^{27}~\cm^{-2}~\s^{-1}$ in the 2014 run.
During years 2010, 2011 and 2014 a Beam Energy Scan (BES) campaign was conducted,
taking data for Au--Au collisions at $CM$ energies per nucleon pair
of 62.4, 39, 27, 19.6, 14.5, 11.5 and 7.7 GeV.
A summary of the data samples of heavy-ion collisions collected over this
period is reported in Tab.~\ref{tab:TableRHICruns}.
Heavy-flavor results were reported by the PHENIX and STAR collaborations
for Au-Au collisions at $\sqrtsNN$ of 39, 62.4, 130 and 200 GeV,
for Cu-Cu collisions at $\sqrtsNN=200~\GeV$, and for $d$Au
collisions at $\sqrtsNN=200~\GeV$.
Also proton-proton ($pp$) collisions at $\sqrt{s}=200~\GeV$ 
were studied as a reference for heavy-ion collisions.

\begin{table}
\centering
\begin{small}
\begin{tabular}{l|c|c|c|c}
    & Year & Collision system & $\sqrtsNN$ & Delivered int. lumi \\
    &      &                  &   (GeV)           & ($\nb^{-1}$) \\
\hline
Run-1  & 2000    & Au--Au & 130 & 0.02 \\[1ex]
Run-2  & 2001/02 & Au--Au & 200 & 0.258 \\
       &         & Au--Au & 19.6 & $0.4 \cdot 10^{-3}$ \\[1ex]
Run-3  & 2002/03 & $d$--Au & 200 & 73 \\[2ex]
Run-4  & 2003/04 & Au--Au & 200 & 3.53 \\
       &         & Au--Au & 62.4 & 0.067 \\[2ex]
Run-5  & 2004/05 & Cu--Cu & 200 & 42.1 \\
       &         & Cu--Cu & 62.4 & 1.5 \\
       &         & Cu--Cu & 22.4 & 0.02 \\[2ex]
Run-7  & 2006/07 & Au--Au & 200 & 7.25 \\[2ex]
Run-8  & 2007/08 & d--Au & 200 & 437 \\[2ex]
Run-10 & 2009/10 & Au--Au & 200 & 10.3 \\
       &         & Au--Au & 62.4 & 0.544 \\
       &         & Au--Au & 39 & 0.206 \\
       &         & Au--Au & 11.5 & $7.8 \cdot 10^{-3}$ \\
       &         & Au--Au & 7.7 & $4.23 \cdot 10^{-3}$ \\[1ex]
Run-11 & 2010/11 & Au--Au & 200 & 9.79 \\
       &         & Au--Au & 27 & 0.063 \\
       &         & Au--Au & 19.6 & 0.033\\[1ex]
Run-12 & 2011/12 & U--U   & 193 & 0.736 \\
       &         & Cu--Au & 200 & 27.0 \\[1ex]
Run-14 & 2013/14 & Au--Au & 200 & 43.9 \\
       &         & Au--Au & 14.5 & 0.044 \\[1ex]
       &         & $^3$He--Au & 204 & 134 \\[1ex]
Run-15 & 2014/15 & $p$--Au & 200 & 1270 \\
       &         & $p$--Al & 200 & 3970 \\
\hline
\end{tabular}
\end{small}
\caption{Summary of heavy-ion data taking periods at RHIC: collision systems, 
$CM$ energies and total delivered integrated luminosity 
(from http://www.rhichome.bnl.gov/RHIC/Runs/).}
\label{tab:TableRHICruns}
\end{table}

The LHC accelerator started operation at the end of year 2009 with 
$pp$ collisions at $\sqrt{s}=0.9~\TeV$.
Subsequently, protons were collided at $CM$ energies of 7~TeV 
(in years 2010-2011), 8~TeV (in 2012), and 13~TeV (in 2015).
Heavy-ion (AA) and proton-nucleus ($p$A) collisions are an integral part of 
the LHC physics program. 
The first two samples of Pb--Pb collisions were collected in years 2010 and 2011
at a $CM$ energy per nucleon-nucleon collision of $\sqrtsNN=2.76~\TeV$.
About $10~\mub^{-1}$ of integrated luminosity were delivered in 2010 
and $166~\mub^{-1}$ in 2011 to the
three LHC experiments that took part in the heavy-ion program, namely 
ALICE (a detector dedicated to heavy-ion collisions), ATLAS and CMS.
The next LHC run with Pb beams was conducted at the end of 2015 at 
$\sqrtsNN=5.02~\TeV$.
In year 2013, $p$Pb collisions were performed at the LHC at 
$\sqrtsNN=5.02~\TeV$. All four main LHC experiments (ALICE, ATLAS, CMS 
and LHCb) took data.  The delivered integrated luminosity was of about 
$30~\nb^{-1}$.

\begin{figure}
\begin{center}
$\vcenter{\hbox{\includegraphics[width=0.47\textwidth]{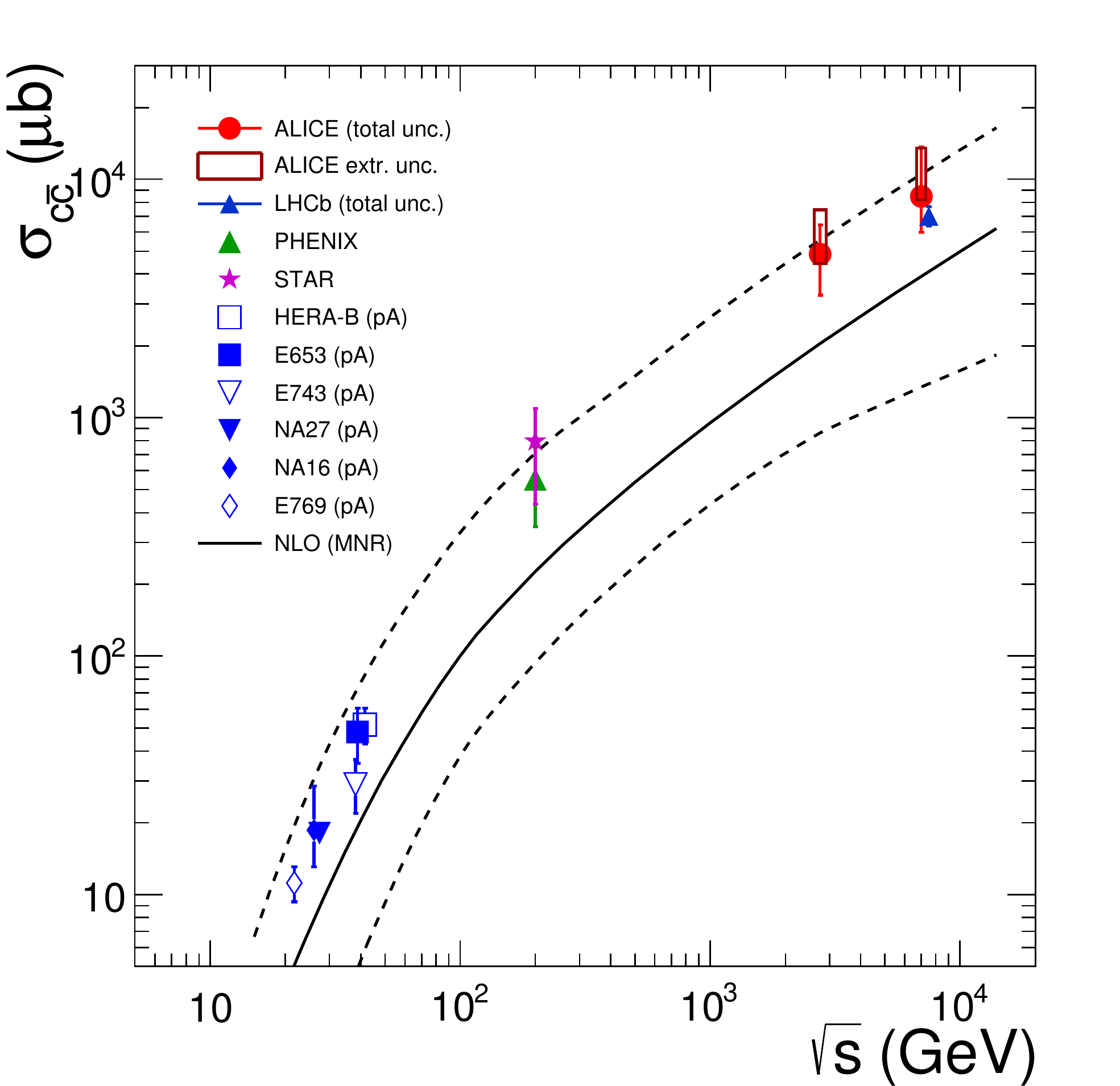}}}$
$\vcenter{\hbox{\includegraphics[width=0.49\textwidth]{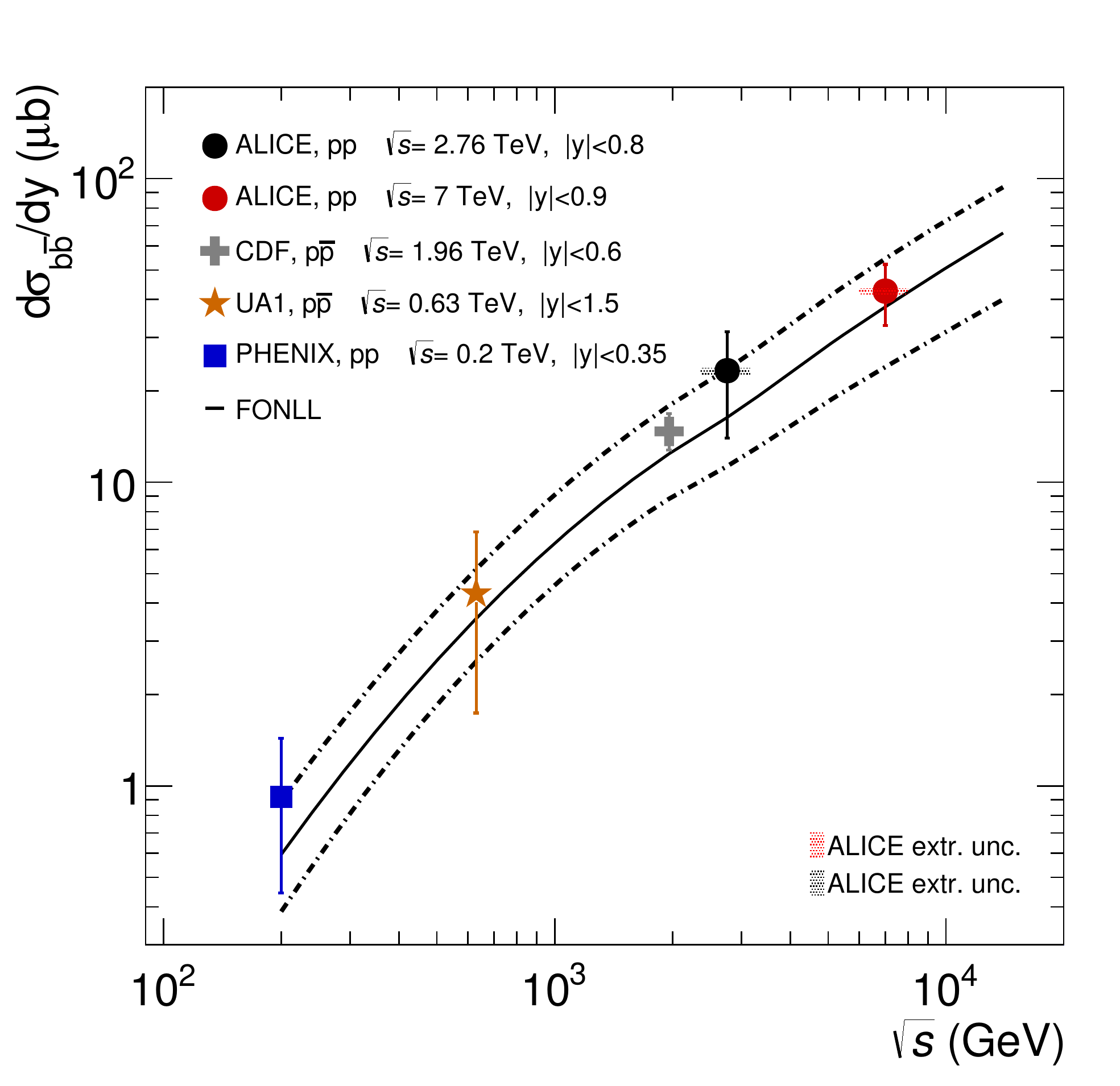}}}$
\caption{
Left: total inclusive charm production cross section as a function of 
$\sqrt{s}$~\cite{Lourenco:2006vw,Adare:2010de,Adamczyk:2012af,Abelev:2012vra,Aaij:2013mga}.
Data from $p$A and deuteron--nucleus ($d$A) collisions 
were scaled assuming no nuclear effects (taken from~\cite{Andronic:2015wma}).
Right: inclusive beauty production cross section per rapidity unit measured at mid-rapidity 
as a function of $\sqrt{s}$ in $pp$ and $p\bar{p}$ 
collisions~\cite{Albajar:1990zu,Acosta:2004yw,Adare:2009ic,Abelev:2014hla,Abelev:2012sca} (taken from~\cite{Abelev:2014hla}).
Results from perturbative QCD calculations and their uncertainties are shown as solid and dashed lines.
}
\label{fig:CScb}
\end{center}
\end{figure}

The increase of the $CM$ energy from fixed-target experiments to 
the colliders RHIC and LHC is reflected in a larger
number of produced particles, a higher temperature of the created 
medium and a longer lifetime of the possible QGP phase.
For example, the energy density of the medium estimated 
with the Bjorken formula for central Au--Au collisions
at a nominal formation time of $1~\fm/c$ 
increases from about $2~\GeV/\fm^{3}$ at $\sqrtsNN=19~\GeV$ 
to about $5~\GeV/\fm^{3}$ at $\sqrtsNN=200~\GeV$~\cite{Adler:2004zn} 
and reaches about $14~\GeV/\fm^{3}$ in central Pb--Pb collisions
at the LHC~\cite{Chatrchyan:2012mb}.
The decoupling time for hadrons at midrapidity increases by about 40\%
from top RHIC to top LHC energy, signaling that at higher
collision energies the system lives longer and expands to a larger 
size at freeze-out~\cite{Aamodt:2011mr}.
Concerning HF hadron production,
the cross section for $c\bar{c}$
and $b\bar{b}$ production in nucleon-nucleon 
collisions increases dramatically with increasing $\sqrtsNN$.
A compilation of the measurements of $\pT$-integrated charm and
beauty production cross section in $pp$ (${p\bar{p}}$) and $p$A
collisions as a function of $\sqrtsNN$ is reported in Fig.~\ref{fig:CScb} 
(taken from~\cite{Andronic:2015wma,Abelev:2014hla}).
As will be discussed in the next section, the experimental 
results are described within uncertainties by calculations within 
perturbative QCD.
The cross section for charm (beauty) production increases
by about a factor of 10 (50) from RHIC to LHC energies.


\subsubsection{Open Heavy-Flavor Production Measurement Techniques}

\hspace{2cm}

Open HF production is measured experimentally in proton--proton,
proton--nucleus and nucleus--nucleus collisions exploiting various 
techniques.
Charm and beauty hadrons can be fully reconstructed from their hadronic
decays, or, alternatively, they can be studied inclusively by measuring 
electrons or muons from their semi-leptonic decays.
Beauty production is also accessed via the inclusive 
$\mathrm{B}\rightarrow \mathrm{J}/\psi+X$ 
decay mode and by reconstructing jets of hadrons produced in the fragmentation
of a beauty quark ($b$-jets).
Experiments featuring vertex detectors with high spatial resolution,
which are available at all four main LHC experiments, as well
as at STAR and PHENIX at RHIC after their recent upgrades,
can exploit the relatively long lifetime of $D$ and $B$ mesons to resolve 
their decay vertex from the interaction point.
$D^{0}$, $D^{+}$ and $D_s^{+}$ mesons decay weakly with mean proper
decay lengths $c\tau$ of approximately 120, 310 and 150~$\mu$m, 
respectively~\cite{Agashe:2014kda}.
Beauty mesons have longer lifetimes than charmed hadrons:
$c\tau \approx 500~\mu$m for $B^{0}$, $B^{+}$ and $B_s^{0}$~\cite{Agashe:2014kda},
which makes it possible to disentangle the charm and beauty contributions in 
the sample of HF decay leptons.
Most of the beauty-hadron decay channels proceed via $b\rightarrow c$ hadron
cascades, giving rise to a topology that contains both a secondary and a
tertiary decay vertex.
In the remainder of this section the key points of the different experimental 
approaches to open HF reconstruction are briefly summarized.

\emph{Inclusive measurements of heavy-flavor decay leptons}.
Open HF production is studied inclusively by measuring electrons 
and muons from semi-leptonic decays of charm and beauty hadrons, which are
characterized by branching ratios (BRs) of the order of 5--15\% depending 
on the hadron species~\cite{Agashe:2014kda}. 
The crucial aspects are the lepton (electron, muon) identification and the 
subtraction of the background due to
leptons not coming from HF hadron decays.\\
\emph{Electrons}: For HF decay electrons, the various experimental
collaborations have exploited different combinations of identification 
techniques, which are effective in different momentum ranges.
The STAR~\cite{Abelev:2006db} and ALICE~\cite{Abelev:2012xe} collaborations 
used specific energy loss (d$E$/d$x$) measurements in combination with  
time-of-flight information at low momenta and with energy and shower shape
from electromagnetic calorimeters at high momenta.
In addition, in ALICE the signal from the Transition Radiation Detector was 
used in $pp$ collisions~\cite{Abelev:2012xe}.
In PHENIX, electron identification was performed using a Ring Imaging Cherenkov 
detector together with the information from an electromagnetic 
calorimeter~\cite{Adare:2010de}.
ATLAS performed a measurement of HF decay electrons in $pp$ collisions 
utilizing clusters in electromagnetic calorimeters matched to charged 
tracks~\cite{Aad:2011rr}.
The background electrons are dominated by photon conversions in the detector 
material and by Dalitz decays of $\pi^0$ and $\eta$ mesons.
Their contribution can be estimated using different techniques and 
subtracted statistically.
Three main approaches to background estimation have been used: 
i) low-invariant-mass $e^+e^-$ pairs (invariant-mass method, utilized by the 
STAR~\cite{Agakishiev:2011mr} and ALICE~\cite{Abelev:2014hla} collaborations); 
ii) Monte Carlo simulation of background electrons from hadron decays (cocktail 
method, used in PHENIX~\cite{Adare:2010de} and ALICE~\cite{Abelev:2012xe}); 
iii) special data taking with additional converter material 
of well defined thickness (converter method exploited in 
PHENIX~\cite{Adare:2010de}).\\
\emph{Muons}: Muons are tracked and identified using spectrometers located 
downstream of absorbers that stop the majority of the hadrons. 
The main background contribution is due to the decay in flight of pions and 
kaons and to hadrons punching through the absorber. 
It can be estimated via cocktail calculations, as done in 
PHENIX~\cite{Adare:2012px} and ALICE~\cite{Abelev:2012qh}.
In ATLAS, the background is discriminated statistically based on a Monte Carlo 
template fit to the distribution of the difference between the track momentum 
measured in the muon spectrometer and that measured in the inner tracker, 
after accounting for energy loss in the calorimeters located in between the 
inner tracker and the muon spectrometer~\cite{Aad:2011rr}.\\
Cocktail subtraction methods use as input the yields and momentum/rapidity 
distributions of all relevant sources of
background leptons, which are mainly pions, kaons and $\eta$ mesons.
Ideally, to reduce the uncertainties, this input should be taken from data 
measured with the same apparatus.

\emph{Dilepton invariant-mass analysis}.
Heavy-flavor production can be studied  via the contribution of simultaneous
semi-leptonic decays of ${D\overline{D}}$ and ${B\overline{B}}$
pairs in the dilepton invariant-mass distribution.
In particular, the open HF contribution is extracted from the
intermediate-mass region, \ie, the region between the $\phi$
and the J/$\psi$ peaks, where the main contributions to the dilepton spectrum
are expected to be open HF decays, thermal radiation from the QGP
and (at SPS energies) Drell-Yan dileptons.
This technique was used in Pb--Pb collisions at the SPS with the 
NA50~\cite{Abreu:2000nj} and NA60~\cite{Arnaldi:2008er} experiments
to measure open-charm and thermal-dimuon production in Pb--Pb and
In--In collisions.
At RHIC, the PHENIX collaboration measured the di-electron production
double differentially in mass and $\pT$ which allowed the separation of 
regions dominated by charm from those dominated by beauty decays and the 
extraction of the ${c\bar{c}}$ and ${b\bar{b}}$ production cross sections 
in $pp$ and $d$Au collisions~\cite{Adare:2008ac,Adare:2014iwg}.

\emph{Beauty measurements from partially reconstructed decays}\\
\emph{Beauty-decay leptons}: the beauty contribution to the sample of 
HF decay leptons can be extracted by exploiting additional 
information. 
In particular, charm and beauty decay leptons can be separated based on
the longer lifetime of beauty hadrons, which results in a larger separation
of the production point of their decay leptons from the interaction 
vertex~\cite{Abelev:2012sca}.
This approach requires high resolution (of the order or better than 100 $\mu$m) 
on the track impact parameter, which is the distance of closest approach of a 
track (\ie, the reconstructed particle trajectory) to the interaction vertex.
Alternatively, the fraction of beauty-decay electrons has
been estimated from  the azimuthal correlations between HF decay 
electrons and charged 
hadrons~\cite{Adare:2009ic,Aggarwal:2010xp,Abelev:2014hla}. 
This method exploits the fact that the width of the near-side peak is larger
for beauty than for charm hadron decays, thus allowing one to disentangle
the corresponding relative contribution to the yield of
HF decay electrons. The main limitation of the beauty
measurement via single electrons (muons) is that the correlation 
between the measured momentum of the lepton and that of the parent
$B$ meson is very broad, especially at low momentum.\\
\emph{Non-prompt J/$\psi$}: the contribution of beauty-hadron decays
to the yield of $J/\psi$ mesons can be measured inclusively by
decomposing the $J/\psi$ yield into its prompt (\ie, $J/\psi$ produced
at the interaction point) and non-prompt (displaced) components.
This is achieved via fits to the distribution of the measured distances
between the interaction vertex and the $J/\psi$ decay vertex.
This approach has been used in $pp$, $p$Pb and Pb--Pb collisions at the LHC by 
ALICE~\cite{Abelev:2012gx,Adam:2015rba}, CMS~\cite{Khachatryan:2010yr} 
and LHCb~\cite{Aaij:2011jh,Aaij:2013zxa}.
The branching ratio of inclusive beauty-hadron decays into $J/\psi$ 
measured at LEP is 1.16\%~\cite{Buskulic:1992wp,Adriani:1993ta,Abreu:1994rk}, 
lower than that into single leptons.
However, as compared to the measurement of beauty-decay leptons, this 
channel provides a more direct measurement of the beauty-hadron kinematics 
due to the narrower correlation between the momentum of the $J/\psi$ and 
that of the parent beauty hadron.

\emph{Fully reconstructed decays of charm and beauty hadrons}\\
This approach allows the reconstruction of HF hadrons exploiting 
hadronic decays of $D$ mesons (and in the future of $\Lambda_{c}$ baryons), 
and $J/\psi$+hadron (with the $J/\psi$
reconstructed from dileptonic decays) decays of $B$ mesons.
The branching ratios are smaller than those in semi-leptonic channels,
but this technique has the advantage of providing access to the 
kinematics of the heavy-flavor hadron.\\
\emph{Charm mesons}: $D$-meson production has been measured down to low
transverse momentum by STAR~\cite{Adamczyk:2012af} and 
ALICE~\cite{ALICE:2011aa,Abelev:2012tca} in Au--Au and Pb--Pb 
collisions at RHIC and LHC, respectively.
The measurement is based on the invariant-mass analysis of fully reconstructed 
decay topologies of the following hadronic decay channels: 
${D}^0\rightarrow {K}^-\pi^+$ (BR=3.88\%), 
${D}^+\rightarrow {K}^-\pi^+\pi^+$ (BR=9.13\%), 
${D}^{*+}\rightarrow {D}^0\pi^+$ (BR=67.7\%), and
${D_s}^+\rightarrow \phi\pi^+\rightarrow {K}^-{K}^+\pi^+$ (BR=2.24\%) 
and
their charge conjugates~\cite{Agashe:2014kda}.
The huge combinatorial background is reduced by selections on the
d$E$/d$x$ and time-of-flight that provide kaon and pion identification.
In addition, the spatial resolution of the ALICE silicon tracker makes it
possible to reconstruct the $D$-meson decay vertex and to apply
geometrical selections on its separation from the interaction 
vertex~\cite{ALICE:2011aa}.
A technique based on geometrical selections on displaced
decay-vertex topologies was used also for the preliminary results
on ${D}^0$-meson production in Pb--Pb collisions recently reported
by CMS~\cite{CMS:2015hca}.\\
\emph{Beauty mesons}: CMS Collaboration recently published
results of the production of $B$ mesons in $p$Pb collisions at the 
LHC in the transverse momentum range 
$10<\pT<60~\GeV/c$~\cite{Khachatryan:2015uja}.
The following decays were reconstructed: 
${B}^+ \rightarrow {J}/\psi+{K^+}$, 
${B}^0 \rightarrow {J}/\psi+{K^{*0}(892)}$ and 
${B_s}^+\rightarrow {J}/\psi+\phi$, all having branching ratios
of about 0.1\%~\cite{Agashe:2014kda}.
Kinematic selections on the displaced decay vertex topologies were applied
to reduce the combinatorial background. \\
\emph{b-jets}: Reconstructed jets associated with beauty hadrons 
(``$b$-jets'') can be identified based on 
kinematic variables related to the relatively long lifetime and large mass 
of beauty hadrons (``$b$-tagging'').
The first measurement of $b$-jets in heavy-ion collisions was performed by 
the CMS collaboration at the LHC~\cite{Chatrchyan:2013exa}.
Measurements of $b$-jets are complementary to those of beauty hadrons,
because they are typically performed in a different momentum range and
because the reconstructed jet energy is closely related to that of the $b$ 
quark.

\subsubsection{Experimental observables}
\label{sec:expobs}
\hspace{2cm}

Interactions of heavy quarks with the constituents of the hot and dense
medium created in nucleus--nucleus collisions are expected to
modify the momentum and angular distribution of HF hadrons
as compared to $pp$ collisions.
The main experimental observables used up to now to study these effects
are the nuclear modification factor $R_{\rm AA}$ and the elliptic flow $v_2$,
which is the coefficient of the second harmonic in the Fourier expansion
of the particle azimuthal distributions.
Further insight into the interaction mechanisms of heavy quarks with the 
medium and the properties of the medium can be provided by measurements of angular
correlations involving HF particles and/or their decay products, and by 
measuring the higher harmonics in the Fourier expansion of the particle
azimuthal distributions.
However, with the currently available experimental samples, these 
more differential observables could only be accessed with limited 
statistical precision in nucleus--nucleus collisions.

The nuclear modification factor $R_{\rm AA}$ is commonly used to study the
modification of the production yield and the momentum distribution
of particles originating from hard (\ie, high-$Q^2$) partonic scattering 
processes in nucleus--nucleus collisions relative to $pp$ collisions.
Since the rate of hard processes is expected to scale with the 
average number of binary 
nucleon--nucleon collisions occurring in the nucleus--nucleus collision,
$\left\langle N_{\rm coll} \right\rangle$, the nuclear modification factor of 
the transverse momentum ($\pT$) distributions is expressed as:
\begin{equation}
\label{eq:Raa}
R_{\rm AA}(\pT)=
{1\over {\left \langle N_{\rm coll}\right\rangle}} \cdot {\mathrm{d} N_{\rm AA}/\mathrm{d}\pT \over \mathrm{d} N_{\rm pp}/\mathrm{d}\pT} = 
{1\over {\left \langle T_{\rm AA} \right\rangle}} \cdot 
{\mathrm{d} N_{\rm AA}/\mathrm{d}\pT \over 
\mathrm{d}\sigma_{\rm pp}/\mathrm{d}\pT}\,,
\end{equation}
where $\mathrm{d} N_{\rm AA}/\mathrm{d}\pT$ is the $\pT$ spectrum measured in 
A--A collisions, $\mathrm{d} N_{\rm pp}/\mathrm{d}\pT$ 
($\mathrm{d}\sigma_{\rm pp}/\mathrm{d}\pT$) is the $\pT$-differential yield 
(cross section) in $pp$ collisions and $\left\langle T_{\rm AA} \right\rangle$ 
is the average nuclear overlap function.
The nuclear overlap function is defined as the convolution of the nuclear 
density profiles 
of the colliding ions in the Glauber model~\cite{glauber,Miller:2007ri} and 
is related to the number of nucleon--nucleon collisions as 
$N_{\rm coll} = \sigma_{\rm NN}^{\rm inel} \cdot T_{\rm AA}$, where 
$\sigma_{\rm NN}^{\rm inel}$
is the inelastic nucleon--nucleon cross section at given $\sqrt{s}$.
In absence of nuclear effects, a value of $R_{\rm AA}=1$ is expected
for HF particles in the whole $\pT$ range, because,
due to their large mass, charm and beauty quarks are produced predominantly
in hard partonic scattering processes.
Thermal production in the QGP, which does not scale with $N_{\rm coll}$,
is expected to be negligible at the medium temperatures reached in heavy-ion 
collisions at RHIC and at the LHC~\cite{BraunMunzinger:2000dv,Zhang:2007dm}.
Parton in-medium energy loss causes a suppression, $\Raa<1$, of hadrons at 
high transverse momenta ($\pT \gtsim 5~\GeV/c$), and is expected to be the 
dominant effect in the nuclear modification factor. The $\Raa$ can therefore 
be used to characterize HQ in-medium energy loss and to infer from it
the corresponding transport coefficients.
At low and intermediate $\pT$ ($\pT \ltsim 5~\GeV/c$), effects other than 
in-medium energy loss are expected to modify the HF hadron production yield, 
leading to values of $\Raa$ different from unity.
These effects are related both to the formation of a hot and dense medium and 
to nuclear effects in the initial state of the collision.
Among the initial-state effects, nuclear modification of the parton 
distribution functions, $k_{\rm T}$-broadening and cold-nuclear-matter energy 
loss due to multiple scatterings of the initial partons could modify the 
production yield and $\pT$ distribution of HF hadrons, as discussed 
in Section~\ref{sssec_pAdata}.
Final-state effects which could modify the hadron spectra in addition to
energy loss are the collective ``radial'' flow of the medium and the 
modification of hadronization in the presence of a medium.
Low-momentum heavy quarks, including those shifted to low momentum  
by parton energy loss, could participate in the collective expansion of the 
system and possibly reach thermalization with the medium as a consequence of 
multiple interactions with the medium 
constituents~\cite{Batsouli:2002qf,Greco:2003vf}. 
In elementary collisions heavy quarks are expected to hadronize mainly 
through fragmentation, while the ample presence of quarks and antiquarks 
in the medium makes the coalescence of charm and beauty quarks with light 
quarks a plausible hadronization mechanism (as discussed in 
Sec.~\ref{sssec_hadro}).  This will impart (part of) the 
large radial and elliptic flow of the quarks from the medium on the heavy quarks 
and introduce a $\pT$-dependent modification to the observed charmed hadron 
spectrum compared to a pure fragmentation scenario (even in the limit where 
heavy quarks do not take part in the collective expansion of the 
medium)~\cite{Greco:2003vf,He:2011qa,Gossiaux:2010yx,Beraudo:2014boa}.
Furthermore, hadronization via coalescence may lead to a modification of the
relative abundance of the different HF hadron species.
In particular, a baryon-to-meson and strange-to-non-strange enhancement 
for charmed hadrons, similar to that observed for light-flavor hadrons, is 
predicted~\cite{Andronic:2003zv,Kuznetsova:2006bh,Sorensen:2005sm,MartinezGarcia:2007hf,Oh:2009zj,He:2012df}.

An observable complementary to the nuclear modification factor $\Raa$
to study the interaction of heavy quarks with the medium is
the anisotropy in the azimuthal distribution of HF hadrons.
Anisotropy in particle momentum distributions originates from the initial 
anisotropy in the spatial distribution of the nucleons 
participating in the collision.
The anisotropy of the azimuthal distribution of the particles
produced in the collision is usually characterized with the Fourier 
coefficients: 
\begin{equation}
\label{eq:vn}
v_n=\langle\cos[n(\varphi-\Psi_n)]\rangle,
\end{equation}
where $\varphi$ is the azimuthal angle of 
the particle, and $\Psi_n$ is the azimuthal angle of the initial state 
symmetry plane for the $n$-th harmonic.
If nuclei were spherically symmetric with a matter density depending only on 
the distance from the centre of the nucleus, the symmetry planes $\Psi_n$ for 
all harmonics would coincide with the reaction plane, \ie, the plane defined 
by the beam direction and the impact parameter of the two colliding nuclei.
For non-central collisions the overlap region of the colliding nuclei 
has an ``almond" shape, and the largest contribution to the
anisotropy is given by the second coefficient $v_2$, which is called 
elliptic flow.
At RHIC and LHC energies, a positive $v_2$ is observed for many different 
hadron species~\cite{Adams:2004bi,Adler:2003kt,Abelev:2014pua} and is 
commonly ascribed to the combination of two mechanisms. 
The first one, which is dominant at low and intermediate momenta
($\pT<6~\GeV/c$), is the build-up of a collective expansion 
through interactions among the medium constituents, which convert the 
initial-state geometrical anisotropy into a momentum anisotropy of the 
final-state particles~\cite{Ollitrault:1992bk}.
The second mechanism is the path length dependence of in-medium parton 
energy loss, which is predicted to give rise to a positive $v_2$ for 
hadrons up to large $\pT$~\cite{Gyulassy:2000gk,Shuryak:2001me}.
This path length dependence is expected to be different for the different 
energy-loss mechanisms: linear for collisional 
processes~\cite{Thoma:1990fm,Braaten:1991we} 
and close to quadratic for radiative processes~\cite{Baier:1996sk}.
Hence, the measurement of $v_2$ at high $\pT$ can constrain the path length 
dependence of parton energy loss.
At low $\pT$, HF hadron $v_2$ offers a unique opportunity to test 
whether also heavy quarks take part in the 
collective expansion and possibly thermalize in the medium.
In addition, at low and intermediate $\pT$, the $D$ and $B$ meson elliptic flow 
(as well as that of their decay leptons) is expected to be sensitive to the 
role of hadronization via recombination, which is predicted to augment 
the $v_2$ of HF hadrons with respect to that of charm and beauty
quarks~\cite{Molnar:2003ff,Greco:2003vf}.

The measurement of higher-order flow coefficients of HF hadrons
was proposed in Ref.~\cite{Nahrgang:2014vza} as a sensitive probe of the degree 
of charm- and beauty-quark thermalization. 
In particular, in this model, the triangular-flow coefficient $v_3$ is 
likely to show an incomplete coupling of the heavy quarks to the bulk
medium as well as the expected mass hierarchy.

The study of angular correlations between HF  hadrons could further 
constrain the models of interactions of heavy quarks with the medium, 
possibly providing valuable information on the path length dependence of 
energy loss and on the relative contributions of collisional and radiative 
processes~\cite{Nahrgang:2013saa,Uphoff:2013rka,Beraudo:2014boa,Cao:2014pka,Renk:2013xaa}.
The medium effects on the angular (de)correlation and momentum imbalance of 
heavy quark-antiquark pairs can be studied via the correlations of two 
HF hadrons, ${D\overline{D}}$ and ${B\overline{B}}$.
However, this measurement is extremely challenging because of the huge data 
samples required to cope with 
the small branching ratios of the hadronic channels through 
which $D$ and $B$ mesons can be fully reconstructed.
In the beauty sector, angular correlations and energy imbalance of $b$-jets, 
the latter quantified, \eg, by measuring the asymmetry 
$A_{\rm J}=(E_{T1}-E_{T2})/(E_{T1}+E_{T2})$ of di-jets emitted
in opposite hemispheres~\cite{Aad:2010bu,Chatrchyan:2011sx},
could become accessible in Pb--Pb collisions with the integrated luminosities 
that will be recorded during the LHC Run-2, \ie, the 2015--2018 data 
taking period.
As an alternative, correlations involving electrons and muons from HF
hadron decays can be studied, such as ${D-e}$, ${e^+-e^-}$, 
${\mu^+-\mu^-}$, ${e-\mu}$.
The interpretation of the results from these observables is, however, 
less straightforward than that of ${D\overline{D}}$ or $b$-jet
correlations, because they contain also angular decorrelation effects due 
to the HF hadron decay kinematics, and because the lepton carries
only a fraction of the parent-hadron momentum.
With the currently available data samples, a measurement of ${e-\mu}$
correlations could be carried out in $d$Au collisions~\cite{Adare:2013xlp}.
In addition, two-particle correlations involving heavy flavors were measured 
utilizing a $D$ meson or a HF decay lepton as trigger particle 
and the hadrons produced in the collision as associated particles.
Such observables have also sensitivity to the HQ production and fragmentation
process, as well as to the presence of collective effects that induce 
correlations between the HF particles and the bulk of hadrons 
produced in the collision.
Measurements of $e$--hadron correlations in Au--Au collisions at RHIC were 
published by PHENIX~\cite{Adare:2010ud}, but with the 
current level of statistical uncertainties no quantitative conclusion can be
drawn on the modification of the azimuthal-correlation shape as compared to
$pp$ collisions.
Preliminary results for $e$--hadron correlations were also reported by STAR 
for Au--Au collisions at RHIC~\cite{Wang:2008kha} and by ALICE 
for Pb--Pb collisions at the LHC~\cite{Thomas:2014cwa}.
The larger data samples of Au--Au and Pb--Pb collisions that will be collected 
in the coming years at RHIC and at the LHC will make it possible to fully
exploit the potential of $e$--hadron and $D$--hadron correlations, 
to provide further constraints for the characterization of HQ 
interactions with the medium.

\subsection{Open Heavy-Flavor Data in pp and p-A (d-A) Collisions}
\label{ssec_element}

\subsubsection{Production in pp and Comparison to pQCD Calculations}
\label{sssec_ppdata}

\hspace{2cm}

In the context of the study of HQ production in heavy-ion
collisions, measurements in $pp$ reactions provide a crucial 
reference to establish a baseline that allows the identification of
modifications in AA reactions, which can be related to the 
formation of a hot and dense medium.
In particular, a $pp$ reference is needed to quantify the modification
of the momentum distribution of HF particles in AA collisions
through the nuclear modification factor $\Raa$.
Also in the studies of two-particle angular correlations, comparisons 
to $pp$ results are mandatory to interpret the measurements in heavy-ion 
collisions.

A key aspect of open HF studies is that their production
in elementary hadronic collisions can be calculated in the framework of 
perturbative QCD (pQCD) down to low $\pT$ with the HQ mass, $m_{Q}$, 
acting as a long distance cutoff~\cite{Combridge:1978kx,Mangano:1991jk}.
The differential cross section for the inclusive production of HF 
hadrons in nucleon--nucleon collisions can be computed using the 
collinear factorization approach as a convolution of three factors:
(i) the parton distribution functions of the incoming nucleons, 
(ii) the partonic hard scattering cross section calculated as a 
perturbative series in the strong coupling constant, and
(iii) the fragmentation process describing the non-perturbative 
transition of a charm (beauty) quark into a heavy hadron.
The latter is modeled by a fragmentation function, which parametrizes the 
fraction of quark energy transferred to the produced hadron, as discussed
in Sec.~\ref{sssec_hadro}. This formalism is implemented at next-to-leading 
order (NLO) accuracy in the general-mass variable-flavor-number scheme, 
GM-VFNS~\cite{Kniehl:2004fy,Kniehl:2005mk} and in the fixed order 
next-to-leading-log (FONLL) resummation approach~\cite{Cacciari:1998it}.
Both of these frameworks are based on the matching of ``massive'' NLO 
calculations in the Fixed-Flavor Number Scheme (FFNS, valid at low $\pT$), 
where heavy quarks are not active partons in the nucleon appearing only in 
the partonic hard scattering process, with ``massless'' calculations in 
the Zero-Mass Variable Flavor Number Scheme (ZM-VFNS, valid for 
$\pT \gg m_{Q}$), where charm and beauty quarks appear also as 
active partons in the nucleon PDFs and their mass can be 
neglected in the partonic hard scattering cross section.

Calculations of inclusive production cross section of HF hadrons
in hadronic collisions are also performed in the leading order 
(LO) approximation within the framework of $k_{\rm T}$-factorization with
unintegrated gluon distribution functions (UGDFs) to account for the 
transverse momenta of the initial 
partons~\cite{Luszczak:2008je,Maciula:2013wg,Catani:1990eg}.

These theoretical frameworks allow for a mostly analytic calculation
of inclusive $\pT$ and $y$ differential cross section of heavy quarks,
HF hadrons and their decay leptons at NLO+NLL (or LO) accuracy.
However, for some particular studies, such as those of HF jets or  
correlations of HF hadrons with other particles produced in the interaction, 
a more complete description of the hadronic final state is needed, which 
can be obtained using Monte Carlo (MC) generators.  General-purpose MC 
generators, such as PYTHIA~\cite{Sjostrand:2006za,Sjostrand:2007gs} or 
HERWIG~\cite{Corcella:2000bw}, can serve this purpose, but are limited to 
LO accuracy. A more versatile description of hard processes can be achieved 
with the MC@NLO~\cite{Frixione:2003ei} and POWHEG~\cite{Frixione:2007nw} 
generators, where a consistent matching of NLO pQCD calculations with parton 
showers is implemented, thus combining the strength of MC generators
(\ie, a complete modelling of the hadronic final state) with NLO accuracy 
in the hard scattering process and leading-logarithm
resummation in the soft/collinear regimes.

These pQCD-based approaches are used in the models of HQ production 
and in-medium energy loss to compute the initial HQ distributions in 
heavy-ion collisions.
In particular, HQ distributions calculated with FONLL are used in 
WHDG~\cite{Wicks:2007am,Wicks:2005gt}, MC@sHQ+EPOS~\cite{Nahrgang:2013saa}, 
TAMU~\cite{He:2014cla}, Djordjevic {\it et al.}~\cite{Djordjevic:2013xoa},
and CUJET 3.0~\cite{Xu:2014tda,Xu:2015bbz}.
Other models make use of MC generators, \eg,  MC@NLO in 
BAMPS~\cite{Uphoff:2011ad}, POWHEG in POWLANG~\cite{Alberico:2011zy}, and
PYTHIA (tuned to fit FONLL predictions for heavy quarks) in 
PHSD~\cite{Song:2015sfa,Song:2015ykw}. 
In the model of Cao {\it et al.}, a LO pQCD calculation was originally 
used to compute the initial HQ distributions~\cite{Cao:2013ita}, while in 
recent studies of angular correlations an improved initialization based 
on MC@NLO was adopted~\cite{Cao:2014pka}.

Since the above-mentioned pQCD calculations and MC generators
provide a fundamental input for the studies of HQ interactions in
the QGP, it is crucial to verify that they can describe the data from $pp$ 
collisions at RHIC and LHC energies.
This data-to-theory comparison consistutes also an important test of our 
understanding of QCD in its perturbative regime.
Furthermore, a solid understanding of open-charm production is important also
for cosmic-ray and neutrino astrophysics, because `prompt' neutrino
flux from charm decays is the dominant background for astrophysical neutrinos 
at energies of the order of 1~PeV~\cite{Bhattacharya:2015jpa,Gauld:2015yia}.
In the following, a selection of measurements of $\pT$ and $y$
differential inclusive production cross sections of open HF in 
$pp$ collisions at RHIC and at the LHC will be presented and compared
to the results of the pQCD calculations and MC generators presented above.
Note that HF production was also measured at the Tevatron 
collider~\cite{Acosta:2004yw} at $CM$ energies in between the RHIC and LHC 
ones. They are not reported in this review, which focuses on the accelerators
and energies at which heavy-ion collisions have been studied.

\begin{figure}
\begin{center}
$\vcenter{\hbox{\includegraphics[width=0.5\textwidth]{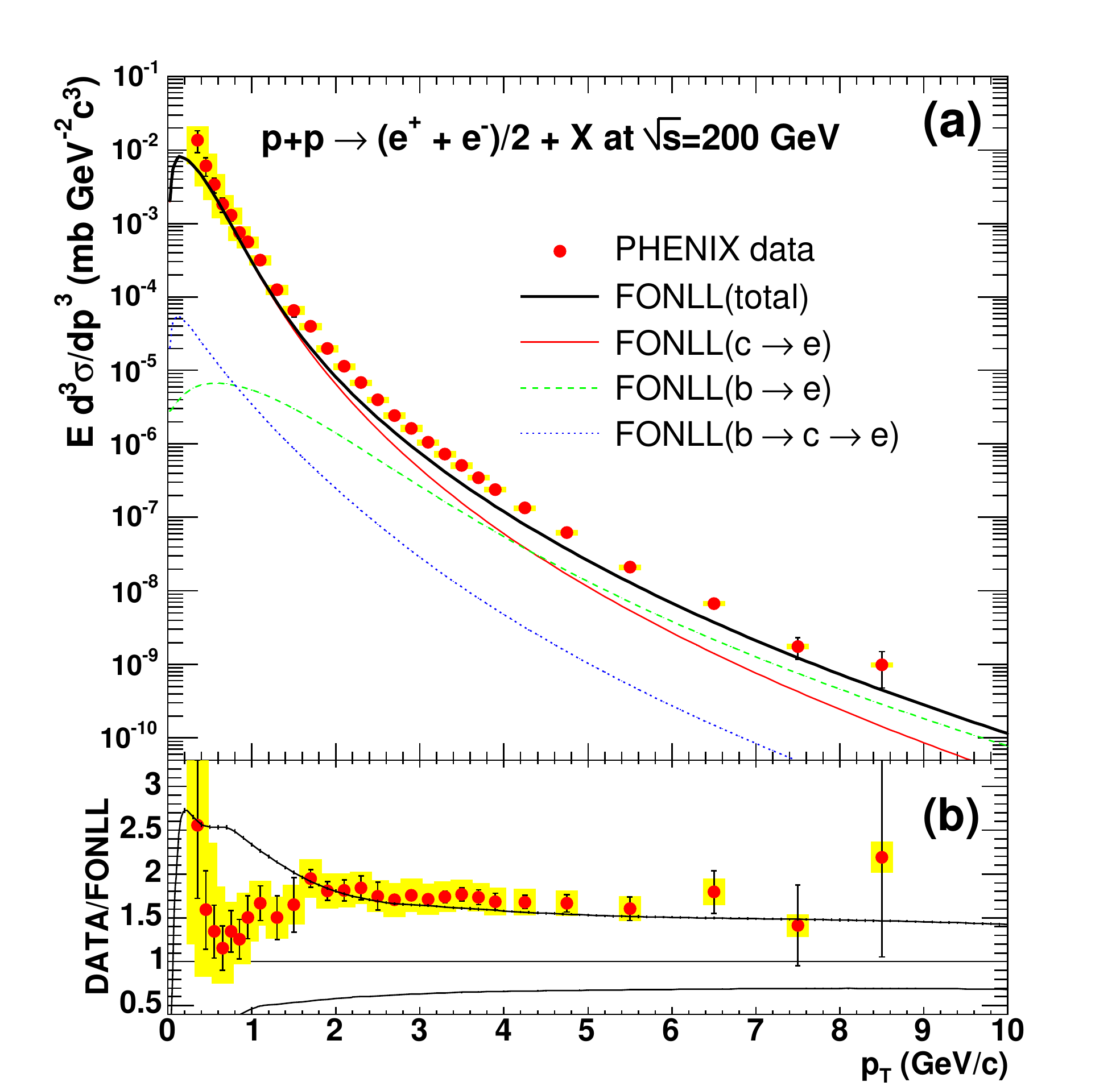}}}$
$\vcenter{\hbox{\includegraphics[width=0.46\textwidth]{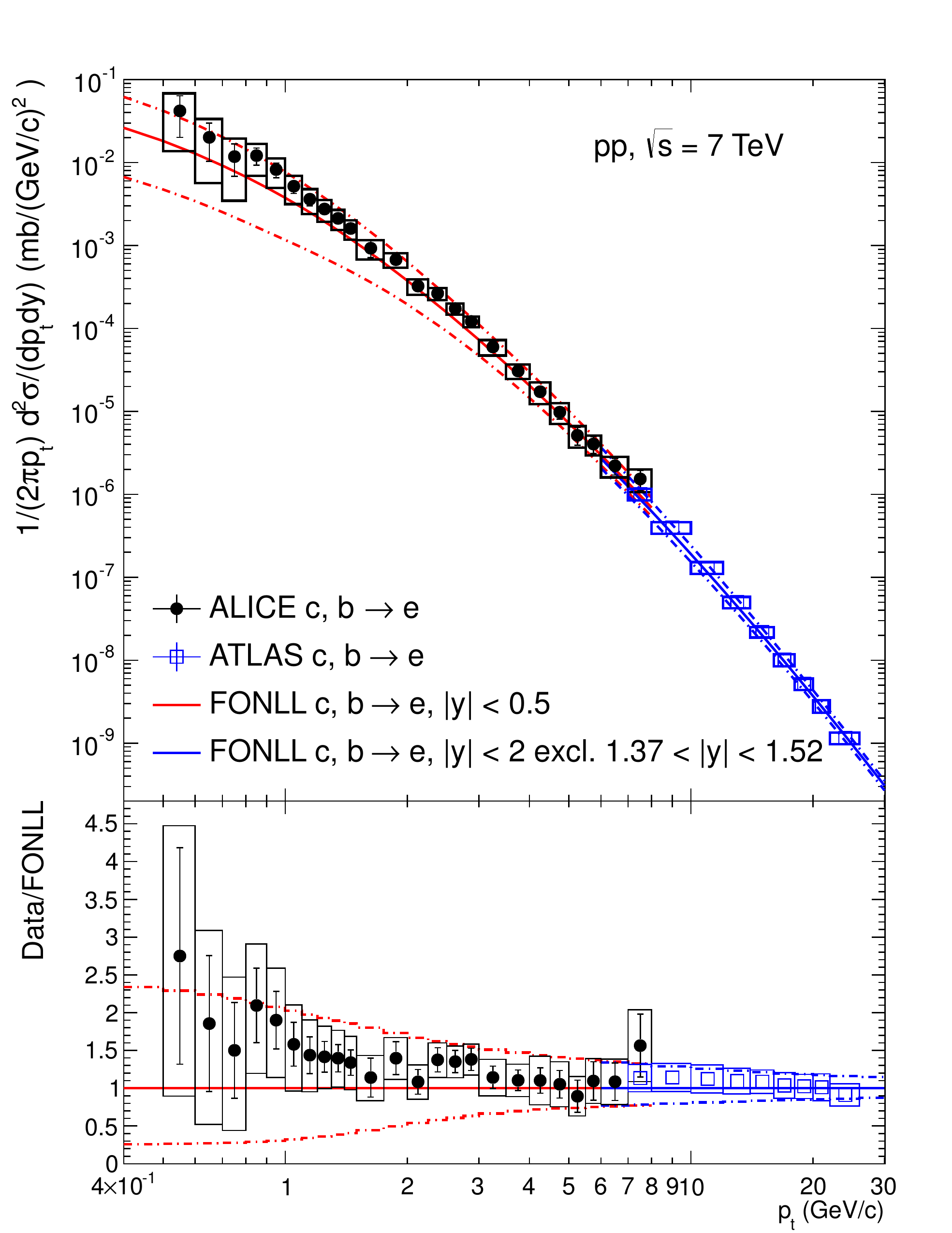}}}$
\caption{Transverse-momentum differential production cross sections of 
HF decay electrons at mid-rapidity. 
Left: results at $\sqrt{s}=200~\mathrm{GeV}$ from PHENIX~\cite{Adare:2006hc}.
Right: results at $\sqrt{s}=7~\TeV$ from ALICE~\cite{Abelev:2012xe} and 
ATLAS~\cite{Aad:2011rr}.
Data are compared to FONLL calculations~\cite{Cacciari:1998it,Cacciari:2012ny}.}
\label{fig:HFEpp}
\end{center}
\end{figure}

In Fig.~\ref{fig:HFEpp} the $\pT$-differential production cross section
of electrons from HF hadron decays at mid-rapidity measured
by the PHENIX collaboration in $pp$ collisions at $\sqrt{s}=200~\GeV$ 
(left panel)~\cite{Adare:2006hc} and by the
ALICE~\cite{Abelev:2012xe} and 
ATLAS~\cite{Aad:2011rr} collaborations at $\sqrt{s}=7~\TeV$ (right panel) is 
shown.
The data are compared to FONLL pQCD calculations~\cite{Cacciari:2012ny}, which 
at both energies can describe the measurements within uncertainties.
The central values of FONLL calculations are obtained using 
CTEQ-6.6 parton distribution functions~\cite{Nadolsky:2008zw}, 
$m_{c}=1.5~\GeV/c^2$ and $m_{b}=4.75~\GeV/c^2$ for the charm and beauty 
quark masses, and QCD
renormalization ($\mu_{\rm R}$) and factorization ($\mu_{\rm F}$) scales 
of $\mu_{\rm R}=\mu_{\rm F}= \sqrt {p^2_{\rm T} + m^2_{q}}$.
The theoretical uncertainty band is defined by varying the
charm and beauty quark masses in the ranges 
$1.3 < m_{c} < 1.7$ GeV/$c^2$ and $4.5 < m_{b} < 5.0$ GeV/$c^2$, 
and the QCD scales in the ranges 
$0.5 <\mu_{\rm R}/\mu_0 < 2$ and $0.5 <\mu_{\rm F}/\mu_0 < 2$ with the 
constraint $0.5 <\mu_{\rm F}/\mu_{\rm R} < 2$.
The outcomes of GM-VFNS and POWHEG, not shown in this figure, agree
with the FONLL predictions within uncertainties (see Ref.~\cite{Klasen:2014dba}
for a systematic comparison of FONLL, GM-VFNS and FONLL at LHC energies).
A similar agreement between HF decay electron measurements and 
pQCD-based calculations is found for $pp$ collisions 
at $\sqrt{s}=2.76~\TeV$~\cite{Abelev:2014gla}.

\begin{figure}
\begin{center}
$\vcenter{\hbox{\includegraphics[width=0.48\textwidth]{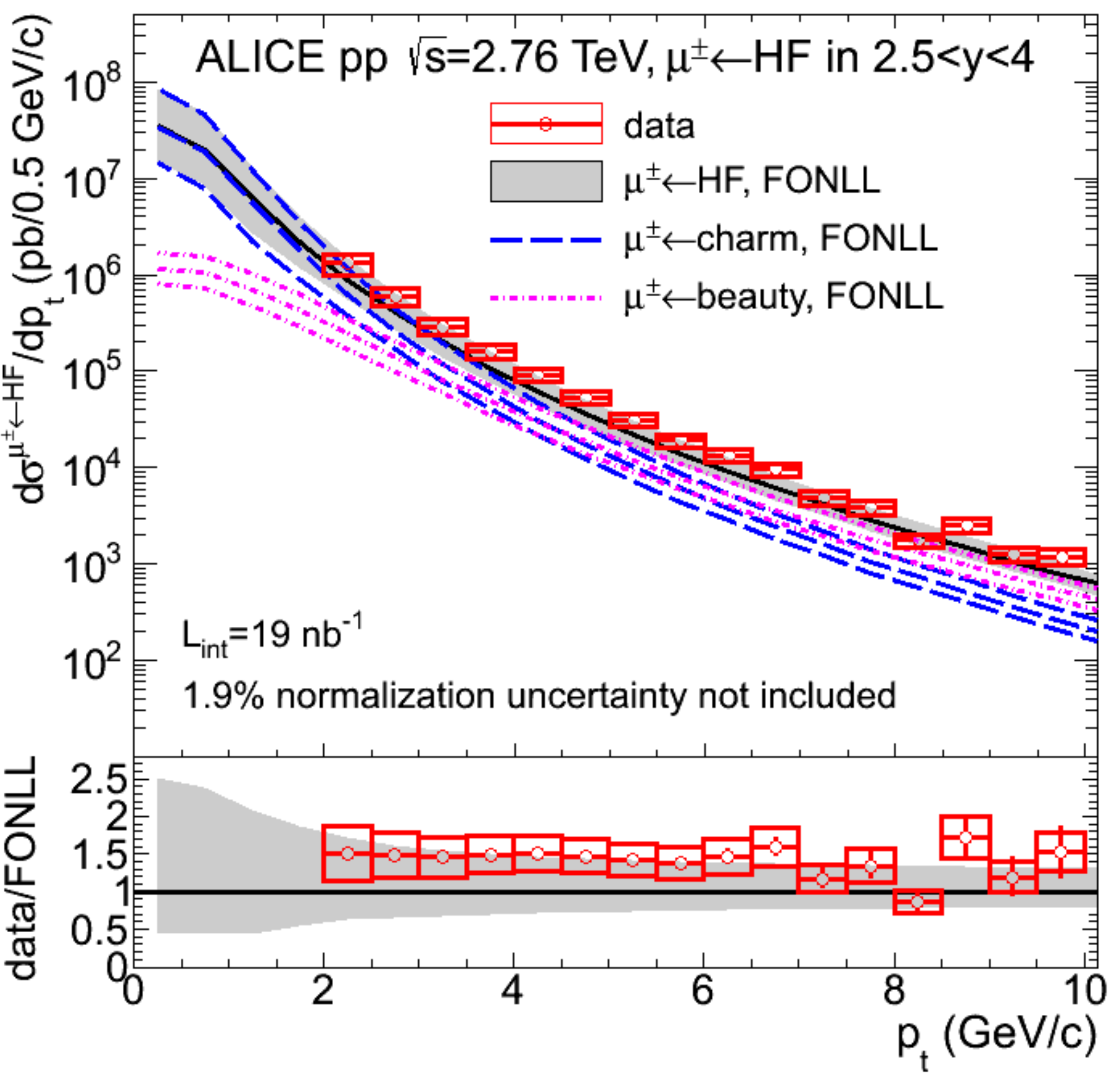}}}$
$\vcenter{\hbox{\includegraphics[width=0.48\textwidth]{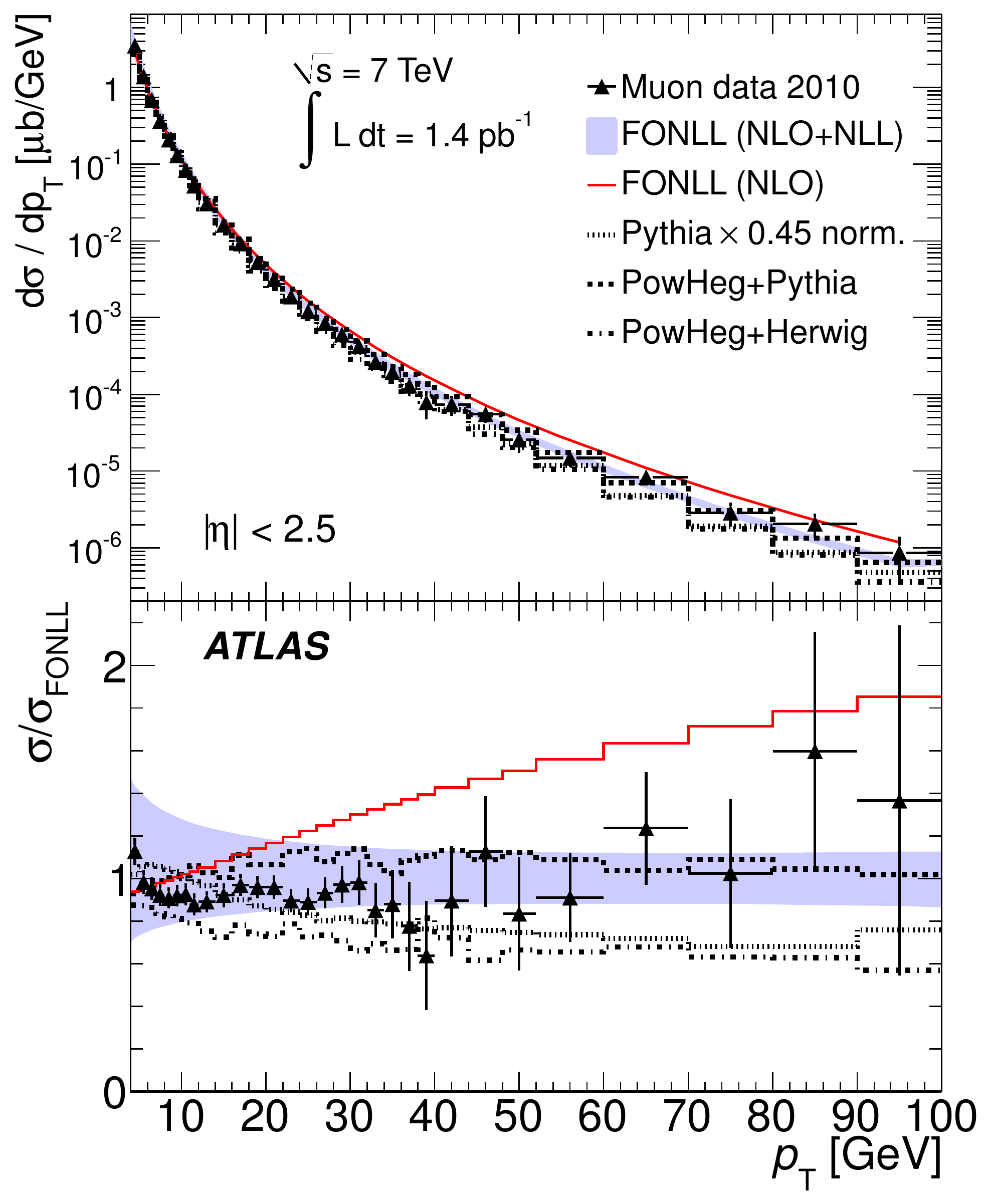}}}$
\caption{Transverse-momentum differential production cross section of 
heavy-flavor decay muons.
Left: results at $\sqrt{s}=2.76~\mathrm{TeV}$ at forward rapidity from
ALICE~\cite{Abelev:2012qh}. 
Right: results at $\sqrt{s}=7~\mathrm{TeV}$ at mid-rapidity from
ATLAS~\cite{Aad:2011rr}.
Data are compared to FONLL~\cite{Cacciari:1998it,Cacciari:2012ny}, 
POWHEG~\cite{Frixione:2007nw} and PYTHIA~\cite{Sjostrand:2006za} predictions.
}
\label{fig:HFMpp}
\end{center}
\end{figure}

Heavy-flavor decay muon production has been measured in $pp$ collisions 
at the LHC by the ATLAS collaboration at mid-rapidity at 
$\sqrt{s}=7~\TeV$~\cite{Aad:2011rr} and by the ALICE collaboration
at forward rapidity ($2.5<y<4$) at $\sqrt{s}=2.76$ and 
7~TeV~\cite{Abelev:2012pi,Abelev:2012qh}.
The measured $\pT$-differential cross sections at forward rapidity at 
$\sqrt{s}=2.76~\TeV$ and at mid-rapidity at $\sqrt{s}=7~\TeV$ are shown in 
the left and right panels of Fig.~\ref{fig:HFMpp}, respectively.
In both rapidity intervals, the predictions from FONLL calculations
are compatible within uncertainties with the measured cross sections.
In the forward-rapidity interval, the results of FONLL calculations are also
shown separately for muons from charm and beauty decays.
The predictions for beauty-decay muons include the contributions of 
muons coming directly from beauty-hadron decays ($B \rightarrow \mu$) and 
of muons from decays of charmed hadrons produced in beauty-hadron decays 
($B \rightarrow D \rightarrow \mu$). 
According to FONLL, the yield of HF decay muons is dominated by
muons from charm decays for $\pT<4-5~\GeV/c$, while at higher $\pT$
the main contribution is from beauty decays.
The mid-rapidity results in the right panel of Fig.~\ref{fig:HFMpp} are 
also compared to predictions obtained with
the POWHEG Monte Carlo generator, interfaced to either PYTHIA or HERWIG for 
the parton shower simulation.
POWHEG+PYTHIA agrees well with the FONLL calculations, while
POWHEG+HERWIG predicts a significantly lower total cross-section.
As pointed out in Ref.~\cite{Aad:2011rr}, less than half of this difference may 
be accounted for by the different HF hadron decay models implemented
in PYTHIA and HERWIG.
Also shown in the right panel of Fig.~\ref{fig:HFMpp} is the outcome
of the PYTHIA event generator (LO plus parton shower), which describes well 
the measured $\pT$ dependence, but overestimates the total cross section by
a factor of about two.
The FONLL(NLO) curve displayed in Fig.~\ref{fig:HFMpp} is the central
value of a FONLL calculation in which the next-to-leading-log 
resummation part was disabled in the pQCD calculation. 
Such a NLO calculation deviates significantly from the measured cross section,
showing sensitivity to the NLL resummation term in the pQCD calculation.
A very good description of the data is also provided by GM-VFNS, as can 
be seen in Fig.~3 of Ref~\cite{Bolzoni:2012kx}.

\begin{figure}
\begin{center}
\includegraphics[width=0.53\textwidth]{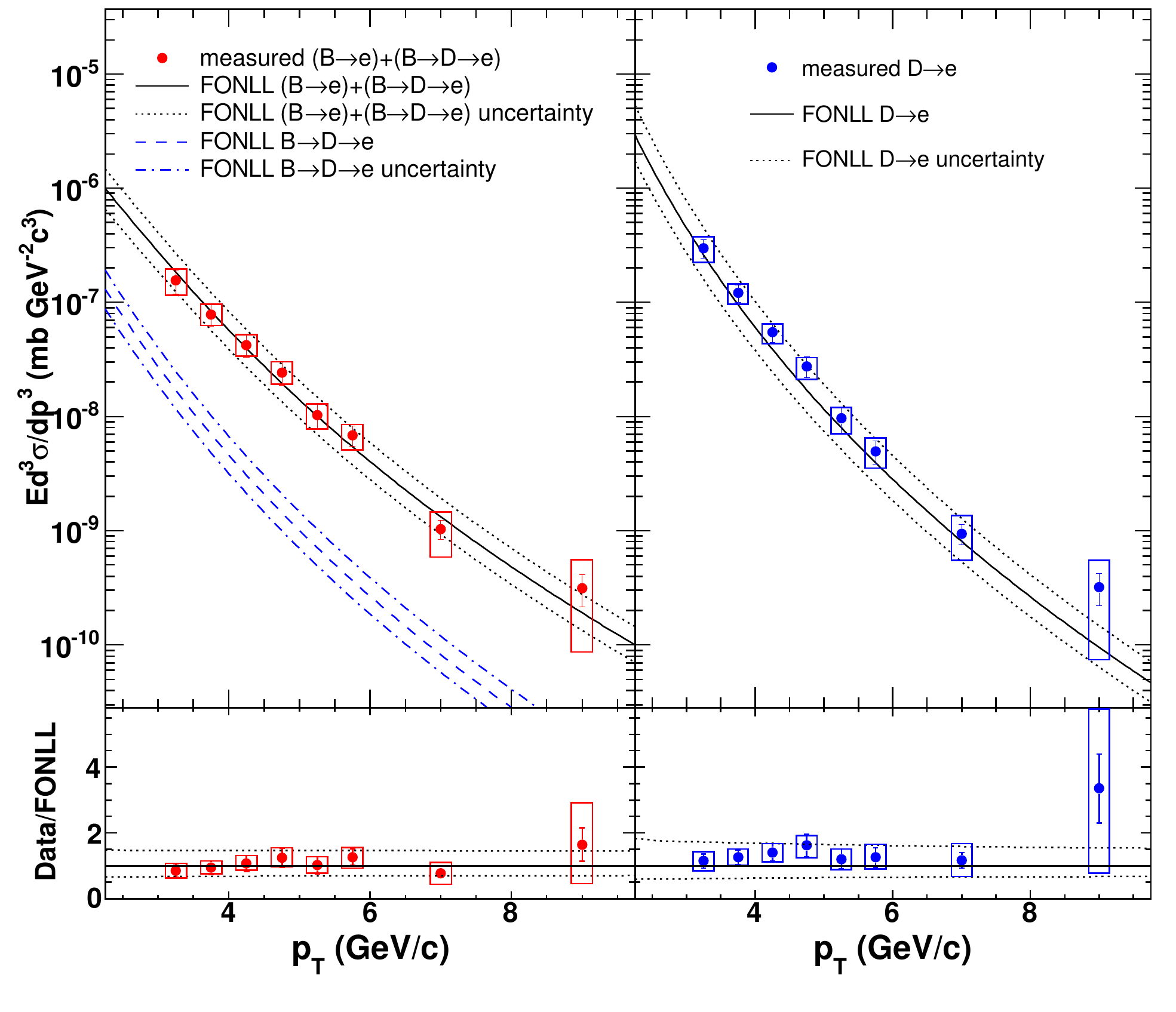}
\includegraphics[width=0.42\textwidth]{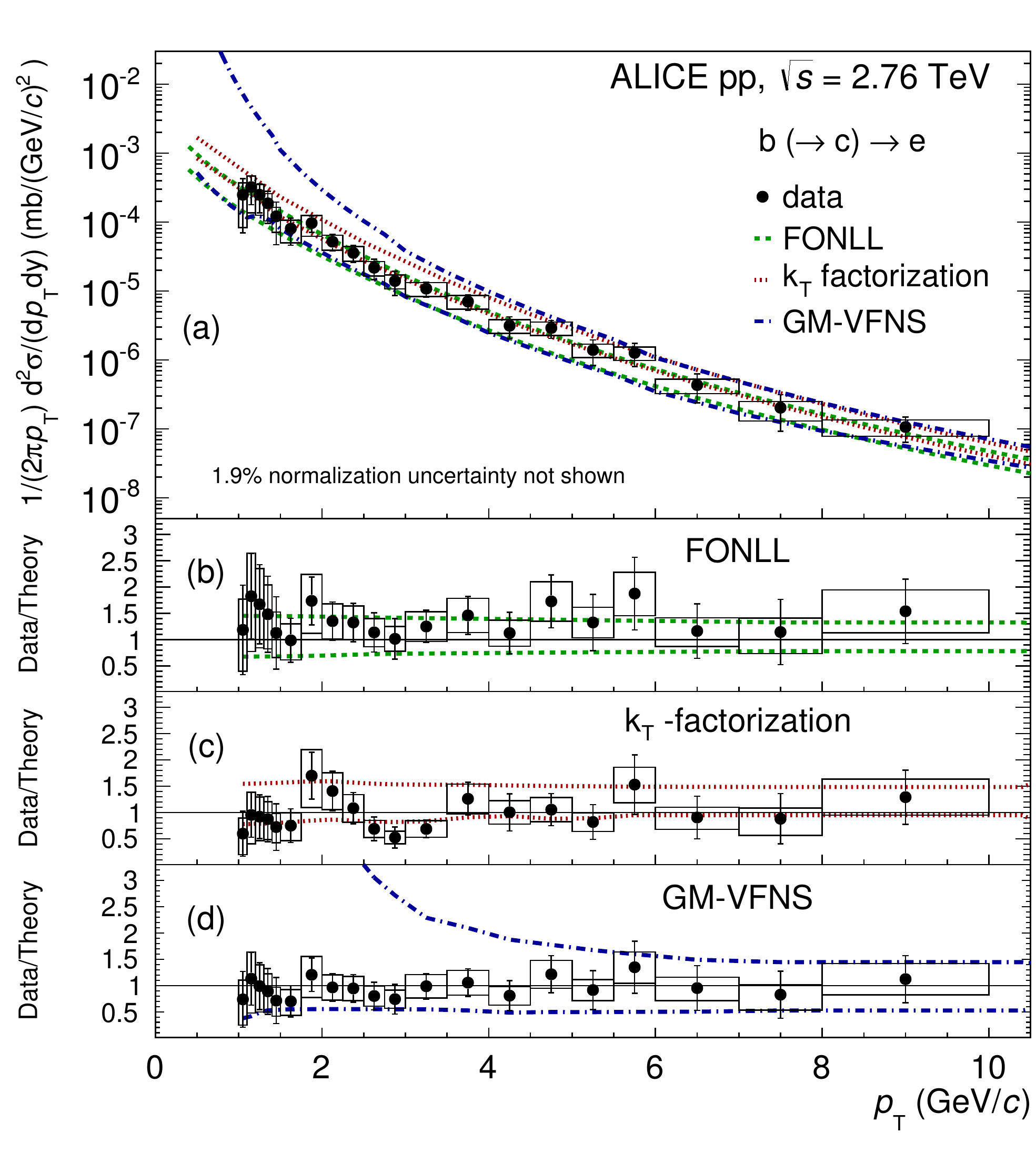}
\caption{Transverse-momentum differential production cross section of beauty-decay 
electrons in $pp$ collisions. 
Left: results at $\sqrt{s}=200~\mathrm{GeV}$ from STAR~\cite{Agakishiev:2011mr}.
Right: results at $\sqrt{s}=2.76~\mathrm{TeV}$ from ALICE~\cite{Abelev:2014hla}.
Data are compared to FONLL~\cite{Cacciari:1998it,Cacciari:2012ny}, GM-VFNS~\cite{Bolzoni:2012kx} 
and LO $k_{\rm T}$-factorization~\cite{Maciula:2013wg} calculations.}
\label{fig:BeautyElpp}
\end{center}
\end{figure}

Electrons from open-charm and -beauty decays could be statistically separated
at RHIC using the ratio $e_B/(e_B + e_D)$ measured from
$e$-hadron azimuthal correlations~\cite{Aggarwal:2010xp} and the measured
HF decay electron cross section.
The invariant cross section of electrons from beauty and charm decays
measured by the STAR Collaboration in $pp$ collisions at $\sqrt{s}=200~\GeV$ 
is shown in the left panel of 
Fig.~\ref{fig:BeautyElpp}~\cite{Agakishiev:2011mr}, together with
FONLL predictions, which describe the data within uncertainties.
At the LHC, in addition to the studies of azimuthal correlations, a 
selection on the electron impact parameter was applied to disentangle
charm and beauty contributions to the measured HF decay 
electrons~\cite{Abelev:2012sca,Abelev:2014hla}.
The ALICE results for beauty-decay electron $\pT$-differential cross section
at $\sqrt{s}=2.76~\GeV$~\cite{Abelev:2014hla} are reported 
in the right panel of Fig.~\ref{fig:BeautyElpp}.
They are compared to the predictions from FONLL, GM-VFNS~\cite{Bolzoni:2012kx} 
and LO $k_{\rm T}$-factorization~\cite{Maciula:2013wg} calculations.
The data and the pQCD calculations are consistent within the experimental and 
theoretical uncertainties.

\begin{figure}
\begin{center}
\begin{tabular}{cc}
$\vcenter{\hbox{\includegraphics[width=0.48\textwidth]{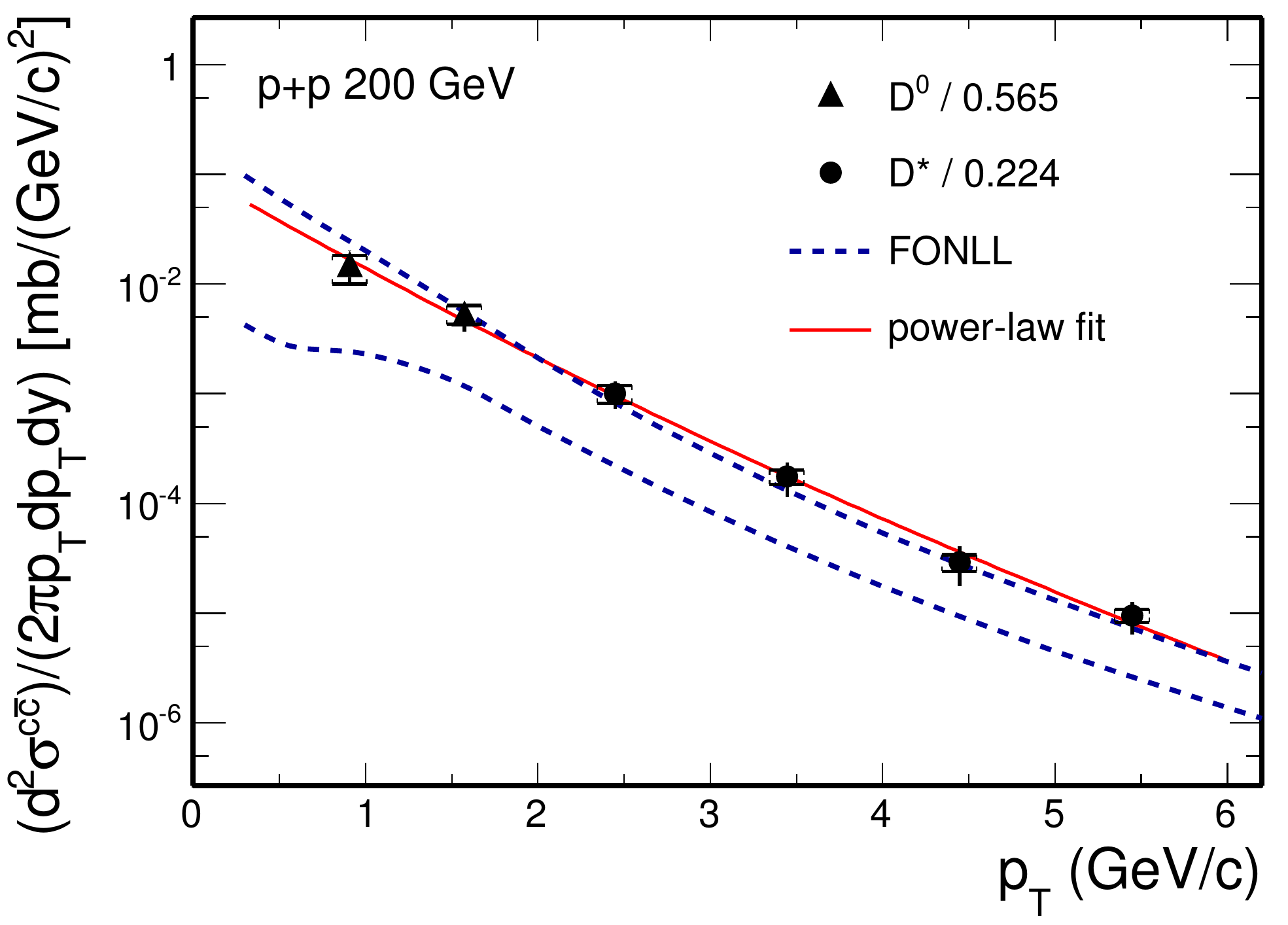}}}$ &
$\vcenter{\hbox{\includegraphics[width=0.48\textwidth]{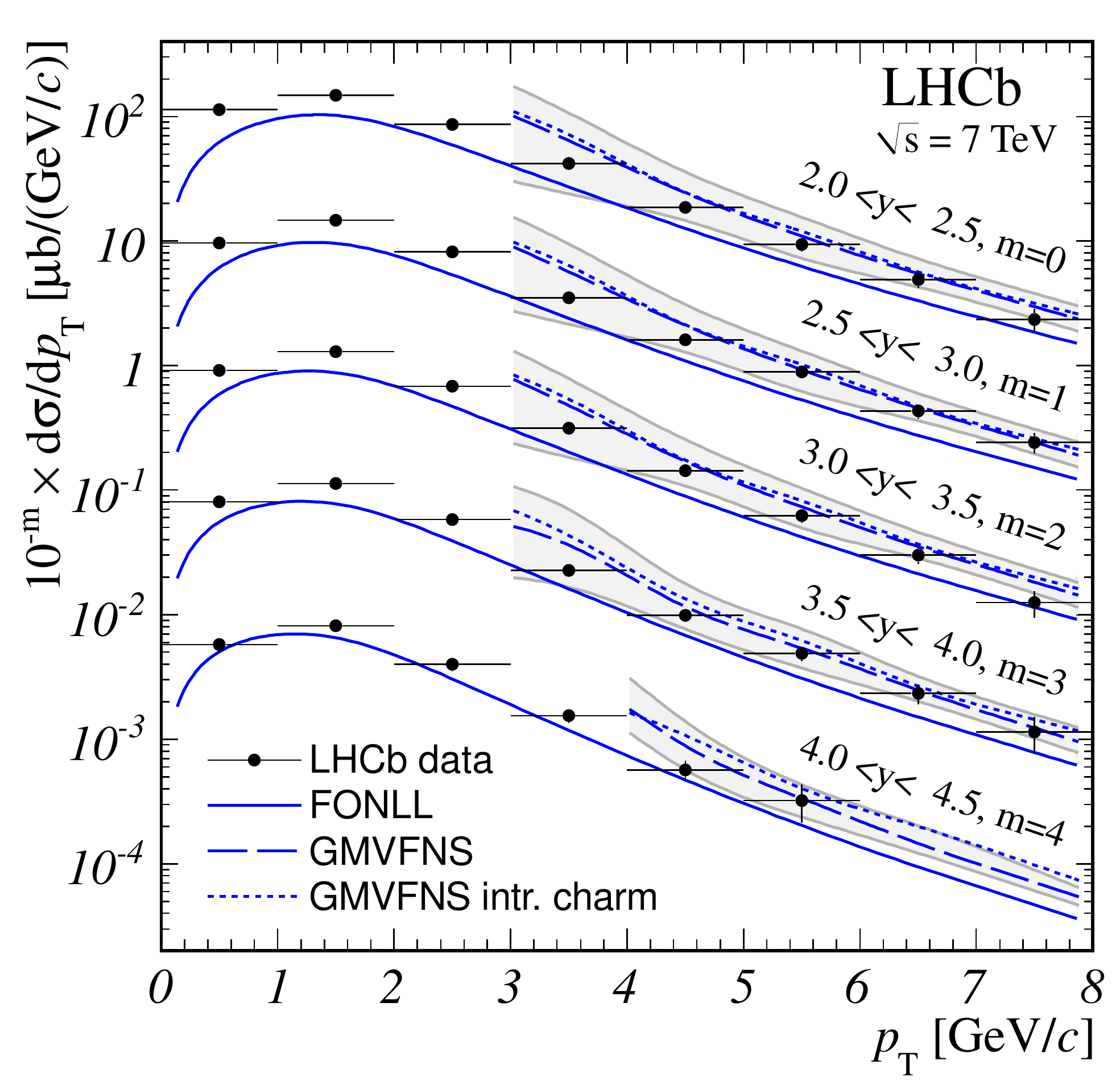}}}$ \\
(a) & (b) \\
\vspace{0.2cm}  \\
\includegraphics[width=0.48\textwidth]{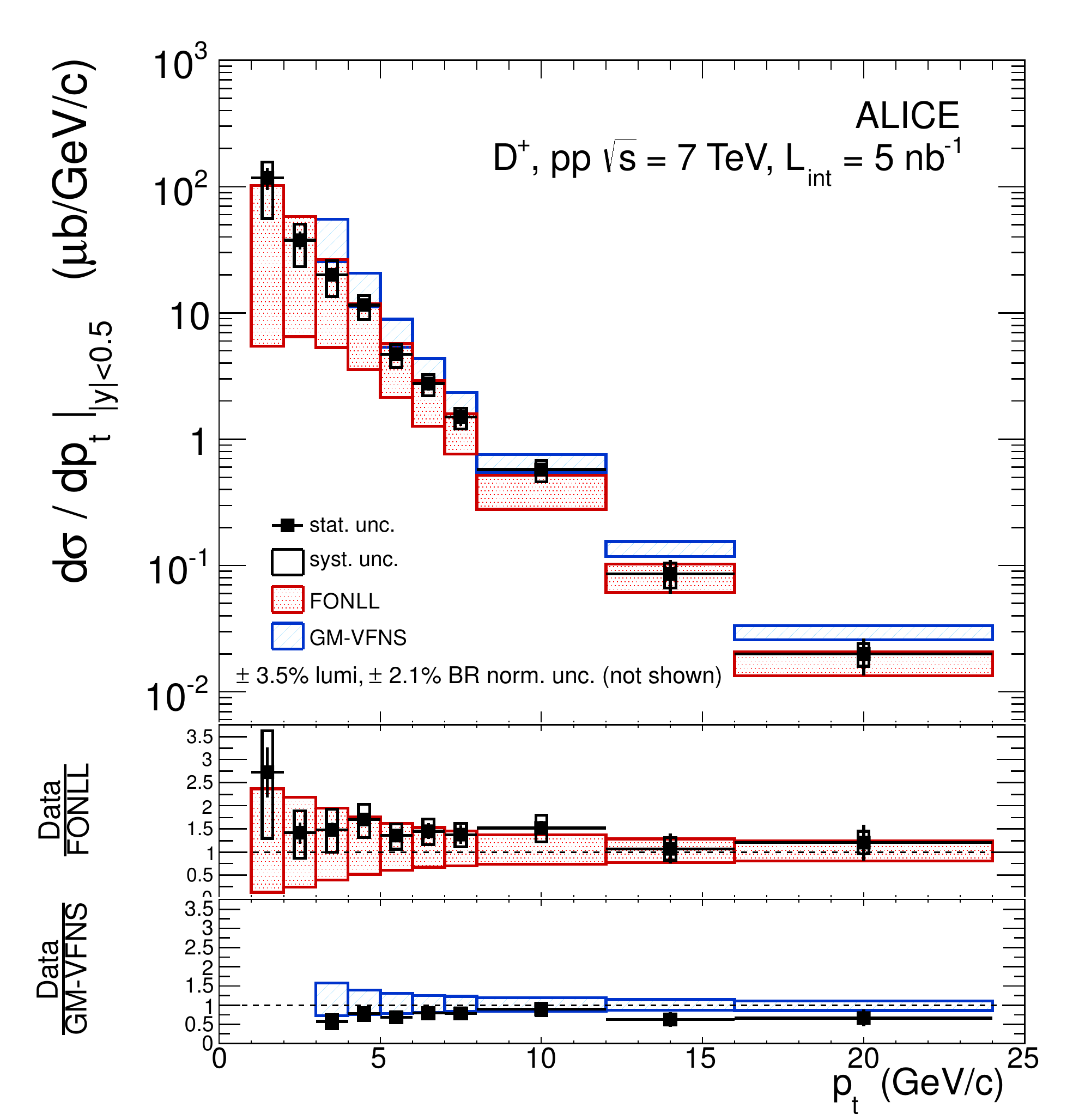} &
\includegraphics[width=0.48\textwidth]{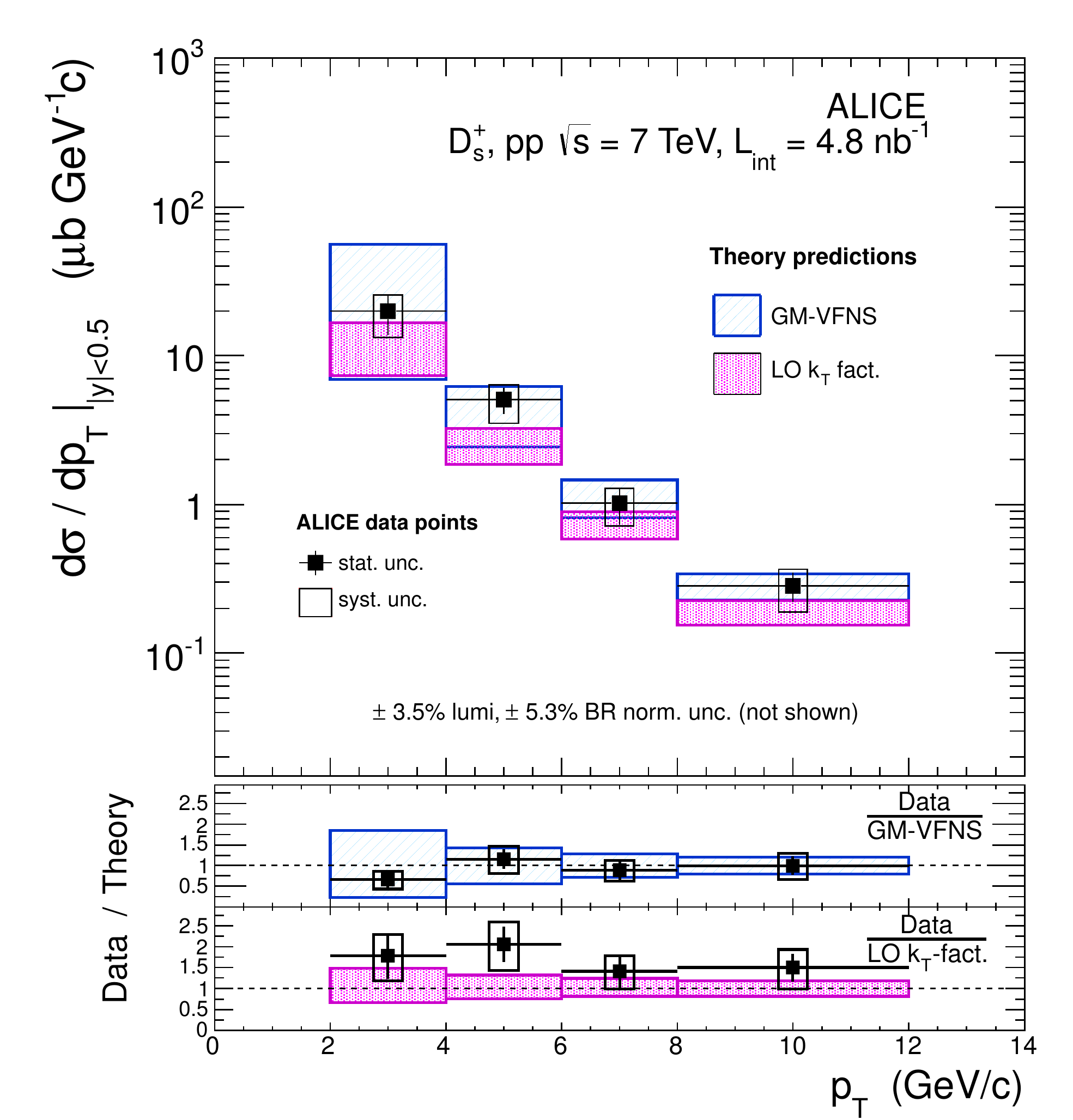}\\
(c) &  (d)\\
\end{tabular}
\caption{Transverse-momentum differential production cross section of prompt $D$ mesons
in $pp$ collisions. 
(a) ${D^0}$ and ${D^{*+}}$ at $\sqrt{s}=200~\mathrm{GeV}$ at 
mid-rapidity from STAR~\cite{Adamczyk:2012af};
(b) ${D^{0}}$ at $\sqrt{s}=7~\mathrm{TeV}$ at forward rapidity 
from LHCb~\cite{Aaij:2013mga};
(c) ${D^{+}}$ at $\sqrt{s}=7~\mathrm{TeV}$ at mid-rapidity
from ALICE~\cite{ALICE:2011aa};
(d) ${D_{s}^{+}}$ at $\sqrt{s}=7~\mathrm{TeV}$ at mid-rapidity
from ALICE~\cite{Abelev:2012tca}.
Data are compared to FONLL~\cite{Cacciari:1998it,Cacciari:2012ny}, GM-VFNS~\cite{Kniehl:2012ti} 
and LO $k_{\rm T}$-factorization~\cite{Maciula:2013wg} calculations.}
\label{fig:Dmespp}
\end{center}
\end{figure}

Open-charm production was measured at RHIC and at the LHC by reconstructing
hadronic $D$-meson decays.
A selection of results on $D$-meson $\pT$-differential cross sections in
$pp$ collisions is presented in Fig.~\ref{fig:Dmespp}.
In the top left panel, Fig.~\ref{fig:Dmespp}(a), the ${D^0}$ and 
${D^{*+}}$ cross sections (scaled by the respective branching fractions 
$f(c\rightarrow{D^0})$ and $f(c\rightarrow{D^{*+}})$ ) measured 
in $pp$ collisions at $\sqrt{s}=200~\GeV$ at mid-rapidity by the STAR 
collaboration~\cite{Adamczyk:2012af} are shown along with FONLL
predictions.
This production cross section is measured for ``inclusive" $D$-meson yields, 
including both the ``prompt" contribution coming from charm-quark fragmentation 
and decays of excited charmed-hadron states and the feeddown contribution 
due to beauty-hadron decays.
The LHCb results on prompt ${D^0}$ production at forward rapidity 
at $\sqrt{s}=7~\TeV$~\cite{Aaij:2013mga} are shown in Fig.~\ref{fig:Dmespp}(b).
The differential cross section is reported as a function of $\pT$
for different rapidity intervals and is compared to pQCD calculations
with the FONLL and GM-VFNS~\cite{Kniehl:2012ti} approaches.
Also shown is the outcome of GM-VFNS calculations using the
the CTEQ-6.5c2 parton densities with intrinsic charm~\cite{Pumplin:2007wg},
showing that the effect of intrinsic charm is expected to be small
in the phase space region of the LHCb measurement.
The LHCb collaboration recently reported results on the $\pT$-differential
production cross section of $D$ mesons at  $\sqrt{s}=13~\TeV$ and found them
to be in agreement with NLO predictions from FONLL, POWHEG, and 
GM-VFNS~\cite{Aaij:2015bpa}.
The measured ratios of the cross sections at 13 and 7~TeV are also found to be
described within uncertainties by FONLL and POWHEG, even though the data
are consistently above the central values of the ratios of pQCD predictions 
in all the considered $\pT$ and $y$ intervals. As argued in 
Refs.~\cite{Mangano:2012mh,Cacciari:2015fta,Gauld:2015yia}, 
ratios of cross sections at different $CM$ energies can be predicted 
through pQCD calculations with an accuracy of a few percent because some 
theoretical parameters (factorization and renormalization scales, quark mass, 
fragmentation fractions) are correlated at different energies and their
uncertainties cancel almost completely in the ratio.
On the other hand, PDF uncertainties do not cancel completely, because of the 
different Bjorken-$x$ range of initial-state partons covered by the 
measurements at the two $CM$ energies, making these ratios
sensitive to the PDFs, and in particular to gluon PDFs at 
small $x$, where they are not yet well constrained by data.
In the bottom panels, (c) and (d), of Fig.~\ref{fig:Dmespp} the ALICE
results on ${D^+}$ and ${D_s^+}$  $\pT$-differential cross sections
at mid-rapidity at $\sqrt{s}=7~\TeV$ are reported together with predictions
from FONLL, GM-VFNS and LO $k_{\rm T}$-factorization.
The NLO pQCD calculations provide also a good description of the 
${D^+}$, ${D^{*+}}$, and ${D_s^+}$ cross sections measured by the ATLAS 
collaboration in $pp$ collisions at $\sqrt{s}=7~\TeV$ in 
the intervals $3.5<\pT<100~\GeV/c$ and $|\eta|<2.1$~\cite{Aad:2015zix}.
A similar conclusion about the ability of pQCD calculations to describe
the measurements at LHC energies is obtained.
At all energies and rapidities the pQCD calculations agree with the
measured cross sections within uncertainties.
Yet, the FONLL predictions obtained with the central values of the calculation 
parameters tend to underestimate the data, with the measured cross sections
lying close to the upper edge of the theoretical uncertainty band.
The tendency of FONLL to underestimate the charm data was also
observed at the Tevatron~\cite{Acosta:2003ax}.
On the other hand, the central value of the GM-VFNS predictions lies 
systematically  above the data.

\begin{figure}
\begin{center}
\begin{tabular}{cc}
$\vcenter{\hbox{\includegraphics[width=0.48\textwidth,bb = 0 8 567 407]{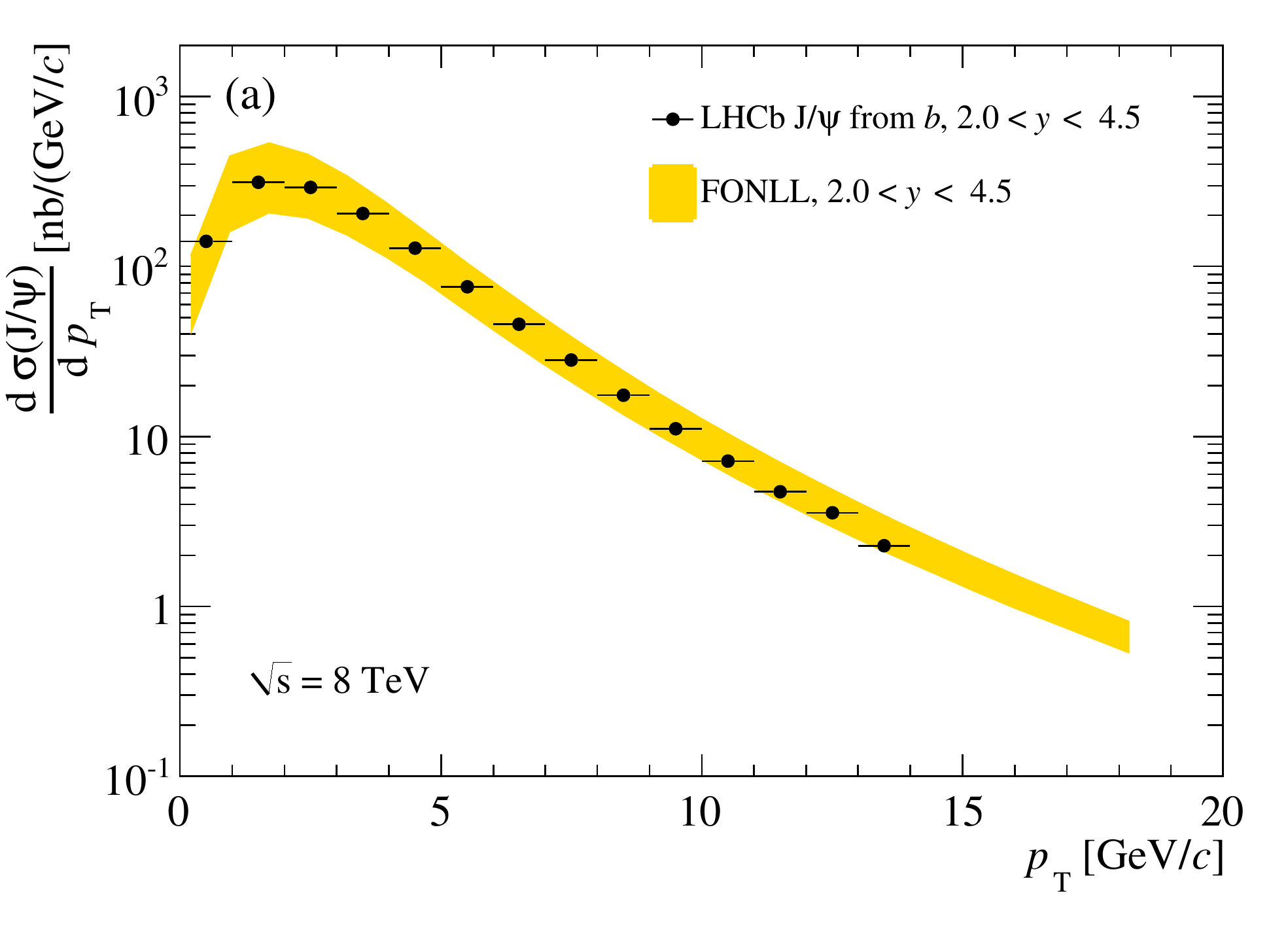}}}$ &
$\vcenter{\hbox{\includegraphics[width=0.48\textwidth,bb = 0 0 567 360]{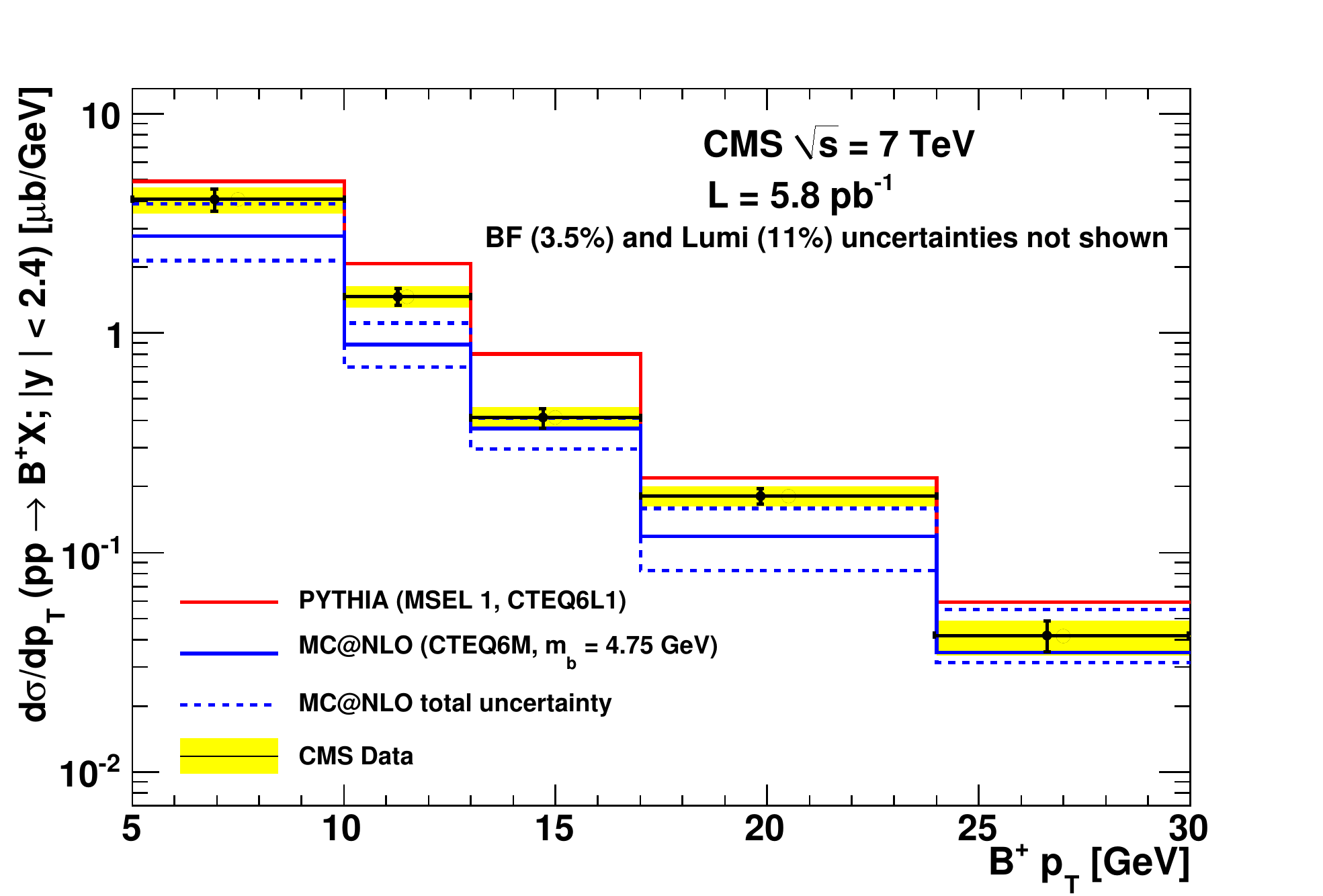}}}$ \\
(a) & (b) \\
\vspace{0.2cm}  \\
\includegraphics[width=0.48\textwidth]{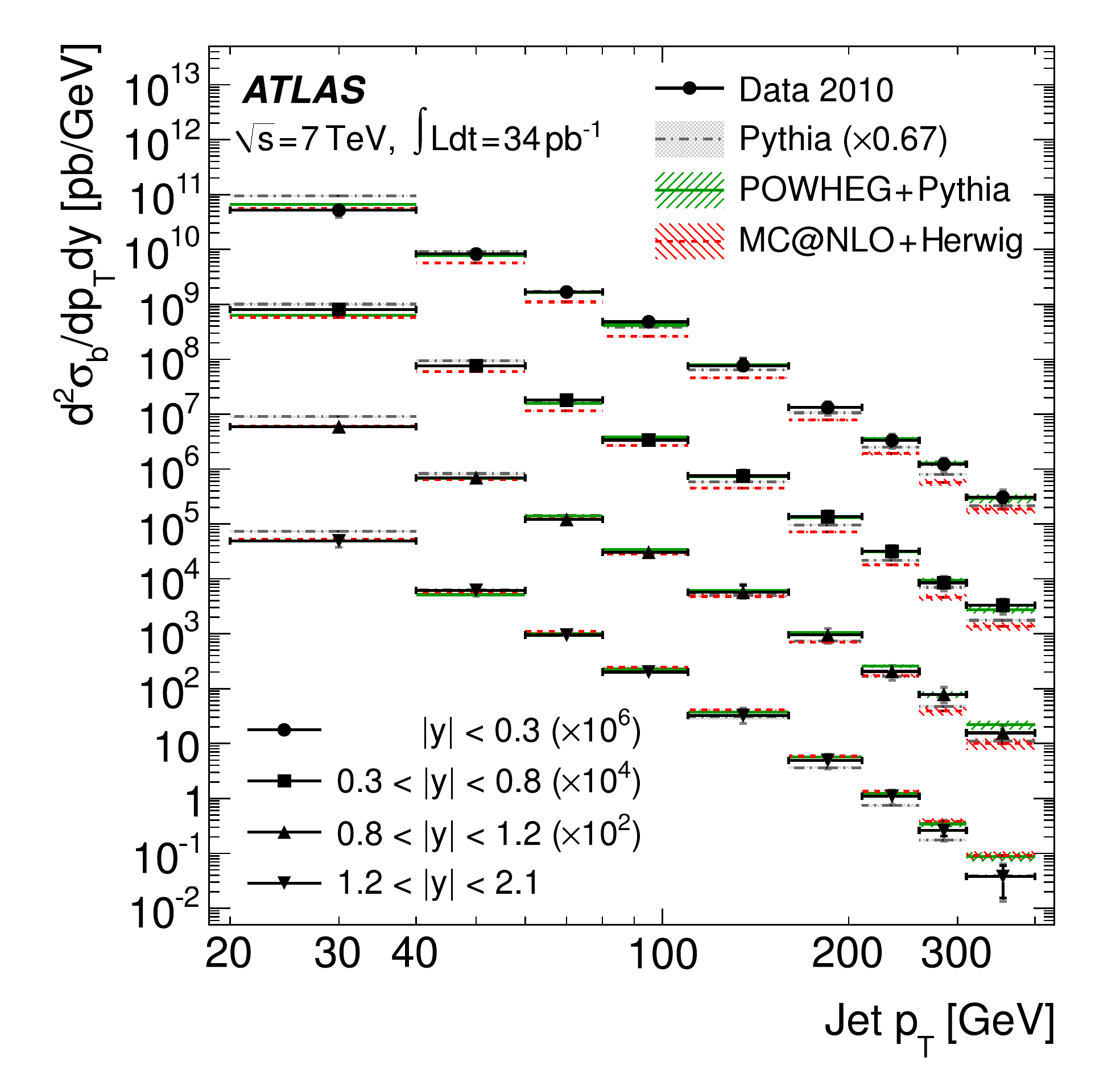} &
\includegraphics[width=0.48\textwidth]{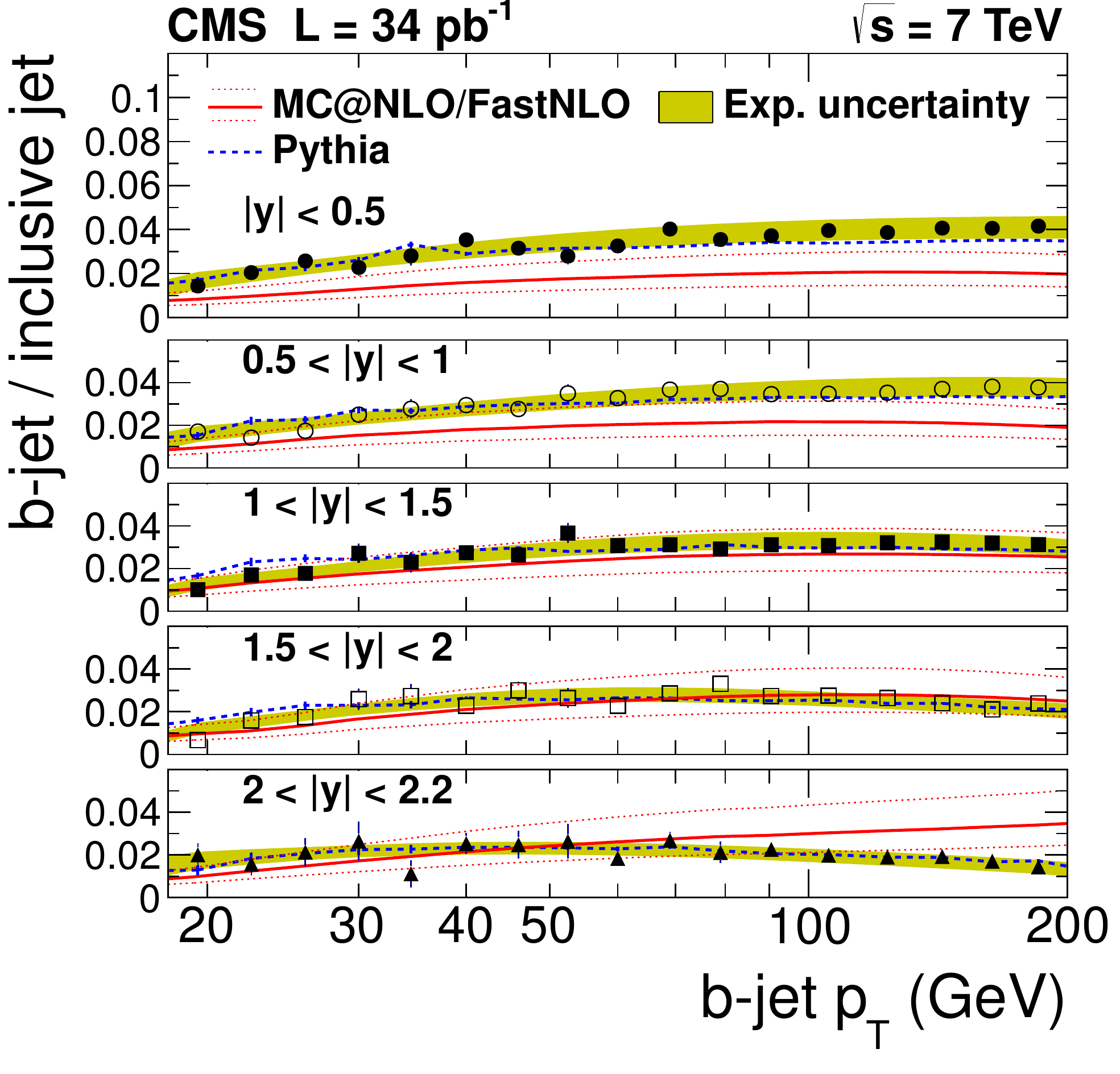}\\
(c) &  (d)\\
\end{tabular}
\caption{Transverse-momentum differential measurements of beauty production at the LHC.
(a) non-prompt $J/\psi$ at $\sqrt{s}=8~\mathrm{TeV}$ at forward
rapidity from LHCb~\cite{Aaij:2013yaa};
(b) ${B^{+}}$ meson at $\sqrt{s}=7~\mathrm{TeV}$ at mid-rapidity 
from CMS~\cite{Khachatryan:2011mk};
(c) $b$-jets at $\sqrt{s}=7~\mathrm{TeV}$ in four rapidity intervals
from ATLAS~\cite{ATLAS:2011ac};
(d) ratio of $b$-jet to inclusive-jet cross section in four rapidity intervals
from CMS~\cite{Chatrchyan:2012dk}.
Data are compared to FONLL~\cite{Cacciari:1998it,Cacciari:2012ny} calculations 
and to MC@NLO~\cite{Frixione:2003ei}, POWHEG~\cite{Frixione:2007nw} and
PYTHIA~\cite{Sjostrand:2006za} MC generators.}
\label{fig:beautypp}
\end{center}
\end{figure}

Figure~\ref{fig:beautypp} displays a selection of results on beauty
production as a function of $\pT$ at the LHC.
In particular, the non-prompt $J/\psi$ cross section at forward rapidity, 
measured by the LHCb collaboration at $\sqrt{s}=8~\TeV$~\cite{Aaij:2013yaa} is 
shown in panel (a);  the ${B^{+}}$-meson cross section at mid-rapidity measured 
by CMS at $\sqrt{s}=7~\TeV$~\cite{Khachatryan:2011mk} is shown in panel (b); and 
panel (c) depicts the cross section
of $b$-jets measured by ATLAS at $\sqrt{s}=7~\TeV$ 
in different rapidity intervals~\cite{ATLAS:2011ac}.
Furthermore, in Fig.~\ref{fig:beautypp}(d), the ratios of $b$-jet to 
inclusive-jet cross sections measured by CMS~\cite{Chatrchyan:2012dk} in
different rapidity intervals are shown as a function of $\pT$.
The data are compared to different theoretical predictions.
FONLL provides a good description of the non-prompt $J/\psi$ data.
A similar agreement with FONLL is observed for the recent measurements of 
non-prompt $J/\psi$ cross section at $\sqrt{s}=13~\TeV$~\cite{Aaij:2015rla}.
However, the measured ratios of the production cross sections 
of $J/\psi$ from beauty-hadron decays at $CM$ energies of
13 and 8 TeV are found to lie systematically above the predictions from FONLL 
calculations. As pointed out above, 
in the calculations of ratios of cross sections at
different energies, the sensitivity to the pQCD scale variation is 
substantially reduced, thus providing some sensitivity to the (mostly gluon) 
PDFs in regions where they are not yet well constrained by 
data~\cite{Cacciari:2015fta,Gauld:2015yia}.
PYTHIA, which has LO+LL accuracy, does not provide a good description of the
measured $B$-meson cross section, which is instead correctly predicted by MC@NLO.
The measurements of $b$-jet cross section can be compared to
predictions from MC event generators featuring a complete 
description of the hadronic final state.
The NLO generators POWHEG (matched to PYTHIA parton shower) and MC@NLO 
(interfaced to HERWIG for the parton shower) are found to describe the data
reasonably well, with POWHEG+PYTHIA providing a slightly better agreement 
with the data across the different rapidity regions (see~\cite{ATLAS:2011ac} 
for details).
The LO+LL predictions from PYTHIA do not predict the correct normalization,
overestimating the measured integrated cross section by a factor of about 1.5,
but they can describe reasonably well the $\pT$ dependence of the $b$-jet 
production cross section.

Additional insight into HQ production and fragmentation can
be obtained from measurements of HF azimuthal correlations,
which help to constrain MC models and to disentangle different
production processes for HF particles.
For example, measurements of azimuthal correlations between $B$ and
$\overline{B}$ hadrons from CMS show a substantial contribution from
production at small opening angles ($\Delta\varphi\approx 0$), which is not 
reproduced by PYTHIA~\cite{Khachatryan:2011wq}. 
This result points to the importance of the ``near"' production (via the gluon 
splitting mechanism) in addition to the ``back-to-back" production (mostly via 
flavor creation).
Preliminary results of $D$-hadron azimuthal correlations from 
ALICE~\cite{Bjelogrlic:2014kia} show that PYTHIA and POWHEG+PYTHIA
provide a reasonable description of the data for this
observable, which is sensitive to charm-quark fragmentation and jet
structure.
The production of $D$ mesons, prompt and non-prompt $J/\psi$, and $\Upsilon$
was also measured by ALICE and CMS as a function of the multiplicity of particles 
produced in the collision~\cite{Abelev:2012rz,Adam:2015ota,Chatrchyan:2013nza}.
The per-event yields of open charm and beauty hadrons and quarkonia are
found to increase with increasing multiplicity.
This trend can be described by models including multiple parton 
interactions (MPI), thus providing sensitivity to the role of MPIs 
at the hard momentum scales relevant for ${c\bar{c}}$ and 
${b\bar{b}}$ pair production.


\subsubsection{Results from p-A (d-A) collisions and Cold-Nuclear-Matter Effects}
\label{sssec_pAdata}

\hspace{2cm}

The study of the properties of the hot and dense medium created in heavy-ion 
reactions requires a quantitative understanding of the effects induced by the 
presence of nuclei in the initial-state of the collisions, as well as, 
possibly, by the relatively high multiplicities of particles in the final state.
Such effects are commonly referred to as cold-nuclear-matter (CNM) effects and
are assessed experimentally  by studying particle production in 
proton--nucleus ($p$A) or deuteron--nucleus ($d$A) collisions.

Heavy-flavor production can be affected by a variety  of CNM effects, 
which include the following.
\begin{itemize}
\item \emph{Modification of the Parton Distribution Functions.} The nuclear 
environment affects the quark and gluon distributions, which are modified 
in bound nucleons compared to those of free nucleons. 
The modification depends on the fractional parton momentum ($x$), on the scale 
of the parton--parton interaction ($Q^2$), and on the atomic mass number (A) 
of the nucleus~\cite{Arneodo:1992wf,Malace:2014uea}.
This effect is commonly studied experimentally via the ratio between the PDF of 
nucleons in nuclei (nPDF) and those of the proton (deuteron), $R^{\rm A}_{i}(x,Q^2)$,
where $i$ is the parton flavor. 
Four regions, depending on the value of $x$, are usually identified:
(i) \emph{Shadowing}, which is a depletion ($R^{\rm A}_{i}<1$) at 
small $x$ ($x<10^{-2}$);
(ii) \emph{Anti-shadowing}, an enhancement ($R^{\rm A}_{i}>1$) at intermediate $x$ 
($10^{-2}<x<10^{-1}$);
(iii) \emph{EMC effect}, a depletion ($R^{\rm A}_{i}<1$) in the valence quark region
($x \sim 10^{-1}$); and 
(iv) an enhancement ($R^{\rm A}_{i}>1$) at large $x$ ($0.8<x<1$) 
associated with the \emph{Fermi motion} of nucleons inside the nucleus.
There is no comprehensive theoretical understanding of the observed pattern
over the entire $x$ range.
The $R^{\rm A}_{i}(x,Q^2)$ values can be calculated using phenomenological
parameterizations based on global fit analyses of lepton--nucleus and 
proton--nucleus data, such as EPS09~\cite{Eskola:2009uj}, 
HKN07~\cite{Hirai:2007sx} and nDS~\cite{deFlorian:2003qf}.
The depletion in the low-$x$ region (shadowing) can be understood as due 
to gluon  phase-space saturation, and it can be described within
the  Colour Glass Condensate (CGC) effective theory, where an initial high-energy 
nucleus is treated as a coherent and dense (saturated) gluonic 
system~\cite{Gelis:2010nm}.
The modification of the PDFs results in a modification of the effective 
partonic luminosity (and consequently of the HQ production cross 
section) in collisions involving nuclei relative to $pp$ collisions.\\
\item \emph{Multiple scattering of partons} in the nucleus before 
and/or after the hard scattering affects the kinematic distribution 
of the produced heavy quarks and/or hadrons.
These multiple collisions lead to transverse-momentum broadening (usually 
denoted Cronin effect)~\cite{Lev:1983hh,Wang:1998ww,Kopeliovich:2002yh} 
and parton energy 
loss~\cite{Gavin:1991qk,Brodsky:1992nq,Vitev:2007ve,Arleo:2012rs}.\\
\item On top of initial-state CNM effects, also
effects in the final state, due to the high multiplicity of particles
produced in $p$A ($d$A) collisions, may be responsible for a modification of
the HF hadron yields and momentum distributions.
The presence of final-state effects in small collision systems is suggested 
by measurements of long-range correlations of charged 
hadrons in $p$Pb collisions at the 
LHC~\cite{CMS:2012qk,Abelev:2012ola,ABELEV:2013wsa,Aad:2012gla}.
These observations can be described by hydrodynamic calculations 
assuming the formation of a medium with some degree of collectivity.
It is still highly debated if such a collective flow is established, and
alternative explanations, based, \eg, on the CGC effective 
theory~\cite{Dusling:2012cg} have been proposed.
It should be pointed out that if a collective motion of the final-state 
particles is established, the medium could also impart a flow on 
HF particles.
Additional indications for the importance of final-state effects
in small collision systems are provided by the larger suppression of the 
$\psi({2S})$ meson with respect to the $J/\psi$ in 
$d$Au collisions at RHIC~\cite{Adare:2013ezl} and $p$Pb collisions at the 
LHC~\cite{Abelev:2014zpa}, see, \eg, Refs.~\cite{Ferreiro:2014bia,Du:2015wha}
for pertinent model calculations. \\
\end{itemize}

\begin{figure}
\begin{center}
$\vcenter{\hbox{\includegraphics[width=0.48\textwidth]{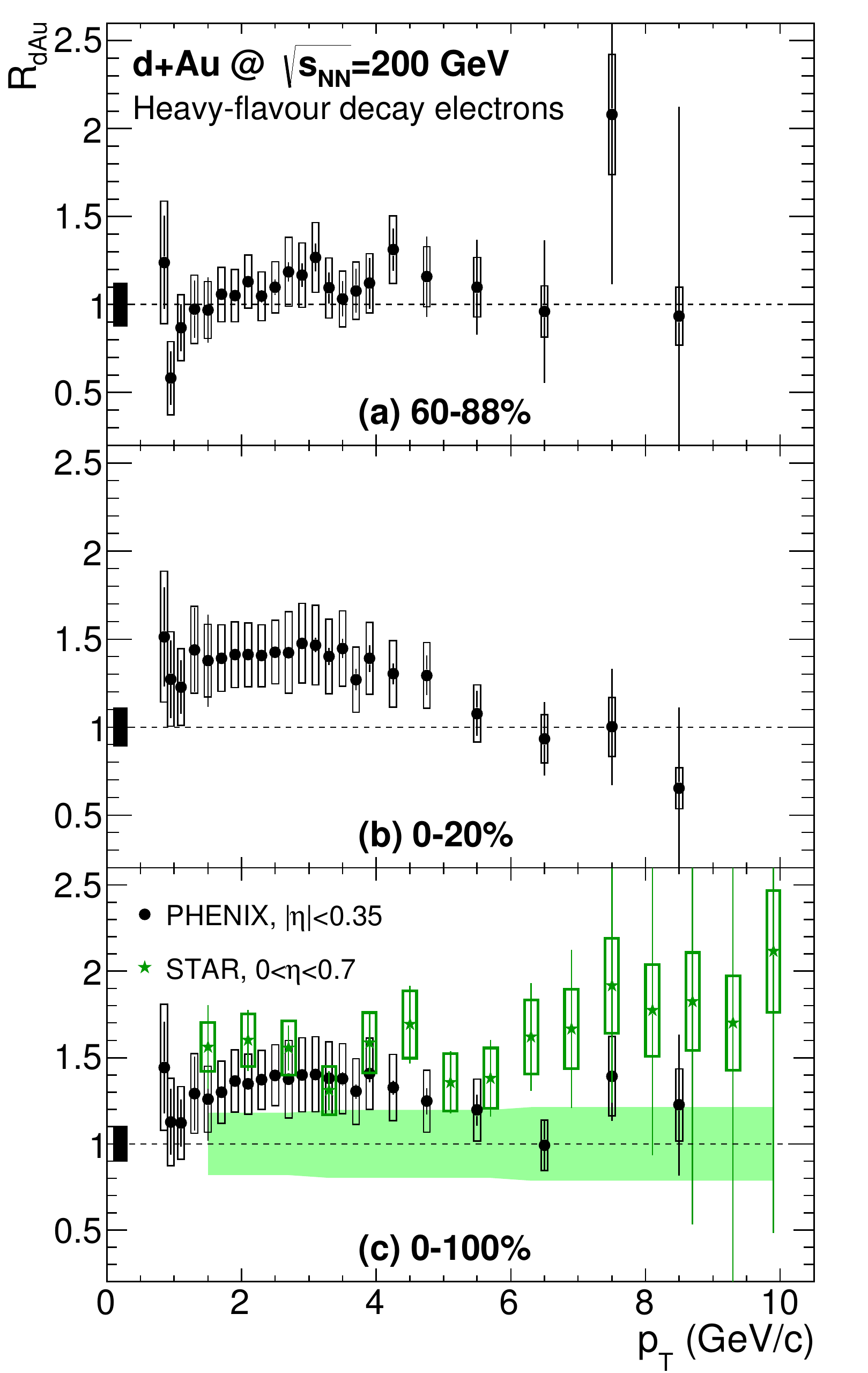}}}$
$\vcenter{\hbox{\includegraphics[width=0.48\textwidth]{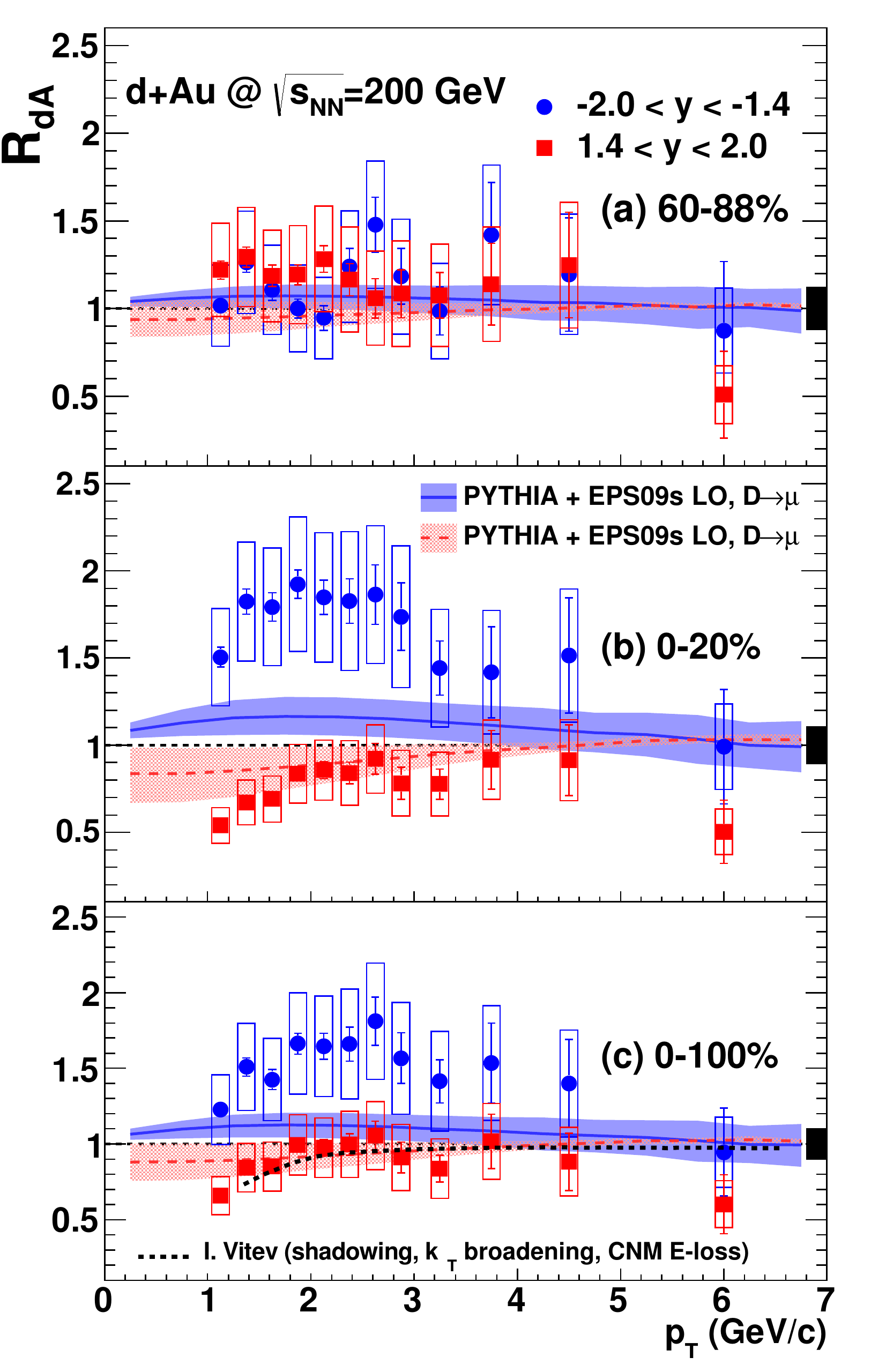}}}$
\caption{Nuclear modification factor of heavy-flavor decay leptons as a 
function of $p_{\rm T}$  in $d$Au collisions at 
$\sqrt{s_{\rm NN}}=200~\mathrm{GeV}$.
Left: heavy-flavor decay electrons at mid-rapidity from 
PHENIX~\cite{Adare:2012yxa} and STAR~\cite{Abelev:2006db}.
Right: heavy-flavor decay muons at backward (Au-going side) and forward 
($d$-going side) rapidity from PHENIX~\cite{Adare:2013lkk}.
PHENIX results are shown for the centrality-integrated sample (0--100\%) and 
for peripheral (60-88\%) and central (0--20\%) collisions separately.
Muon data are compared to predictions based on nuclear modification of
the PDF (EPS09s nPDF~\cite{Helenius:2012wd}); and to a theoretical calculation 
including shadowing, $k_{\rm T}$-broadening and cold-nuclear-matter energy loss 
effects~\cite{Vitev:2007ve}.}
\label{fig:RdAuLeptonsRHIC}
\end{center}
\end{figure}

Heavy-flavor production was studied at RHIC and at the LHC
in $d$Au and $p$Pb collisions at $\sqrtsNN=200~\GeV$ and $5.02~\TeV$, 
respectively.
At RHIC, the measurements were carried out by STAR~\cite{Abelev:2006db} and 
PHENIX~\cite{Adare:2012yxa,Adare:2013lkk} in the semi-leptonic channel.
In particular, HF decay electrons were measured at mid-rapidity
($|\eta|<0.35$ for PHENIX and $0<\eta<7$ for STAR) and HF decay muons 
at forward (deuteron-going direction) and backward (Au-going) rapidity 
with PHENIX.
The PHENIX collaboration also reported results for peripheral and central 
collisions.
The resulting nuclear modification factor, $R_{d\rm Au}$, of HF decay 
leptons is shown in Fig.~\ref{fig:RdAuLeptonsRHIC} as a function of $\pT$ for 
multiplicity integrated (bottom panels), central (middle panels) and 
peripheral (top panels) collisions.
The mid-rapidity results (left panels) show a slight enhancement
of the production with respect to binary-scaled $pp$ collisions, with a
mild dependence on collision centrality.
The data can be described in terms of nuclear PDFs (anti-shadowing) and 
$k_{\rm T}$-broadening (Cronin enhancement).
The possible development of a collective flow in the HQ sector
is also expected to lead to values of $R_{d\rm Au}$ larger than unity
in the $\pT$ range covered by the measurements. 
An approximate description of the measured values of nuclear modification 
factor can be obtained employing a blast-wave function, with parameters
extracted from fits to the light-hadron spectra, to determine the momentum 
distribution of HF hadrons in $d$Au collisions~\cite{Sickles:2013yna}.
Recently, the outcome of calculations in the POWLANG transport setup
at RHIC energies were published~\cite{Beraudo:2015wsd}.
In this model, it is assumed that in $p$A and $d$A collisions a hot and 
deconfined medium is formed.
The relativistic Langevin equation is used to follow the propagation of 
charm and beauty quarks in the hydrodynamically expanding medium until 
hadronization, which is modeled by combining each heavy quark with a light
parton from the medium to form color-singlet objects (strings), which are
fragmented with PYTHIA to produce the final-state hadrons.
The model can describe the midrapidity data within uncertainties, 
suggesting that, within
such a framework, the enhancement of $R_{d\rm Au}$ of HF decay electrons
observed at RHIC reflects the radial flow acquired by the parent-$D$ and -$B$ 
mesons.

In the right panels of Fig.~\ref{fig:RdAuLeptonsRHIC} the results 
on the HF muon $R_{d\rm Au}$ at forward and backward rapidities are 
reported. 
They show a pronounced centrality dependence, with similar $R_{d\rm Au}$ values 
in peripheral collisions at forward and backward rapidity, but significant
differences in central collisions, namely a suppression at 
forward rapidity ($d$-going direction) and an enhancement at backward rapidity
(Au-going direction).
The data at forward rapidity are described both by the model of Vitev 
{\it et al.}~\cite{Vitev:2007ve}, which includes shadowing, $k_{\rm T}$ broadening 
and CNM energy loss, and by pQCD calculations including EPS09 
nPDFs~\cite{Helenius:2012wd}. 
The results at backward rapidity cannot be described by only considering  
nPDF effects, suggesting that other mechanisms are at work.

The ${b\bar{b}}$ production cross section was determined at 
mid-rapidity by the PHENIX collaboration from the invariant-mass and $\pT$ 
distributions of $e^+e^-$ pairs.  The result, 
$\sigma_{{b\bar{b}}}^{d\rm Au}=1.37\pm0.28\mathrm{(stat)}\pm0.46\mathrm{(syst)}~\mb$,
is consistent with binary scaling of the cross section measured in $pp$ 
collisions~\cite{Adare:2014iwg}.
The modification due to CNM effects on HF production at mid-rapidity 
is expected to be small as compared to the quoted uncertainties of the 
measurement.

At LHC energies, charm and beauty production was studied by measuring 
HF decay electrons (ALICE~\cite{Adam:2015qda}), $D$ mesons 
(ALICE~\cite{Abelev:2014hha}), $B$ mesons (CMS~\cite{Khachatryan:2015uja}), 
and $b$-jets (CMS~\cite{Khachatryan:2015sva}) at mid-rapidity; 
at forward ($p$-going side) and backward (Pb-going) rapidity, beauty was 
studied via measurements of non-prompt $J/\psi$ by the LHCb 
collaboration~\cite{Aaij:2013zxa}.
Preliminary results were also reported by the ALICE collaboration for 
HF decay muons at forward and backward rapidity~\cite{Li:2014dha} 
and beauty-decay electrons at mid-rapidity~\cite{Li:2014dha}.

\begin{figure}
\begin{center}
\includegraphics[width=0.48\textwidth]{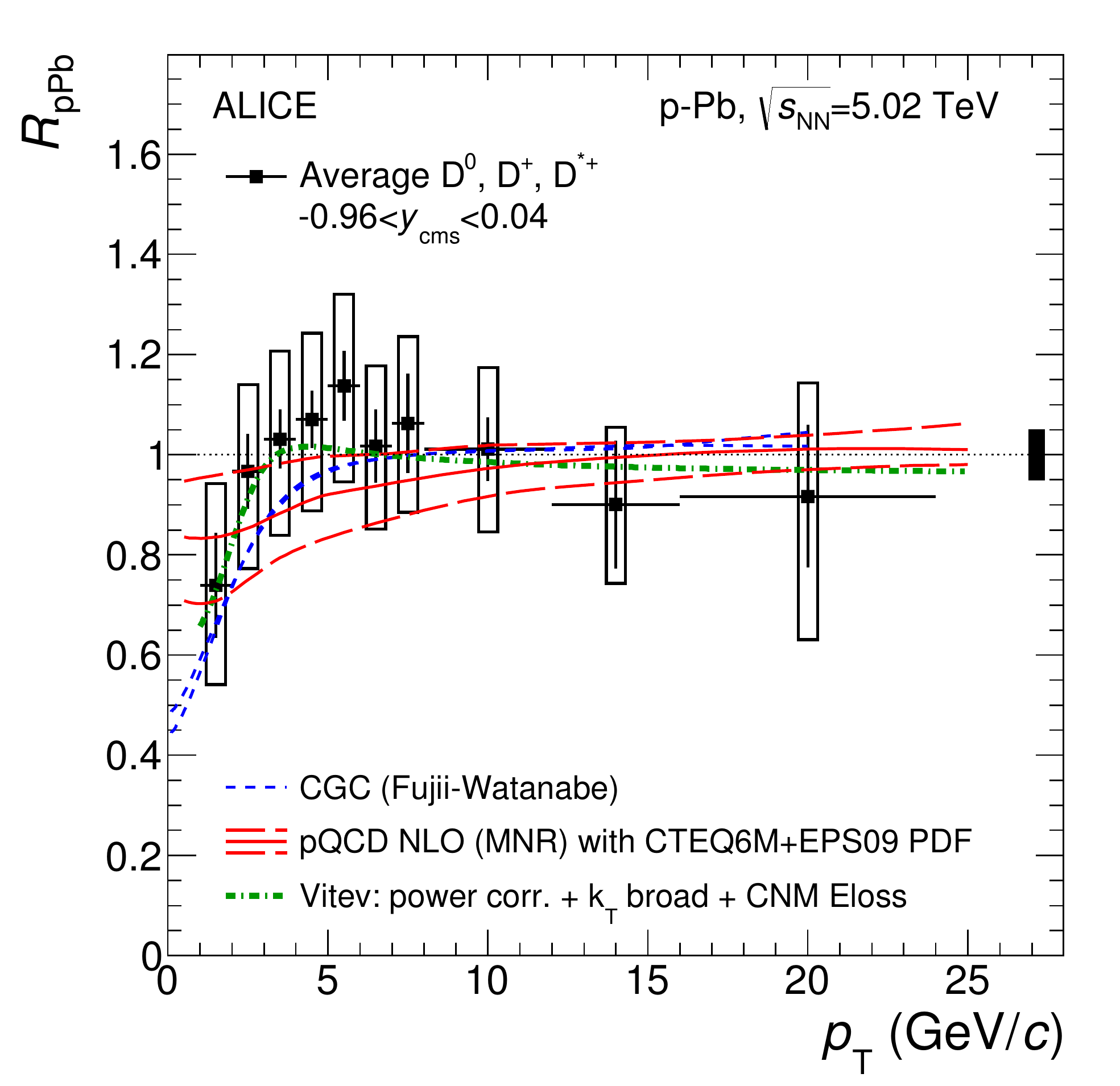}
\includegraphics[width=0.48\textwidth]{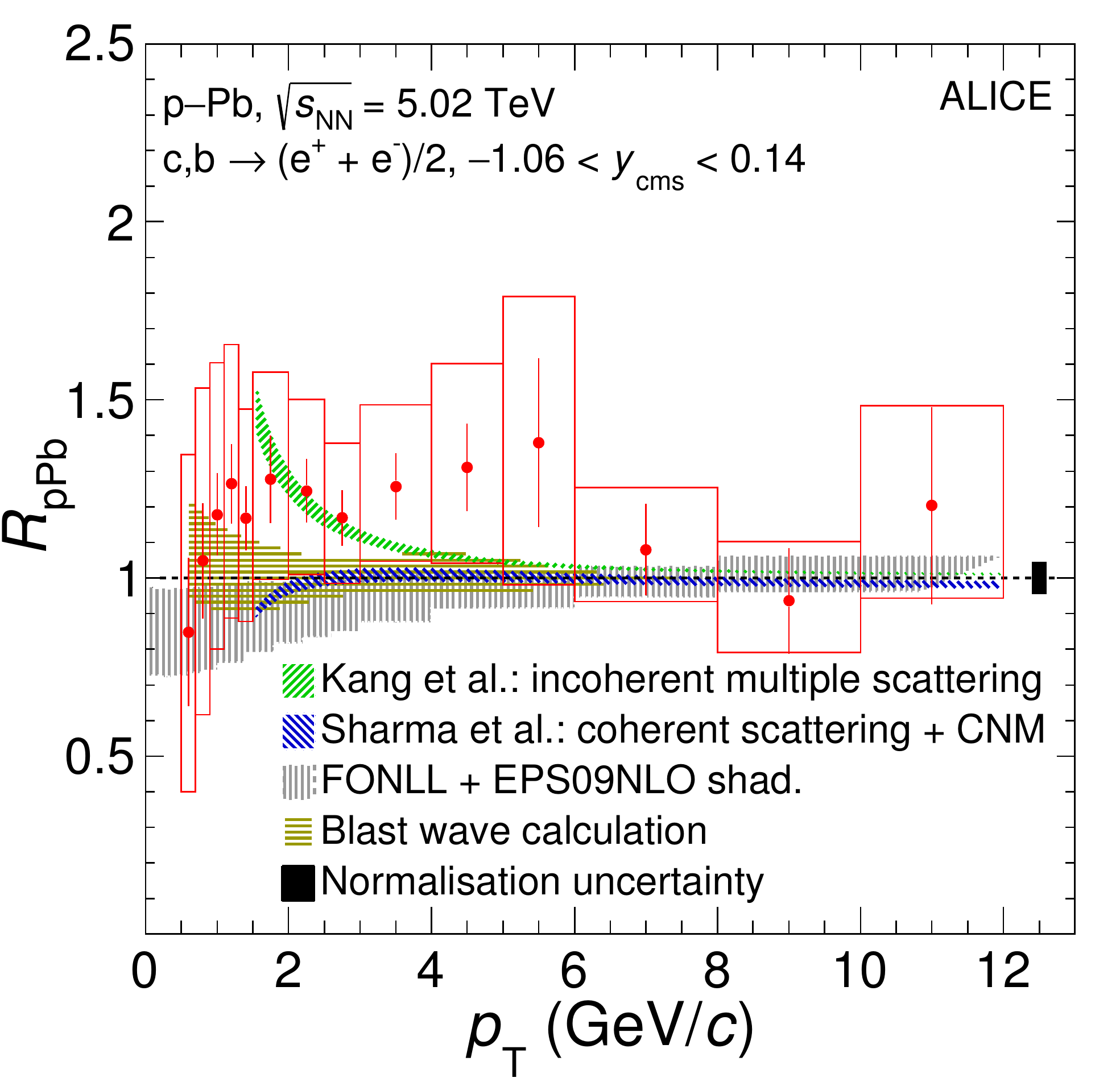}
\caption{Nuclear modification factor of $D$ mesons~\cite{Abelev:2014hha} (left) 
and HF decay electrons~\cite{Adam:2015qda} (right) as a function of 
$p_{\rm T}$  in $p$Pb collisions at 
$\sqrt{s_{\rm NN}}=5.02~\mathrm{TeV}$.
Data are compared to calculations including CNM 
effects: NLO pQCD calculations with EPS09 nPDF~\cite{Eskola:2009uj}, 
Color Glass Condensate~\cite{Fujii:2013yja}, shadowing, $k_{\rm T}$-broadening 
and CNM energy loss~\cite{Sharma:2009hn}, incoherent multiple 
scatterings~\cite{Kang:2014hha} and a blast-wave ansatz for a collectively
expanding medium~\cite{Sickles:2013yna}.}
\label{fig:RpADLHC}
\end{center}
\end{figure}

The nuclear modification factor $R_{p\rm Pb}$ of prompt $D$ 
mesons~\cite{Abelev:2014hha} and HF decay electrons~\cite{Adam:2015qda} 
measured by the ALICE collaboration as a function of $\pT$ at mid-rapidity 
is shown in Fig.~\ref{fig:RpADLHC}. 
The reference $pp$ cross sections at $\sqrt{s}=5.02~{\rm TeV}$ were obtained 
with a pQCD-based energy scaling of the $\pT$-differential cross sections 
measured at $\sqrt{s}=7~\rm{TeV}$~\cite{Averbeck:2011ga}.
No significant difference among different $D$-meson species 
(${D^0}$, ${D^+}$, ${D^{*+}}$ and ${D_s^{+}}$)
is observed~\cite{Abelev:2014hha}.
The average $R_{p\rm Pb}$ of ${D^0}$, ${D^+}$ and ${D^{*+}}$ 
mesons is reported in the left panel of Fig.~\ref{fig:RpADLHC}. 
Within uncertainties, the measured $D$-meson $R_{p\rm Pb}$ is compatible with 
unity, indicating small ($<10-20\%$) CNM effects for $\pT>2~\GeV/c$.
The $D$-meson $R_{p\rm Pb}$ is compared to model calculations including CNM 
effects, namely NLO pQCD calculations with ESP09 nPDF~\cite{Eskola:2009uj}, 
calculations based 
on CGC effective theory~\cite{Fujii:2013yja} and 
predictions including shadowing, $k_{\rm T}$ broadening and CNM energy 
loss~\cite{Sharma:2009hn}.
The data are fairly well described by the above mentioned models, which 
consider only initial-state effects.
The effect of the possible formation of a collectively expanding medium, as 
calculated with the blast-wave approach of~\cite{Sickles:2013yna}, is expected 
to be small for $D$ mesons in multiplicity-integrated collisions. 
The first results from two different transport models assuming the formation of 
a hot and deconfined medium in $p$Pb collisions at the LHC, which modifies 
the propagation and hadronization of heavy quarks, were recently 
published~\cite{Xu:2015iha,Beraudo:2015wsd}
In these frameworks, a small bump in the $R_{p\rm Pb}$ at low/intermediate $\pT$ 
due to radial flow is predicted, possibly accompanied by a moderate 
($<20-30\%$) suppression at high $\pT$, due to in-medium energy loss.
The models describe the data within uncertainties, even though the
measured $R_{p\rm Pb}$ disfavors a suppression larger than 15--20\% in the
transverse momentum interval $5<\pT<10~\GeV/c$.
However, the current uncertainties on the experimental and theoretical side 
do not allow us to discriminate between scenarios with only CNM effects 
or with CNM and hot medium effects.

The HF decay electron result, shown in the right panel
of Fig.~\ref{fig:RpADLHC}, is also consistent with unity within uncertainties 
over the whole $\pT$ range of the measurement. 
Given the large systematic uncertainties, the measured $R_{p\rm Pb}$ is also 
compatible with an enhancement in the transverse-momentum interval
$1<\pT<6~\GeV/c$, as observed at midrapidity in $d$Au collisions at
$\sqrtsNN=200~\GeV$ (see Fig.~\ref{fig:RdAuLeptonsRHIC}).
Similarly to what is observed for $D$ mesons, the data are described within 
uncertainties by pQCD calculations with nPDFs~\cite{Eskola:2009uj}, 
and by a model including shadowing, $k_{\rm T}$ broadening and CNM energy 
loss~\cite{Sharma:2009hn}, both of them predicting a small suppression
at low $\pT$.
Calculations based on incoherent multiple scatterings~\cite{Kang:2014hha} 
or on a blast-wave modeling of the possible establishment of collective 
flow~\cite{Sickles:2013yna} predict an enhancement in the nuclear 
modification factor at low $\pT$.
The uncertainties of the current measurements do not allow a discrimination
among these different theoretical approaches.

\begin{figure}
\begin{center}
\includegraphics[width=0.48\textwidth]{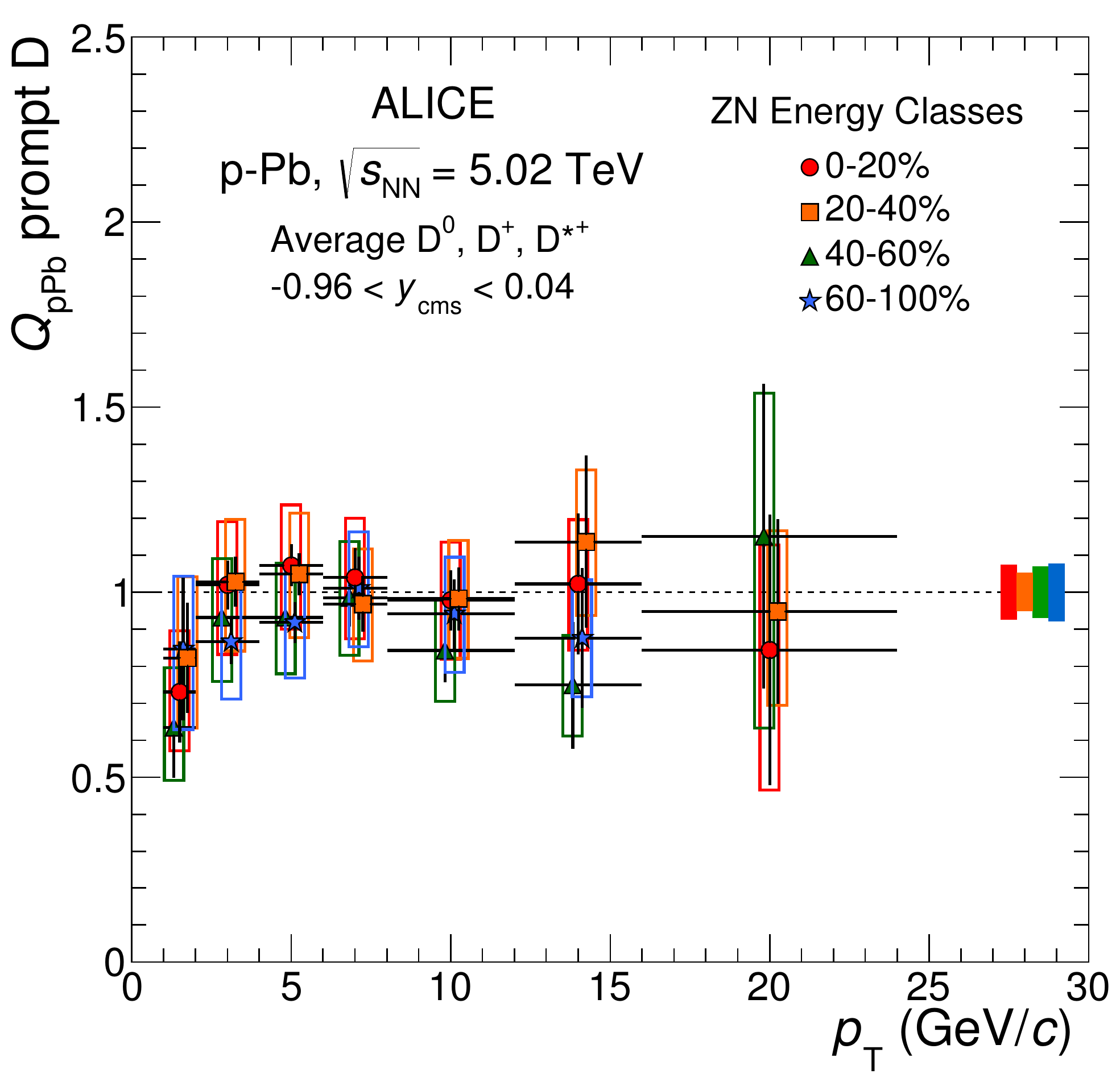}
\caption{Nuclear modification factor of $D$ mesons (average of
${D^0}$, ${D^+}$ and ${D^{*+}}$) as a function of $p_{\rm T}$ in different 
centrality classes for $p$Pb collisions at 
$\sqrt{s_{\rm NN}}=5.02~\mathrm{TeV}$~\cite{Adam:2016mkz}. 
The notation $Q_{p\rm Pb}$ is used for the nuclear modification factor 
in centrality classes to emphasize the possible presence of potential biases 
in the centrality estimation (see Ref.~\cite{Adam:2014qja})}.
\label{fig:QpADLHC}
\end{center}
\end{figure}

Results on $D$-meson production as a function of the $p$Pb collision 
centrality were recently published~\cite{Adam:2016mkz}.
In Fig.~\ref{fig:QpADLHC}, the nuclear modification factor, $Q_{p\rm Pb}$,
of prompt $D$ mesons (average of ${D^0}$, ${D^+}$ and ${D^{*+}}$) is shown
as a function of $p_{\rm T}$ for four different centrality classes,
defined from the energy deposited in the neutron zero-degree 
calorimeters.
This selection provides the least biased estimation of the
collision geometry, as discussed in detail in Ref.~\cite{Adam:2014qja}.
The results indicate that charm-hadron production is compatible with binary 
scaling of the $pp$ reference in all the considered centrality classes.
In particular, no evidence of a substantial modification of $D$-meson 
production with respect to $pp$ collisions is observed in the 20\% most 
central collisions, in which the multiplicity of produced particles is 
comparable to that in peripheral nucleus-nucleus collisions at RHIC/LHC
energies.
Nevertheless, considering the current statistical and systematic 
uncertainties, centrality-dependent effects of the order of 10\% 
cannot be excluded.

\begin{figure}
\begin{center}
$\vcenter{\hbox{\includegraphics[width=0.48\textwidth]{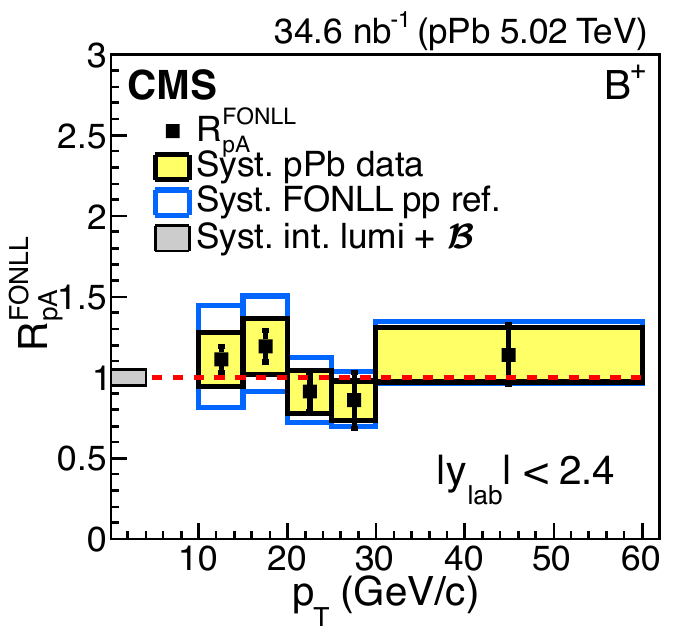}}}$
$\vcenter{\hbox{\includegraphics[width=0.48\textwidth]{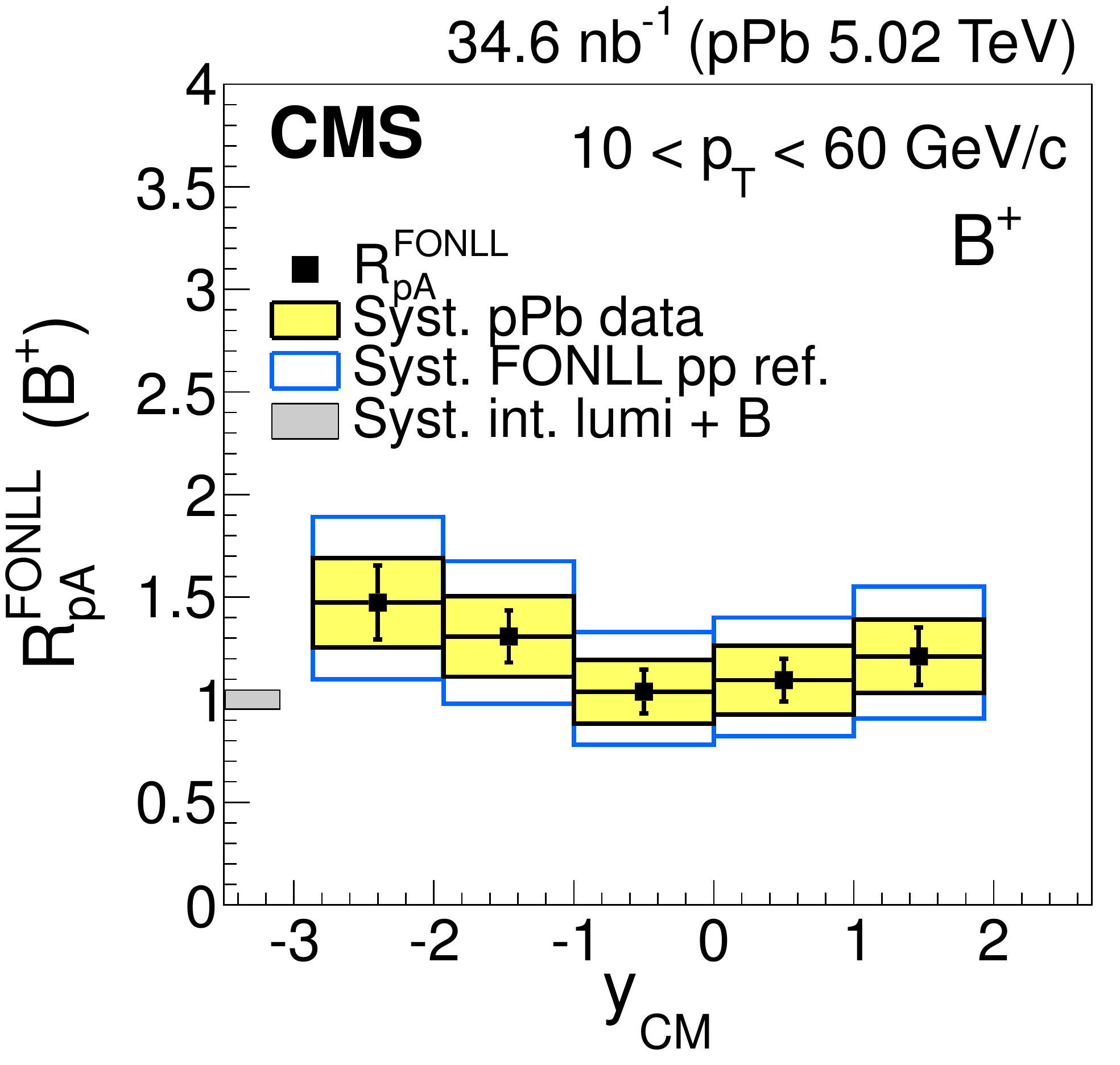}}}$
\caption{Nuclear modification factor of ${B^+}$ mesons as a function of
$\pT$ (left) and rapidity (right)  in $p$Pb collisions at
$\sqrt{s_{\rm NN}}=5.02~\mathrm{TeV}$~\cite{Khachatryan:2015uja}.
The $pp$ reference is taken from FONLL pQCD calculations.}
\label{fig:RpABLHC}
\end{center}
\end{figure}

The production of beauty mesons, namely ${B^0}$, ${B^+}$ and 
${B_s^0}$, was measured by the CMS 
collaboration~\cite{Khachatryan:2015uja} in $p$Pb collisions as a function
of transverse momentum in the range $10<\pT<60~\GeV/c$.
The production of ${B^+}$ mesons was also studied as a function of 
rapidity.
The ${B^+}$ nuclear modification factor, computed using FONLL pQCD 
calculations as the $pp$ reference, is shown in Fig.~\ref{fig:RpABLHC} as a 
function of $\pT$ (left panel) and rapidity (right panel).
No significant modification of $B$-meson production is observed in $p$Pb 
collisions compared to the binary-scaled FONLL reference over the measured 
$\pT$ range. 

\begin{figure}
\begin{center}
$\vcenter{\hbox{\includegraphics[width=0.46\textwidth]{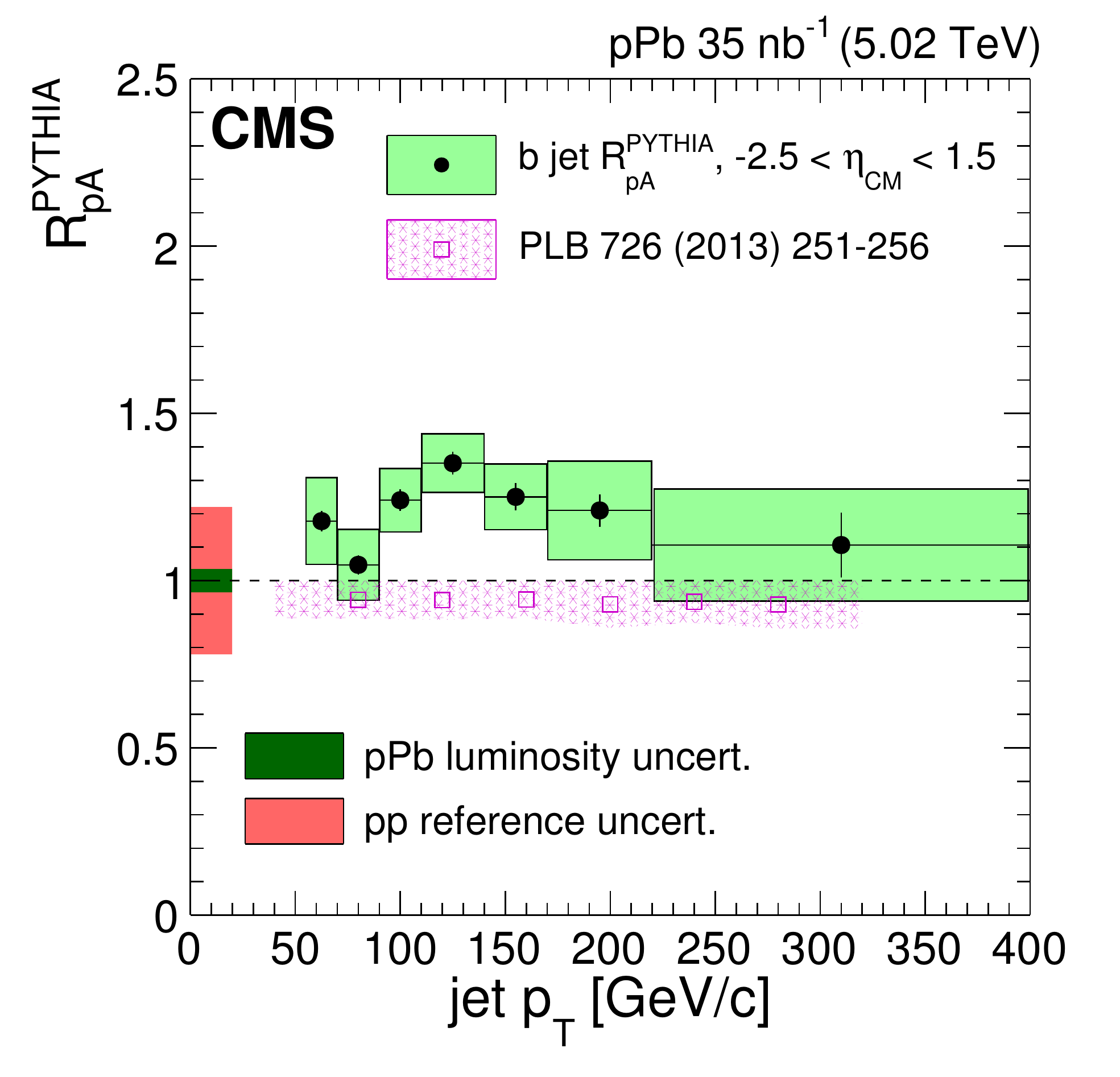}}}$
$\vcenter{\hbox{\includegraphics[width=0.5\textwidth]{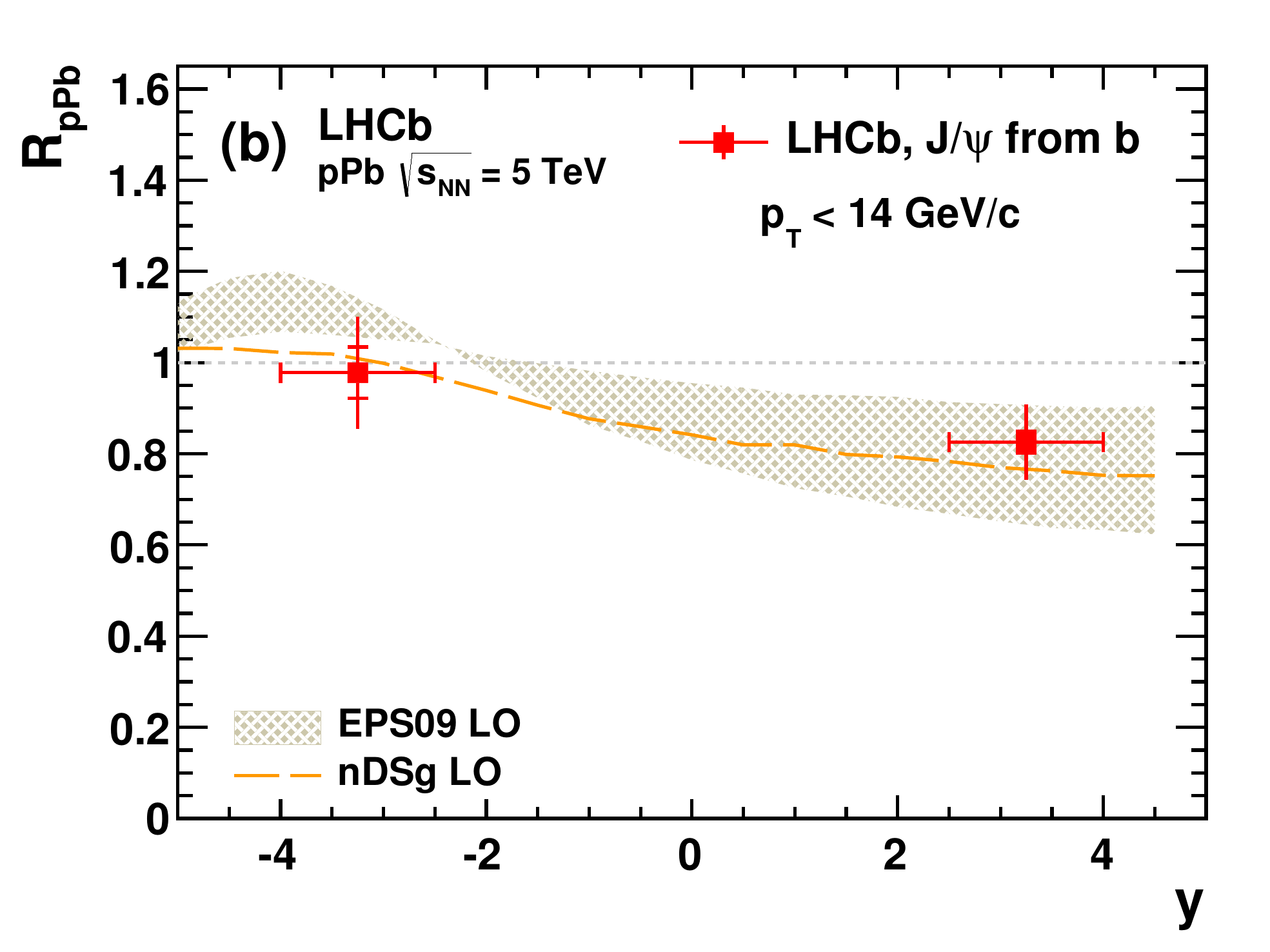}}}$
\caption{Nuclear modification of beauty production in $p$Pb collisions 
at $\sqrt{s_{\rm NN}}=5.02~\mathrm{TeV}$.
Left: $R_{p\rm Pb}$ of $b$-jets as a function of $\pT$ measured by
CMS~\cite{Khachatryan:2015sva}, compared to predictions of a model 
including initial-state energy loss~\cite{Huang:2013vaa}.
Right:  $R_{p\rm Pb}$ of $J/\psi$ from beauty-hadron decays as a function of $y$ 
measured by LHCb~\cite{Aaij:2013zxa}, compared to predictions 
of models with nuclear modification of the PDFs~\cite{delValle:2014wha}.
}
\label{fig:RpAbjetjpsiLHC}
\end{center}
\end{figure}

The CMS collaboration recently reported measurements of $b$-jet 
$\pT$-differential cross section and nuclear modification factor in $p$Pb 
collisions at $\sqrt{s_{\rm NN}}=5.02~\mathrm{TeV}$ in the 
jet transverse-momentum and pseudorapidity intervals  $55<\pT<400~\GeV/c$ and 
$2.5 < \eta_{\rm CMS}< 1.5$~\cite{Khachatryan:2015sva}.
As there are no $pp$ data available at the $CM$ energy of $p$Pb 
collisions, the $pp$ reference for the $R_{\rm pPb}$ calculation was obtained 
from PYTHIA simulations.
The discrepancies between PYTHIA and data observed at $\sqrt{s}=2.76$ and 7 TeV 
were accounted for in the systematic uncertainty.
The resulting $R_{p\rm Pb}^{\rm PYTHIA}$ as a function of $\pT$ is shown in the 
left panel of Fig.~\ref{fig:RpAbjetjpsiLHC}; it is compatible with unity 
within uncertainties,
especially considering the 22\% uncertainty on the PYTHIA reference.
The data are described within uncertainties by a pQCD model that includes 
modest initial-state energy-loss effects~\cite{Huang:2013vaa}.
Overall, the conclusions from the $b$-jet studies agree with those drawn from the 
measurements of $B$ meson production reported above.
Future measurements of $B$ mesons and $b$-jets in $pp$ collisions at 
$\sqrt{s}=5.02~\TeV$ are expected to provide a substantial reduction of the 
uncertainties on the $pp$ reference, thus enabling a more definite assessment 
of possible modifications of beauty production in $p$Pb collisions.

Beauty production in $p$Pb collisions at the LHC was also studied by the 
LHCb collaboration by measuring non-prompt $J/\psi$ at large rapidities 
($2<y_{\rm lab}<4.5$) down to $\pT=0$~\cite{Aaij:2013zxa}.
The $pp$ reference at $\sqrt{s}=5.02~\TeV$ was obtained via an interpolation of 
the measurements at $\sqrt{s}$ values of 2.76, 7 and 8 TeV.
The $\pT$-integrated $R_{p\rm Pb}$ as a function of rapidity is shown in the 
right panel of Fig.~\ref{fig:RpAbjetjpsiLHC} together with theoretical 
calculations.
The nuclear modification factor is compatible with unity at backward rapidity 
(Pb-going side).
In the forward rapidity region the data show a modest suppression relative to 
the binary-scaled $pp$ reference.
Perturbative QCD calculations at LO including EPS09 or nDSg nPDF 
parameterizations~\cite{delValle:2014wha} describe
the data well at forward rapidity, while in the backward region the agreement 
is not as good.

As pointed out in Ref.~\cite{Gauld:2015lxa}, the nuclear 
modification of the PDFs at small $x$ can be better constrained by means
of a data-to-theory comparison of the $\pT$-differential $D$-meson 
forward-to-backward ratio, \ie, the ratio of the $D$-meson cross sections in 
symmetric intervals at backward and forward rapidity,
in the kinematical region accessible with the LHCb apparatus.
In this ratio, the theoretical uncertainties due to pQCD scales and 
quark mass partially cancel, as well as some of the contributions to the 
experimental systematic uncertainties, thus providing improved sensitivity to 
CNM effects on the PDFs.

Further insight into possible modifications of HF production in
$d$Au and $p$Pb collisions can be obtained from measurements of azimuthal
correlations.
The angular correlation between HF hadrons reflects the
correlation between HQ pairs and is therefore sensitive to their
production mechanisms.
In proton(deuteron)--nucleus collisions, the CGC effective theory predicts,
in addition to a reduction of the overall particle yield, a broadening 
and suppression of the away-side peak (at $\Delta\varphi=\pi$)
in the two-particle azimuthal 
correlations~\cite{Gelis:2010nm,Marquet:2007vb,Lappi:2012nh}.
A depletion of the away-side yields in two-particle azimuthal correlations
is also expected to be induced by CNM energy loss and multiple scattering 
processes in the initial and final state~\cite{Kang:2011bp}. 
These effects could also affect HF angular correlations.
The azimuthal correlation between HF decay electrons at mid-rapidity
($\pT > 0.5~\GeV/c$, $|\eta|< 0.35$) and HF decay muons 
at forward rapidity ($\pT > 1~\GeV/c$, $1.4<\eta<2.1$) were measured
by the PHENIX collaboration in $pp$ and $d$Au collisions at 
$\sqrt{s}=200~\GeV$~\cite{Adare:2013xlp}.
The selection of a muon at forward rapidity allows to probe a low-$x$ 
range in the Au nucleus, where saturation effects are predicted to occur.
A suppression of the away-side peak ($\Delta\varphi=\pi$) is observed in $d$Au
collisions compared to $pp$, indicating that the charm-quark pair kinematics
is modified in the cold nuclear medium.
Preliminary results of $D$-hadron azimuthal correlations in $p$Pb collisions at 
the LHC were reported by the ALICE collaboration~\cite{Bjelogrlic:2014kia}. 
With the current level of uncertainties, no conclusion can be drawn
on a possible modification of $D$-hadron correlations with respect to $pp$ 
collisions.

Measurements of two-particle angular correlations in $p$A collisions can 
also be exploited to study whether also charm and beauty show $v_2$-like 
double-ridge long-range angular correlations in ``small systems'', as 
observed for 
light hadrons~\cite{CMS:2012qk,Abelev:2012ola,ABELEV:2013wsa,Aad:2012gla}.
An intriguing possible hint for a non-zero HF $v_2$ in high-multiplicity
$p$Pb collisions is provided by the ALICE measurements of angular 
correlations between a muon at forward rapidity and hadrons at 
midrapidity~\cite{Adam:2015bka}.
A positive $v_2$ is observed in the 20\% highest-multiplicity $p$Pb interactions
for muon tracks up to $\pT \approx 4~\GeV/c$.
Therefore, in the transverse momentum interval $\pT>2~\GeV/c$, where the 
inclusive muon yield is expected to be dominated by HF hadron decays, the data 
may support a finite value of HF $v_2$.

In summary of this section, the measurements in proton(deuteron)--nucleus collisions
show small to moderate modifications of HF production as compared to the
binary-scaled $pp$ references.
The magnitude of this modification depends on rapidity, $\pT$, and 
(at least at RHIC energy) collision centrality.
The measured values of charm and beauty nuclear modification factors can be 
described by models including CNM effects: nuclear modification of the PDFs, 
(gluon) saturation, transverse-momentum broadening, and initial-parton 
energy loss.
The uncertainties on the current experimental results, together with the large 
uncertainties on the parameterized nuclear PDFs at small Bjorken-$x$ prevent 
a more conclusive theoretical interpretation of the data, making it difficult 
to address in a more quantitative way the role of the various CNM effects 
in the initial and final state.


\subsection{Results from A-A Collisions and Model Comparisons}
\label{ssec_AAdata}
Charm and beauty quarks are sensitive probes of the properties of the
hot and dense medium created in heavy-ion collisions at ultra-relativistic 
energies.
They are predominantly produced in hard-scattering processes occurring
in the early stages of the collision, characterized by time scales
shorter than the expected formation times of the QGP medium.
Thermal production in the medium is expected to be small or negligible at
the temperatures attained in heavy-ion collisions at RHIC and at the 
LHC~\cite{BraunMunzinger:2000dv,Zhang:2007dm}.
Their total yield is therefore essentially set by the yield in
$pp$ collisions, which, as discussed in Section~\ref{sssec_ppdata}, is 
described by pQCD calculations within current uncertainties, further modified 
by nuclear corrections to the PDFs (see Section~\ref{sssec_pAdata}) and scaled 
by the number of binary nucleon--nucleon collisions ($N_{\rm coll}$) 
occurring in the nucleus--nucleus interaction.
The initially produced heavy quarks interact with the constituents of the 
medium through the exchange of energy and momentum.
At sufficiently high $\pT$ ($\pT \gtsim 5(15)~\GeV/c$ for charm (bottom)), the 
main effect is 
that heavy quarks lose energy while traversing the medium (although the precise 
characterization of the transition to the energy-loss regime is still an open 
question and is expected to depend on collision energy, centrality, quark flavor, 
etc.). The energy loss can occur via both inelastic (radiative)
and elastic (collisional)
processes, resulting in a suppression of the yield of HF hadrons 
(and their decay leptons) at high $\pT$ as compared to the binary-scaled $pp$ 
reference (recall Secs.~\ref{sssec_qgp} and \ref{sssec_qgp-rad}).  
The interest in HF studies in the high-$\pT$ regime is mostly related to the 
predicted color-charge and quark-mass dependence of in-medium parton energy 
loss, expected to lead to a hierarchy where  beauty quarks lose less energy 
than charm quarks, and the latter less energy than light quarks and gluons.
At lower $p_T$ ($\pT\ltsim5(15)~\GeV/c$ for charm (bottom)), measurements of HF 
particles are sensitive to other aspects of the interactions of charm and beauty 
quarks with the medium. Low-momentum heavy quarks, including those shifted to low 
momentum by energy loss, are expected to couple to the collective expansion 
(flow) of the system and approach local thermal equilibrium with the 
medium~\cite{Batsouli:2002qf,Greco:2003vf}. 
It is also predicted that a significant fraction of low- and 
intermediate-momentum heavy quarks hadronizes via recombination with 
other quarks from the medium~\cite{Greco:2003vf,Andronic:2003zv} (as
discussed in Sec.~\ref{sssec_hadro}).
These questions can be addressed via the study of the nuclear modification
factor, $\Raa$, and of the azimuthal anisotropy, in particular the elliptic flow 
$v_2$, of HF hadrons and their decay leptons at low and 
intermediate $\pT$ (smaller than roughly five times the HQ mass).

\begin{figure}
\begin{center}
$\vcenter{\hbox{\includegraphics[width=0.46\textwidth]{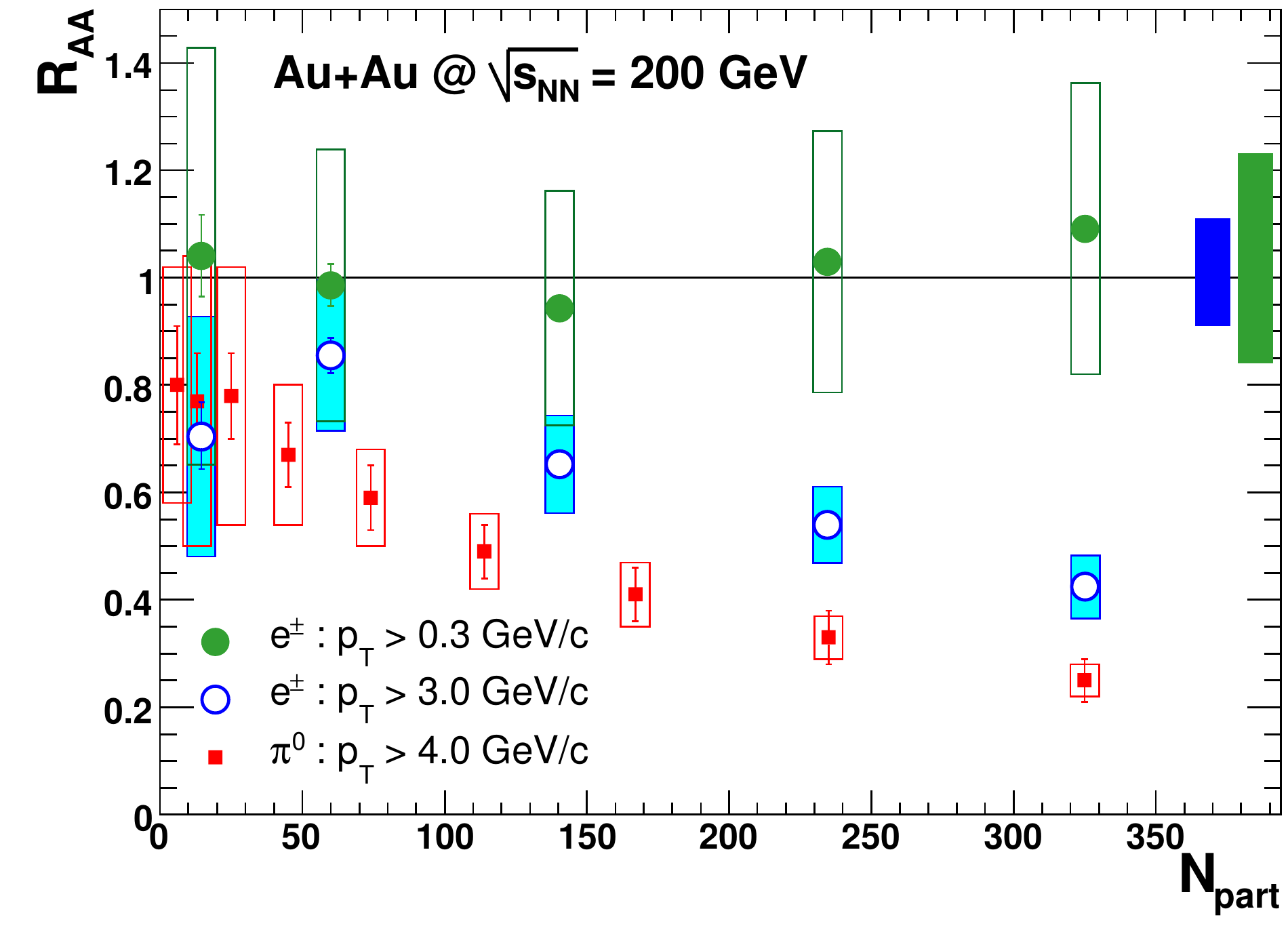}}}$
$\vcenter{\hbox{\includegraphics[width=0.49\textwidth]{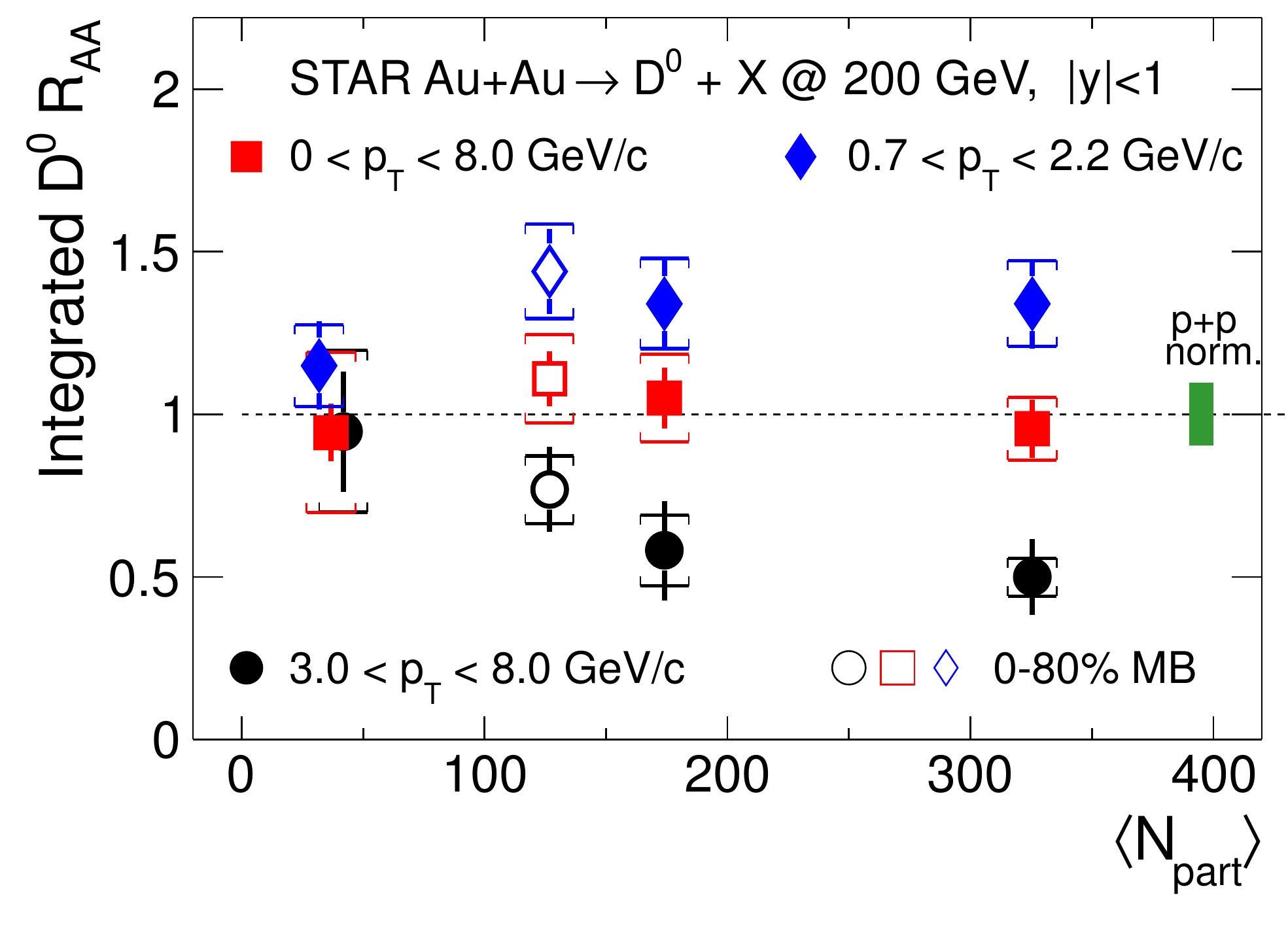}}}$
\caption{Centrality dependence of the nuclear modification factor $\Raa$ of HF 
decay electrons (left, from PHENIX~\cite{Adare:2006nq}) and ${D^0}$ mesons (right, 
from STAR~\cite{Adamczyk:2014uip})
at mid-rapidity in different $\pT$ intervals.
Also shown in the left panel is the $\Raa$ of $\pi^0$ with 
$\pT>4~\GeV/c$.}
\label{fig:RaaCentRHIC}
\end{center}
\end{figure}

All final-state effects described above, which are a consequence of the 
interactions of the heavy quarks with the hot and dense medium, influence 
the momentum distribution of charm (beauty) quarks, but they have little or no 
effect on the total yield of HF hadrons.
For this reason, the $\pT$-integrated yield is expected to be consistent with 
$N_{\rm coll}$ scaling of the yield measured in $pp$ reactions, apart
from effects due to nuclear modifications of the PDFs affecting the production
process.
This is confirmed by the results shown in Fig.~\ref{fig:RaaCentRHIC} where
the nuclear modification factor of HF decay 
electrons~\cite{Adare:2006nq} and ${D^0}$ mesons~\cite{Adamczyk:2014uip}
measured in Au--Au collisions at a $CM$ energy of 
$\sqrtsNN=200~\GeV$ per nucleon pair, are shown as a function of the 
collision centrality, expressed in
terms of the number of participant nucleons $N_{\rm part}$, for different 
$\pT$ intervals.
The production of HF decay electrons with $\pT>0.3~\GeV/c$,
which measures a large part of the total charm production yield,
and of ${D^0}$ mesons with $0<\pT<8~\GeV/c$, are consistent with a scaling 
with the number of binary collisions (corresponding to $\Raa=1$) within 
experimental uncertainties, as expected for initial HQ production from 
hard-scattering processes.
In contrast, the yield at high $\pT$ shows a clear suppression  ($\Raa<1$), 
which increases from peripheral to central collisions, following
the qualitative expectation from in-medium charm (and beauty) quark energy loss.

At present, HF production has not been measured down to $\pT=0$ 
in Pb--Pb collisions at the LHC.
Hence, no results of total charm (beauty) production in heavy-ion collisions 
at LHC energies are currently available to check the binary scaling.
It is worth to point out that the nuclear modification of the PDF, which
has a small effect on the total charm production at RHIC energies, is
expected to induce a $\sim$20\% reduction of initial charm production at
$\sqrtsNN=2.76~\TeV$,, according to NLO calculations~\cite{Mangano:1991jk} 
with the EPS09 parameterization of nPDFs~\cite{Eskola:2009uj}.
This is due to the smaller values of Bjorken-$x$ probed at the LHC, \ie, 
a larger expected shadowing, as compared to lower collision energies.
Hence, the $\pT$-integrated $\Raa$ of charm hadrons is expected to be 
lower than unity at LHC energies.

In the remainder of this section, the results from  measurements at RHIC 
and at the LHC will be briefly summarized, focusing on the $\pT$ dependence 
of the nuclear modification factor $\Raa$ and the elliptic flow $v_2$.
The overview of the experimental results is organized as follows.
In the first subsection, the results from measurements of HF decay leptons 
at different collision energies are reviewed. Then, the $D$-meson 
measurements at RHIC and at the LHC are discussed and compared to results 
for light-flavor hadrons.
Finally, the experimental results in the beauty sector from the LHC experiments
are reported together with a comparison of the nuclear modification factors 
of charm and beauty hadrons.
The presentation of the experimental results is accompanied by a discussion of 
their consequences for the characterization of the produced QCD matter and the 
estimation of its transport coefficients. The main ingredients of the 
different theoretical model calculations utilized in this discussion are 
summarized in Table~\ref{tab_models}.

\begin{table}[!t]
\label{tab_models}
\scriptsize
\centering
\begin{tabular}{ccccccc}
\hline
Model & Heavy-quark & nPDFs & Medium    & Quark-medium & Hadroni- & Hadron   \\
      & production  &       & modelling & interactions & zation              & phase\\
\hline
\multicolumn{7}{c}{Transport models}\\
\hline
BAMPS                 & MC@NLO & No    & Boltzmann & Boltzmann & frag & no \\
\cite{Uphoff:2011ad,Uphoff:2012gb,Uphoff:2014hza} & & & parton 3+1D & pQCD coll+rad  &      & \\
\\
Cao {\it et al.}/Duke & MC@NLO & EPS09 & Hydro 2+1D & Langevin & frag+ & yes \\ 
\cite{Cao:2013ita,Cao:2014pka,Cao:2015hia} & & & viscous & coll+pQCD rad & reco  &  \\
\\
MC@sHQ+EPOS & FONLL & EPS09 & Hydro 3+1D & Boltzmann & frag+ & no \\
\cite{Gossiaux:2008jv,Gossiaux:2010yx,Nahrgang:2013xaa} & & & (EPOS) & pQCD coll+rad & reco  & \\
\\
PHSD & PYTHIA & EPS09 & off-shell parton& off-shell trans  & frag+ & yes \\ 
\cite{Song:2015sfa,Song:2015ykw} & & & transport & pQCD coll & reco  & \\
\\
POWLANG & POWHEG & EPS09 & Hydro 2+1D & Langevin & string- & no\\
\cite{Alberico:2011zy,Alberico:2013bza,Beraudo:2014boa} & & & viscous & pQCD coll& reco& \\
\\
TAMU & FONLL & EPS09 & Hydro 2+1D & Langevin & frag+ & yes\\
\cite{He:2011qa,He:2011zx,He:2014cla} & & & ideal & T-mat coll & reco  & \\
\\
\hline
\multicolumn{7}{c}{Energy-loss models}\\
\hline
\\
AdS/CFT (HG) & FONLL & No & Glauber & AdS/CFT & frag & no \\
\cite{Horowitz:2007su,Horowitz:2011wm} & & & no hydro & drag & & \\
\\
CUJET 3.0 & FONLL & No & Hydro 2+1D & rad+coll & frag & no \\
\cite{Xu:2014tda,Xu:2015bbz} & & & viscous & \\
\\
Djordjevic {\it et al.} & FONLL & No & Glauber & rad+coll+ & frag & no \\
\cite{Djordjevic:2013pba,Djordjevic:2014tka} & &  & no hydro & magn.~mass & &\\
\\
Vitev {\it et al.} & non-zero mass & No & Glauber+ & rad+ & frag & no \\
\cite{Adil:2006ra,Sharma:2009hn} & VFNS & & 1D Bjorken exp & in-med dissoc & &\\
\\
WHDG & FONLL & No & Glauber & rad+coll  & frag & no \\
\cite{Wicks:2007am,Wicks:2005gt} & &  & no hydro & \\
\hline
\end{tabular}
\caption{Overview of the main features of models of heavy-quark in-medium energy loss
and transport; see Sec.~\ref{sec_theo} for more details. The non-standard acronyms are:
coll=collisional, rad=radiative, frag=fragmentation, reco=recombination, dissoc=dissociation, 
exp=expansion.}
\end{table}

\subsubsection{Heavy-flavor decay leptons}
\hspace{2cm}

Measurements of HF decay leptons in heavy-ion collisions
have been carried out at RHIC and at the LHC by the STAR, PHENIX, ALICE and ATLAS 
collaborations at mid- and forward rapidity, exploiting both  
semi-electronic and the semi-muonic decay modes.

The first measurements of HF decay electron spectra in Au--Au
reactions at RHIC were performed by PHENIX on the data 
sample recorded during the first RHIC run at a $CM$ energy of $\sqrtsNN=130~\GeV$ 
per nucleon pair~\cite{Adcox:2002cg}.
The limited precision of the measurement prevented from drawing conclusions
about possible modifications of charm production in heavy-ion collisions
relative to a binary-scaled $pp$ reference.
The larger data sample of Au--Au collisions at $\sqrtsNN=200~\GeV$
collected in year 2001 allowed the measurement of the centrality
and $\pT$ dependence of the yield of electrons from HF hadron decays 
at mid-rapidity~\cite{Adler:2004ta,Adler:2005xv}.
The measurements of the nuclear modification factor of 
HF decay electrons, reported in~\cite{Adler:2005xv}, showed
a substantial suppression for $\pT>~2\GeV/c$ in central Au--Au collisions
relative to the expectation based on binary scaling of the yields measured in 
$pp$ collisions.
The magnitude of the observed suppression turned out
to be compatible with that measured for neutral pions and
was larger than expected in radiative-energy loss calculations,
posing a challenge to this class of models.
More precise measurements were carried out by the 
PHENIX~\cite{Adare:2006nq,Adare:2010de} and STAR~\cite{Abelev:2006db} 
collaborations utilizing the larger data samples of Au--Au collisions
at $\sqrtsNN=200~\GeV$ collected during the RHIC run-4.
In addition, PHENIX published results from a sample of Cu--Cu collisions at 
$\sqrtsNN=200~\GeV$ collected in run-5~\cite{Adare:2012px} and in Au--Au 
collisions at $\sqrtsNN=62.4~\GeV$ from run-10~\cite{Adare:2014rly}.
The STAR collaboration published a study of $v_2$ in Au--Au collisions
at different energies utilizing the run-10 data samples.
Recently, the PHENIX collaboration reported first results of separated
yields of single electrons from charm- and beauty-hadron decays from the 
sample of Au--Au collisions at $\sqrtsNN=200~\GeV$ collected in run-11
after the installation and commissioning of the vertex 
detector~\cite{Adare:2015hla}.
During the first run at the LHC (years 2009-2013), HF decay muon 
$\Raa$ and $v_2$ were measured at forward rapidity by the ALICE 
collaboration~\cite{Abelev:2012qh,Adam:2015pga}. 
Preliminary results for HF decay electrons and muons at mid-rapidity
were reported by ALICE~\cite{Sakai:2013ata} and ATLAS~\cite{ATLAS-HFM}.

\begin{figure}
\begin{center}
\includegraphics[width=0.48\textwidth]{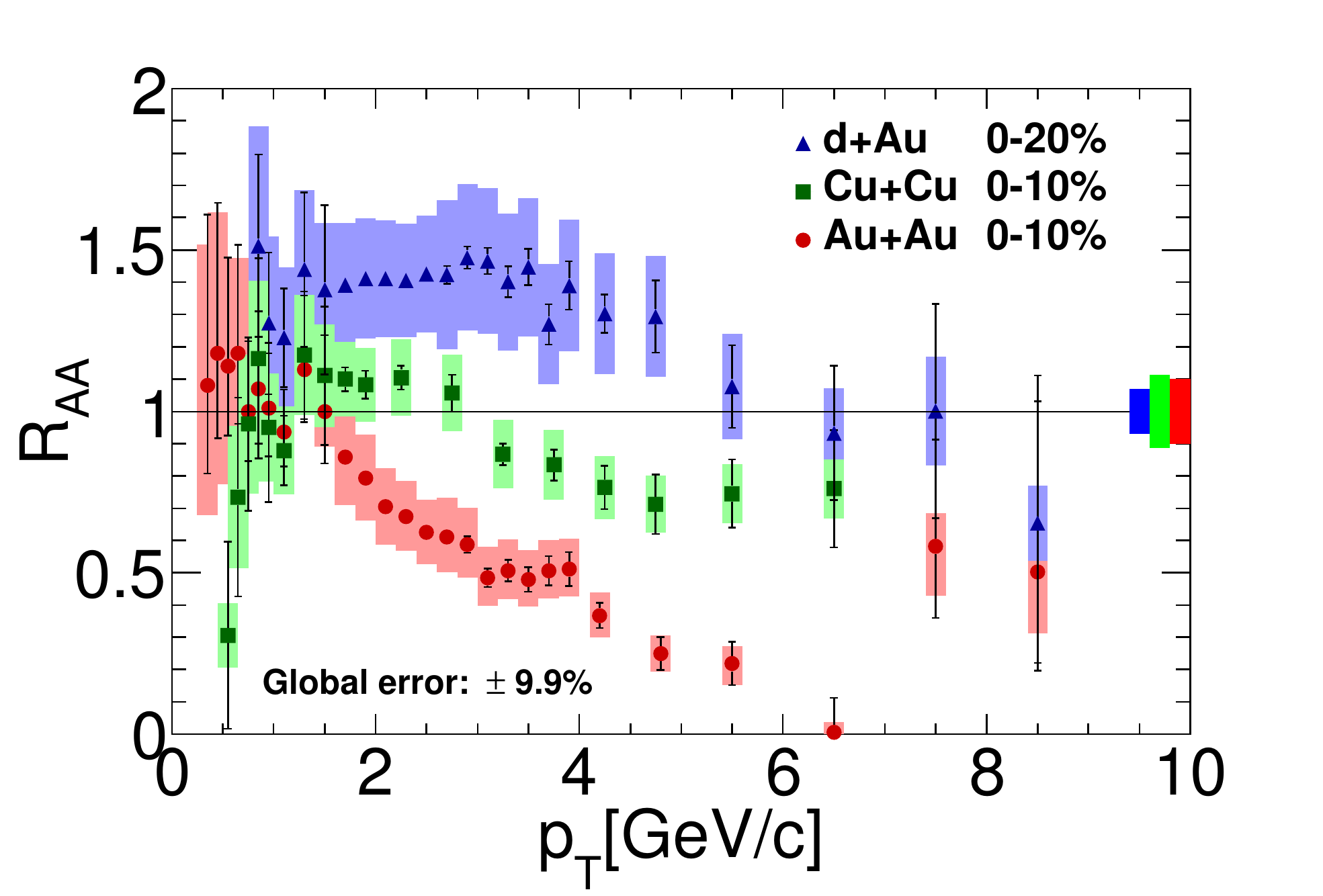}
\includegraphics[width=0.48\textwidth]{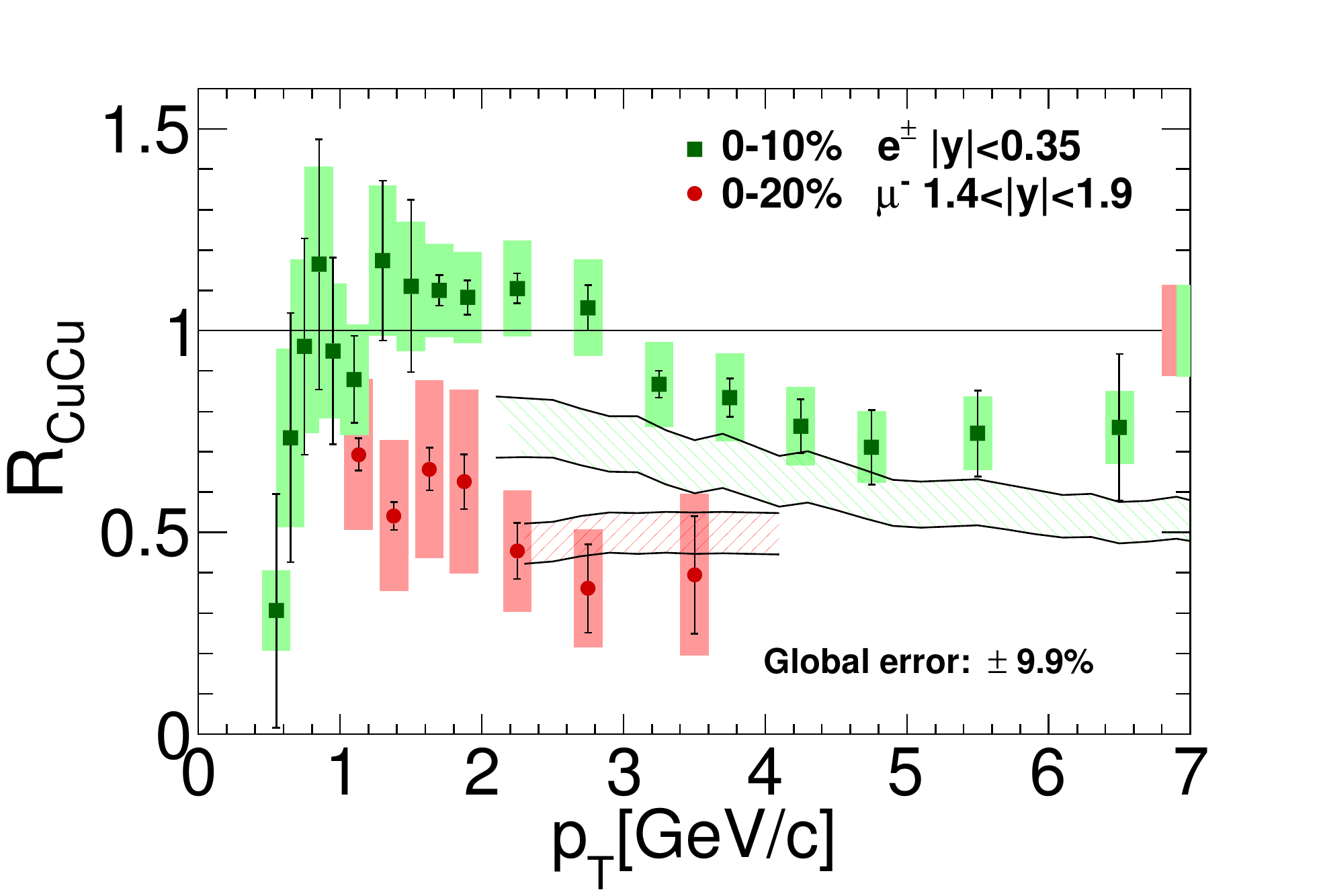}
$\vcenter{\hbox{\includegraphics[width=0.48\textwidth]{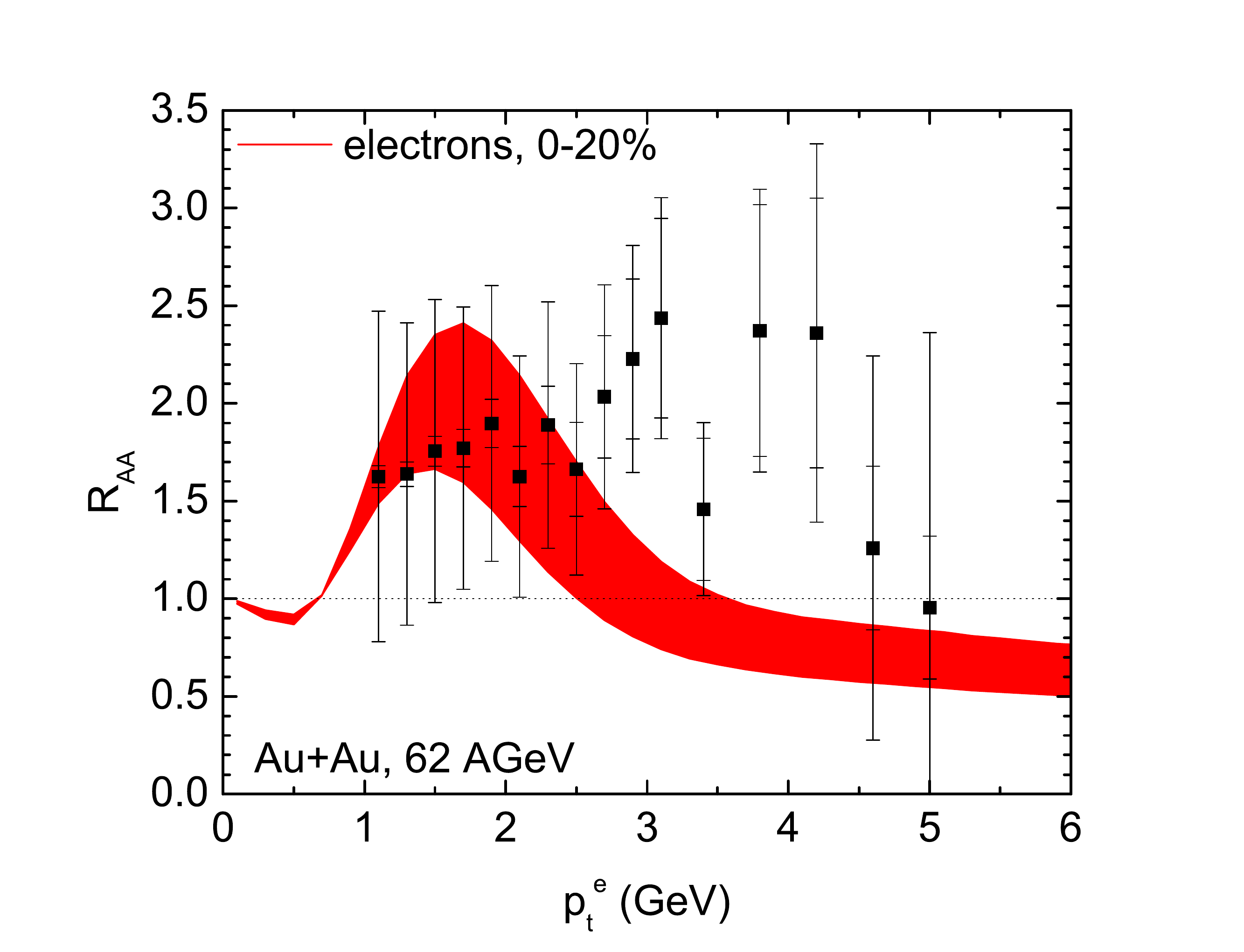}}}$
$\vcenter{\hbox{\includegraphics[width=0.48\textwidth]{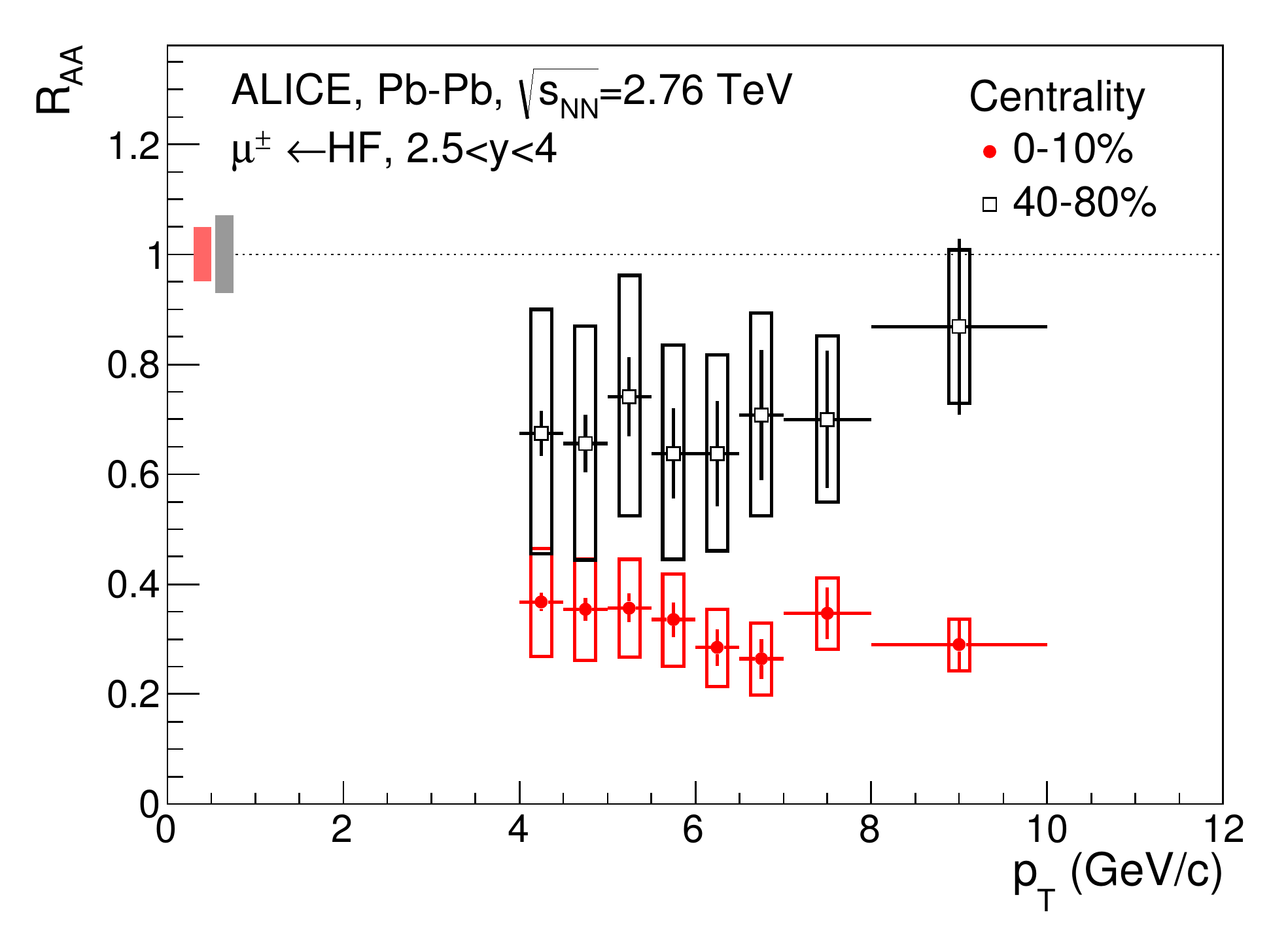}}}$
\caption{Nuclear modification factor of HF decay leptons as
a function of $\pT$ at RHIC and at the LHC.
Top left: Heavy-flavor decay electrons at mid-rapidity in central 
$d$Au~\cite{Adare:2012yxa}, Cu--Cu~\cite{Adare:2013yxp} and Au--Au 
collisions~\cite{Adare:2010de} at $\sqrtsNN = 200~\GeV$.
Top right:  Heavy-flavor decay leptons at mid- (electrons) and 
forward (muons) rapidity in central Cu--Cu collisions at 
$\sqrtsNN = 200~\GeV$~\cite{Adare:2012px} compared to model 
calculations~\cite{Sharma:2009hn}.
Bottom left: Heavy-flavor decay electrons at mid-rapidity in central 
Au--Au collisions at $\sqrtsNN = 62.4~\GeV$~\cite{Adare:2014rly},
compared to the TAMU model calculation~\cite{He:2014epa}.
Bottom right: Heavy-flavor decay muons at forward rapidity in central and 
peripheral Pb--Pb collisions at $\sqrtsNN = 2.76~\TeV$~\cite{Abelev:2012qh}.}
\label{fig:RaaHFLPt}
\end{center}
\end{figure}

A selection of results on the $\pT$ dependence of the nuclear
modification factor of HF decay leptons for different colliding
systems and collision energies and centralities is collected  
in  Fig.~\ref{fig:RaaHFLPt}.
In the top-left panel, the results on HF decay electron $\Raa$ at 
midrapidity in central Au--Au, Cu--Cu and $d$Au collisions at 
$\sqrtsNN=200~\GeV$ are compared~\cite{Adare:2013yxp}.
A substantial suppression of the yield of electrons with $\pT>1.5~\GeV/c$
is observed in the 10\% most central Au--Au collisions~\cite{Adare:2010de}.
The suppression increases with increasing $\pT$, reaching a factor of about 
four at $\pT=4~\GeV/c$.
The results from $d$Au collisions~\cite{Adare:2012yxa}, showing a nuclear 
modification factor consistent or larger than unity, provide clear evidence 
that the high-$\pT$ suppression observed in central Au--Au collisions is a
final-state effect due to the formation of a hot and dense medium.
In the 10\% most central Cu--Cu collisions a moderate suppression of 
the HF decay electron yield is observed for $\pT >3~\GeV/c$. 
The magnitude of the suppression is smaller than that observed in central
Au--Au collisions, as expected from the smaller size of the system created
in the collisions of the lighter Cu nuclei.

In Cu--Cu collisions at $\sqrtsNN=200~\GeV$, HF decay muons were 
measured at forward rapidity ($1.4<y<1.9$)~\cite{Adare:2012px}.
The results are reported in the top-right panel of Fig.~\ref{fig:RaaHFLPt}
and compared to the mid-rapidity results.
Open HF production is found to be significantly
more suppressed at forward rapidity than at midrapidity.
The magnitude of the suppression observed in Cu--Cu collisions
at forward rapidity is comparable to that in central Au--Au
collisions at midrapidity.
This observation suggests that in-medium energy loss is not the only 
mechanism responsible for the observed suppression, as the medium size
and density are larger in Au--Au than in Cu--Cu collisions.
Other nuclear effects, such as gluon shadowing at small Bjorken-$x$ or partonic 
energy loss in cold nuclear matter, can  be relevant for the description 
of the forward-rapidity results.
The gross features of the data are caught by the Vitev {\it et al.}
model calculations~\cite{Sharma:2009hn}, shown in the top-right 
panel of Fig.~\ref{fig:RaaHFLPt}, which includes in-medium energy loss due to 
gluon radiation and in-medium hadron formation and dissociation, as well 
as cold-nuclear matter effects such as shadowing and $k_{\rm T}$-broadening.

The HF decay electron $\Raa$ in the 20\% most central Au--Au
collisions at $\sqrtsNN=62.4~\GeV$ is shown in the bottom-left panel of
Fig.~\ref{fig:RaaHFLPt}~\cite{Adare:2014rly}.
Since a sample of $pp$ collisions at RHIC is not available for this 
$CM$ energy, the $pp$ reference was taken from ISR data.
The $\Raa$ is found to be consistently larger than unity and
no suppression is observed in the measurement $\pT$ range.
In contrast to the HF results, the $\pi^0$ measurements at this lower 
collision energy show a suppression that increases with 
centrality~\cite{Adare:2012uk}.
Measurements at this lower collision energy offer the possibility to
study HF production in a situation in which the initial temperature
of the medium is reduced as compared to top RHIC energy, while still 
encompassing the transition region.
This could shed light on the question whether the HQ coupling
to the medium is primarily driven by an increasing temperature (or energy 
density), or by an increase in coupling strength in the pseudo-critical region
of the chiral/deconfinement transition~\cite{He:2014epa}.
The measured HF decay electron nuclear modification factor can
be described within uncertainties by the TAMU model~\cite{He:2014epa}, shown 
as a red band in the bottom-left panel of Fig.~\ref{fig:RaaHFLPt}.
In this approach, the $\Raa$ pattern at $\sqrtsNN=62.4~\GeV$ emerges
from the interplay of initial- and final-state effects, in particular the
partial thermalization of heavy quarks in the hot medium (starting
from a softer initial-production spectrum than at $\sqrtsNN=200~\GeV$)
and a Cronin enhancement, which is known to become more pronounced toward 
lower collision energies.

In the bottom-right panel of Fig.~\ref{fig:RaaHFLPt}, the $\Raa$ of
HF decay muons at forward rapidity ($2.5<y<4$) is shown for central
and peripheral Pb--Pb collisions at the LHC energy of $\sqrtsNN=2.76~\TeV$.
A strong suppression, by a factor of 3-4, is observed in the 10\% most central
collisions in the measurement $\pT$ range ($4<\pT<10~\GeV/c$), without a 
significant $\pT$ dependence within uncertainties.
A smaller suppression is observed for peripheral (40--80\%) collisions. 
The higher collision energy allowed a precise determination of the nuclear 
modification factor in the momentum interval $\pT>6~\GeV/c$, which was
accessible with limited statistical precision at RHIC energies.
In this high-$\pT$ interval, according to the central value of FONLL pQCD 
calculations, the dominant contribution to the HF muon yield is 
due to beauty-hadron decays.

\begin{figure}
\begin{center}
$\vcenter{\hbox{\includegraphics[width=0.48\textwidth]{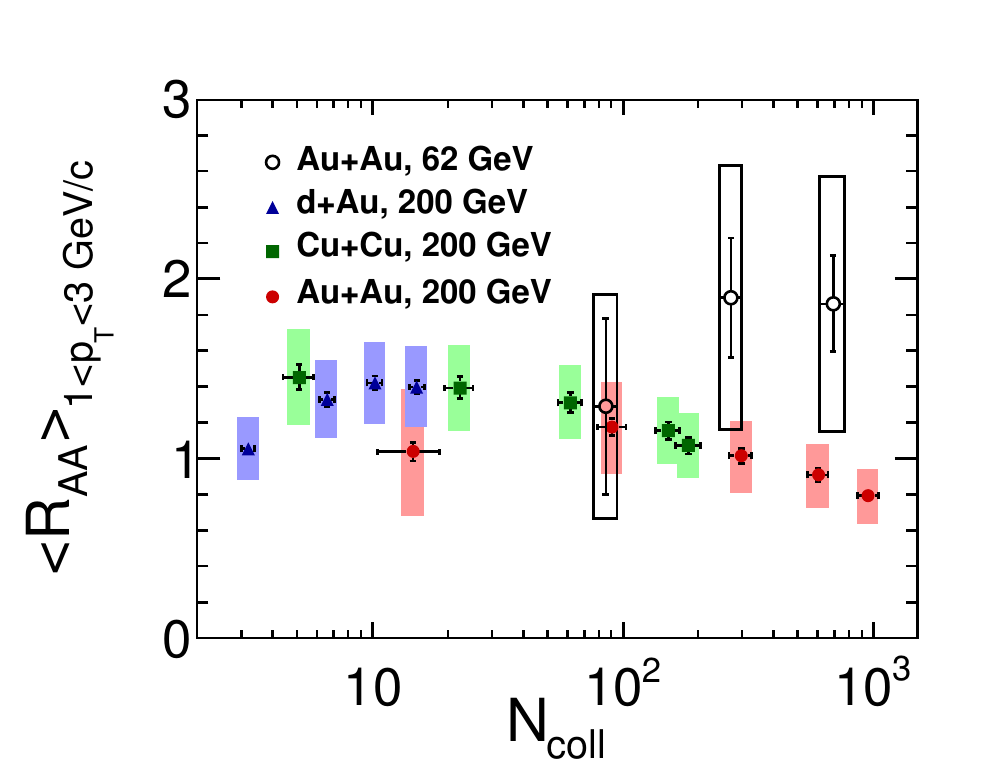}}}$
$\vcenter{\hbox{\includegraphics[width=0.48\textwidth]{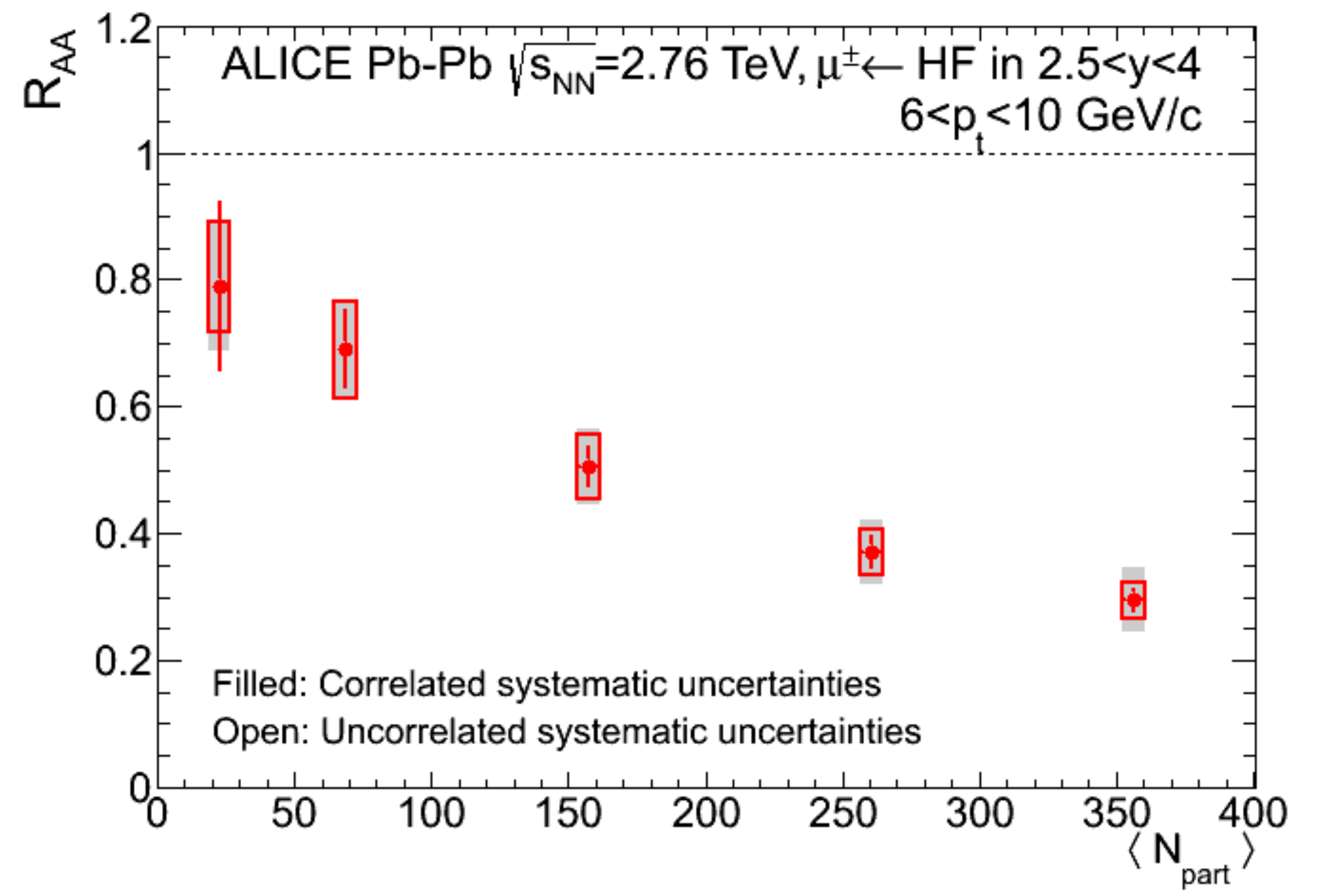}}}$
\caption{System size dependence of HF decay lepton $\Raa$.
Left: Heavy-flavor decay electrons in $1<\pT<3~\GeV/c$
in $d$Au, Cu--Cu and Au--Au collisions at RHIC energies
as a function of the average number of nucleon--nucleon collisions
~\cite{Adare:2013yxp,Adare:2014rly}.
Right: Heavy-flavor decay muons in $6<\pT<10~\GeV/c$ in Pb--Pb
collisions at the LHC as a function of the average number of participant
nucleons~\cite{Abelev:2012qh}.
}
\label{fig:RaaHFLcent}
\end{center}
\end{figure}

The system size dependence of the nuclear modification factor can be
investigated by comparing results at different centralities and in different
colliding systems.
It was observed in Ref.~~\cite{Adare:2013yxp} that when results from $d$Au, Cu--Cu 
and Au--Au collisions are compared in centrality intervals of comparable 
system size (\ie, similar average number of participant nucleons or
binary collisions), similar trends and magnitudes are found for the $\Raa$ 
as a function of $\pT$.
This is illustrated in the left panel of Fig.~\ref{fig:RaaHFLcent},
where the HF decay electron $\Raa$ in the range $1<\pT<3~\GeV/c$ from
$d$Au, Cu--Cu and Au--Au collisions is compiled as a function of 
$\left\langle N_{\rm coll} \right\rangle$.
The data at $\sqrtsNN=200~\GeV$ indicate a common trend among the three 
different systems, showing an enhancement which increases with increasing 
system size at low $\left\langle N_{\rm coll} \right\rangle$, followed by 
suppression for larger system sizes.
This common trend suggests that the enhancement and suppression effects are 
dependent on the size of the colliding system and the produced medium
and are the result of the interplay between CNM (nPDF, Cronin enhancement) 
and hot-medium (energy-loss, radial flow) effects.
The results at $\sqrtsNN=62.4~\GeV$, also shown in the left panel of 
Fig.~\ref{fig:RaaHFLcent}, suggest that at this lower collision energy the 
competition among different effects (Cronin enhancement, flow 
and energy loss) favors HF enhancement over suppression, 
consistently with previous observations of an increased Cronin enhancement 
with decreasing collision energy~\cite{Adler:2006xd}.
Studies of the system size dependence in Pb--Pb collisions at the LHC are shown
in the right panel of Fig.~\ref{fig:RaaHFLcent}, which reports
the $\Raa$ of HF decay muons for $2.5<y<4$ and $6<\pT<10~\GeV/c$, 
where, according to FONLL calculations, the beauty contribution is expected to 
be dominant.
A trend of increasing suppression with increasing centrality is observed, 
qualitatively similar to that found in Au--Au collisions at RHIC, suggesting 
that HQ in-medium energy loss dominates over other cold and hot medium effects.

The nuclear modification factor of electrons from HF hadron decays 
cannot be compared directly to that of light hadrons at the same $\pT$,
because of the kinematics of the semi-leptonic decay.
According to PYTHIA simulations~\cite{Sjostrand:2006za}, HF 
decay electrons with $\pT>3~\GeV/c$ originate to a large extent from the 
decay of $D$ mesons with $\pT>4~\GeV/c$, with the beauty contribution expected 
to be small at low $\pT$ according to FONLL calculations.
For this reason, in the left panel of Fig.~\ref{fig:RaaCentRHIC},
the $\Raa$ of HF decay electrons with $\pT>3~\GeV/c$
is compared to that of $\pi^{0}$  for $\pT>4~\GeV/c$.
In this intermediate $\pT$ range, the data suggest a smaller suppression of 
HF hadrons as compared to light-flavor mesons.
In addition, in order to draw conclusions on the parton energy loss
starting from the measurements of charm and light-flavor hadron $\Raa$, one
should also consider the effects of the different momentum distributions
of the initially produced charm quarks as compared to light quarks and gluons,
their different fragmentation functions into hadrons, as well as the
different initial-state effects on light and heavy quarks 
(\eg, the different Cronin enhancement of hadrons with different 
mass~\cite{Adler:2006xd,Adare:2012yxa}).
This will be discussed in more detail in the next sub-section where the
$D$-meson measurements are presented and discussed.

\begin{figure}
\begin{center}
$\vcenter{\hbox{\includegraphics[width=0.48\textwidth]{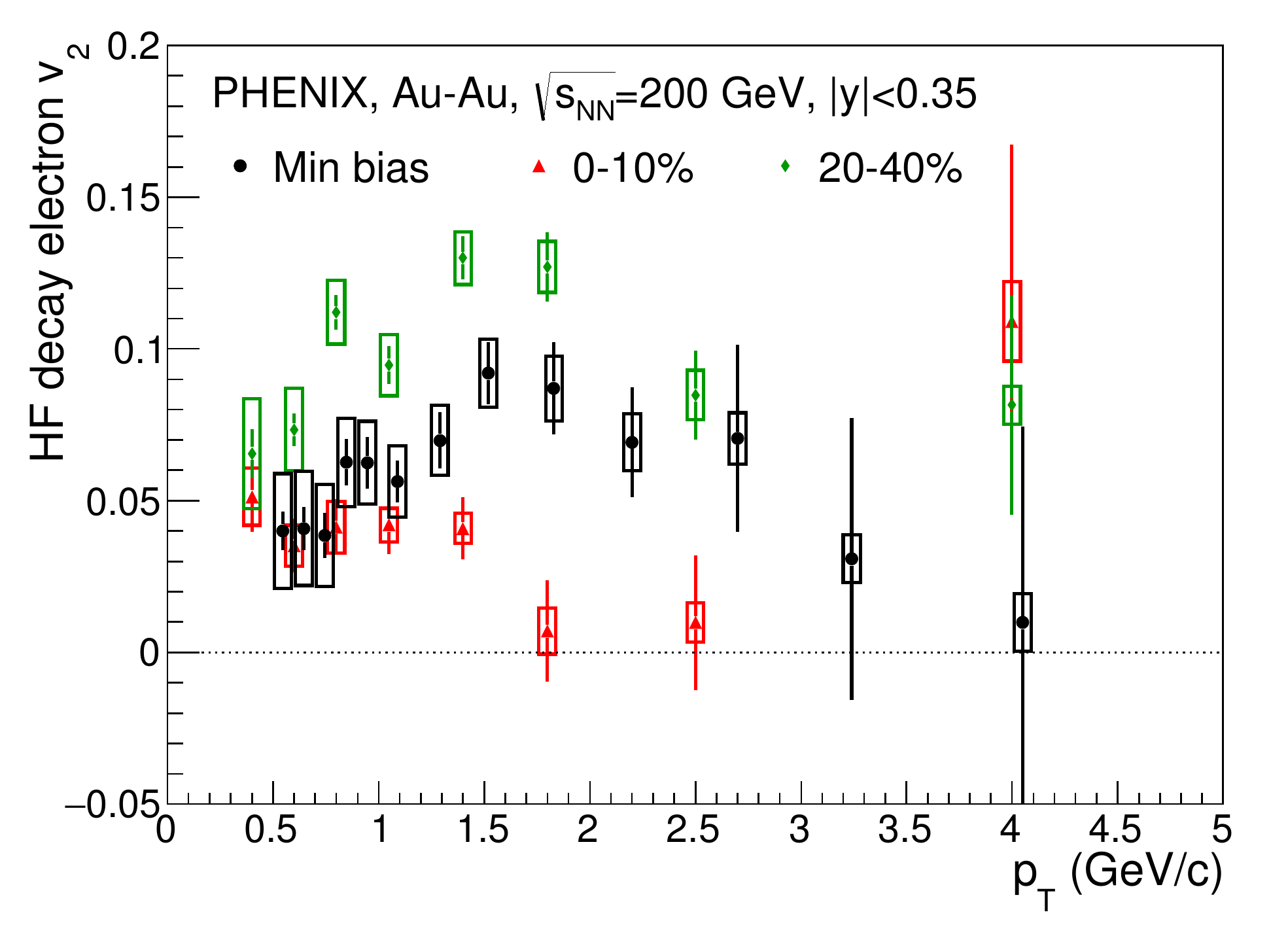}}}$
$\vcenter{\hbox{\includegraphics[width=0.48\textwidth]{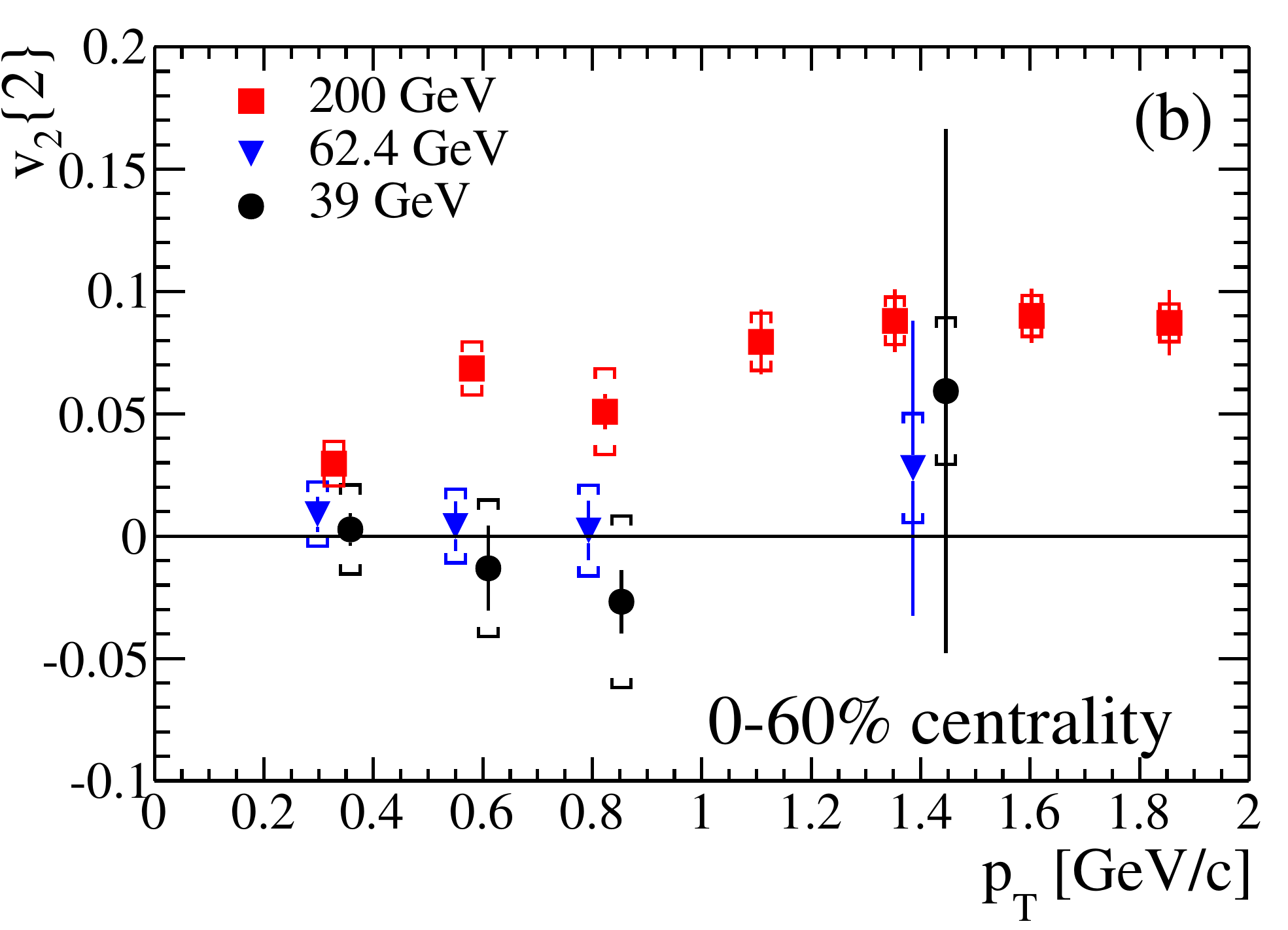}}}$
\caption{Heavy-flavor decay electron $v_2$ in Au--Au collisions as function of 
$\pT$.
Left: PHENIX results for different centrality intervals at 
$\sqrtsNN=200~\GeV$~\cite{Adare:2010de}.
Right: STAR results at three different collision 
energies~\cite{Adamczyk:2014yew}.}
\label{fig:v2HFL}
\end{center}
\end{figure}

Further insight into the interaction of heavy quarks with the medium is
provided by the measurements of elliptic flow and their comparison
to model calculations.
The results of the measurements of HF decay electron $v_2$ at mid-rapidity 
as a function of $\pT$ in Au--Au collisions at RHIC are shown in 
Fig.~\ref{fig:v2HFL}.
The left panel reports the results of the PHENIX collaboration
for collisions at $\sqrtsNN=200~\GeV$ in different centrality
intervals~\cite{Adare:2010de}.
The largest $v_2$ is observed in semi-peripheral collisions (20--40\% shown
in the plot and 40-60\% not shown in the plot), for which the initial
geometrical anisotropy is largest.
The elliptic flow is found to be larger than zero in the interval 
$0.5<\pT<2.5~\GeV/c$, with a maximum value of about 0.1 at 
$\pT \approx 1.5~\GeV/c$.
Note that a non-zero $v_2$ of HF decay electrons does not necessarily 
imply a non-zero $v_2$ of charm (beauty) quarks usually associated with 
heavy quarks taking part in the collective expansion of the medium.
A significant contribution to HF electrons at low and intermediate 
$\pT$ may arise from the decays of charm (and beauty) hadrons produced 
via the recombination of a heavy quark with a light quark from the medium.
Due to the light-quark collective flow, HF hadrons produced with this 
recombination mechanism can acquire a non-zero $v_2$ also in the case in 
which charm quarks have vanishing elliptic flow~\cite{Greco:2003vf}\footnote{Recall,
however, that a sharp separation between diffusion and coalescence effects 
is somewhat academic since hadronization should smoothly emerge from the
interactions that a heavy quark undergoes when approaching $T_{pc}$ from 
above (cf.~the discussion in Sec.~\ref{sssec_hadro}).}.
In the right panel of Fig.~\ref{fig:v2HFL} the STAR measurements of HF 
decay electron $v_2$ at three different collision energies are compared 
for the 0--60\% centrality class~\cite{Adamczyk:2014yew}. The results at 
$\sqrtsNN=200~\GeV$ are compatible with the measurement by the PHENIX 
collaboration in the same centrality class (see Ref.~\cite{Adamczyk:2014yew}
for the comparison).
At lower collision energies, $\sqrtsNN=39$ and $62.4~\GeV$, the $v_2$
values are smaller and consistent with zero within uncertainties.
Also the PHENIX collaboration reported a measurement of HF decay
electron $v_2$ at $\sqrtsNN=62.4~\GeV$ in the interval $1<\pT<4~\GeV/c$ .
The central values are non-zero but lower than at 200 GeV, and consistent 
with the STAR results at lower $\pT$.
However, the large uncertainties prevent from drawing firm conclusions
on the dependence of $v_2$ on the collision energy.

\begin{figure}
\begin{center}
\includegraphics[width=0.48\textwidth]{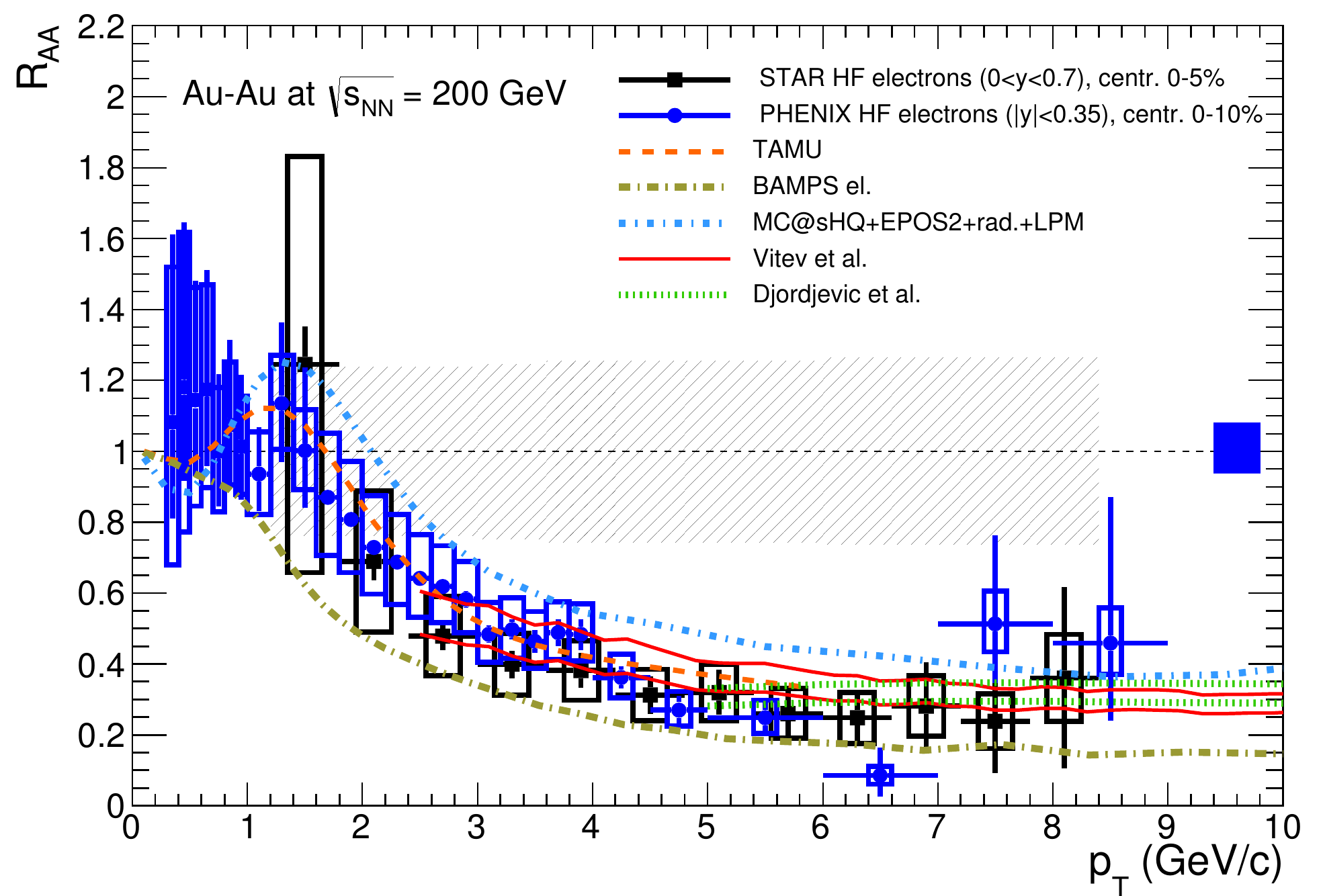}
\includegraphics[width=0.48\textwidth]{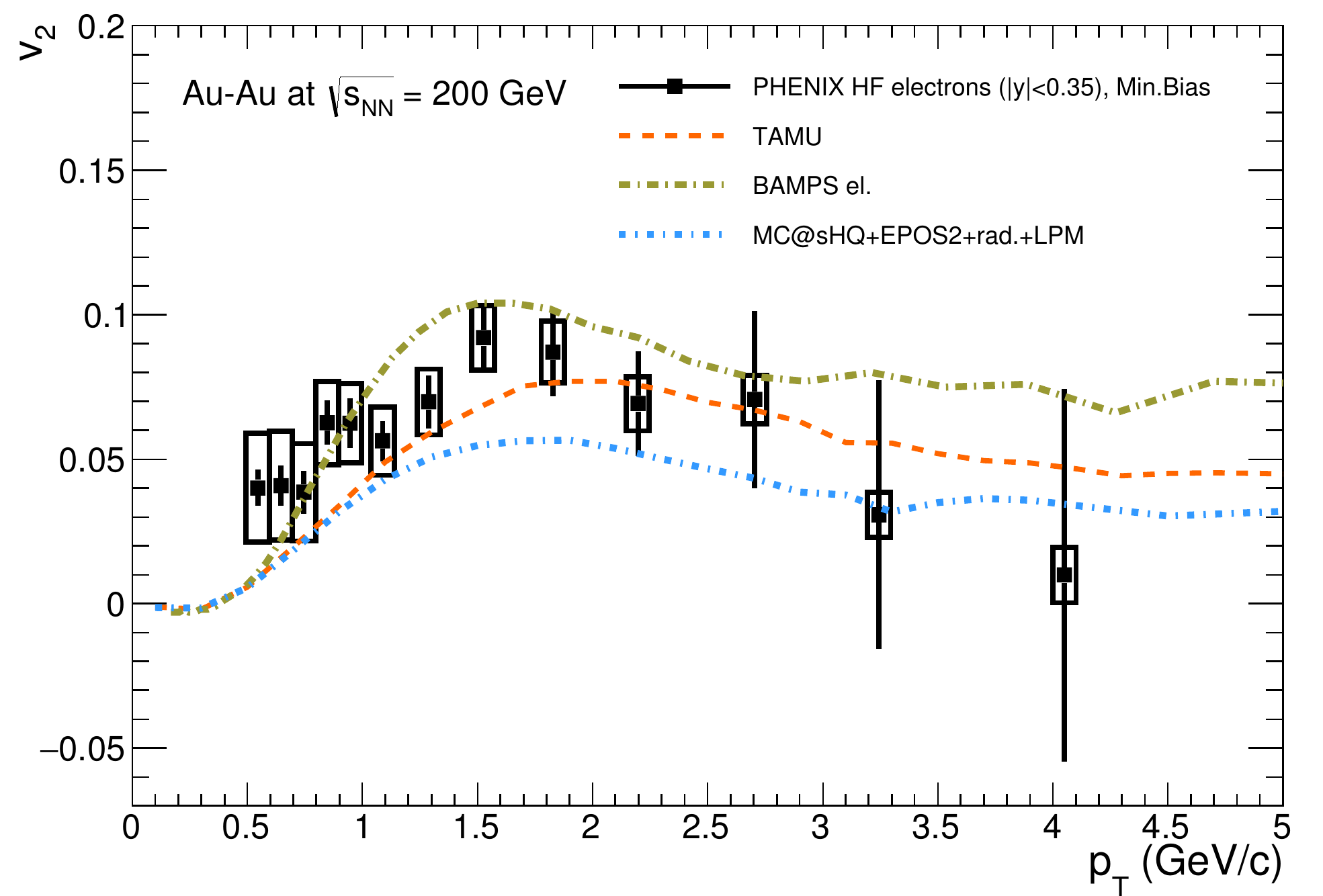}
\caption{Heavy-flavor decay electron $\Raa$ in central and $v_2$
in minimum-bias Au--Au collisions at RHIC from PHENIX~\cite{Adare:2010de} and
STAR~\cite{Abelev:2006db,Adare:2010de} compared to model predictions:
MC@sHQ+EPOS~\cite{Gossiaux:2008jv,Nahrgang:2013xaa}, 
TAMU~\cite{He:2011qa,He:2014cla}, 
BAMPS~\cite{Uphoff:2011ad,Uphoff:2012gb,Uphoff:2014hza}, 
Vitev et al~\cite{Sharma:2009hn} and 
Djordjevic et al.~\cite{Djordjevic:2014tka} 
(taken from~\cite{Andronic:2015wma}).}
\label{fig:Raav2modelsRHIC}
\end{center}
\end{figure}

A number of theoretical model calculations are available for 
the elliptic flow coefficient and the nuclear modification factor
of HF decay electrons at RHIC energies.
In Fig.~\ref{fig:Raav2modelsRHIC}, taken from~\cite{Andronic:2015wma}, a 
comprehensive comparison of the outcome of model calculations to the
measurements is shown.
Overall, the  HF decay electron $\Raa$ in central Au--Au collisions 
measured by the PHENIX~\cite{Adare:2010de} and STAR~\cite{Abelev:2006db} 
collaborations and the $v_2$ measured by PHENIX~\cite{Adare:2010de} in 
minimum-bias (MB) collisions are fairly well described by available model 
calculations. In some of the models the quark-medium coupling (represented 
by the medium density/temperature and interaction cross section)
is tuned to describe the $\Raa$ of pions (Djordjevic et al., WHDG, 
Vitev et al.) or electrons (BAMPS) at RHIC energies.

\begin{figure}
\begin{center}
\includegraphics[width=0.48\textwidth]{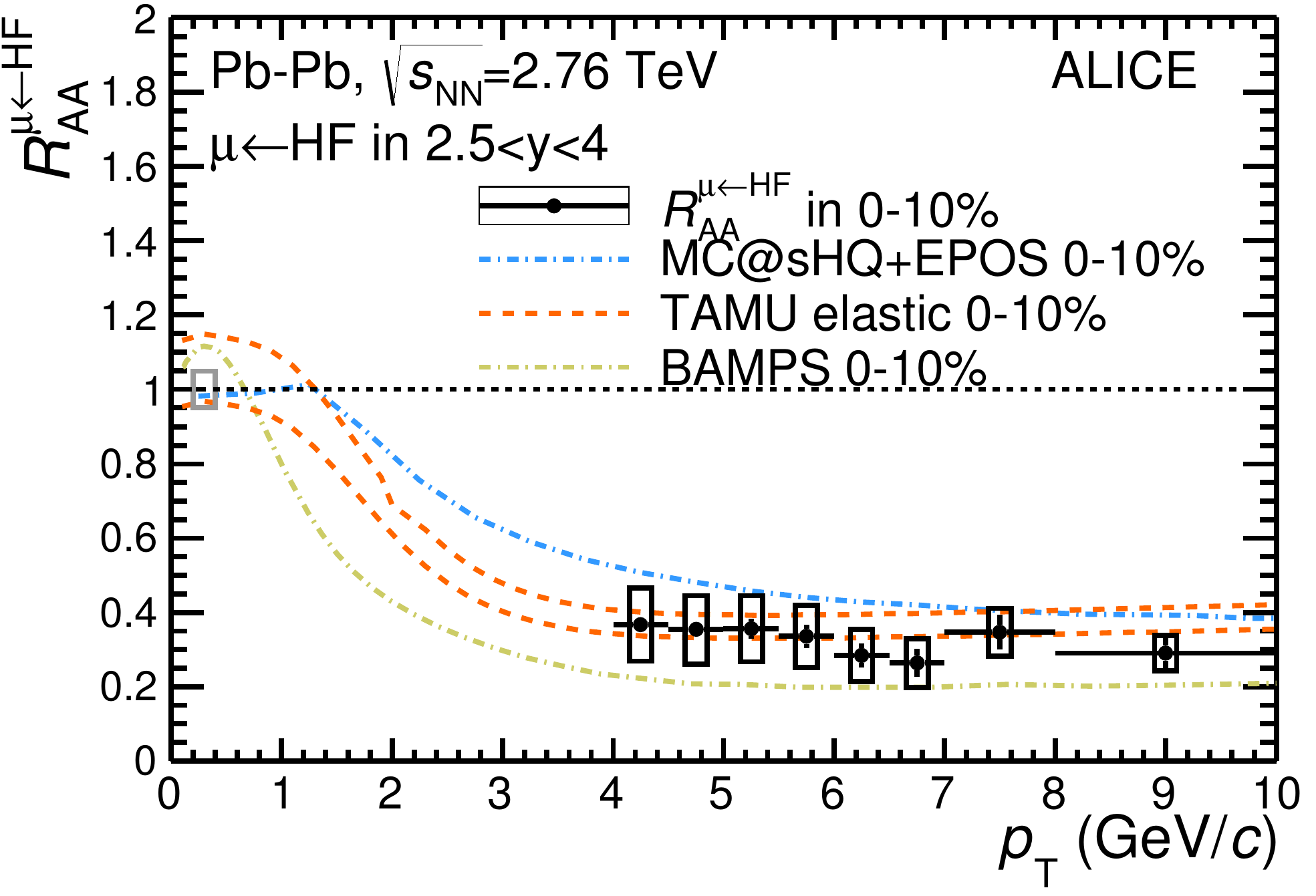}
\includegraphics[width=0.48\textwidth]{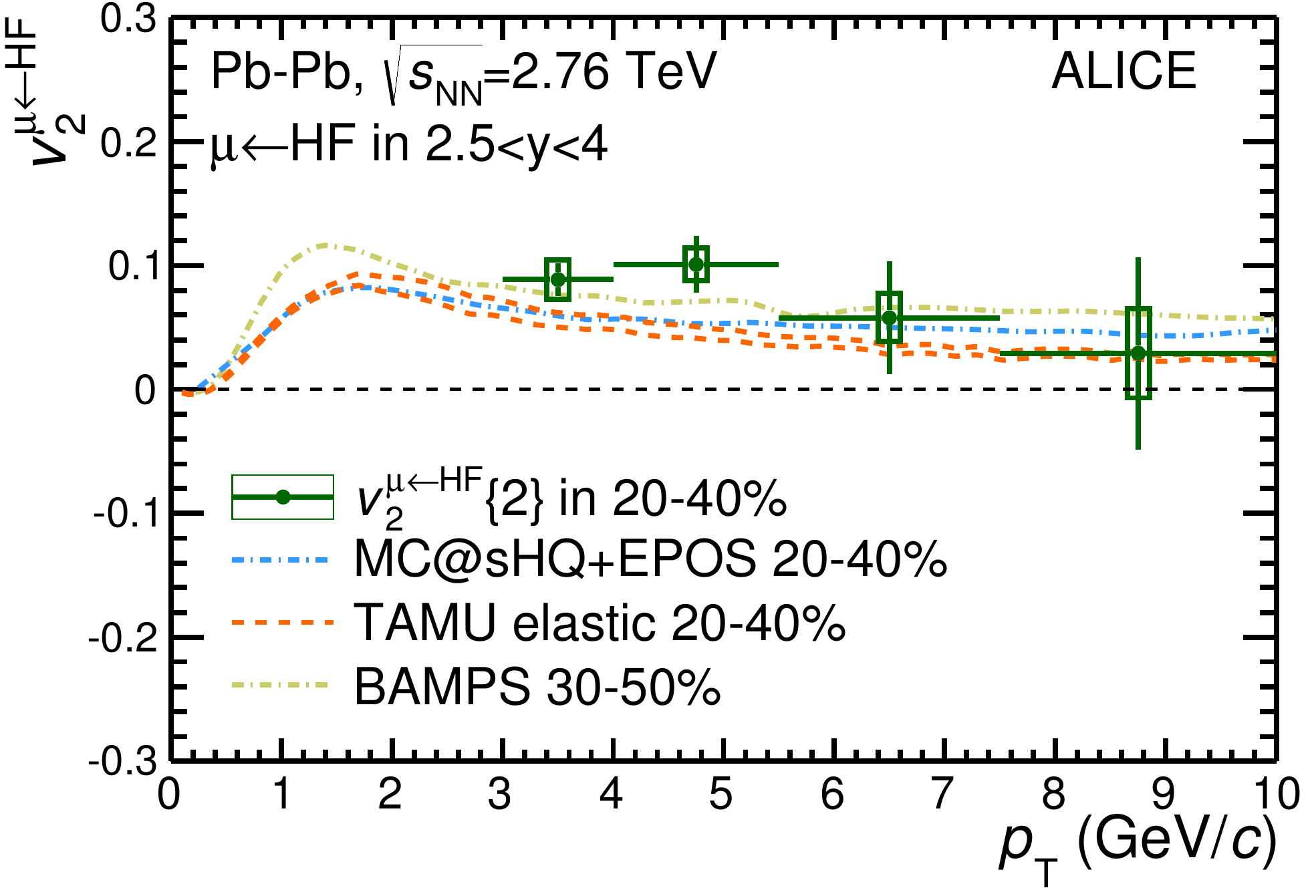}
\caption{Heavy-flavor decay muon $\Raa$ in central and $v_2$ 
in semi-central Pb--Pb collisions at the LHC~\cite{Adam:2015pga} as a 
function of $\pT$ compared to transport model predictions:
MC@sHQ+EPOS~\cite{Nahrgang:2013xaa}, TAMU~\cite{He:2014cla} and 
BAMPS~\cite{Uphoff:2011ad,Uphoff:2012gb,Uphoff:2014hza}.}
\label{fig:Raav2HFM}
\end{center}
\end{figure}

The ALICE Collaboration recently published results on the
elliptic flow of HF decay muons at forward rapidity in the 
interval $3<\pT<10~\GeV/c$ for three different centrality classes of 
Pb--Pb collisions at the LHC~\cite{Adam:2015pga}.
The $\pT$ dependence of $v_2$ in the 20--40\% class is shown in the
right panel of Fig.~\ref{fig:Raav2HFM}.
A positive $v_2$ is observed with a significance larger than $3\,\sigma$.
The data are compared to the predictions of the 
MC@sHQ+EPOS~\cite{Nahrgang:2013xaa}, TAMU~\cite{He:2014cla} and 
BAMPS~\cite{Uphoff:2011ad,Uphoff:2012gb} models.
All the three models predict a substantial suppression of the high-$\pT$
yield (left panel of Fig.~\ref{fig:Raav2HFM}) and a positive $v_2$
(right panel of Fig.~\ref{fig:Raav2HFM}), approximately consistent 
with what is observed in the data.
However, the BAMPS and MC@sHQ+EPOS models, which give a good description of
the measured $v_2$, tend to underestimate and overestimate the $R_{\rm AA}$,
respectively, while the TAMU model describes the $R_{\rm AA}$, but slightly 
underestimates the elliptic flow.
This indicates that it is challenging to simultaneously describe
the strong suppression of HF decay muons at high-$\pT$
in central collisions and their azimuthal anisotropy in semi-central collisions.

\begin{figure}
\begin{center}
\begin{minipage}{0.45\textwidth}
\includegraphics[width=\textwidth]{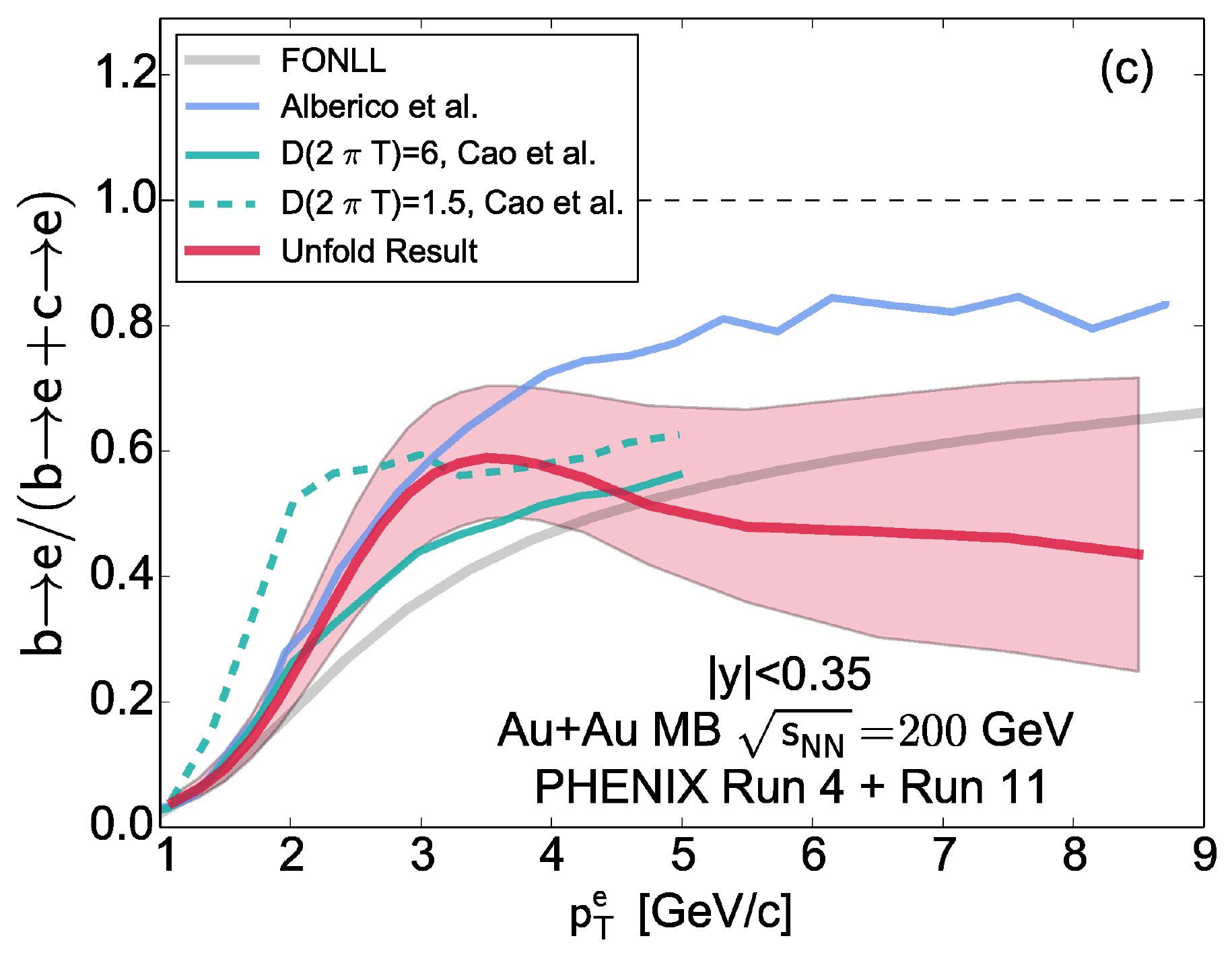}
\includegraphics[width=\textwidth]{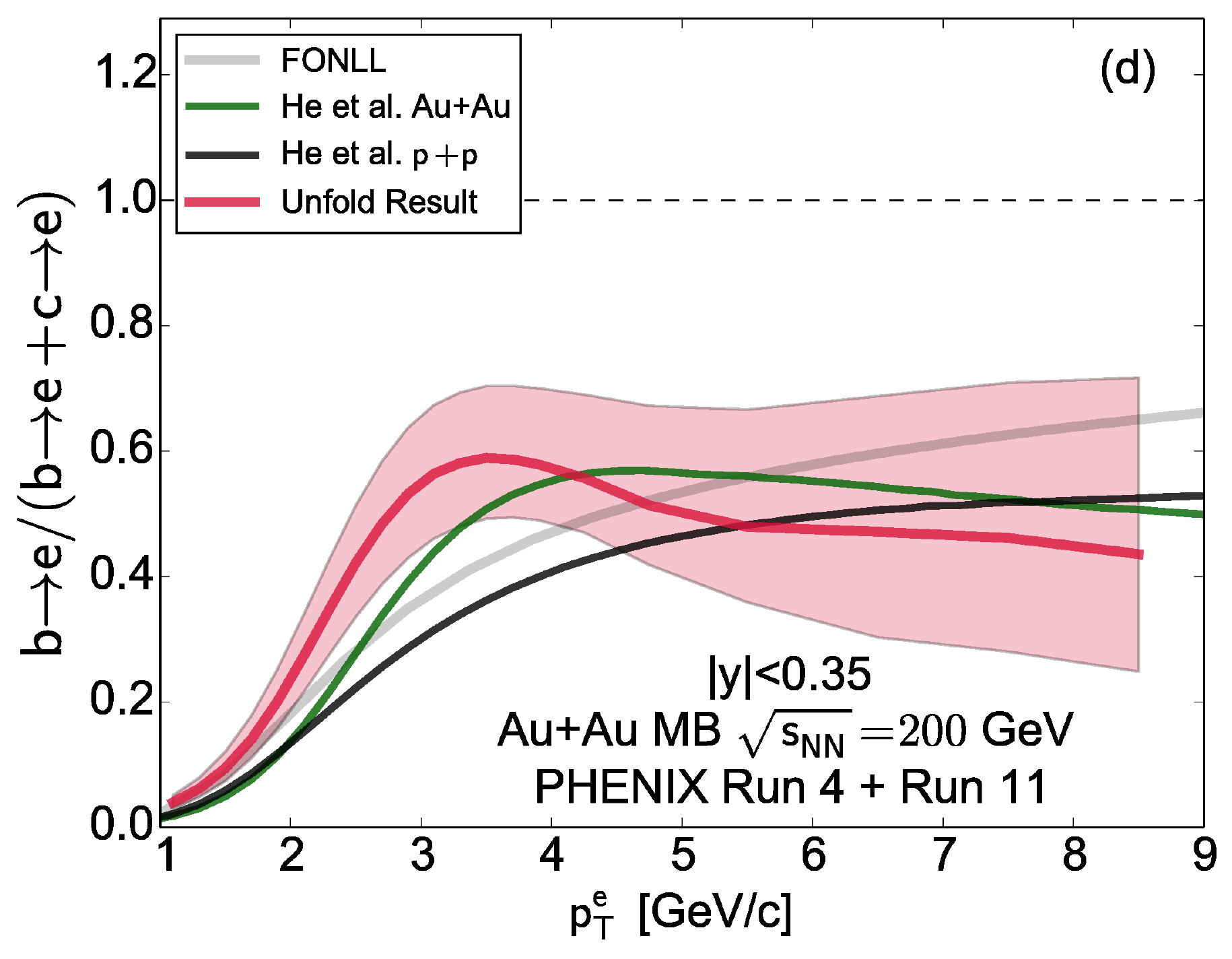}
\caption{Fraction of electrons from beauty-hadron decays in MB 
Au--Au collisions as a function of 
$\pT$ from the DCA-based unfolding analysis by the PHENIX 
collaboration~\cite{Adare:2015hla} compared 
to model calculations.}
\label{fig:bfracAuAuRHIC}
\end{minipage}
\begin{minipage}{0.49\textwidth}
\includegraphics[width=\textwidth]{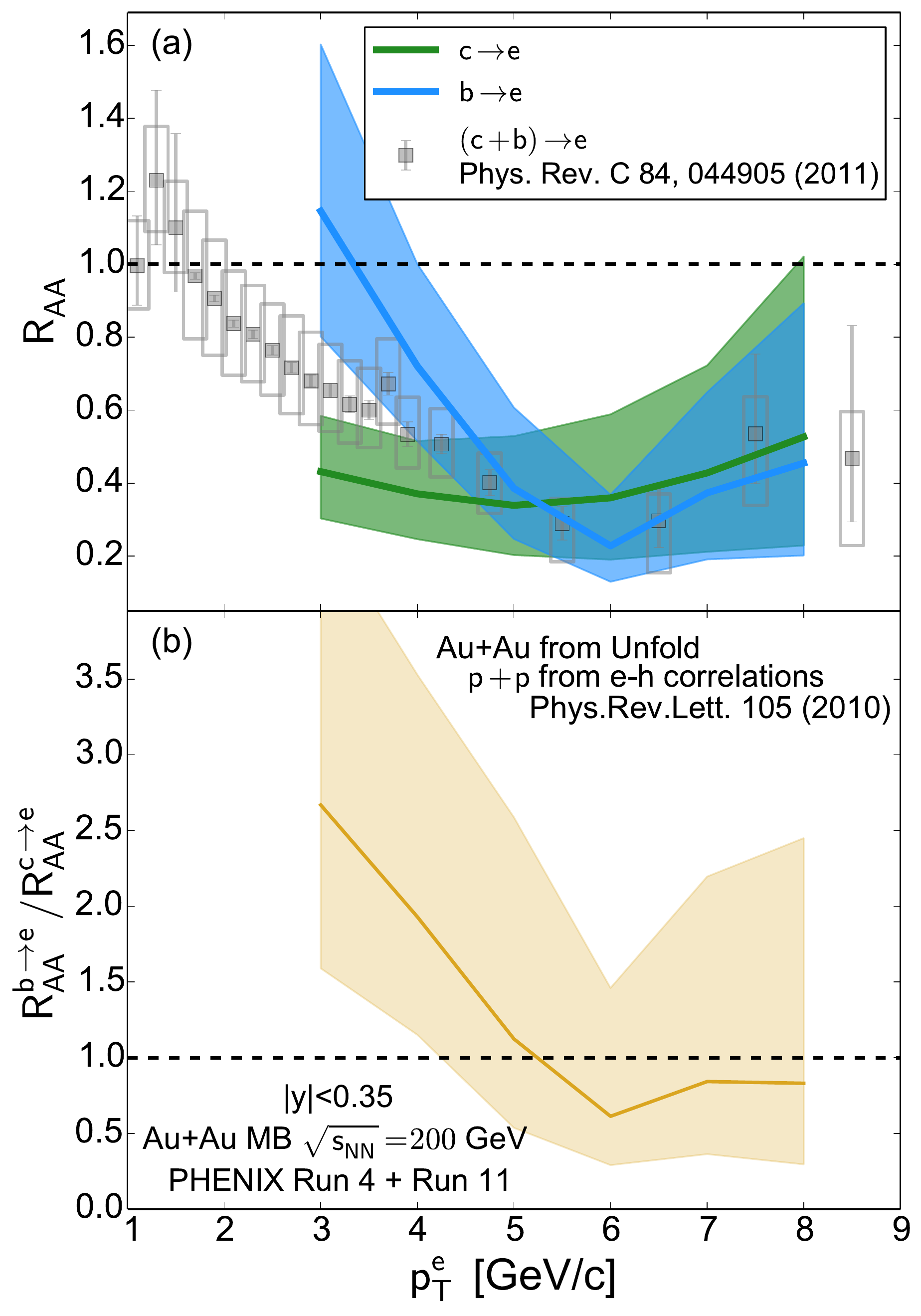}
\caption{Nuclear modification factor of electrons from charm and beauty
hadron decays in MB Au--Au collisions from the DCA-based unfolding 
analysis by the PHENIX collaboration~\cite{Adare:2015hla}}
\label{fig:RaacbeRHIC}
\end{minipage}
\end{center}
\end{figure}

A major progress in the HF decay lepton studies can be provided
by the separation of the contributions of charm and beauty hadron decays.
The PHENIX collaboration recently published the measurements performed on a 
sample of MB Au--Au collisions at $\sqrtsNN=200~\GeV/c$ 
recorded in 2011 with their new vertex detectors (VTX)~\cite{Adare:2015hla}.
The enhanced vertexing capabilities allow the separation of 
beauty and charm decay electrons based on the shape of the measured 
distributions of the distance of closest approach (DCA) of the tracks to the
interaction vertex.
An unfolding procedure was utilized to infer the parent charm-
and beauty-hadron yields as a function of $\pT$ starting from the measured
electron yield as a function of $\pT$ and DCA.
The extracted fraction of beauty-decay electrons in the HF electron 
yield ($\frac{b \rightarrow e}{c \rightarrow e + b\rightarrow e}$) 
as a function of $\pT$ is shown as a red band in Fig.~\ref{fig:bfracAuAuRHIC} 
and compared to the expectation from theoretical calculations.
In particular, predictions from FONLL pQCD calculations (corresponding to
no medium effects) and from three transport models, namely 
POWLANG~\cite{Beraudo:2014boa}, Duke (also denoted as Cao, Qin, Bass
or Cao {\it et al.}) for two different 
values of the $D_s(2\pi T)$ parameter governing the HQ-medium 
coupling~\cite{Cao:2012jt}, and TAMU~\cite{He:2014cla} are displayed.
Comparisons to other model calculations, not shown in 
Fig.~\ref{fig:bfracAuAuRHIC}, can be found in Ref.~\cite{Adare:2015hla}.
The models are in reasonable agreement with the extracted bottom 
electron fraction within the relatively large uncertainties.
From these results on the beauty-fraction in Au--Au collisions, the 
nuclear modification factors of electrons from 
charm and beauty hadron decays were disentangled using (i) the
additional information on the beauty-electron fraction 
in $pp$ collisions extracted from the the angular correlation analysis performed 
by the STAR collaboration~\cite{Aggarwal:2010xp} and (ii) the measured $R_{\rm AA}$ 
of HF decay electrons in Au--Au collisions.
The resulting $R_{\rm AA}$'s are reported in Fig.~\ref{fig:RaacbeRHIC} together
with the ratio $R_{\rm AA}^{ b \rightarrow e} / R_{\rm AA}^{ c \rightarrow e}$.
The electrons from beauty-hadron decays are found to be less
suppressed than those from charm-hadron decays at a $1\,\sigma$ level
in the range $3 <\pT<4~\GeV/c$.

Preliminary results on beauty-decay electron $R_{\rm AA}$ in central 
Pb--Pb collisions at the LHC, also exploiting the different DCA shapes
of charm and beauty decay electrons, were reported by the ALICE
Collaboration~\cite{Volkl:2014zha}.
A suppression of the yield relative to the binary-scaled $pp$ cross section
($R_{\rm AA}<1$), albeit with sizeable uncertainties, is observed for 
$\pT>3~\GeV/c$, consistent with in-medium energy loss of beauty quarks.

\subsubsection{Charm hadrons}
\hspace{2cm}

The production of $D$ mesons was measured $\pT$ differentially in Au--Au 
collisions at RHIC by the STAR collaboration~\cite{Adamczyk:2014uip} and in 
Pb--Pb collisions at the LHC
with ALICE~\cite{ALICE:2012ab,Abelev:2013lca,Abelev:2014ipa,Adam:2015nna,Adam:2015sza,Adam:2015jda}.
Preliminary results on ${D^{0}}$-meson nuclear modification factor
in Pb--Pb collisions at the LHC were recently reported by 
CMS~\cite{CMS:2015hca} and they are consistent with the ALICE results.
All these measurements were carried out at midrapidity ($|y|<1$ 
in the case of STAR and CMS and $|\eta|<0.8$ for ALICE).
The STAR collaboration measured the ${D^{0}}$-meson yield in the 
transverse-momentum range $0<\pT<8~\GeV/c$ using a data sample of 
$\sim 8.2 \cdot 10^8$ MB-triggered  
events and $\sim 2.4 \cdot 10^8$ events in the 0-10\% centrality interval
recorded during run-10 and run-11~\cite{Adamczyk:2014uip}.
Preliminary results were reported on ${D^{0}}$ production
in U--U collisions at $\sqrtsNN=193~\GeV$~\cite{Ye:2014eia}.
Recently, STAR reported preliminary results on ${D_{s}^{+}}$-meson
$\Raa$ and $v_2$ in Au--Au collisions at top RHIC energy~\cite{Nasim:2015qbe},
which, although limited by statistics, show similar features as
the ALICE measurements at the LHC~\cite{Adam:2015jda}.
ALICE obtained the first results on ${D^{0}}$-, ${D^{+}}$- and
${D^{*+}}$-meson nuclear modification factors in Pb--Pb collisions at
$\sqrtsNN=2.76~\TeV$ by analyzing a sample of  $13\times 10^6$ collisions
in the centrality range of 0--80\% collected in 2010~\cite{ALICE:2012ab}.
This sample allowed the measurement of the $D$-meson $\Raa$ in the momentum
interval $2<\pT<16~\GeV/c$ in the 0--20\% and 40--80\% centrality classes.
Using the larger data sample of  $16.4 \cdot 10^6$ central (0--10\%) 
and $9.0 \cdot 10^6$ semi-peripheral (30--50\%) Pb--Pb collisions recorded in 
year 2011, the $\Raa$ of ${D^{0}}$, ${D^{+}}$, and 
${D^{*+}}$ mesons could be measured with improved precision in a wider 
transverse-momentum interval ($1<\pT<36~\GeV/c$ for the 10\% most central 
collisions)~\cite{Adam:2015sza,Adam:2015nna}.
In addition, the $D$-meson elliptic flow~\cite{Abelev:2013lca,Abelev:2014ipa}
and the ${D_{s}^{+}}$-meson nuclear modification 
factor~\cite{Adam:2015jda} were measured for the first time.
Due to the small size of the data sample of $pp$ collisions collected at 
$\sqrt{s}=2.76~\TeV$, the $pp$ reference for the $\Raa$ at the LHC was obtained 
via a $\sqrt{s}$-scaling of the measurements at $\sqrt{s}=7~\TeV$.
The scaling factor and its uncertainty were obtained from FONLL calculations 
of the $D$-meson $\pT$ differential cross section at $\sqrt{s}=2.76$ and 
$7~\TeV$~\cite{Averbeck:2011ga}.

\begin{figure}
\begin{center}
$\vcenter{\hbox{\includegraphics[width=0.48\textwidth]{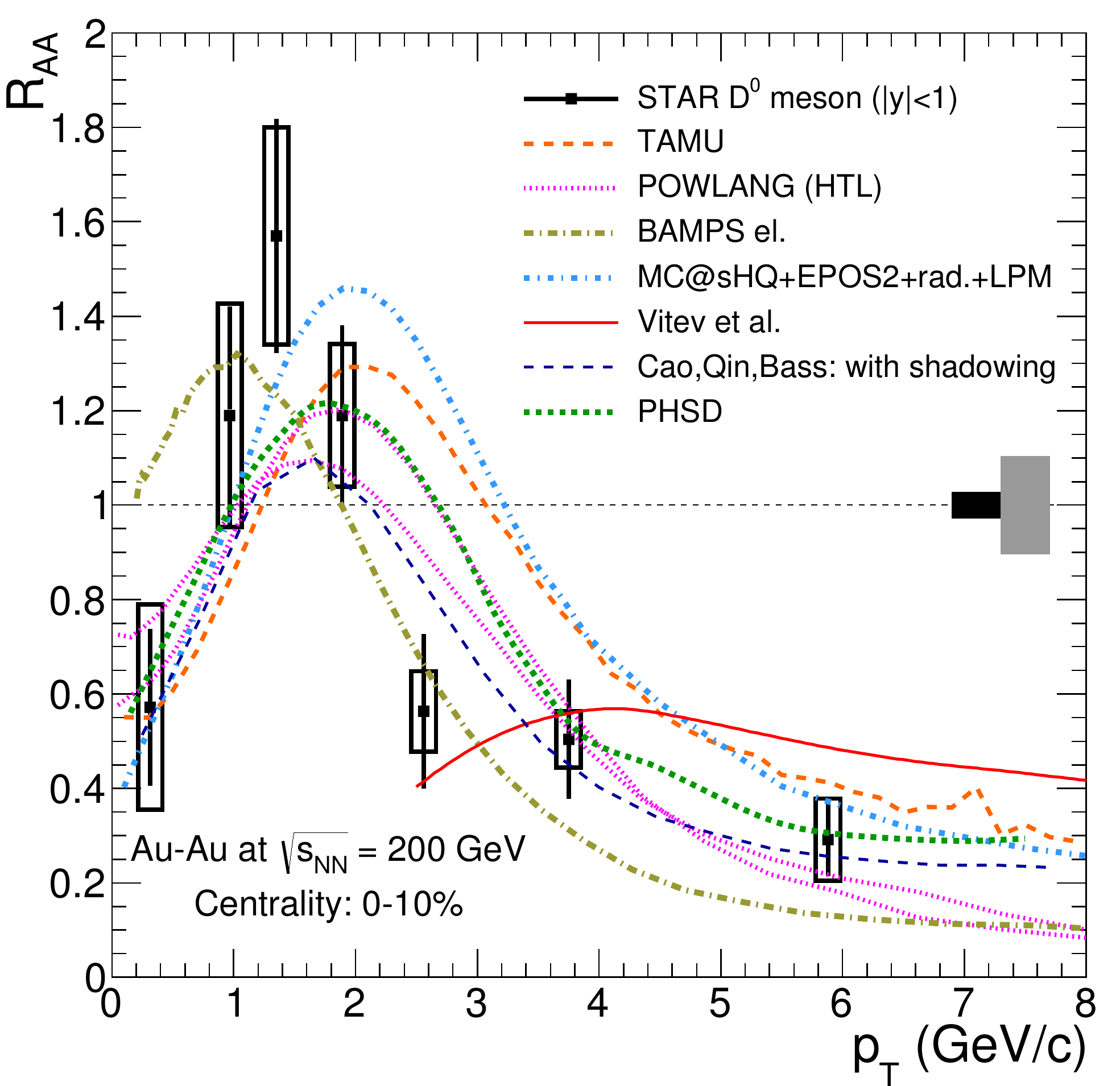}}}$
$\vcenter{\hbox{\includegraphics[width=0.48\textwidth]{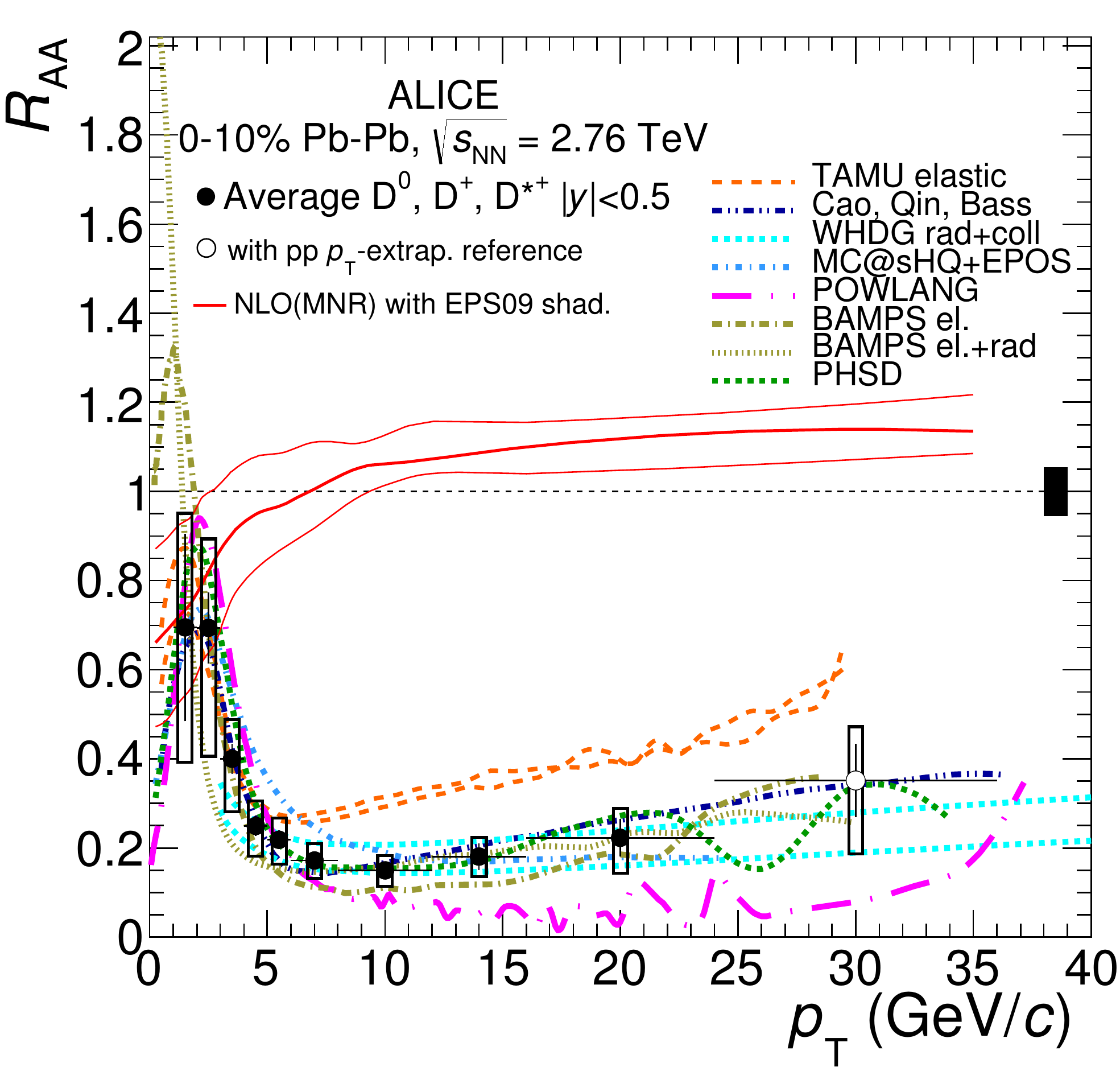}}}$
\includegraphics[width=0.48\textwidth]{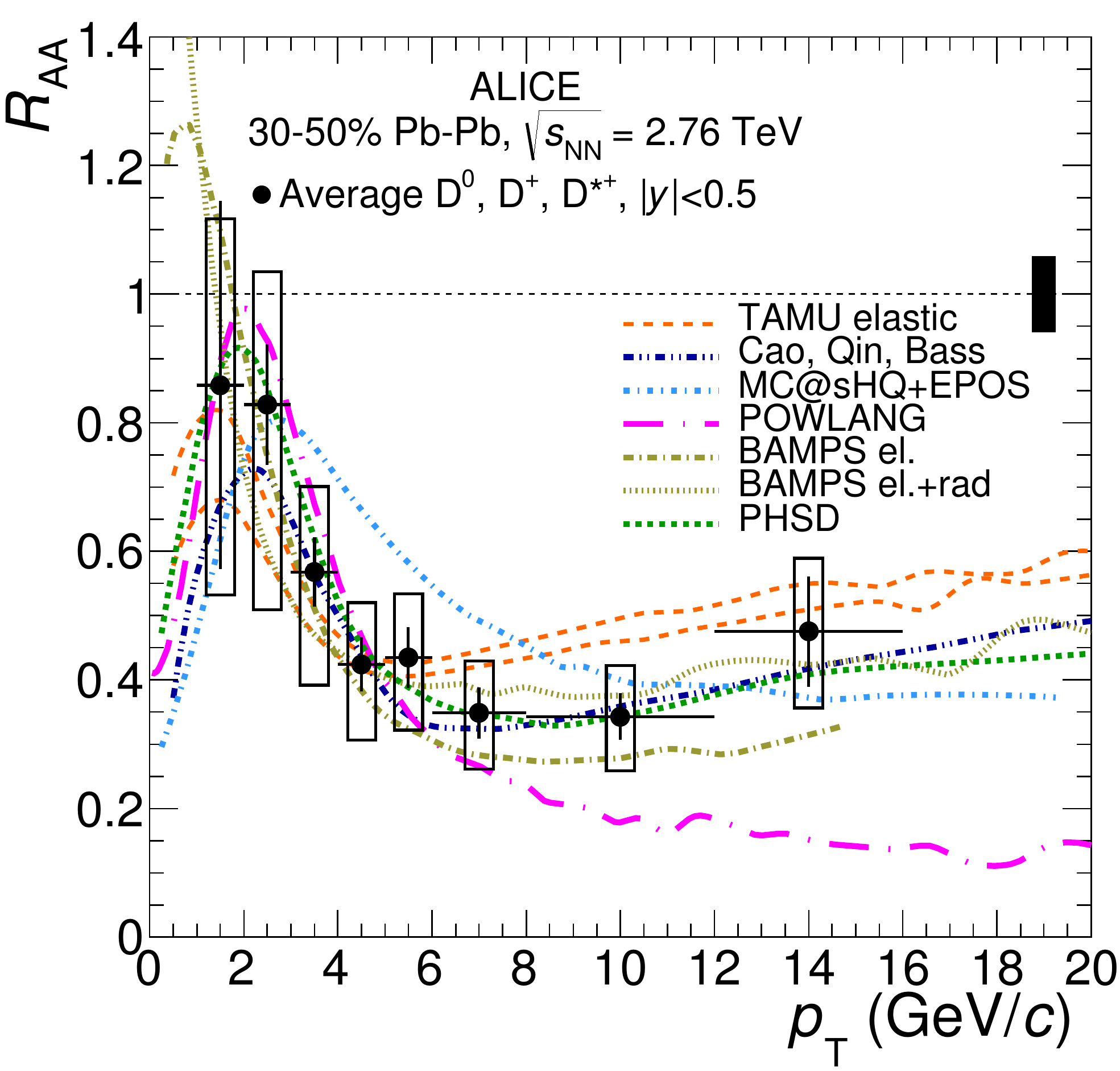}
\includegraphics[width=0.48\textwidth]{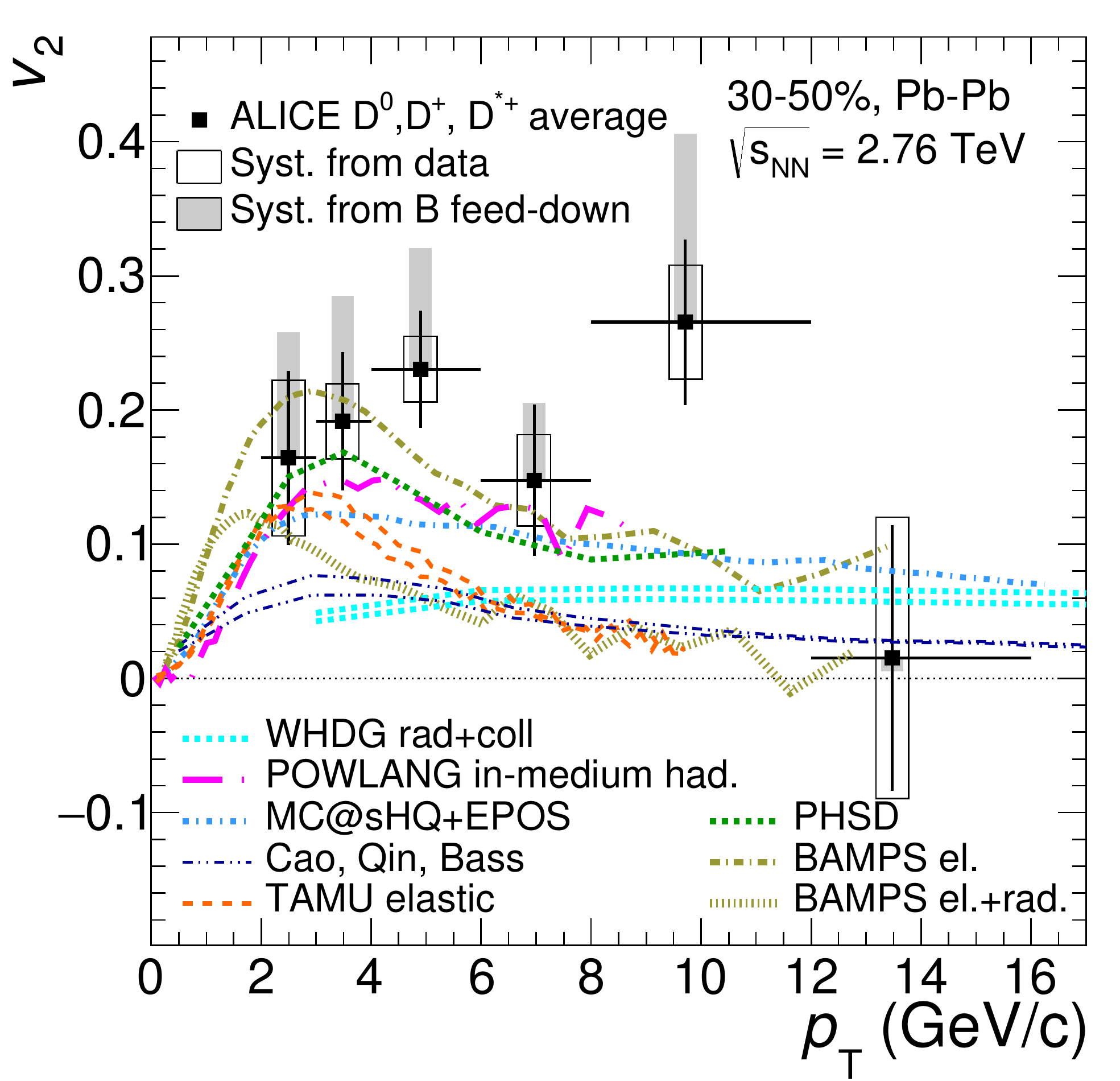}
\caption{$D$-meson nuclear modification factor and elliptic flow as a function of $\pT$ 
compared to model 
calculations~\cite{He:2011qa,He:2014cla,Gossiaux:2010yx,Nahrgang:2013xaa,Beraudo:2014boa,Cao:2013ita,Wicks:2005gt,Horowitz:2011gd,Horowitz:2011wm,Sharma:2009hn,Uphoff:2014hza,Song:2015sfa,Song:2015ykw}
Top left:  ${D^0}$-meson $\Raa$ in central Au--Au collisions 
at $\sqrtsNN=200~\GeV$~\cite{Adamczyk:2014uip} (taken from Ref.~\cite{Andronic:2015wma}).
Top Right: $D$-meson (average of ${D^0}$, ${D^+}$ and 
${D^{*+}}$) $\Raa$ in the 10\% most central Pb--Pb 
collisions at $\sqrtsNN=2.76~\TeV$~\cite{Adam:2015sza} (taken from Ref.~\cite{Dainese:2016ofs}).
Bottom left: $D$-meson $\Raa$ in semi-central (30--50\%) Pb--Pb collisions at 
$\sqrtsNN=2.76~\TeV$~\cite{Adam:2015sza}.
Bottom right: $D$-meson $v_2$ in semi-central (30--50\%) Pb--Pb collisions at 
$\sqrtsNN=2.76~\TeV$~\cite{Abelev:2013lca,Abelev:2014ipa} (taken from Ref.~\cite{Dainese:2016ofs}).}
\label{fig:Raav2DLHCRHIC}
\end{center}
\end{figure}

A selection of results on the $D$-meson nuclear modification factor and elliptic 
flow is shown in Fig.~\ref{fig:Raav2DLHCRHIC}.
In the top-left panel, the transverse-momentum dependence of the $\Raa$ of
${D^0}$ mesons in the 10\% most central Au–-Au collisions at 
$\sqrtsNN = 200~\GeV$ from the STAR experiment~\cite{Adamczyk:2014uip} is 
compared to the outcome of various model 
calculations~\cite{He:2011qa,Gossiaux:2010yx,Beraudo:2014boa,Cao:2013ita,Sharma:2009hn,Uphoff:2014hza}.
The data show a structure in transverse-momentum which is characterized by an
increase of $\Raa$ with increasing $\pT$ for  $\pT<1.5~\GeV/c$, a maximum at 
$\pT$ around $1.5~\GeV/c$, where a value $\Raa>1$ is measured, followed by a 
decrease.
For $\pT>3~\GeV/c$, a clear suppression relative to the binary-scaled $pp$ cross 
section is observed.
A similar trend is observed in the preliminary results from U--U 
collisions~\cite{Ye:2014eia}.
The $\Raa$ measured by the STAR collaboration in the interval $0<\pT<3~\GeV/c$ 
is described qualitatively, and to some extent also quantitatively, by the 
models that include interactions in an expanding fluid-dynamical medium, 
causing energy 
loss and radial flow (TAMU~\cite{He:2011qa}, BAMPS~\cite{Uphoff:2014hza}, 
Cao {\it et al.}~\cite{Cao:2013ita}, MC@sHQ~\cite{Gossiaux:2010yx}, 
POWLANG~\cite{Beraudo:2014boa}, and PHSD~\cite{Song:2015sfa}). 
In these models, the $\Raa$ shape at low $\pT$ is the effect of the collective 
flow on the light and charm quarks and of the contribution of the 
recombination mechanism to charm-quark hadronization.
In some of this models, \eg, TAMU~\cite{He:2011qa}, POWLANG~\cite{Beraudo:2014boa} 
and Duke~\cite{Cao:2015hia}, the effect of hadronization via
recombination, which converts low- and intermediate-$\pT$ charm quarks into 
$D$ mesons, is rather crucial to describe the data at low and intermediate $\pT$.
In the TAMU~\cite{He:2011zx} and PHSD models~\cite{Song:2015sfa} 
the contribution to the $v_2$ due to $D$-meson rescattering in the hadronic phase 
is found to be significant (also for the $\Raa$ in the PHSD model).
The model by Vitev {\it et. al.}~\cite{Sharma:2009hn}, which includes 
CNM and hot QGP (in-medium energy loss and meson dissociation) effects, 
is consistent with the data in the region of its applicability, $\pT > 3~\GeV/c$ 
(since the medium's transverse collective expansion is neglected).

The prompt $D$-meson $\Raa$ (average of ${D^{0}}$, ${D^{+}}$, and
${D^{*+}}$ nuclear modification factors) measured with 
ALICE~\cite{Adam:2015sza} in central (0--10\%) Pb--Pb collisions at 
$\sqrtsNN=2.76~\TeV$ is shown in the top-right panel 
of Fig.~\ref{fig:Raav2DLHCRHIC} together with a selection of model 
predictions~\cite{He:2014cla,Nahrgang:2013xaa,Beraudo:2014boa,Cao:2013ita,Uphoff:2014hza,Wicks:2005gt,Horowitz:2011gd,Horowitz:2011wm}.
In particular, only models for which simultaneous predictions for $v_2$ 
are available were included in this plot (other predictions compared to data 
are deferred to the left panel of Fig.~\ref{fig:RaaDandPiVsModels}).
The prompt $D$-meson yield at high $\pT$ is found to be strongly suppressed with 
respect to the binary-scaled $pp$ reference.
In the interval $3<\pT<10~\GeV/c$, the suppression increases 
($\Raa$ decreases) with increasing $\pT$.
The maximal suppression is observed around $\pT=10~\GeV/c$, where the yields 
are reduced by a factor of 5--6 relative to the binary-scaling expectation 
value.
For $\pT> 10~\GeV/c$, the $\Raa$ appears to increase (decreasing suppression) 
with increasing $\pT$, even though the large uncertainties prevent a 
conclusion on the trend of the nuclear modification factor at high $\pT$.
A significant suppression, $\Raa < 0.5$, is observed for $D$ mesons with 
$\pT > 25~\GeV/c$.
Since no significant modification of the $D$-meson production is observed in 
$p$Pb collisions for $\pT>2~\GeV/c$, the strong suppression of the $D$-meson 
yields observed for $\pT>3~\GeV/c$  cannot be explained in terms of 
CNM effects and therefore is predominantly due to final-state 
effects induced by the hot and dense medium created in the collisions.
This is also supported by the fact that the data cannot be described by the
the outcome of a NLO pQCD calculation~\cite{Mangano:1991jk} including only
the initial-state effects related to the nuclear modification of the 
PDF~\cite{Eskola:2009uj} (``NLO with EPS09" curve in 
Fig.~\ref{fig:Raav2DLHCRHIC}), which instead was able to reproduce the 
measured $D$-meson nuclear modification factor in $p$Pb collisions 
(see Fig.~\ref{fig:RpADLHC}).
On the other hand, all the models including interactions of charm quarks with 
an hot and dense partonic medium provide in general a reasonable description 
of the observed $\Raa$.

\begin{figure}
\begin{center}
$\vcenter{\hbox{\includegraphics[width=0.48\textwidth]{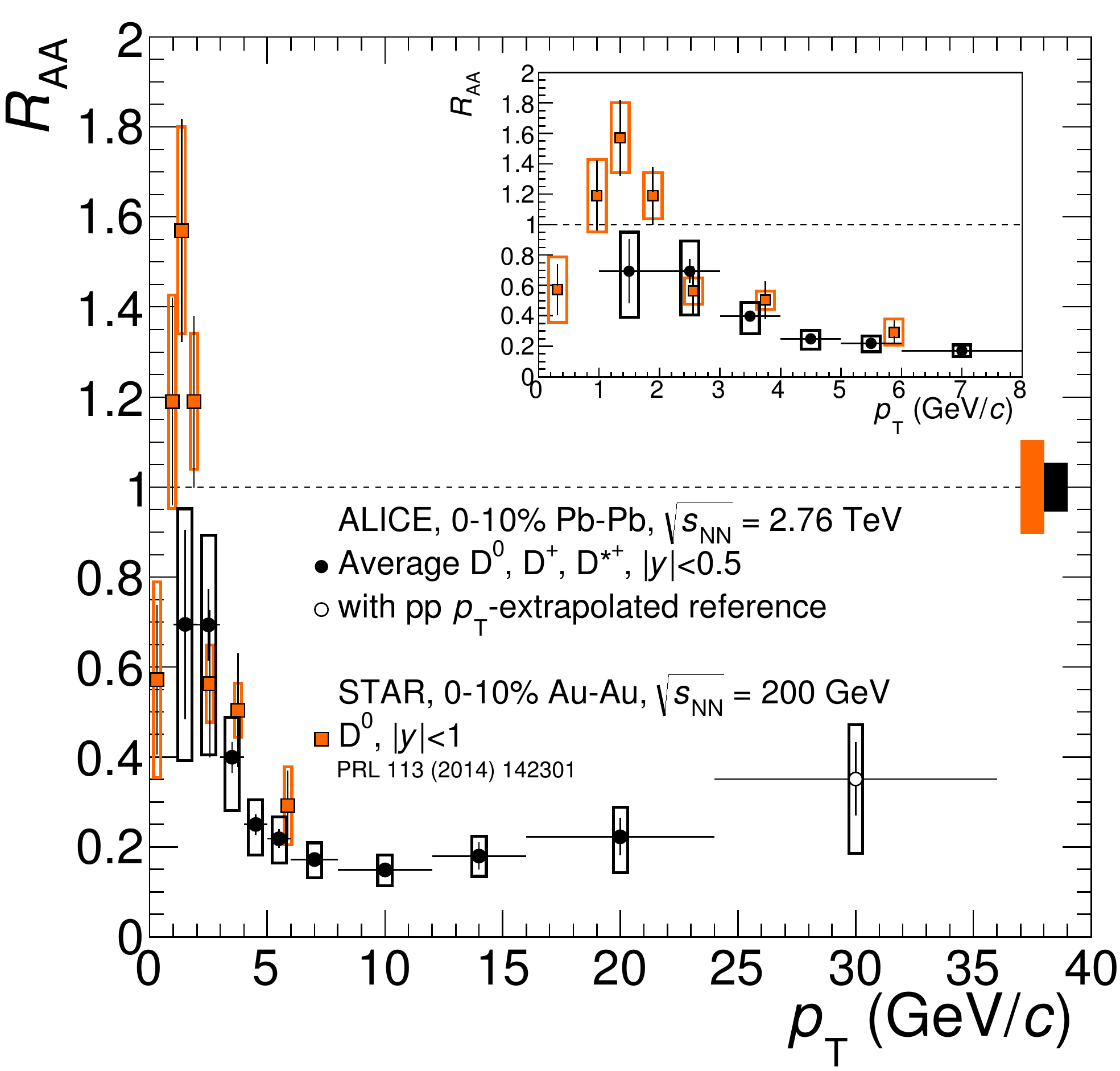}}}$
$\vcenter{\hbox{\includegraphics[width=0.48\textwidth]{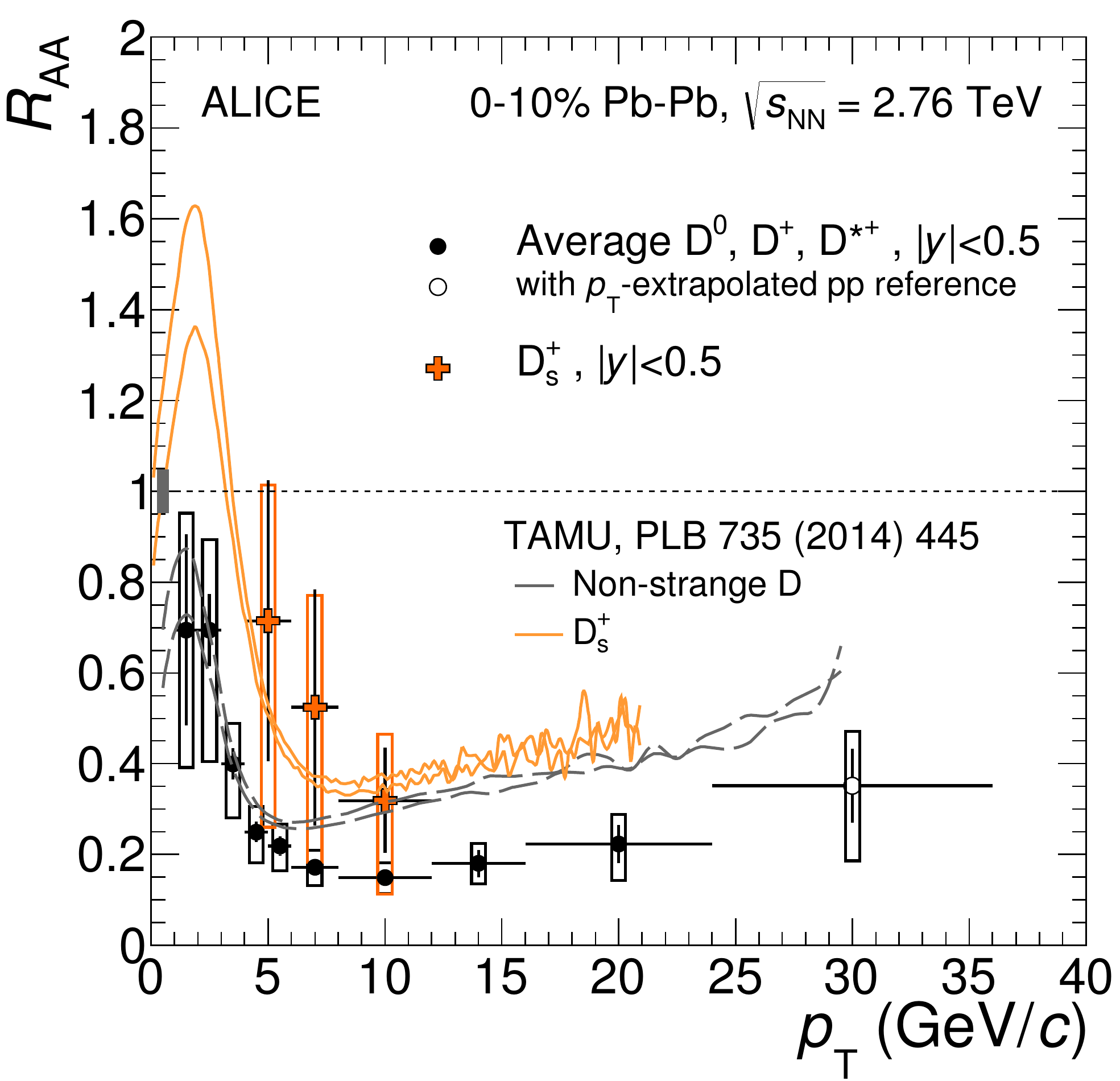}}}$
\caption{
Left: comparison between the $\Raa$ of ${D^0}$ mesons in central 
(0--10\%) Au--Au collisions at $\sqrtsNN=200~\GeV$~\cite{Adamczyk:2014uip} 
and of $D$ mesons (average of ${D^0}$, ${D^+}$ and 
${D^{*+}}$) in central Pb--Pb collisions at 
$\sqrtsNN=2.76~\TeV$~\cite{Adam:2015sza}.
Right: ${D_s^{+}}$~\cite{Adam:2015jda} and average $D$-meson~\cite{Adam:2015sza} 
$\Raa$ in central Pb--Pb collisions at $\sqrtsNN=2.76~\TeV$ compared
to predictions from the TAMU model~\cite{He:2014cla}.}
\label{fig:RaaDsD}
\end{center}
\end{figure}

To illustrate the evolution of the nuclear modification factor with $\sqrt{s}$, 
the $D$-meson $\Raa$ measured in the 10\% most central Au--Au and Pb--Pb 
collisions at $\sqrtsNN = 200$ and $2.76~\TeV$
are compared in the left panel of Fig.~\ref{fig:RaaDsD}.
At high $\pT$ ($\pT>3~\GeV/c$), where the nuclear modification factor is 
expected to be dominated by the effect of in-medium parton energy 
loss, the $\Raa$ values measured at the two energies are compatible
within uncertainties.
However, as pointed out in~\cite{Adam:2015sza}, this
does not necessarily imply a similar charm-quark energy loss 
and medium density at the two collision energies, since
the nuclear modification factor is also sensitive
to the slope of the $\pT$ spectra of the hard-scattered partons.
Therefore, the combined effect of a denser medium and harder 
initial $\pT$ spectra at the LHC could result in similar values of $\Raa$ 
as at RHIC energies (see, \eg, Ref.~\cite{Baier:2002tc}).
At lower $\pT$, the $\Raa$ measured at $\sqrtsNN=2.76~\TeV$ is lower
than the one at $\sqrtsNN=200~\GeV$ and does not show a `bump-like' trend 
with a rise, a maximum and a successive decrease with increasing $\pT$.
However, due to the large uncertainties and the coarser binning at low 
$\pT$, no firm conclusion can be drawn.
In this comparison, it has to be considered that the $\Raa$ at low 
and intermediate $\pT$ is the result of the interplay of different effects 
occurring in the initial and final state. 
Therefore, a different $\Raa$ trend at different $CM$ energies 
could arise from a different role of initial-state effects and radial flow.
As far as initial-state effects are concerned, with increasing $\sqrtsNN$ one 
expects a stronger reduction of the HQ production yields at low $\pT$ 
due to nuclear shadowing (due to the smaller values of Bjorken-$x$ being 
probed~\cite{Eskola:2009uj}) and a less pronounced Cronin peak at 
intermediate $\pT$~\cite{Wang:1998ww,Vogt:2001nh}.
On the other hand, the radial flow of the medium at LHC energies is about 10-20\% 
larger than at RHIC~\cite{Abelev:2008ab,Abelev:2012wca}.
However, this does not necessarily imply that
the bump-like structure observed in the $\Raa$ trend at low $\pT$ at
RHIC energy should become more pronounced with increasing collision energy.
The stronger radial flow effect could be counter-balanced in the $\Raa$
by the different shape of the reference spectra in $pp$ collisions at 
different $\sqrt{s}$.
In this respect, is interesting to notice in the top-left and top-right 
panels of Fig.~\ref{fig:Raav2DLHCRHIC} that a reasonable description of the 
low and intermediate $\pT$ data at both the collision energies is obtained 
with the models that include nuclear modification of the PDF, charm-quark 
interactions with the medium constituents, hydrodynamical medium expansion 
and hadronization via recombination, such as TAMU~\cite{He:2014cla, He:2011qa}, 
POWLANG~\cite{Beraudo:2014boa}, Duke~\cite{Cao:2013ita} and 
MC@sHQ+EPOS~\cite{Gossiaux:2010yx,Nahrgang:2013xaa}.
The BAMPS model~\cite{Uphoff:2014hza}, which does not 
include nuclear modification of the PDFs, predicts for LHC energies at low 
$\pT$ (where shadowing is relevant) a value of $\Raa$ larger than that
observed in the data.

In the bottom panels of Fig.~\ref{fig:Raav2DLHCRHIC} the $\Raa$ (left) 
and $v_2$ (right) of prompt $D$ mesons (average of ${D^{0}}$, 
${D^{+}}$, and ${D^{*+}}$) in semi-central (30--50\%) Pb--Pb
collisions at $\sqrtsNN=2.76~\TeV$ are shown and compared to model predictions.
The nuclear modification factor indicates that the $D$-meson yield
is suppressed in the 30--50\% centrality class in the measured $\pT$
range with respect to the binary-scaled $pp$ reference~\cite{Adam:2015sza}.
This reduction of the yield in Pb--Pb collisions is smaller, by a factor
of about two, than in the 10\% most central collisions, as expected due to
the decreasing medium density, size and lifetime from central to peripheral 
collisions.
A positive elliptic flow is measured for prompt $D$ 
mesons in the centrality class 30--50\%~\cite{Abelev:2013lca}.
In particular, in the interval $2<\pT<6~\GeV/c$ the measured $v_2$ is found to
be larger than zero with $5.7\,\sigma$ significance.
A positive $v_2$ of ${D^0}$ mesons was also observed in the 10--30\%
centrality class~\cite{Abelev:2014ipa}. 
These results indicate that the interactions with the medium constituents 
transfer information on the azimuthal anisotropy of the 
system to the charmed particles~\cite{Abelev:2013lca}.
It also suggests that low-momentum charm quarks take part in the 
collective expansion of the medium, even though, with the current 
uncertainties, the possibility that the observed $D$-meson $v_2$ is completely 
due to the light-quark contribution in a scenario with hadronization via 
recombination cannot be ruled out. 
A positive $v_2$ is also observed for $\pT>6~\GeV/c$, which is likely to 
originate from the path-length dependence of the partonic energy loss, 
although the large uncertainties do not allow for a firm conclusion.

As already pointed out in the discussion of the HF decay lepton results,
the simultaneous comparison of the measured $\Raa$ and $v_2$ to 
theoretical model calculations constrains the description of the 
interactions of heavy quarks with the medium, possibly providing  
sensitivity to the relative contributions of elastic (collisional) and 
inelastic (radiative) processes, and to the path length dependence of 
in-medium parton energy loss.
Overall, the observed elliptic flow is qualitatively described by the models
that include both charm-quark energy loss in a spatially anisotropic medium 
and momentum gain processes transferring elliptic flow produced  
through the system expansion to charmed particles.
The WHDG model does not include a hydrodynamical description of the
medium expansion, so that the anisotropy results only from path length 
dependent energy loss (the models of Djordjevic {\it et al.} and Vitev 
{\it et al.} do not provide a calculation for $v_2$; their $\Raa$ is compared 
to the data in Fig.~\ref{fig:RaaDandPiVsModels} below).
The models that include only collisional energy loss (TAMU, POWLANG, BAMPS-elastic
and PHSD) provide in general a good description of the $v_2$, but
tend to overestimate (TAMU) or underestimate (POWLANG, BAMPS-elastic) 
the $\Raa$ in central and/or semi-peripheral collisions.
On the other hand, models including both radiative and collisional energy 
loss (Cao {\it et al.}, BAMPS rad+el and WHDG) describe the $\Raa$ in central 
collisions well, but tend to underestimate the elliptic flow at low $\pT$.
This may be a consequence of the fact that the inclusion of radiative 
processes reduces the weight of elastic interactions, which 
are more effective in building up the azimuthal momentum anisotropy.
The MC@sHQ+EPOS model, which also includes both collisional and
radiative processes, can describe the measured $\Raa$ and $v_2$ within
uncertainties at low ($\pT<2~\GeV/c$) and high ($\pT>6-8~\GeV/c$) transverse 
momenta at different collision energies and centrality, but it tends to 
overestimate $\Raa$ and underestimate $v_2$ in the intermediate $\pT$ region.
From this discussion, it emerges that the role of the different interaction 
mechanisms, in particular radiative and collisional energy loss, is not yet 
completely clarified, even though the  data-to-theory comparison 
suggests that both of these contributions are relevant. 
Finally, models including hadronization of charm quarks from recombination with 
light quarks from the medium (TAMU, Cao {\it et al.}, MC@sHQ+EPOS, POWLANG and 
PHSD) predict a more pronounced radial flow peak in the low-$\pT$ $\Raa$ and
a larger $v_2$, due to the light-quark contribution, thus providing a better
description of the data at low $\pT$ (see, \eg, the discussion
in Ref.~\cite{Beraudo:2014boa}). 
In summary, this comprehensive data-to-theory comparison comparison reiterates 
the challenges for theoretical models to simultaneously describe the 
measured $D$-meson $\Raa$ and $v_2$ at different collision energies and 
centralities.
This indicates that the current data from RHIC and LHC have the potential to 
better constrain the description of the interactions of charm quarks with 
the medium constituents and their hadronization mechanism.

In the right panel of Fig.~\ref{fig:RaaDsD}, the nuclear modification
factor of $D$ mesons (average of ${D^0}$, ${D^+}$ and 
${D^{*+}}$) in central (0--10\%) Pb--Pb collisions at 
$\sqrtsNN=2.76~\TeV$ is compared to that of ${D_s^{+}}$ 
mesons~\cite{Adam:2015jda} and to the corresponding predictions from
the TAMU model~\cite{He:2012df,He:2014cla}.
This comparison is meant to address the expected effect of hadronization via 
quark recombination in the partonic medium on the relative abundances of 
strange and non-strange $D$-meson species.
An enhancement of the ${D_s}^+$ yield relative to that of non-strange 
$D$ mesons at low and intermediate momenta is expected in nucleus--nucleus 
collisions as compared to $pp$ interactions, if the dominant process for $D$-meson 
formation  is in-medium hadronization of charm quarks via 
recombination with light quarks, due to the large abundance of strange quarks 
in the QGP~\cite{Rafelski:1982pu,Koch:1986ud,Andronic:2003zv,Kuznetsova:2006bh,He:2012df}.
In particular, the $\pT$ dependence of this comparison has been suggested as a 
tool to map out the relative importance of recombination processes~\cite{He:2012df}.
In the three $\pT$ intervals, in which the ${D_s}^+$ yield could be
measured in Pb--Pb collisions, the central values of its $R_{\rm AA}$ are found 
to be higher than those of non-strange $D$ mesons, although compatible within 
uncertainties.
Even though part of the systematic uncertainty is correlated
between strange and non-strange $D$ mesons\cite{Adam:2015jda}, the current 
uncertainties prevent a conclusion on the expected modification of the 
relative abundance of charm-hadron species due to hadronization via 
recombination.
Among the various models of open-charm production in heavy-ion collisions, 
TAMU is the only one providing a quantitative prediction for the ${D_s}^+$-meson 
nuclear modification factor.
The measured $\Raa$ is described within uncertainties by this prediction,
where the $\pT$ dependent enhancement of ${D_s}^+$ mesons relative to that of 
non-strange $D$ mesons is a consequence of the recombination of charm quarks 
with thermally equilibrated strange quarks in the QGP.
As discussed in Sec.~\ref{sssec_hadro}, an enhanced production of ${D_s}^+$ 
mesons (as well as $\Lambda_{\rm c}$ baryons, etc.) in heavy-ion collisions 
due to recombination entails a reduction of charm quarks available for
hadronization into non-strange meson species.
This ``chemistry effect" should therefore be considered in the interpretation 
of the comparison of the nuclear modification factors of non-strange $D$ mesons 
and light-flavor hadrons (\eg, pions), which is discussed in the next paragraphs.

\begin{figure}
\begin{center}
$\vcenter{\hbox{\includegraphics[width=0.48\textwidth]{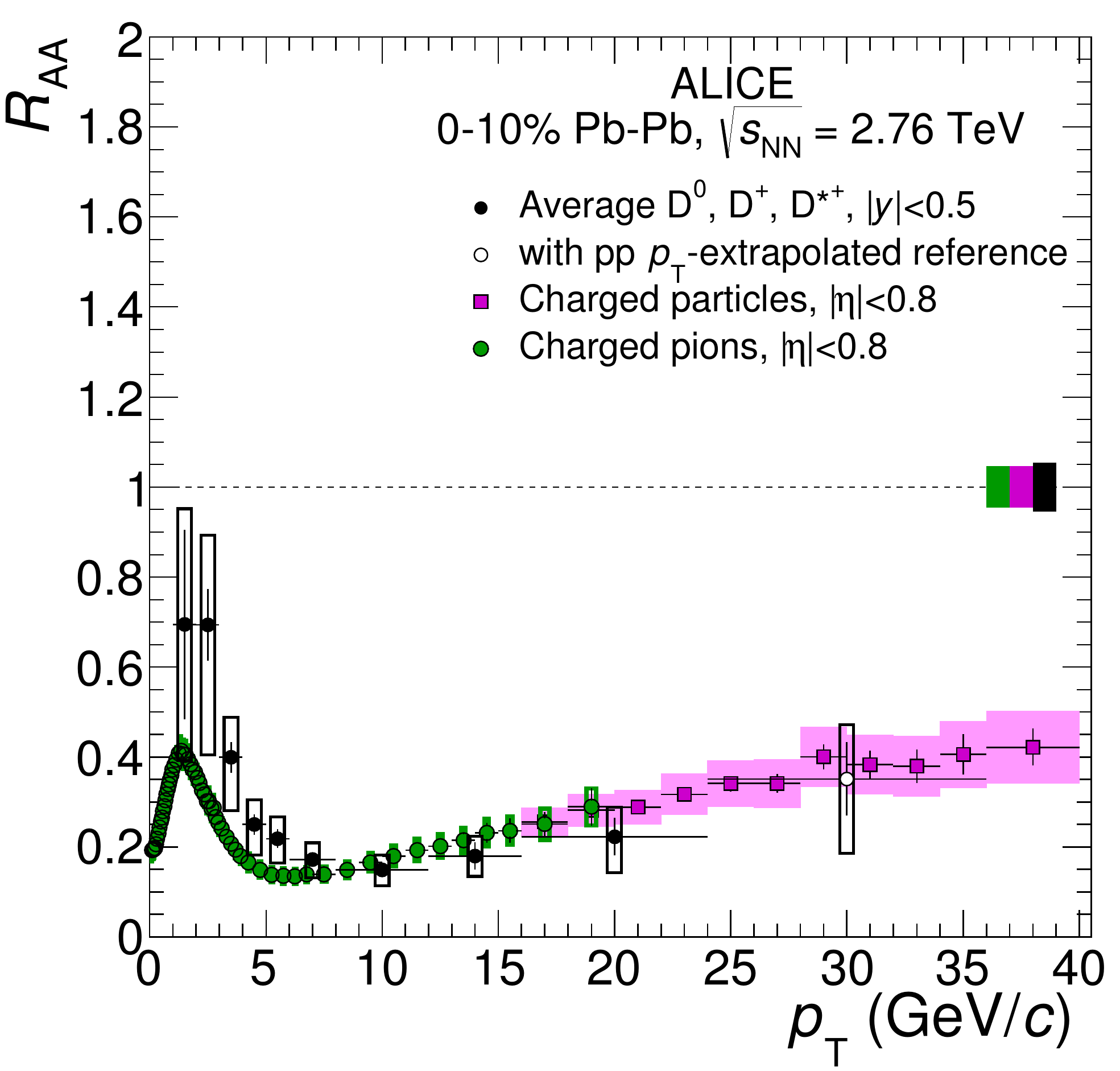}}}$
$\vcenter{\hbox{\includegraphics[width=0.48\textwidth]{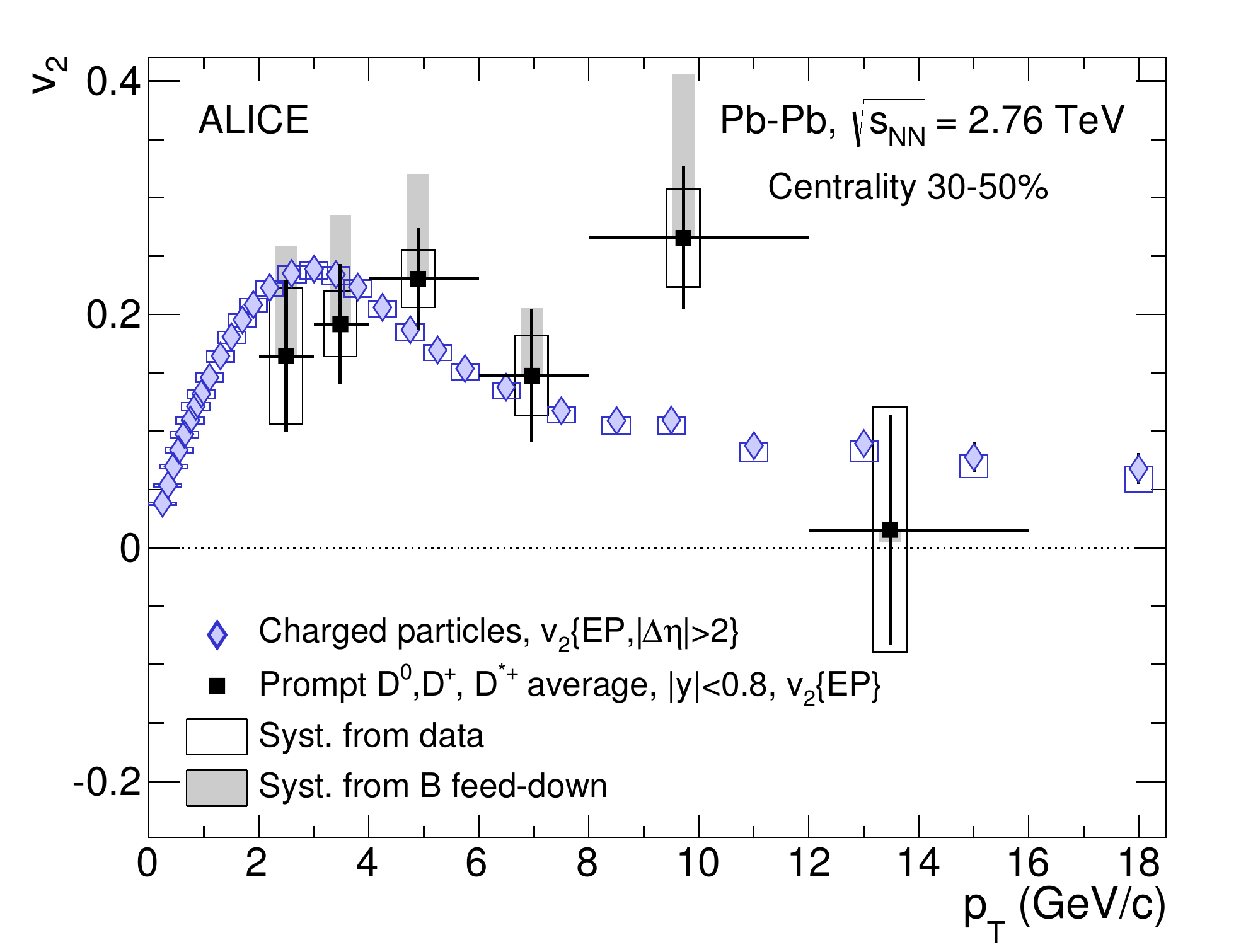}}}$
\caption{$D$-meson $\Raa$ in central~\cite{Adam:2015sza} and $v_2$ in semi-peripheral~\cite{Abelev:2013lca}
Pb--Pb collisions at $\sqrtsNN=2.76~\TeV$ compared to results for charged pions~\cite{Abelev:2014laa}
and charged particles~\cite{Abelev:2012hxa,Abelev:2012di}.}
\label{fig:Raav2DandPi}
\end{center}
\end{figure}

It is interesting to compare both the nuclear modification factor and 
the elliptic flow of $D$ mesons with those of light-flavor particles,
as done in Fig.~\ref{fig:Raav2DandPi}.
This could provide some additional insight into the interactions of partons 
with the medium constituents and on the degree
of equilibration of charm quarks in the collectively expanding system.
In the right panel of Fig.~\ref{fig:Raav2DandPi}, the elliptic
flow coefficients of $D$ mesons~\cite{Abelev:2013lca} and charged 
particles~\cite{Abelev:2012di} measured in Pb--Pb 
collisions at $\sqrtsNN=2.76~\TeV$ in the 30--50\% centrality class
are compared.
The magnitude and $\pT$ trend of $v_2$ are observed to be similar (within
uncertainties of about 30\%) for charmed and light-flavor hadrons, which 
dominate the charged-particle sample.
A similar observation was made for the ${D^0}$-meson and 
charged-particle $v_2$ in the 0--10\% and 10--30\% centrality 
classes~\cite{Abelev:2014ipa}.
The current uncertainties do not allow a conclusion on whether the
$D$-meson elliptic flow follows the mass dependence predicted by
hydrodynamical calculations, which would suggest a full thermalization of 
charm quarks with the medium.
The comparison of the nuclear modification factor of $D$ mesons and 
light-flavor particles (pions), mostly originating from gluon fragmentation 
at LHC energies, was long proposed as a test for the expected color-charge 
and quark mass dependence of in-medium parton energy 
loss~\cite{Gyulassy:1990ye,Baier:1996sk,Dokshitzer:2001zm,Armesto:2003jh}.
This comparison is shown in the left panel of Fig.~\ref{fig:Raav2DandPi}
for the 10\% most central Pb--Pb collisions at $\sqrtsNN=2.76~\TeV$, where
the $\Raa$ of $D$ mesons, pions 
(in the interval $1<\pT<20~\GeV/c$)~\cite{Abelev:2014laa} and charged 
particles (in $16<\pT<40~\GeV/c$)~\cite{Abelev:2012hxa} are collected.
At high $\pT$ ($\pT>8-10~\GeV/c$) all light-flavor hadron species are 
found to be equally suppressed in Pb--Pb collisions and the particle 
ratios are compatible with those in vacuum~\cite{Abelev:2014laa}, so that the
charged particle $\Raa$ can be used in this comparison in place of the 
pion $\Raa$ at high $\pT$.
The nuclear modification factors of $D$ mesons and light-flavor hadrons are 
found to be consistent for $\pT>6~\GeV/c$.
At lower $\pT$ ($\pT<6~\GeV/c$), the $\Raa$ of $D$ mesons tends to be 
slightly higher than that of pions. 
Since, as pointed out in~\cite{Adam:2015sza}, the systematic uncertainties of 
$D$-meson yields are mainly correlated across $\pT$ intervals, the current
data provide a hint for $\Raa^{\rm D}>\Raa^{\pi}$ (at about a 1\,$\sigma$ level)
at intermediate and low $\pT$.
An interpretation of this potential difference between the $D$-meson and 
pion $\Raa$ at in terms of different in-medium parton energy 
loss of charm quarks, light quarks and gluons is, however, not straightforward 
because in this $\pT$ range the $\Raa$ is sensitive to other initial- and 
final-state effects, which could have rather different weights in the light 
and charm sectors.
As pointed out in Ref.~\cite{Djordjevic:2013pba}, similar values of
the $D$-meson and pion $\Raa$ could originate from the interplay 
of the color-charge and quark-mass dependent energy loss with the 
different $\pT$ distributions in the $pp$ reference and different 
fragmentation functions of charm quarks as compared to light quarks and 
gluons. In addition, at LHC energies the pion yield at $\pT\sim 3-4~\GeV/c $ 
could still have a significant component from soft production (due to 
the strong radial flow) which does not scale with the number of binary 
nucleon--nucleon collisions (contrary to the $D$-meson yield, modulo
shadowing).
A priori, initial-state effects, radial flow and hadronization via recombination 
(see the above remark about the hadro-chemistry in the charm sector) can
affect the $\Raa$ of $D$ mesons and pions quite differently.

\begin{figure}
\begin{center}
\includegraphics[width=0.48\textwidth]{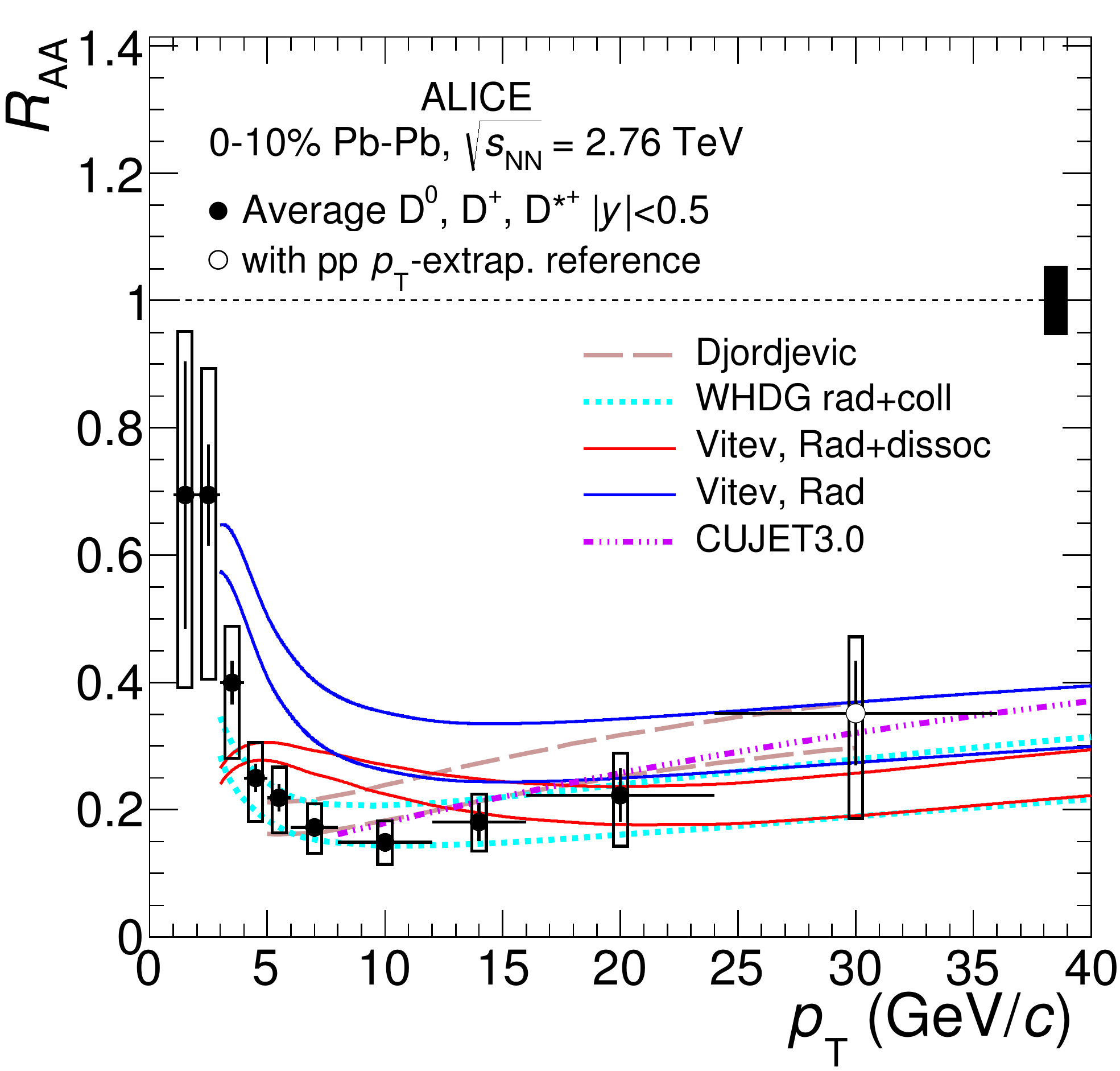}
\includegraphics[width=0.48\textwidth]{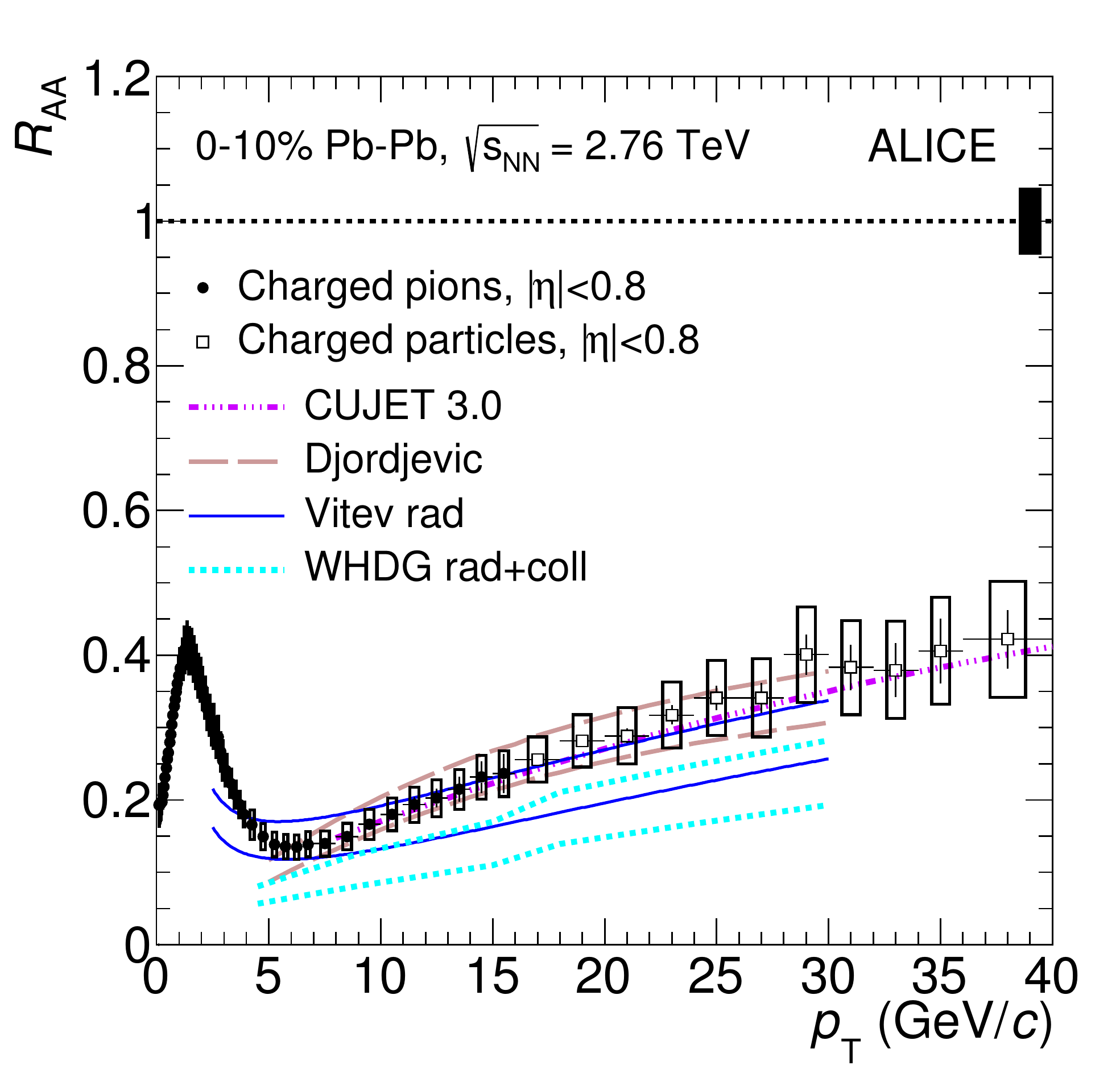}
\caption{$D$-meson~\cite{Adam:2015sza}, pion~\cite{Abelev:2014laa} and 
charged-particle~\cite{Abelev:2012hxa} $\Raa$ as a 
function of $\pT$ in the 10\% most central Pb--Pb collisions at 
$\sqrtsNN=2.76~\TeV$, compared to predictions from pQCD energy loss 
models: Djordjevic~\cite{Djordjevic:2014tka}, 
CUJET 3.0~\cite{Xu:2014tda,Xu:2015bbz}, WHDG with radiative and collisional
energy loss~\cite{Wicks:2005gt,Horowitz:2011gd,Horowitz:2011wm} and Vitev 
with radiative and dissociation processes~\cite{Sharma:2009hn}.}
\label{fig:RaaDandPiVsModels}
\end{center}
\end{figure}

Four models provide a calculation for the nuclear modification factors of 
$D$ mesons and pions (charged particles) namely 
Djordjevic {\it et al.}~\cite{Djordjevic:2014tka},  
CUJET3.0~\cite{Xu:2014tda,Xu:2015bbz},
WHDG~\cite{Wicks:2005gt,Horowitz:2011gd,Horowitz:2011wm} and 
Vitev {\it et al.}~\cite{Sharma:2009hn}.
In Fig.~\ref{fig:RaaDandPiVsModels}, the outcome of these model 
calculations is compared to the measured $\Raa$ of $D$ mesons (left) and 
pions/charged particles (right) in central Pb--Pb collisions at the LHC.
The models Djordjevic {\it et al.}, WHDG and CUJET3.0 include both radiative 
and collisional energy loss.
The WHDG calculations tend to overpredict the measured suppression of the 
pion $\Raa$ while describing the $D$-meson one within experimental and 
theoretical uncertainties. 
For the Vitev {\it et al.} model, two different implementations are considered: 
the first one (labelled as 'Vitev rad') includes only radiative energy loss;
the second one ('rad+dissoc') considers in addition the effect of in-medium 
formation and dissociation of HF hadrons.
The in-medium formation and dissociation process is not considered as being 
relevant for pions due to their much longer formation time.
These model calculations can describe the measured
light-flavor $\Raa$ within uncertainties, while for $D$ mesons a better 
agreement with the data is obtained 
when the in-medium dissociation mechanism is included in the calculation, 
indicating the relevance of this effect in the Vitev {\it et al.} approach. 
However, within this model an 
overestimation of the measured $D$-meson $\Raa$ toward lower momenta, 
$\pT\ltsim12~\GeV/c$, persists.
The Djordjevic {\it et al.} and CUJET3.0 models can describe both the pion and 
$D$-meson $\Raa$ results (as well as their ratio, see Ref.~\cite{Adam:2015sza}) 
over the full $\pT$ interval of the calculations ($\pT>5$ and 8~GeV/$c$, respectively).
In these models, which include collisional and radiative energy loss,
the nuclear modification factors of $D$ mesons and light-flavor hadrons turn 
out to be similar as a consequence of 
the interplay among (i) the larger energy loss of gluons with respect to that 
of charm quarks (mainly due to the larger color coupling factor),  
(ii) the different amount of gluon and light-quark contributions to the 
observed pion yield in $pp$ and Pb--Pb collisions, and 
(iii) the harder $\pT$ distribution and fragmentation of 
charm quarks with respect to those of gluons and light quarks.

\subsubsection{Beauty production}
\hspace{2cm}

Besides the measurements of leptons from beauty-hadron decays described
above, beauty production in Pb--Pb collisions at the LHC was studied at 
mid-rapidity through the measurements of non-prompt $J/\psi$ carried out 
by the ALICE~\cite{Adam:2015rba} and CMS~\cite{Chatrchyan:2012np} 
collaborations.  These measurements cover the low- and intermediate-$\pT$ 
regions, $1.5<\pT^{J/\psi}<30~\GeV/c$, although with large uncertainties at 
low $\pT$.  At higher momenta, $\pT>80~\GeV/c$, beauty production could be 
studied by CMS via the measurement of $b$-jets~\cite{Chatrchyan:2013exa}.

\begin{figure}
\begin{center}
\includegraphics[width=0.48\textwidth]{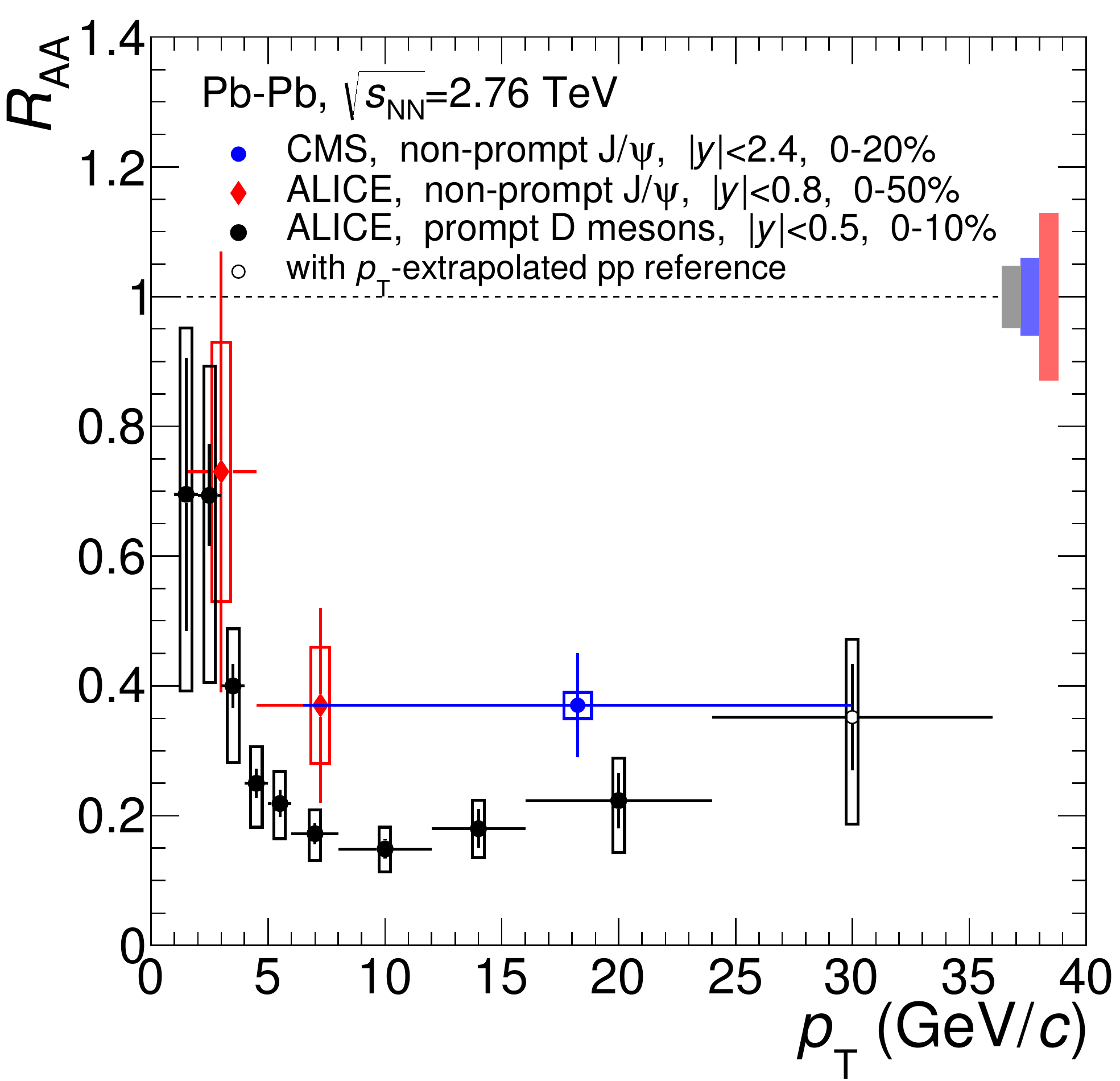}
\includegraphics[width=0.48\textwidth]{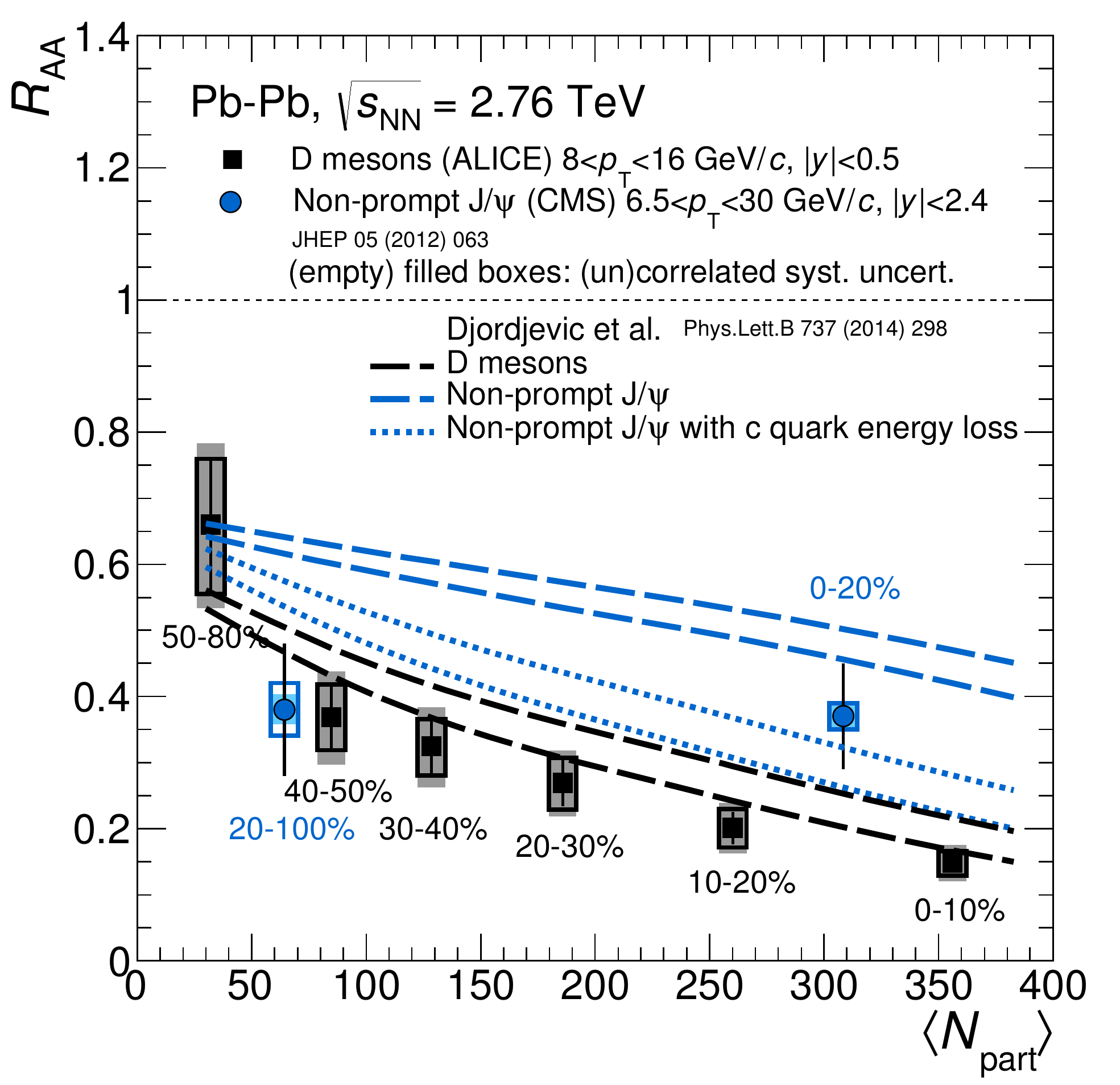}
\caption{Left: non-prompt $J/\psi$ $\Raa$ as a function of $\pT$ at mid-rapidity
in Pb--Pb collisions at $\sqrtsNN=2.76~\TeV$ compared to the prompt 
$D$-meson $\Raa$~\cite{Adam:2015sza}.
Results from ALICE~\cite{Adam:2015rba} at low $\pT$ in the 0--50\% centrality 
class and CMS~\cite{Chatrchyan:2012np} at high $\pT$ in the 0--20\% centrality 
class are reported.
Right: non-prompt $J/\psi$~\cite{Chatrchyan:2012np} and prompt $D$-meson~\cite{Adam:2015nna}
$\Raa$ as a function of centrality compared to model predictions including 
radiative and collisional energy loss~\cite{Djordjevic:2014tka}.
For non-prompt $J/\psi$ additional model calculations are shown in which the 
beauty-quark interactions are calculated using the charm-quark mass~\cite{Andronic:2015wma}.}
\label{fig:RaaJpsiB}
\end{center}
\end{figure}

The $\Raa$ of non-prompt $J/\psi$ in Pb--Pb collisions at $\sqrtsNN=2.76~\TeV$ 
is shown in the left panel of Fig.~\ref{fig:RaaJpsiB} as a function of $\pT$.
The ALICE results (at low $\pT$, $1.5<\pT^{{\rm J}/\psi}<10~\GeV/c$) are obtained
for the centrality class 0--50\%~\cite{Adam:2015rba}, while the CMS results 
at higher $\pT$ ($6.5<\pT^{{\rm J}/\psi}<30~\GeV/c$) are for the 20\% most central 
collisions~\cite{Chatrchyan:2012np}.
A clear suppression as compared to the binary-scaled $pp$ reference is observed 
in central collisions for $\pT^{{\rm J}/\psi}>6.5~\GeV/c$, with 
$\Raa=0.37 \pm 0.08\mathrm{(stat)} \pm0.02\mathrm{(syst)}$.
A suppression is also observed in the intermediate $\pT$ interval,
covering the range $4.5<\pT^{{\rm J}/\psi}<10~\GeV/c$, albeit with larger 
uncertainties.
A larger data sample is needed at lower $\pT$, where the uncertainties on the
current result do not allow to draw any conclusion.
Nevertheless, the results at intermediate and high $\pT$ indicate that the
beauty quarks are substantially affected by interactions with the 
constituents of the hot and dense medium, which induce
a significant modification of their momentum distributions 
in heavy-ion collisions as compared to those observed in $pp$ interactions.
The nuclear modification factor of non-prompt $J/\psi$ is compared to that 
measured for prompt $D$ mesons (average of ${D^0}$, ${D^+}$ and 
${D^{*+}}$) in central (0--10\%) Pb--Pb collisions~\cite{Adam:2015sza}.
This comparison is meant to test the expected quark-mass dependence of 
in-medium energy loss~\cite{Dokshitzer:2001zm,Armesto:2003jh,Armesto:2005iq}.
The suppression of non-prompt $J/\psi$ seems to be weaker than that of $D$ mesons
at high and intermediate $\pT$, although the uncertainties on the 
measurements reported in Fig.~\ref{fig:RaaJpsiB} prevent from drawing
strong conclusions.
In the discussion of this comparison of $\Raa$ magnitudes, it is worth noting 
that the $\pT$ of the $J/\psi$ is shifted to lower 
momenta with respect to that of the parent $B$ meson, due to the decay kinematics.
The average $\pT$ of the parent $B$ mesons in the highest J/$\psi$ transverse 
momentum interval measured by CMS is about $11~\GeV/c$.
For a more direct comparison of $D$ and $B$ nuclear modification, the $D$-meson
$\Raa$ was measured by the ALICE Collaboration~\cite{Adam:2015nna} in the 
interval $8<\pT<16~\GeV/c$, which provides a significant overlap with the
$\pT$ distribution of $B$ mesons decaying to $J/\psi$ particles with 
$6.5<\pT^{{\rm J}/\psi}<30~\GeV/c$ (for which 70\% of the parent $B$ mesons 
are estimated to have transverse momenta in the range $8<\pT<16~\GeV/c$).
In the right panel of Fig.~\ref{fig:RaaJpsiB}, the centrality dependence
of the nuclear modification factors of $D$ mesons and non-prompt $J/\psi$
in the chosen $\pT$ intervals are compared.
The $D$-meson $\Raa$ values in the centrality classes 0--10\% and 10--20\% 
are lower than that of non-prompt $J/\psi$ mesons in the centrality class 
0--20\%.
The significance of this difference is, however, smaller than $3\,\sigma$
considering the statistical and systematic uncertainties~\cite{Adam:2015nna}.
A preliminary measurement of non-prompt $J/\psi$ production performed on the 
larger data sample of Pb--Pb collisions recorded in 2011 was reported by the 
CMS Collaboration~\cite{CMS:2012vxa}.
The non-prompt $J/\psi$ nuclear modification factor in $|y|<1.2$ 
is measured as a function of centrality using finer centrality intervals 
and the same $\pT$ interval ($6.5<\pT^{{\rm J}/\psi}<30~\GeV/c$) of the published 
result.
Considering this measurement, the $\Raa$ of non-prompt $J/\psi$ is
larger than that of $D$ mesons in the 0--10\% and 10--20\% centrality
classes with a significance of about $3.5\,\sigma$~\cite{Adam:2015nna}.

The experimental observation of 
$\Raa^D<\Raa^{J/\psi \leftarrow B}$ alone does not allow to 
conclude on the predicted difference between the in-medium energy loss 
of charm and beauty quarks.
In analogy to the comparison of charm and light-flavor hadron nuclear
modifications discussed above, effects other than quark-mass dependent 
parton energy loss could contribute to differences in the $\Raa$: 
(i) the different $\pT$ 
distributions of the initially produced charm and beauty quarks (which are
steeper for charm than for beauty), and 
(ii) the different shapes of the fragmentation functions (which is harder 
for beauty than for charm quarks) as well as (iii) recombination contributions.
Therefore, for a proper interpretation of the experimental results,  
the measured $\Raa$ of $D$ and $B$ mesons (via non-prompt $J/\psi$) should be 
compared with the outcome of model calculations including HQ 
production, in-medium propagation and hadronization.
Essentially all available models predict 
\mbox{$\Raa^{D} < \Raa^{B}$} in the momentum range $\pT<20~\GeV/c$,
where the quark masses are not negligible with respect to their 
momenta~\cite{Wicks:2005gt,Horowitz:2011gd,Horowitz:2011wm,Armesto:2005iq,Adil:2006ra,Sharma:2009hn,Buzzatti:2012dy,He:2012xz,He:2014cla, Gossiaux:2012ya,Uphoff:2012gb,Alberico:2013bza,Cao:2013ita,Lang:2012cx}.
In the right panel of Fig.~\ref{fig:RaaJpsiB}, the data are compared
to the calculations by Djordjevic {\it et al.}~\cite{Djordjevic:2014tka},
which include both radiative and collisional processes and consider 
dynamical scattering centers in the medium.
Note that, as discussed above, this model can describe the similarity of 
the $D$-meson and pion $\Raa$.
The model describes well the centrality dependence of the $D$-meson nuclear
modification factor in the high-$\pT$ range considered in 
Fig.~\ref{fig:RaaJpsiB} and predicts a smaller suppression of non-prompt
$J/\psi$ mesons as compared to $D$ mesons, in qualitative agreement with
the CMS result for the most central collisions.
Care has to be taken in the data-to-model comparison for the 20--100\%
class, as the centrality interval is very broad.
The preliminary CMS results in finer centrality intervals
from the 2011 Pb--Pb sample~\cite{CMS:2012vxa}, are well described
by the calculations of  Djordjevic {\it et al.} (see Fig.~4 in 
Ref.~\cite{Djordjevic:2014tka}).
In order to study the origin of the difference in the nuclear modification
factors of $D$ and $B$ mesons in this model, the $\Raa$ of non-prompt 
$J/\psi$ was also computed using the charm-quark mass value in the 
calculation of the in-medium interactions of beauty quarks.
The outcome of this test case, depicted as dotted lines in the right panel 
of Fig.~\ref{fig:RaaJpsiB}, shows a substantially lower $\Raa$ of 
non-prompt $J/\psi$, close to that of $D$ mesons, as compared to the case
in which the beauty-quark mass is used in the calculation (dotted lines in 
Fig.~\ref{fig:RaaJpsiB}).
This indicates that, in this model, the large difference in the $\Raa$ of 
$D$ mesons and non-prompt $J/\psi$ arises predominantly from the mass 
dependence of quark-medium interactions and is only moderately affected 
by the different production and fragmentation kinematics of charm and beauty 
quarks.
Similar conclusions are derived by performing the same test with the 
MC@sHQ+EPOS and TAMU models (see~Refs.~\cite{Adam:2015nna,Andronic:2015wma}).

\begin{figure}
\begin{center}
\includegraphics[width=0.48\textwidth]{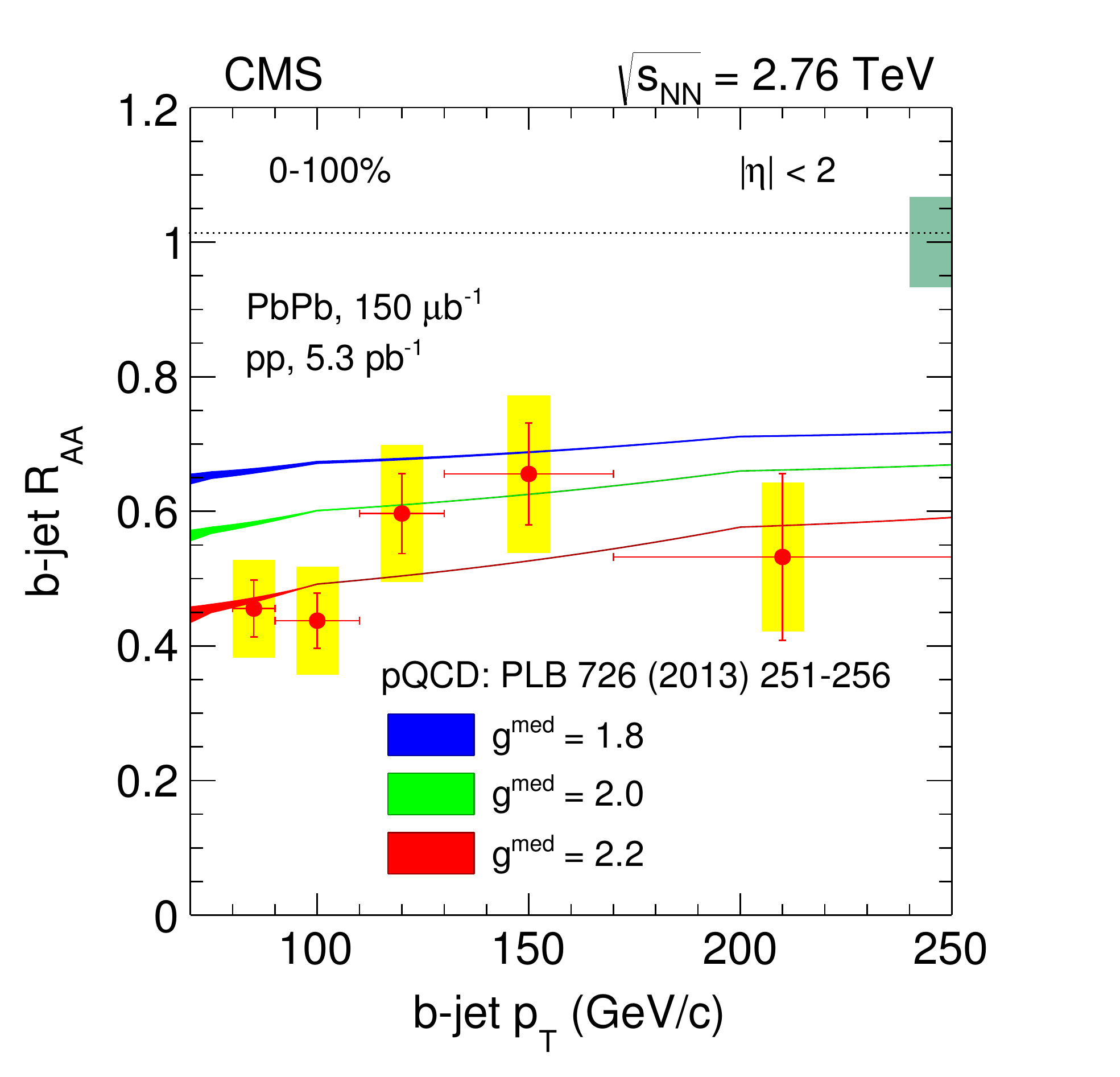}
\includegraphics[width=0.48\textwidth]{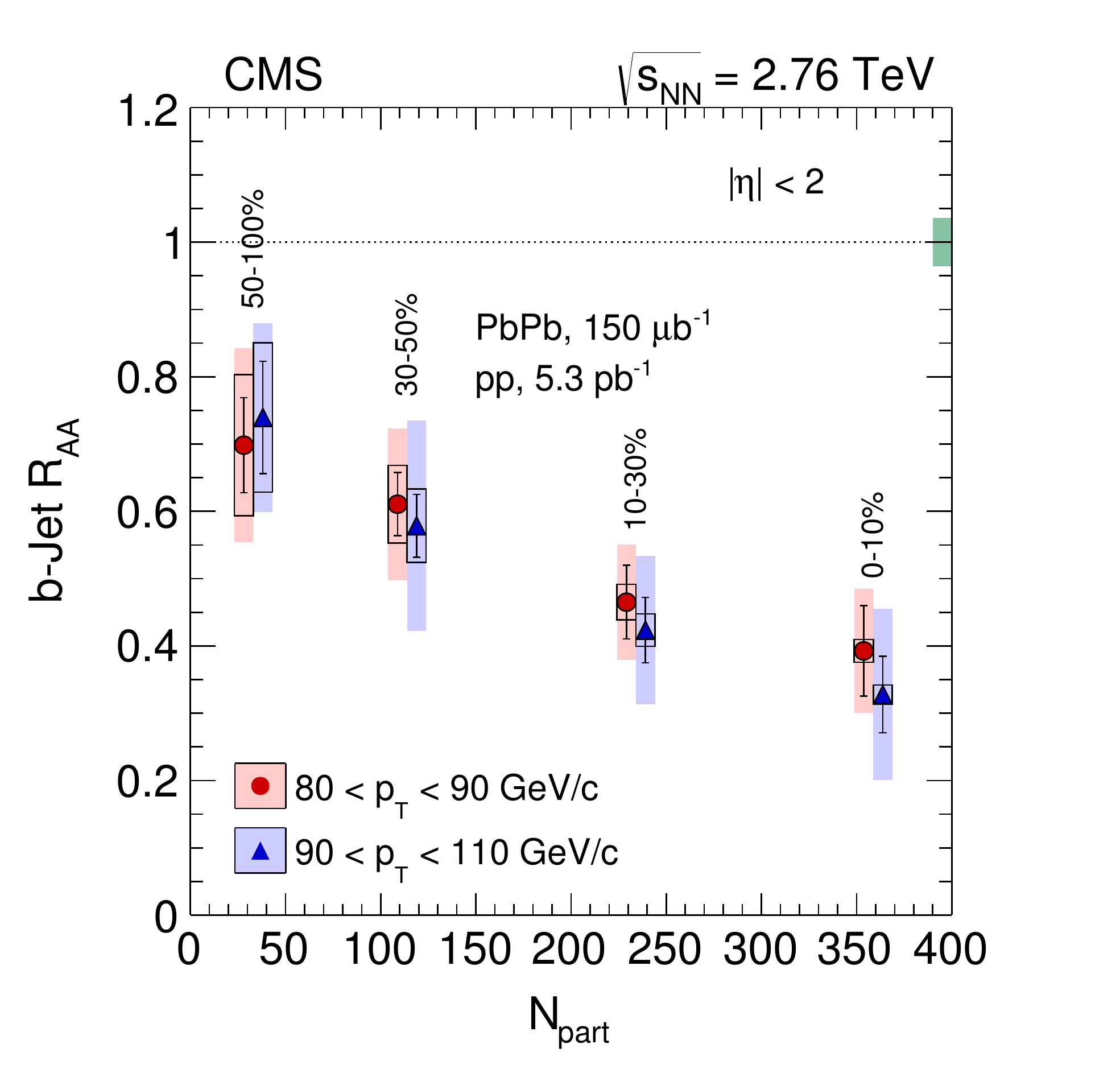}
\caption{Nuclear modification factor of $b$-jets in
Pb--Pb collisions at $\sqrtsNN=2.76~\TeV$~\cite{Chatrchyan:2013exa}.
Left: centrality-integrated $\Raa$ as a function of jet $\pT$ compared
to pQCD energy-loss calculations~\cite{Huang:2013vaa}.
Right: $\Raa$ as a function of centrality for two different
jet $\pT$ intervals.}
\label{fig:bjetRaa}
\end{center}
\end{figure}

A complementary approach to study beauty-quark interactions with the medium
is provided by measurements of $b$-jets.
Assuming that the quark hadronizes outside the medium, the jet energy 
should be, in first approximation, the sum of the energy of the beauty quark 
after its interaction with the medium and of the energy transferred by the 
quark to the medium that remains inside the jet cone. 
The nuclear modification factor of $b$-jets is shown in Fig.~\ref{fig:bjetRaa}.
In the left panel, the centrality-integrated $\Raa$ is displayed
as a function of $\pT$ and compared to the pQCD-based calculations
by Vitev {\it et al.} including radiative and collisional energy 
loss~\cite{Huang:2013vaa}.
The data show a significant suppression, almost independent of $\pT$ in
the measured range, of the $b$-jet yield relative to the 
$pp$ expectation, indicating parton energy loss in the hot and dense medium.
The measured $\Raa$ is described by the Vitev {\it et al.} model using values
of the jet-medium coupling parameter in the range $g^{\rm med}=1.8-2$, 
similar to the value found for inclusive jets.
In the right panel of Fig.~\ref{fig:bjetRaa}, the $b$-jet $\Raa$ is 
shown as a function of centrality, expressed in terms of the number of 
participant nucleons, $N_{\rm part}$, for two $\pT$ intervals.
A smooth decrease of $\Raa$ with increasing centrality is observed, reaching
a suppression of a factor of about 2.5 in the most central collisions.
In the $\pT$ range covered by these measurements the nuclear modification
of the $b$-jet yield is found to be compatible, within the sizeable systematic 
uncertainties, with that of inclusive jets~\cite{Aad:2012vca,CMS:2012kxa}.
This observation challenges models based on the strong-coupling limit, \eg, 
within the AdS/CFT correspondence~\cite{Horowitz:2007su}, in which 
quark-mass dependent effects persist up to large $\pT$.
On the other hand, in pQCD-based models the quark mass effects are
expected to be small at large $\pT$~\cite{Huang:2013vaa}.
Nevertheless, even though quark mass related effects may not play a role 
in the high-momentum interval where the measurement was carried out, 
a difference between the $\Raa$'s of $b$-jets and inclusive jets could have been 
expected due to the color charge dependence of energy loss, since inclusive 
jets at LHC energies should be dominated by gluon jets up to very large $\pT$.
In this respect, it should be considered that at LHC energies a sizeable
fraction of beauty quarks are produced by splitting of gluons into 
$b\bar{b}$ pairs (gluon splitting)~\cite{Banfi:2007gu}.
As pointed out in Ref.~\cite{Andronic:2015wma}, in the case of $b$-jets
at very high $\pT$, a significant part of the in-medium path length is likely 
to be covered by the parent gluon.
For example, the formation (coherence) time of beauty quarks with an energy of 
150 GeV is of about 1~fm/$c$, implicating that the very early 
(hot and dense) stages of the medium evolution are probed by the
parent gluon, and not by the beauty quark.
Note that, following the arguments of coherence time for HQ
pairs from gluon splitting~\cite{Andronic:2015wma}, the 
fact that the medium could be probed by the parent gluon is relevant
only for beauty quarks at high $\pT$.
In the case of charm (for which the contribution of gluon splitting
is smaller, $\approx$\,10-20\%, than for 
beauty~\cite{Mueller:1985zz,Mangano:1992qq}), and for 
beauty at not-too-high $\pT$ ($\pT\ltsim50~\GeV/c$, for which the coherence time 
is small because it is not increased by a large Lorentz boost) it is 
reasonable to assume that HF hadrons (and jets) probe the 
interactions of the heavy quarks with the medium.


\subsection{Summary of Model-to-Data Comparisons and Implications for HF Transport}
\label{ssec_impl}
In this section we summarize and discuss the current experimental results in
light of their theoretical interpretations, organized into two regimes of high 
and low transverse momentum following the expectation of different prevailing 
processes. We augment this discussion with an attempt to identify a 
transverse-momentum scale which possibly delineates the predominantly elastic 
and radiative interaction regimes, accompanied by schematic estimates of the 
pertinent HF transport coefficients. Since theoretical calculations of HF 
transport predict appreciable momentum depedencies of the coefficients (recall 
Fig.~\ref{fig_A}), a separation into regions of ``low" and ``high" $\pT$ may 
be considered a minimal accounting procedure of this aspect, while keeping
in mind that the dominant effects may also arise from rather different 
temperatures (\eg, around $T_{\rm pc}$ vs. the early hottest phases). 


\vskip 0.5 cm

At high $\pT$, the initial HQ spectra are far above the equilibrium limit of the
 medium in heavy-ion collisions, and thus the interactions chiefly probe energy 
{\em loss} mechanisms of charm and beauty particles. This can also be gleaned 
from the Langevin process, eq.~(\ref{langevin}), where the dominant mechanism 
becomes a momentum degradation by a force $dp_j/dt =-\Gamma(p,T) p_j$. In a 
simplified form, the friction coefficient, 
$\Gamma\simeq A \sim n_p \sigma_{Qp}^{\rm trans}$, is a product of medium 
density and HQ-parton transport cross section, where the latter is expected 
to be in the perturbative regime with a mild temperature dependence. Since 
$n_p\propto T^3$, the energy loss is presumably dominated by the hottest phases 
in a heavy-ion collision. The RHIC and LHC data on the nuclear modification 
factor of $D$ mesons, non-prompt $J/\psi$ and HF decay leptons all show a strong 
suppression of the high-$\pT$ HF yield in semi-/central heavy-ion collisions
relative to binary scaled $pp$ cross sections, 
clearly associated with final-state effects. For high-$\pT$ $D$-mesons in 
central Pb--Pb collisions at the LHC, the BAMPS, Duke and MC@sHQ+EPOS models
describe the 
ALICE data well, while the elastic approaches of POWLANG and TAMU over- and 
under-predict the suppression, respectively. Taking Fig.~\ref{fig_A} around 
$p$$\simeq$10\,GeV and for $T$$\simeq$0.3-0.4\,GeV as 
guidance (with MC@sHQ+EPOS typically requiring a moderate $K$ factor for 
elastic-only calculations), we estimate an ``average" high-$p_T$ friction 
coefficient of $A\simeq 0.2\pm0.08$/fm to describe the data (this probably 
includes part of the systematic uncertainty due to, \eg, the different medium 
evolutions employed in the transport models). The POWLANG and TAMU coefficients 
with $A(p$=10\,GeV)$\simeq$0.25-0.3/fm and 0.07-0.1/fm, respectively, roughly 
bracket the estimated uncertainty. A conversion of the HQ transport coefficients 
into 
a jet-quenching parameter, $\hat{q}$, as commonly adopted for light-parton 
energy loss, has been suggested in Ref.~\cite{Gubser:2006nz} 
in terms of the transverse-momentum broadening per unit path length, leading to 
$\hat{q} \equiv \hat{q_\perp} = 4TE_p^2A/p$ ($E_p$: HQ energy). This translates 
into an estimate of 
$\hat{q}\simeq 2.8\pm 1.1$\,GeV$^2$/fm at $T$$\simeq$0.35\,GeV, or 
$\hat{q}/T^3\simeq13\pm5$, 
which is more stable in temperature. This result tends to be larger than the values 
extracted from light-parton jet quenching~\cite{Burke:2013yra}, possibly indicative 
of non-perturbative physics (and/or larger contributions from elastic interactions) 
which could still play a role for a charm quark at $\pT\simeq10$\,GeV.

The measurements of the azimuthal momentum anisotropy suggest a positive $v_2$ 
at high $\pT$, but they do not have enough precision yet to yield meaningful 
constraints on transport coefficients or the path length dependence of parton 
energy loss.


The beauty sector, currently accessed via non-prompt $J/\psi$ mesons and $b$-jets
at the LHC, shows less suppression in the $\Raa$ than charm in a $\pT$ interval 
around 
$10~\GeV/c$. This observation is consistent with the expectation of quark-mass
dependent energy loss (gluon Bremsstrahlung off the heavier beauty quarks is 
suppressed). However, the measured $\Raa$ and $v_2$ of HF decay leptons at both 
RHIC and the LHC also show fair agreement with the BAMPS, MC@sHQ+EPOS and TAMU 
transport calculations, up to $p_t$\,$\simeq$\,10\,GeV. 
Since TAMU does not inlcude radiative contributions, this suggests that the 
$\pT$$\simeq$10\,GeV regime for beauty is still in the realm of elastically 
dominated interactions. The generic expectation for the transition from elastic 
to radiative regimes is a scaling with mass. Thus, based on an estimate for the 
transition regime around $\pT^{\rm trans}$$\simeq$15\,GeV for beauty, one would 
deduce $\pT^{\rm trans}$$\simeq$5\,GeV for charm. The latter value is supported
in the discussion of the low-$\pT$ regime below.

The color-charge dependence of parton energy loss has been one of the original 
motivations for high-$\pT$ HF measurements. The measured $\Raa$'s of $D$ mesons 
and pions, which are expected to be sensitive to the different coupling of 
gluons and quarks with the medium, agree within current uncertainties at RHIC 
and LHC energies. The $\Raa$'s of $b$-jets and inclusive jets are also observed 
to be compatible within uncertainties at very high momenta, $\pT>80$\,GeV. 
Hence, an experimental evidence of the 
color-charge dependence of energy loss remains elusive from the current data.
According to model calculations, the similarity of the $\Raa$'s for $D$ mesons 
and pions could result from the combined effect of a color-charge dependent 
energy loss and the softer $\pT$ distribution and fragmentation function of 
gluons relative to charm quarks. 
To scrutinize this question, comparisons of $B$-, $D$- and $\pi$-meson data
over a larger range in $\pT$ will be needed, at both RHIC and the LHC.
However, care has to taken when going down to $\pT$'s where effects of radial
flow and/or recombination set in. In the beauty sector these could become 
relavant at transverse momenta as high as 15\,GeV. Likewise, at very high
$p_T$, the gluon splitting contribution to $b$-quarks is expected to become
sizable (implying that part of the energy loss is suffered by the parent 
gluon).

\vskip 0.5 cm

The low-$\pT$ region is particularly interesting because of its sensitivity
to a low-momentum transport coefficient of the QCD medium, \ie, the HF diffusion
coefficient $D_s$. In addition, the approach toward equilibrium does not merely
induce a suppression of the spectra, but is expected to produce non-monotonic 
structures (like a ``flow bump" in the HF hadron $\Raa$'s) whose quantitative 
features (such as the flow bump's height and location in $\pT$) are especially 
revealing for both medium evolution and the transport properties of the embedded 
HF particles. Measurements of $\Raa$ and $v_2$ of different HF hadron species 
are expected to quantify the degree to which charm and beauty particles participate 
in the collective expansion of the system, thus directly reflecting their 
coupling strength to the medium. This includes the effects of in-medium
hadronization, which is a manifestation of the HQ-medium coupling in the 
vicinity of $T_{\rm pc}$ (through hadronic pre-/resonant states). Experimentally, 
the $D$-meson $\Raa$ at RHIC energy shows a pronounced maximum around  
$\pT$$\simeq$1-2~GeV, which was predicted by models including strong elastic 
interactions of charm quarks in an expanding QGP, together with hadronization 
via in-medium heavy-light quark recombination. The same models also describe 
the low-$\pT$ $D$-meson $\Raa$ measured at the LHC, where the ``radial-flow 
bump" is less pronounced (\eg, due to the harder initial spectra and stronger 
shadowing) and/or not yet resolved. The approximate agreement of these models 
extends to the $\Raa$ of the HF decay leptons at both accelerators.
Importantly, a fair description also emerges of the low-$\pT$ elliptic flow, 
which is measured to be at the $\sim$10-15\% level for $D$ mesons (slightly
smaller for HF leptons) in non-central collisions at RHIC and the LHC,
with still rather significant uncertainties.
The data favor scenarios in which the heavy quarks pick up substantial 
collectivity from the expanding QGP, which is then
further augmented by hadronization through recombination (by several tens of
percent), and by another 10-20\% in the hadronic phase. The relative importance 
of the hadronization process in generating $v_2$ is another reflection of a 
strong HQ-medium coupling in the vicinty of $T_{\rm pc}$.   
At first sight, the underlying HF transport coefficients appear to be quite 
different among current models. For example, comparing the low-momentum 
region of the MC@sHQ+EPOS (or Nantes) and TAMU models from Fig.~\ref{fig_A} 
left, an appreciable discrepancy is found. The discrepancy is, however, 
smallest at the lowest temperatures, less than a factor of 2. It is further 
mitigated if one recalls that the charm-quark mass close to $T_{\rm pc}$ in 
the MC@sHQ+EPOS model ($m_c$=1.5\,GeV) is smaller than in the TAMU model (where 
the in-medium mass grows as $T_{\rm pc}$ is approached from above, reaching up 
to $m_c$$\simeq$1.8\,GeV). In the conversion from $A$ to the spatial diffusion
coefficient $D_s$, the $m_Q$ dependence of the former is approximately 
divided out, cf.~eq.~(\ref{Ds}), which can be recast as $D_s\simeq T/m_QA(p=0)$.  
This suggests that the coupling strength around $T_{\rm pc}$ is most important 
in building up the HF $v_2$, which is not surprising since the bulk $v_2$ is 
largest from this stage on~\cite{Rapp:2008zq,Das:2015ana} (the importance of 
recombination processes in the hadronization of the heavy quarks further
corroborates this point). More quantitatively from Fig.~\ref{fig_A}, taking 
$T$=0.2\,GeV, one finds $D_s (2\pi T)$\,$\simeq$\,2-3 for the Nantes 
coefficients and $D_s (2\pi T)$\,$\simeq$\,3-4 for the TAMU one. For the 
PHSD model, which also reproduces the $D$-meson $\Raa$ and $v_2$ at the LHC 
reasonably well, the underlying spatial diffusion constant is somewhat 
larger, $D_s (2\pi T)$\,$\simeq$\,5-6 (with $m_c$=1.5\,GeV).  
However, the concrete implementation of heavy-light quark recombination differs 
significantly in the current models, both technically as well as in its
impact on observables (the TAMU implementation tends to give larger effects
from recombination than the one in MC@sHQ+EPOS, Duke and PHSD). This contributes 
significantly to the uncertainty in the model interpretations of the experimental 
data (such as the extraction of the transport coefficient).  


Promising tools to discern different types of hadronization mechanisms and 
the nature of the interactions near $T_{\rm pc}$ are the $\Raa$ and $v_2$ of 
HF hadrons with different quark composition and different mass. In particular, 
the production of mesons carrying a strange quark ($D_s^+$ and $B_s^0$) and 
of baryons ($\Lambda_c^+$, $\Xi_c^+$, $\Xi_c^0$, $\Lambda_b^0$) at low and 
intermediate $\pT$ has been suggested to encode pertinent information. The 
first measurement of the ${D_s^+}$ $\Raa$ at the LHC and a subsequent 
preliminary result at RHIC provide a hint for the relevance of recombination 
of charm with thermal strange quarks, although the current uncertainties do not 
allow for definite conclusions. A stronger signal for recombination of charm 
quarks, with anti-charm quarks, comes from the observed $J/\psi$ 
$\Raa$~\cite{Abelev:2012rv,Abelev:2013ila} and $v_2$~\cite{ALICE:2013xna}
at the LHC. The larger data samples that will be collected in the next years 
will be most valuable to shed further light on the relevance heavy-light and 
heavy-heavy recombination processes.

Let us finally return to question of elastic vs.~radiative HQ interactions 
in the QGP. The TAMU transport approach with non-perturbative elastic 
interactions only approximately describes the $\Raa$ and $v_2$ of $D$-mesons 
up to $\pT$$\simeq$5\,GeV (with significant deviations above), and the HF 
lepton observables, with a substantial beauty component, out to 
$\pT$$\simeq$10\,GeV.
This may serve as an initial estimate of the regime where elastic interactions
dominate the HQ coupling to the QCD medium. Interestingly, the BAMPS and 
MC@sHQ+EPOS approaches seem to do best in a simultaneous description of $\Raa$ 
and $v_2$ toward higher $\pT$ when the elastic HQ interactions in the QGP are 
maximized relative to the radiative ones. The former appear to be more effective 
in transfering collectivity from the expanding medium on heavy quarks than path 
length differences probed by energy loss mechanisms. On the other hand, theoretical
calculations suggest that radiative contributions are important at high $\pT$, 
\eg, for the observed mass dependence of energy loss. Hence, future measurements
of the high-$\pT$ $v_2$ will provide crucial information. Scattering processes 
of 2$\to$3 and 3$\leftrightarrow$3 type~\cite{Liu:2006vi} may also play a role in 
future improvements of the theoretical models.


\vskip 0.5 cm


\vskip 0.5 cm


\section{Conclusions and Outlook}
\label{sec_sum} 
The measurements of open heavy-flavor production in heavy-ion collisions at RHIC 
and the LHC have remarkably progressed in recent years. The pioneering measurements
of semi-leptonic HF decays at RHIC are now augmented by $D$-meson measurements at 
both colliders, as well as beauty particles through their 
$J/\psi$ decay products and 
$b$-jets. The nuclear modification factor exhibits large deviations from unity, 
and the accompanying elliptic flow substantial non-zero values, for all measured
HF particles (data on beauty $v_2$ are not available yet), while control
experiments in small systems yield $R_{p{\rm A}} \simeq R_{d\rm{A}}\simeq 1$. 
The main features of the data are in general understood and corroborate that 
HF particles acquire a substantial collective flow at low and intermediate momenta 
which requires a strong coupling to the QCD medium including recombination as 
hadronization mechanism. At the same time, high-$\pT$ HF particles are significantly 
degraded in energy (beauty particles less than charm particles), which is most 
naturally associated with parton energy loss in the QGP. 
 
Detailed comparisons of the experimental observations with various theoretical 
model calculations have been instrumental in deducing the above qualitative 
insights. Here, we have attempted to take the next step by extracting 
semi-quantitative estimates for the HF transport coefficients. In the low-$\pT$ 
regime, the heavy-quark diffusion coefficient appears to lie in the range of 
$D_s (2\pi T)$\,$\simeq$\,2-5 at temperatures of around $T$$\simeq$0.2\,GeV, which 
are arguably the most relevant ones to obtain the experimentally measured large 
charm-$v_2$ values. This result is not far from current estimates within 
quenched lattice QCD. The pertinent thermalization timescale of 3-6\,fm/$c$ 
suggests that low-$\pT$ charm particles are close to being thermalized in 
central AA collisions at RHIC and the LHC, and probably decouple slighlty below 
$T_{\rm pc}$. High-$\pT$ suppression suggests an energy loss transport coefficient 
of approximately $\hat{q}/T^3$\,$\simeq$\,10-15 for a 10\,GeV charm quark at 
temperatures of $T$$\simeq$0.35\,GeV, which tends to be larger than for light 
partons. We also deduced that elastic scattering dominates the transport of 
charm (beauty) particles for $\pT\ltsim5(15)$\,GeV, and may 
remain important until much larger momenta.     

The above estimates call for a more rigorous quantification of their uncertainties, 
including those caused by the medium evolution, hadronization process, and initial 
conditions. However, given that the current (statistical and systematic) 
uncertainties of the experimental data are still rather large in many cases, and 
that some observables (like beauty $v_2$ and heavy-baryon spectra) are not 
available yet, we see an excellent potential to further narrow down the latitude 
in the phenomenological modeling of HF transport in AA collisions at RHIC and the 
LHC. A base set of precision data on the $\Raa$ and $v_2$ of $D$ and $B$ mesons, 
charm-strange mesons and charm baryons down to low $\pT$ would go a long way 
toward disentangling different mechanisms, identifying their microscopic origin 
and obtaining reliable and accurate results for the temperature and momentum 
dependence of the charm and beauty transport coefficients in QCD matter. 
Accurate low-$\pT$ measurements, including heavy baryons, are also essential
to assess the total production yields of charm and beauty in URHICs, which are, 
\eg, crucial for quantifying regeneration contributions in the quarkonium 
sector; current electron and $D$-meson measurements at RHIC show consistency 
with binary scaling within a $\sim$30\% accuracy. 
The experimental requirements for the above measurements are very likely to be 
met in the coming years through the larger data samples that, augmented with 
advanced detection methods, will become available at RHIC and the LHC.     
 
Continued theoretical efforts will be required to firmly root the phenomenological
analyses in finite-temperature QCD. Lattice results for the HF diffusion coefficient
in full QCD are much anticipated and can serve as a benchmark for effective-theory
calculations of this quantity. Such calculations provide a natural bridge between
lattice-QCD and experiment, and can, in fact, draw constraints from a broad class 
of quantities that can be computed with good precision on the lattice, including
heavy-quark free energies, correlation functions and susceptibilities. The latter,
\eg, are a well-known diagnostic tool for the prevalent degrees of freedom in the QCD
medium, \ie, whether charm propagates as individual quarks or as part of hadronic
states. One is then automatically led into the quarkonium sector, which we did not 
discuss in the present review, but which obviously bears close connections to
open heavy flavor, both theoretically and phenomenologically.        

Future issues on the experimental side also include the role of HF transport in 
small colliding systems, at smaller (and higher) colliding energies, as well as 
correlation observables. For the former, $p$A collisions at the LHC found 
possible indications for a non-zero $v_2$, while the $\Raa$ appears to be only 
mildly modified, paralleling the more accurate observations for light-flavor hadrons. 
This could develop into quite a challenge for model calculations. 
At lower collision energies, in Au-Au reactions at $\sqrt{s}$=62\,GeV, evidence has
been found that the $\Raa$ of semileptonic decay electrons is larger than at
$\sqrt{s}$=200\,GeV and 2.76\,TeV, where the high-$\pT$ suppression is comparable.
In how far this finding is due to initial-state effects (\eg, softer initial spectra 
and/or stronger Cronin effect) or final-state effects needs to be clarified, thereby  
providing further insights into the impact of the QCD transition region on HF
transport. High-$\pT$ HF correlations hold, in principle, the potential for 
disentagling different mechanisms of energy loss, \ie, how the energy dissipated 
by a heavy quark migrates into the medium. Current model calculations suggest that 
the discrimination power between, \eg, elastic and radiative mechanisms requires a 
rather high experimental precision, which may be achievable with future LHC runs.    

In conclusion, heavy-flavor physics in heavy-ion collisions remains one of the 
most promising areas in QCD matter research. In particular, it simultaneously 
incorporates two critical aspects, namely  
(i) taking advantage of upcoming precision data to quantitatively extract transport 
properties of the QCD medium while, 
(ii) taking advantage of controlled theoretical approaches to illuminate the 
in-medium properties of the fundamental QCD force in a non-perturbative regime. 
We believe that this opportunity will enable decisive progress in understanding the
inner workings of the medium that filled the Universe in the first $\sim$10 
microseconds of its existence.     
 


\vspace{1cm}

\noindent
{\bf Acknowledgments}\\
We thank Andrea Beraudo, Andrea Dainese and Min He for helpful comments on the 
ms., and Min He, Davide Caffarri and Diego Stocco for help with the figures. 
This work has been supported by the US National Science Foundation under grant 
no.~PHY-1306359.



\end{document}